\documentclass[aps,prb,twocolumn,showpacs,groupedaddress,floatfix,nofootinbib]{revtex4-1}
\usepackage{graphicx}
\usepackage{amsmath}
\usepackage{epsfig}
\usepackage{helvet}
\usepackage{amssymb}
 
\newcommand{\tr}{\mbox{tr}}
\newcommand{\bra}[1]{\mbox{$\langle #1 |$}}
\newcommand{\ket}[1]{\mbox{$| #1 \rangle$}}
\newcommand{\braket}[2]{\mbox{$\langle #1  | #2 \rangle$}}
\newcommand{\proj}[1]{\mbox{$|#1\rangle \!\langle #1 |$}}

\begin{document}

\title{Entanglement renormalization and gauge symmetry}

\author{L. Tagliacozzo}
\affiliation{School of Mathematics and Physics, the University of Queensland, QLD 4072, Australia}

\author{G. Vidal}
\affiliation{School of Mathematics and Physics, the University of Queensland, QLD 4072, Australia}

\date{\today}

\begin{abstract}
A lattice gauge theory is described by a redundantly large vector space that is subject to local constraints, and can be regarded as the low energy limit of an extended lattice model with a local symmetry. We propose a numerical coarse-graining scheme to produce low energy, effective descriptions of lattice models with a local symmetry, such that the local symmetry is exactly preserved during coarse-graining. Our approach results in a variational ansatz for the ground state(s) and low energy excitations of such models and, by extension, of lattice gauge theories. This ansatz incorporates the local symmetry in its structure, and exploits it to obtain a significant reduction of computational costs.
{ We test the approach in the context of a $Z_2$ lattice gauge theory formulated as the low energy theory of  a specific regime of the toric code with a magnetic field,}  for lattices with up to $16\times 16$ sites ($16^2\times 2 = 512$ spins) on a torus.
We reproduce the well-known ground state phase diagram of the model, consisting of a deconfined and spin polarized phases separated by a continuous quantum phase transition, and obtain accurate estimates of energy gaps, ground state fidelities, Wilson loops, and several other quantities.
\end{abstract}


\maketitle

\section{Introduction}
\label{sec:Intro}

Gauge invariance is one of the most important concepts in modern physics. 
It is at the core of the equivalence principle of general relativity\cite{Misner1973} as well as
 an essential ingredient in the quantum field formulation of the standard model of particle physics\cite{Peskin1995}.
 More broadly, gauge theories are useful in several research areas, including condensed matter, nuclear and high energy physics.

Wilson's lattice gauge theory\cite{Wilson1974LGT} was proposed in order to study quantum chromodynamics QCD non-perturbatively.
By replacing continuous space-time by a discrete lattice, it offers both a specific regularization scheme and a convenient starting point for numerical studies.
 Through large scale Monte Carlo computations, lattice gauge theory has provided, among other results, numerical evidence of QCD confinement\cite{bali_qcd_2000} and 
of the presence of a chiral condensate \cite{jlqcd_collaboration_determination_2009}, the determination from first principles of the  value of the quark masses
  \cite{bazavov_full_2009} and numerical evidences of the formation of quark-gluon plasma at high temperatures\cite{ruester_phase_2005}.
In spite of their incontestable importance in lattice gauge theory calculations, Monte Carlo sampling techniques can not be applied to systems with large chemical potential due to the fermionic sign problem. It is therefore desirable to explore alternative approaches.
 
In the last few years, tensor network algorithms have received increasing attention as variational, non-perturbative methods to study spin lattice systems. The simplest tensor network variational ansatz, the matrix product state\cite{Affleck1988,Fannes1992,Ostlund1995} (MPS), is the basis of White's extremely successful density matrix renormalization group\cite{White1992,White1993,Schollwock2005} (DMRG) algorithm to compute the ground state of spin chains. 
Several generalizations of the MPS to two and larger spatial dimensions exist both for spin and fermionic systems, such as the projected entangled-pair states\cite{Verstraete2004, Jordan2008, Verstraete2008, Gu2008, Gu2009, Murg2009, Xie2009, Chen2009, Zhao2010, Kraus2010, Corboz2010fPEPS, Barthel2009,sierra_density_1998, maeshima_vertical_2001,nishio_tensor_2004}(PEPS), also referred to as tensor product states, and the multi-scale entanglement renormalization ansatz\cite{Vidal2007ER, Vidal2008MERA, Cincio2008, Evenbly20092DIsing, Evenbly2009Alg, Giovannetti2009, Evenbly2010Kagome, Corboz2009fMERA, Pineda2010} (MERA), both of which lead to scalable simulations. 
Since tensor network algorithms do not suffer from the sign problem, they are suitable to study problems beyond the reach of Monte Carlo techniques. Recent calculations include frustrated antiferromagnets\cite{Murg2009, Evenbly2010Kagome, Mezzacapo2010} and interacting fermions\cite{Li2010,Corboz2010fPEPS} in two spatial dimensions.
 
The goal of this paper is to explore the use of tensor network techniques within the Hamiltonian formulation of lattice gauge theory. We will consider the simplest non-trivial case, namely $Z_2$ lattice gauge theory in two spatial dimensions\cite{Wegner1971, Wilson1974LGT, Horn1979, Kogut1979RevModPhys}, with a Hamiltonian that contains the usual kinetic and potential terms for the gauge field and no fermionic matter. The ground state phase diagram of the theory, which we aim to reproduce, is already well understood, due e.g. to previous studies using Monte Carlo techniques.
Therefore the aim of the present paper is not to uncover new physics, but rather to describe a new approach and to confirm its validity in a well-known, simple context. The merit of the present strategy resides in that it can be generalized to more complex settings beyond the reach of Monte Carlo techniques.

In this work we will describe (i) a coarse-graining scheme for the lattice model specifically designed to preserve and exploit its local $Z_2$ symmetry; (ii) a variational ansatz for the ground state(s) and low energy states of theory; and (iii) numerical results to illustrate the potential of the approach.
We will regard lattice gauge theories from the broader and more recent perspective of topological order\cite{wen_topological_1995}. The $Z_2$ lattice gauge theory can be studied as the low energy sector of a spin model, namely Kitaev's toric code model\cite{Kitaev2003ToricCode}, in a suitable limit (see also Refs. \onlinecite{Sachdev_1991, Wen_1991,Sachdev_1991b,Sachdev_2008}). The hopping term of the gauge field is implemented by deforming the toric code with a magnetic field, as investigated in Refs. \onlinecite{Trebst2007BreakDown,hamma_2008}. In the deformed toric code model, the $Z_2$ group is a local symmetry of the spin Hamiltonian, and states are not forced to be gauge invariant. In particular, excited states exist that are gauge covariant. Such states can be understood as describing the presence of static matter in specific locations of the lattice. Our variational ansatz can also represent such states.

\subsection{Previous work}

Previous related work can be divided into two categories. On the one hand, Sugihara\cite{Sugihara2005} has investigated the use of a MPS to describe the ground state of $Z_2$ lattice gauge theory in quasi-one dimensional systems, both directly or through a mapping to the quantum Ising model. We notice that no attempt is made to adapt the MPS, primarily a tensor network for one-dimensional systems (where $Z_2$ lattice gauge theory is trivial) to the presence of the symmetry.
 
On the other hand, Schuch, Cirac and Perez-Garcia\cite{Schuch2010}, as well as Swingle and Wen\cite{Swingle2010}, have recently considered a PEPS representation of a state of a two dimensional lattice with topological order and have investigated how to extract topological information from it,   
while Chen et al.\cite{Chen2010Symmetry} have stressed the importance of explicitly preserving certain symmetry in a PEPS in order to represent topological order.  
In contrast with the present work, where we propose and benchmark an algorithm to compute the ground state(s) of a two dimensional system with topological order (in some limiting regime) or, relatedly, of $Z_2$ lattice gauge theory, none of these previous contributions explains how to actually obtain, given a Hamiltonian of a topologically ordered system, a tensor network representation of its ground state(s).

\begin{figure}[t]
  \includegraphics[width=7.5cm]{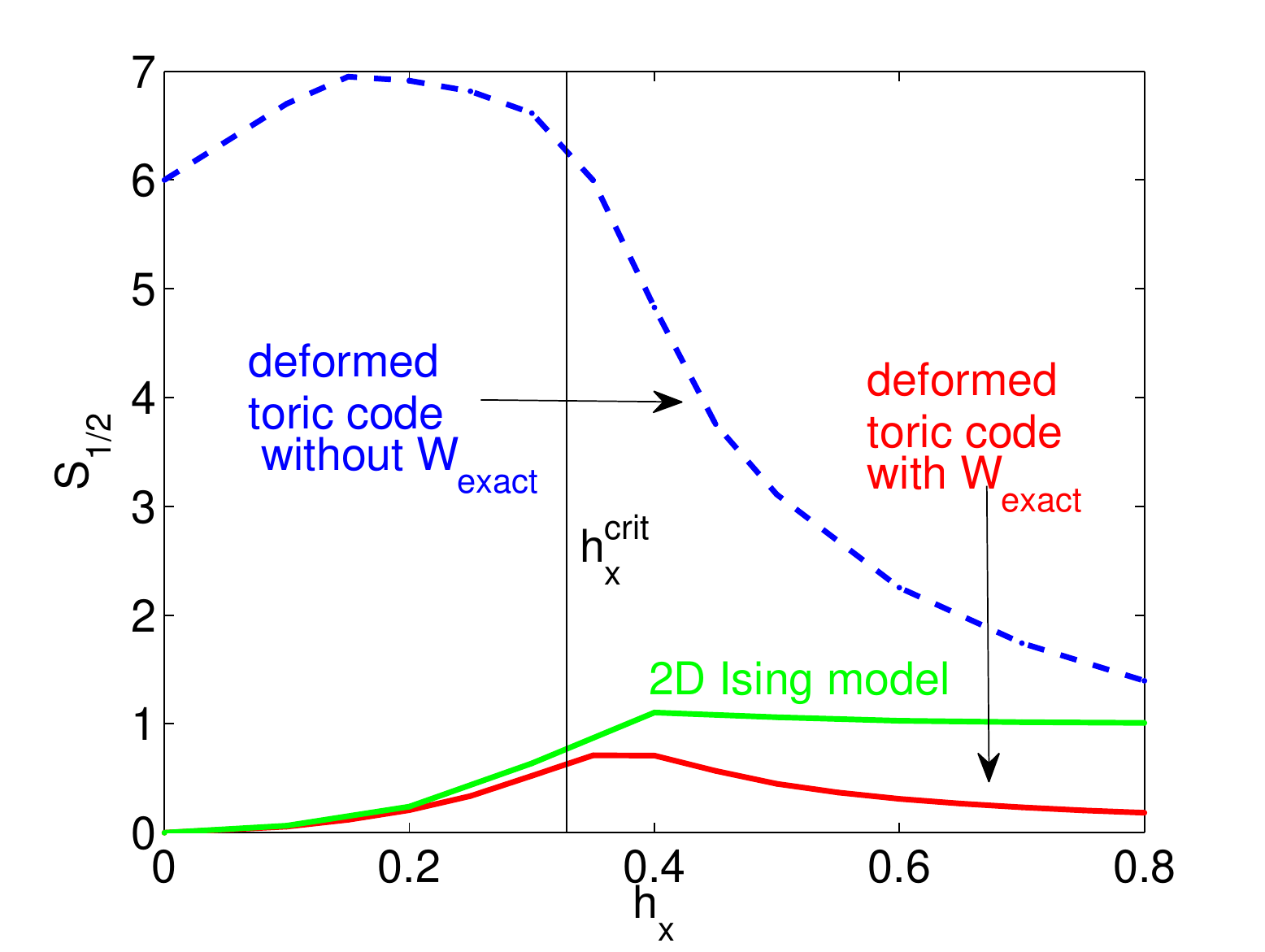}
\caption{Entanglement entropy of half a $4\times 4$ ($32$ spin) lattice $\mathcal{L}$ in the ground state $\ket{\Phi_{+,+}}$ of $H_{\text{TC}}^{\text{x}}$, as a function of the magnetic field $h_x$. The critical magnetic field $h_x^{\text{crit}} = 0.3285(1)$\cite{Hamer2000} is denoted by a vertical line. The blue (upper) curve is a lower bound to the entanglement entropy of the deformed toric code model without $W_{\text{exact}}$. For comparison, the green (middle) curve corresponds to the entanglement entropy of the ground state of the quantum Ising model on a $4\times 4$ ($16$ spin) lattice with periodic boundary conditions and no vacancy. 
Finally, the red (lower) curve corresponds to the entanglement left in the ground state of $H_{\text{TC}}^{\text{x}}$ after applying the analytical transformation $W_{\text{exact}}$. Notice that the amount of entanglement is very similar to that of the Ising model. Similar reduction of the entanglement entropy was observed by several authors in the context of global symmetries \cite{mcculloc_2002,Sierra1997505}
Thus, by implementing a local version of the duality transformation to the Ising model, the analytical transformation $W_{\text{exact}}$ maps the robustly entangled ground state $\ket{\Phi_{{+,+}}}$ of the deformed toric code model to a significantly less entangled ground state.}
\label{fig:Entanglement}
\end{figure} 

\begin{figure}[t]
  \includegraphics[width=8.5cm]{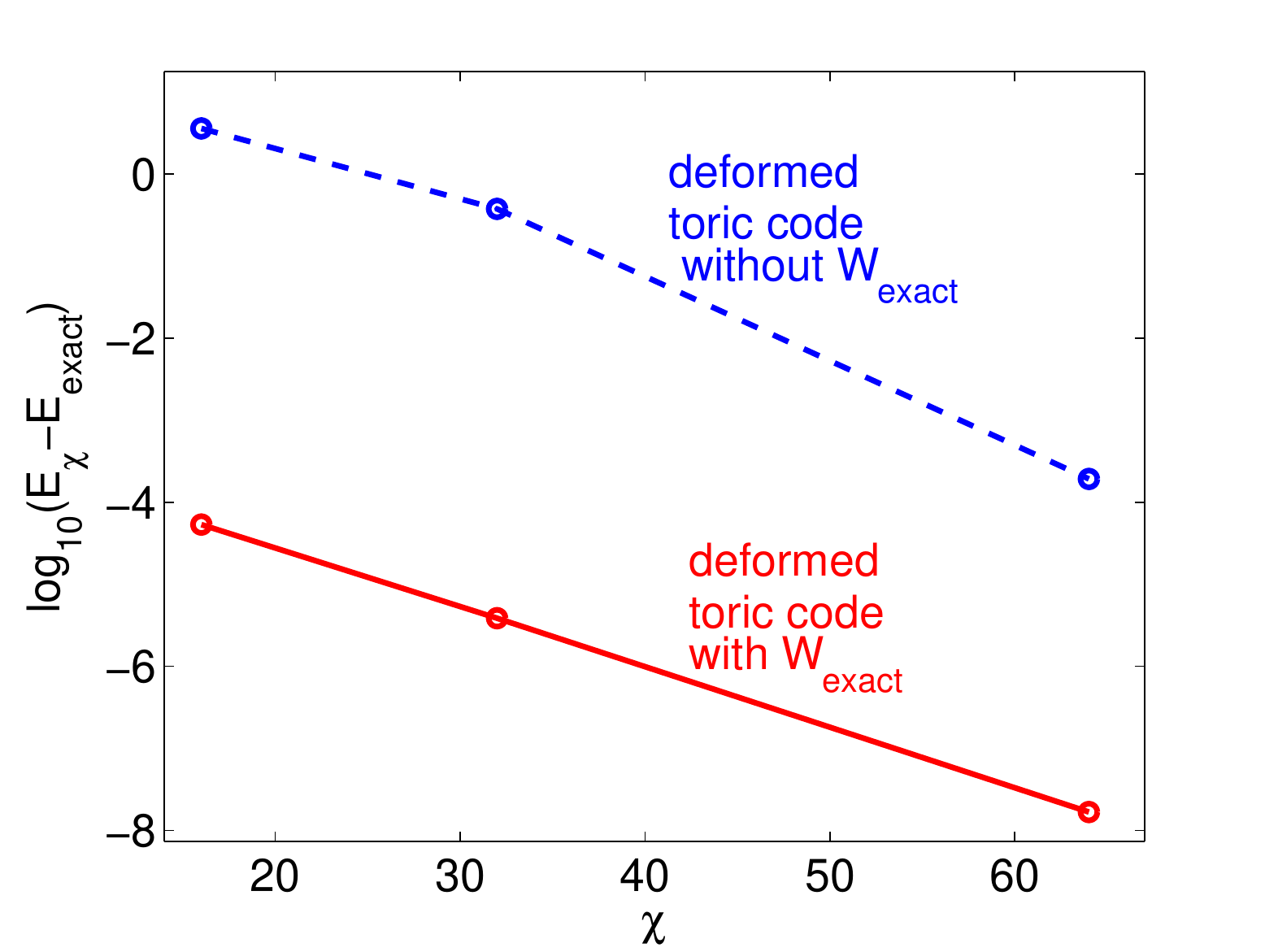}
\caption{
Accuracy in the ground state energy of the deformed toric code in a $4\times 4$ lattice, as a function of the refinement parameter $\chi$ (see Sect. \ref{sec:bench}), for magnetic field $h_x=0.1$ in $H^{\text{x}}_{\text{TC}}$ of Eq. \ref{eq:Hhx}. { On the $4\times 4$ lattice we obtain the exact ground state energy by directly diagonalizing the full Hamiltonian and  use it to asses the precision of the results as a function of the computational cost.} The computational cost grows monotonically with $\chi$. Both curves correspond to using a tree tensor network (see Sect. \ref{sec:coarse}) as a means to numerically coarse-grain the system. However, the lower curve was obtained by first applying $W_{\text{exact}}$ (in this work we refer as the \textit{hybrid ansatz} to the combination of $W_{\text{exact}}$ and a tree tensor network, see Sect. \ref{sec:coarse}). For a fixed value of $\chi$, which roughly corresponds to the same computational cost, the hybrid ansatz leads to about four more digits of accuracy in the ground state energy. }
\label{fig:Bare}
\end{figure}

\subsection{Guide}

The paper is organized in several sections. Sects. \ref{sec:DTC}-\ref{sec:ER} are entirely devoted to introducing background material, whereas Sect. \ref{sec:coarse} describes the proposed approach. Sect. \ref{sec:bench} presents benchmark results, and Sect. \ref{sec:discussion} concludes with a discussion of the proposed technique and of a number of possible generalizations to be pursued in future work.

In more detail, Sect. \ref{sec:DTC} reviews the toric code and its magnetic field deformation, Sect. \ref{sec:Z2} reviews two-dimensional $Z_2$ lattice gauge theory and its duality with the two-dimensional quantum Ising model, and Sect. \ref{sec:ER} contains a short introduction to coarse-graining transformations based on entanglement renormalization and to the MERA. A reader with sufficient previous knowledge on the toric code, $Z_2$ lattice gauge theory and entanglement renormalization may decide to skip Sects. \ref{sec:DTC}-\ref{sec:ER} in a first reading and jump directly to Sect. \ref{sec:coarse}, which describes our proposal --- possibly returning to the Sect. \ref{sec:DTC}-\ref{sec:ER} to seek clarification on nomenclature.

A reader who simply wants to obtain an overall picture of the approach may prefer to read the \textit{summary} below and then have a look at the benchmark results of Sect. \ref{sec:bench}.

\subsection{summary}

As explained in Sect. \ref{sec:coarse}, the lattice model is coarse-grained by means of a transformation that breaks into two parts $W_{\text{exact}}$ and $W_{\text{num}}$. The first part $W_{\text{exact}}$ can be viewed as a local version of the duality transformation to the quantum Ising model. The spin model is divided into blocks of spins, and within each block a duality transformation to the Ising model is implemented. This produces two types of spins: at the boundary between blocks, spins which have not yet been mapped to the Ising model (referred to as \textit{constrained} spins) and in the interior of each block, spins that have been mapped to the Ising model (referred to as \textit{free} spins). The second part $W_{\text{num}}$ of the coarse-graining transformation replaces the free spins inside a block with a single, effective \textit{free} spin. The resulting effective spin model contains \textit{constrained} spins interspersed with \textit{free} spins. 

Importantly, the local $Z_2$ symmetry, which acts on the constrained spins only, is \textit{exactly preserved} during the coarse-graining. In addition, thanks to applying duality mappings to the Ising model only within each block of spins, the coarse-graining transformation is completely \textit{local}: any local operator of the original model is mapped into a local operator of the effective model. 
Finally, composition of coarse-graining transformations reduces a finite system to a small number of spins, which can be addressed with exact diagonalization. As it is costumary in tensor network algorithms, a \textit{variational ansatz} for e.g. the ground state of the model is then obtained by regarding the coefficients that characterize the coarse-graining transformation as variational parameters.

Thus, one of the highlights of the approach is that it produces a tensor network representation of the wave function of the ground state (and low energy states) of the model, from which the expectation value of arbitrary local observables can be computed. We notice here that, depending on the choice of transformation $W_{\text{num}}$, namely depending on whether $W_{\text{num}}$ incorporates \textit{disentanglers} or not, our approach can consider arbitrarily large systems or is restricted to small lattices. In the first case, one can study the renormalization group flow to a fixed point (in preparation). In the second case, on which we will concentrate here
in addition to local observables one can also evaluate non-local order parameters, Wilson loops and ground state fidelities, see Sect. \ref{sec:bench}.

As it is costumary in most tensor network algorithms, entanglement is central to the present discussion. The more entangled a system is, the more costly it is to simulate it with a tensor network ansatz. An important aspect of our work is that it highlights the potential role of duality transformations in transforming a strongly entangled ground state into a weakly entangled one, which is then suitable for tensor network algorithms. In particular, the toric code for small magnetic field has a robustly entangled ground state that is mapped into an Ising model with large magnetic field, whose ground state is only weakly entangled, see Fig. \ref{fig:Entanglement}. A key of our approach is then to be able to map the model to its less entangled dual while preserving locality. This is precisely the role of $W_{\text{exact}}$. After the entanglement in the ground state has been reduced significantly, as illustrated in Fig. \ref{fig:Entanglement}, $W_{\text{num}}$ can proceed to coarse-grain the system by discarding high energy degrees of freedom.
The reduction of entanglement obtained with $W_{\text{exact}}$ significantly decreases computational costs. As a result, with a fixed computational cost (as parameterized by some refinement parameter $\chi$, see Sect. \ref{sec:bench}), use of $W_{\text{exact}}$ produces an improvement of four orders of magnitude in the estimate of the ground state energy in a $4\times 4$ lattice, as shown in Fig \ref{fig:Bare}.
As a side remark the entanglement structure of the same lattice gauge theory we consider here has also been studied  in Ref. \onlinecite{Buividovich2008141}.

\section{The deformed toric code}
\label{sec:DTC}

In this section we briefly review the toric code model, an exactly solvable model proposed by Kitaev in Ref. \onlinecite{Kitaev2003ToricCode}. We also review a deformation of the toric code model obtained by adding a magnetic field on the $\hat{x}$ direction, as analysed by Trebst et al. in Ref. \onlinecite{Trebst2007BreakDown} and by Hamma et al. in Ref. \onlinecite{hamma_2008}, for which an exact solution no longer exists. The deformed toric code model has a local $Z_2$ symmetry and will be used throughout this work to illustrate the proposed coarse-graining transformation.

\subsection{Toric code}
\label{sec:DTC:TC}

We consider the toric code model \cite{Kitaev2003ToricCode} on a square lattice $\mathcal{L}$ made of $L\times L$ sites and with periodic boundary conditions. Recall that in this model a spin-$1/2$ degree of freedom sits on each of the $2L^2$ links of the lattice, with total vector space 
\begin{equation}
	\mathbb{V}_{\text{TC}}\cong \left(\mathbb{C}_2\right)^{\otimes 2L^2} 
\end{equation}
and Hamiltonian
\begin{equation}
	H_{\text{TC}} \equiv -J_{e}\sum_{s} A_s - J_{m}\sum_p B_p,
	\label{eq:HTC}
\end{equation}
with $J_{e},J_{m} > 0$. Here the star operator $A_s = \prod_{j\in s} \sigma_{j}^x$ acts on the spins adjacent to site $s$ and the plaquette operator $B_p = \prod_{j\in p} \sigma_j^{z}$ acts on all the spins surrounding plaquette $p$, where $\sigma^x$ and $\sigma^z$ are Pauli matrices, see Fig. \ref{fig:ABcc}. 

\subsubsection{Ground states and topological sectors}

\begin{figure}[t]
  \includegraphics[width=6cm]{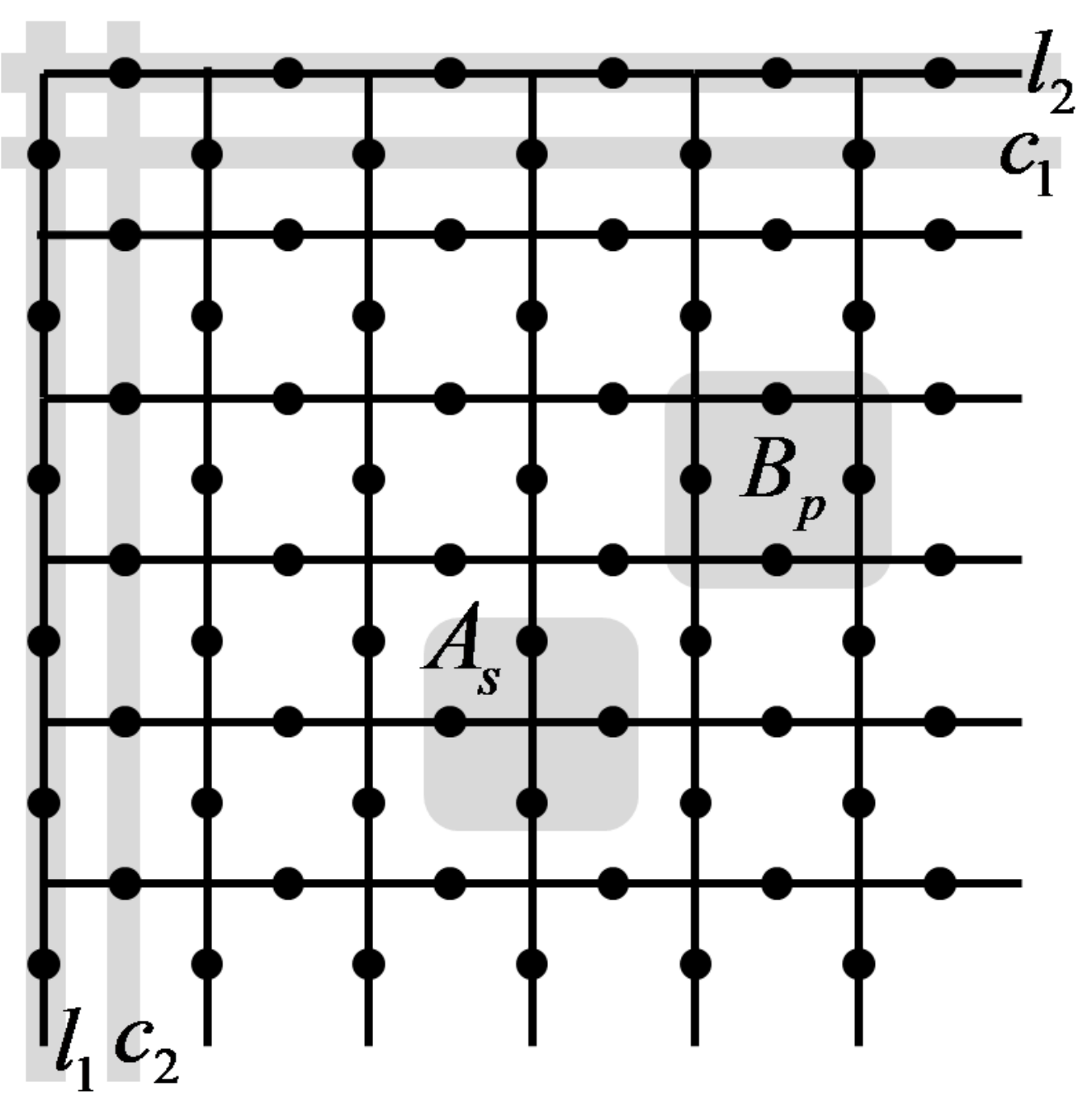}
\caption{
In the toric code model, spin 1/2 degrees of freedom sit at the edges of a square lattice $\mathcal{L}$. Here we consider a lattice $\mathcal{L}$, with periodic boundary conditions on both directions (torus). A star operator $A_s$ acts on the four
spins surrounding site $s\in \mathcal{L}$, whereas a plaquette operator $B_p$ acts on the four spins spins surrounding a plaquette $p$. The non-contractible cuts $c_1$ and $c_2$ and the non-contractible loops $l_1$ and $l_2$ are the support of operators $X_1$ and $X_2$ in Eq. \ref{eq:X} and of operators $Z_1$ and $Z_2$ in Eq. \ref{eq:Z}, respectively.}
\label{fig:ABcc}
\end{figure}

All operators $A_s$ and $B_p$ commute with each other and the ground state subspace of Hamiltonian $H_{\text{TC}}$ consists of the states $\ket{\xi}\in \mathbb{V}_{\text{TC}}$ that simultaneously fulfill the \textit{star constraints}
\begin{equation}
	A_s\ket{\xi} = \ket{\xi},~~~~~\forall s \in \mathcal{L},
	\label{eq:star}
\end{equation}
as well as the \textit{plaquette constraints}
\begin{equation}
	B_p\ket{\xi} = \ket{\xi},~~~~~\forall p \in \mathcal{L},
	\label{eq:plaquette}
\end{equation}
and thus have energy $-L^2(J_{e}+J_{m})$.
Notice that, on the torus, there are $L^2$ star operators $A_s$ and $L^2$ plaquette operators $B_p$ fulfilling
\begin{equation}
	\prod_{s\in\mathcal{L}}A_s = \text{I},~~~~~~~~\prod_{p\in\mathcal{L}}B_p = \text{I}.
	\label{eq:constrained}
\end{equation}
Therefore Eqs. \ref{eq:star} and \ref{eq:plaquette} only represent $L^2-1$ independent constraints each, that is, a total of $2L^2-2$ constraints in the space of $2L^2$ spin-1/2 sites. As a result, there are $2^2=4$ linearly independent ground states. Let us introduce the operators
\begin{equation}
	X_1 \equiv \prod_{j\in c_1} \sigma_j^x,~~~~~~	X_2 \equiv \prod_{j\in c_2} \sigma_j^x,
	\label{eq:X}
\end{equation}
where $c_1$ and $c_2$ denote the two non-contractible cuts of Fig. \ref{fig:ABcc}. Operators $X_1$ and $X_2$ commute with each other and with $H_{\text{TC}}$, and cannot be expressed as products of $A_s$'s and $B_p$'s. The eigenvectors $\ket{\Phi_{v_1,v_2}}$ of $X_1$ and $X_2$, 
\begin{equation}
X_1 \ket{\Phi_{v_1,v_2}} =v_1\ket{\Phi_{v_1, v_2}},~~~~ X_2 \ket{\Phi_{v_1,v_2}} =v_2\ket{\Phi_{v_1, v2}},
\end{equation}
where $v_1,v_2 = \pm 1$, form a basis of the ground state subspace. We refer to $\{\ket{\Phi_{+,+}}$, $\ket{\Phi_{+,-}}$, $\ket{\Phi_{-,+}}$, $\ket{\Phi_{-,-}} \}$ as the ground states of the four \textit{topological sectors} of the model, and we label each topological sector by the pair $(v_1,v_2)$. 

For later reference, we also introduce two operators
\begin{equation}
	Z_1 \equiv \prod_{j\in l_1} \sigma_j^z,~~~~~~	Z_2 \equiv \prod_{j\in l_2} \sigma_j^z,
	\label{eq:Z}
\end{equation}
where $l_1$ and $l_2$ denote the two non-contractible loops of Fig. \ref{fig:ABcc}.

\subsubsection{Electric charges and magnetic vortices}

Excited states of the toric code Hamiltonian $H_{\text{TC}}$ are characterized by violations of the star and plaquette constraints, Eqs. \ref{eq:star} and \ref{eq:plaquette}. We say that state $\ket{\xi}$ contains an \textit{electric charge} on site $s$ if $A_s\ket{\xi} = - \ket{\xi}$. Similarly, we say that state $\ket{\xi}$ contains a \textit{magnetic vortex} (or magnetic monopole) in plaquette $p$ if $B_p\ket{\xi} = - \ket{\xi}$. 

Operator $\sigma_j^z$ acting on a ground state $\ket{\Phi_{v_1,v_2}}$ produces an excited state with a pair of electric charges sitting on the two sites $s$ and $r$ connected by link $j$. Indeed, since for site $s$ (equivalently, for site $r$) we have $A_{s} \sigma^z_j = -\sigma^z_j A_{s}$, it follows that
\begin{equation}
	A_{s}\left(\sigma_j^z \ket{\Phi_{v_1,v_2}}\right) = -\sigma_j^z A_{s}\ket{\Phi_{v_1,v_2}}= - \left(\sigma_j^z \ket{\Phi_{v_1,v_2}}\right).
	\label{eq:pairEC}
\end{equation}
State $\sigma_j^z \ket{\Phi_{v_1,v_2}}$ has an energy $4J_{e}$ above the ground state energy, since each violation of a star constraint increases the energy by $2J_{e}$. 
More generally, any other state that fulfils Eqs. \ref{eq:star} and \ref{eq:plaquette} except in two sites, again corresponding to the presence of two electric charges, has energy $4J_{e}$ above the ground state. This implies that a pair of electric charges created locally by acting with $\sigma_j^z$ on the ground state can be separated in space without a change in energy. We say that electric charges are \textit{deconfined}. On the other hand, operator $\sigma_j^x$ acting on a ground state $\ket{\Phi_{v_1,v_2}}$ produces a pair of magnetic vortices (monopoles) sitting on the two plaquettes $p$ and $q$ that contain link $j$. For plaquette $p$ (equivalently, for plaquette $q$) we have $B_{p} \sigma^x_j = -\sigma^x_j B_{p}$, and therefore
\begin{equation}
	B_{p}\left(\sigma_j^x \ket{\Phi_{v_1,v_2}}\right) = -\sigma_j^x B_{p}\ket{\Phi_{v_1,v_2}}= - \left(\sigma_j^x \ket{\Phi_{v_1,v_2}}\right).
	\label{eq:pairMV}
\end{equation}
State $\sigma_j^x \ket{\Phi_{v_1,v_2}}$ has an energy $4J_{m}$ above the ground state energy. Any other state with just two magnetic vortices has energy $4J_{m}$ and we say that magnetic vortices are also \textit{deconfined}.

\subsection{Hamiltonian deformation of the toric code}
\label{sec:DTC:DTC}

The toric code Hamiltonian $H_{\text{TC}}$ in Eq. \ref{eq:HTC} is exactly solvable. Here we will be interested in a non-solvable deformation of the toric code obtained by introducing a magnetic field in the $\hat{x}$ direction, \cite{Trebst2007BreakDown,hamma_2008}
\begin{equation}
	H_{\text{TC}}^{\text{x}} \equiv -J_{e}\sum_{s} A_s - J_{m}\sum_p B_p - h_x\sum_{j} \sigma_j^x.
	\label{eq:Hhx}
\end{equation}
The magnetic field lifts the ground state degeneracy. Notice, however, that since $\sum_{j} \sigma_j^x$ also commutes with the operators $X_1$ and $X_2$, Hamiltonian $H_{\text{TC}}^{\text{x}}$ still decomposes into four sectors $(v_1,v_2)$. The ground state of any of the sectors may no longer fulfill the plaquette constraints, indicating the presence of pairs of magnetic vortices, for which $\sum_{j} \sigma_j^x$ also acts as a hopping term. 


The limit $J_{e} \gg J_{m}, h_x$, where low energy states fulfill the star constraints of Eq. \ref{eq:star} and therefore contain no electric charges, is well understood\cite{Trebst2007BreakDown} (it corresponds to the $Z_2$ lattice gauge theory\cite{Kogut1979RevModPhys}, which in turn is dual to the quantum Ising model, as reviewed in Sect. \ref{sec:Z2}). In this regime, the ground state phase diagram of $H_{\text{TC}}^{\text{x}}$ contains two phases, one in which pairs of magnetic vortices are deconfined, and another with a Bose condensate of magnetic vortices.

\subsubsection{Deconfined phase}

For small values of the magnetic field $h_x$, $h_x \ll 1$, the model is in a \textit{deconfined phase}, in which a pair of (dressed) magnetic vortices created locally can be separated an arbitrary distance incurring only a finite energy penalty. [Notice that in this work the term 'deconfined' refers to magnetic vortices, and not to the electric charges, which by construction are not present in the low energy sector of the model.]

The deconfined phase is a topologically ordered phase with four nearly degenerate ground states on the torus, one for each sector $(v_1,v_2)$. The sector $(+,+)$ has the smallest energy and the energy separation $\Delta$ to another sector vanishes exponentially fast with the linear size $L$ of the lattice\cite{Kitaev2003ToricCode},
\begin{equation}
	\Delta \approx e^{-L/\xi},
	\label{eq:Delta}
\end{equation}
where $\xi$ is a finite length scale that vanishes for $h_x=0$, i.e. $\Delta = 0$ for the undeformed toric code.

\begin{figure}[t]
  \includegraphics[width=6cm]{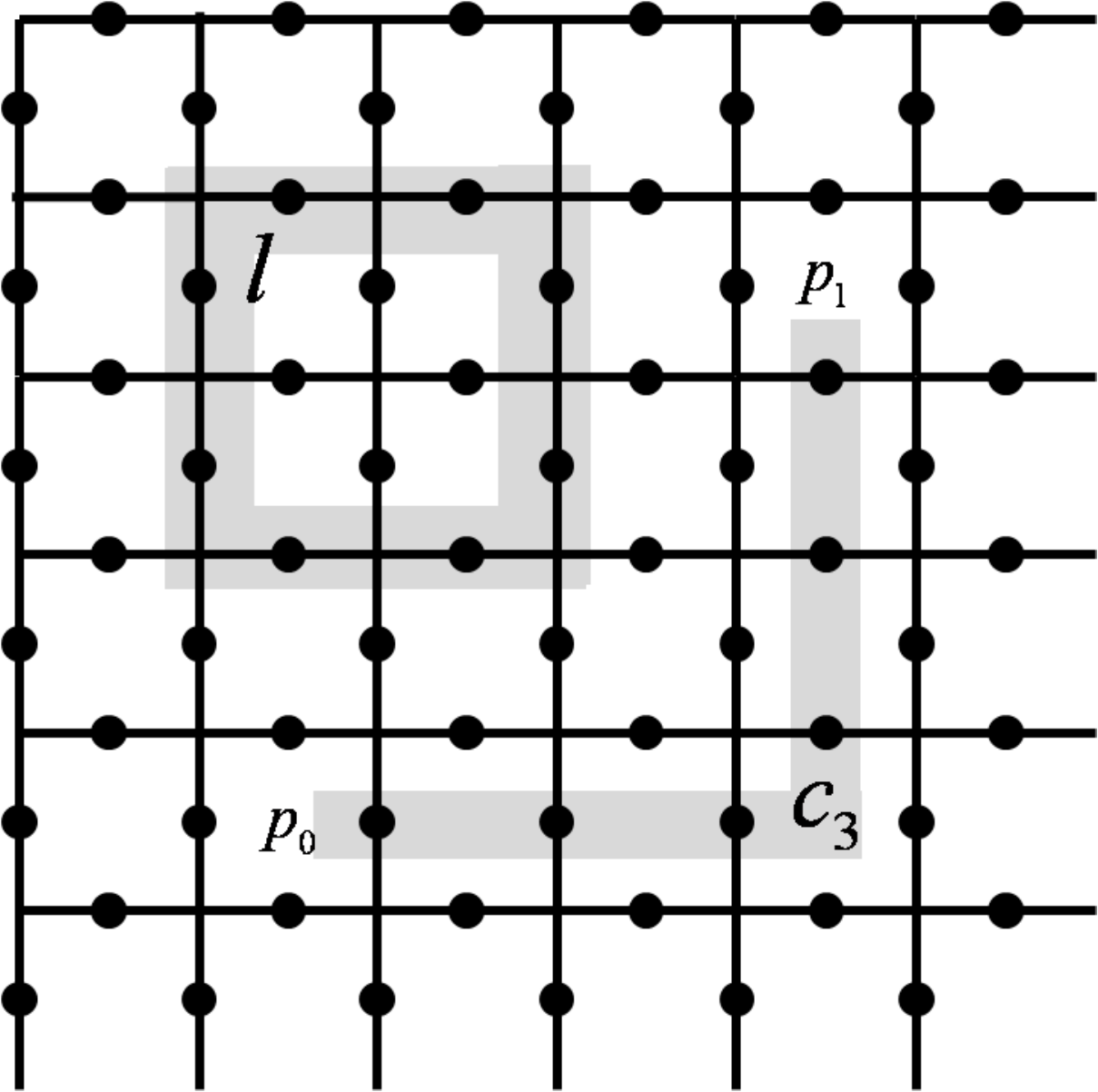}
\caption{
Cut $c_3$ connects two plaquettes $p_0$ and $p_1$ that are as far apart in $\mathcal{L}$ as possible. Example of loop $l$ that has a length of eight spins, $|l|=8$, and encloses an areas of four plaquettes, $\Sigma(l) = 4$. }
\label{fig:X3}
\end{figure}

To further characterize this phase, let us introduce the string operator\cite{kadanoff_1971,Kogut1979RevModPhys,Frohlich:1987er,digiacomo-1995-349}
\begin{equation}
	X_3 \equiv \prod_{j\in c_{3}} \sigma_j^x,
	\label{eq:X3}
\end{equation}
where $c_3$ is the cut of Fig. \ref{fig:X3} connecting a pair $(p_0,p_1)$ of plaquettes that are as distant as possible in the torus. Since $X_3$ anticommutes with $B_{p_0}$ and $B_{p_1}$, and commutes with the rest of plaquette terms, an argument  similar to that in Eq. \ref{eq:pairMV} above shows that this operator acting on a ground state $\ket{\Phi_{v_1,v_2}}$ for $h_x = 0$ (undeformed toric code) produces a state with a pair of magnetic vortices / monopoles   sitting on plaquettes $p_0$ and $p_1$. This state is orthogonal to the ground state and therefore, for $h_x=0$, the ground state expectation value of $X_3$ vanishes,
\begin{equation}
	\langle X_3 \rangle = 0.
	\label{eq:X30}
\end{equation}
In the thermodynamic (or large $L$) limit, $\langle X_3 \rangle$ vanishes for the whole deconfined phase and it is used as a (non-local) disorder parameter.

Alternatively, the deconfined phase can also be characterized by the scaling of the expectation value of Wilson loops\cite{Wilson1974LGT}. For every contractible loop $l$ on $\mathcal{L}$, see Fig. \ref{fig:X3}, we can define a \emph{Wilson loop} operator $Z[l]$ by
\begin{equation}
	Z[l] \equiv \prod_{j \in l} \sigma^{z}_j.
	\label{eq:Wilson}
\end{equation}
Operator $Z[l]$ amounts to the product of operators $B_p$ for all plaquettes $p$ contained in the interior of loop $l$. Therefore, for any ground state $\ket{\Phi_{v_1,v_2}}$ for $h_x =0$ (undeformed toric code), we have the expectation value
\begin{equation}
	\langle Z[l] \rangle = 1.
	\label{eq:Wilson0}
\end{equation}
More generally, in the deconfined phase Wilson loops obey a \textit{perimeter law} \cite{Wilson1974LGT} , in the sense that they decay as
\begin{equation}
	\langle Z[l] \rangle \approx e^{-\alpha p(l)}, 
	\label{eq:WilsonP}
\end{equation}
where $p(l)$ is the length of the loop $l$, and where $\alpha \geq 0$ vanishes for $h_x = 0$, recovering Eq. \ref{eq:Wilson0} for the undeformed toric code.
{ The behaviour of the Wilson loop with respect to its length is related to the dependence of the potential between two static electric charges with their mutual distance, which is in turn induced by their interaction with the gauge fields.
In the deconfined phase this potential is such that the force between the two charges does not increase with their distance. This translate into a decay of the Wilson loop proportional to the perimeter of the loop, as can be found in standard text-books about lattice gauge theories \cite{Montvay:1994cy} and in the original papers on the subject \cite{Wilson1974LGT,Kogut:1974ag,Kogut:1980sg,Marchesini:1981kt}.}

\subsubsection{Spin polarized phase}

For large values of $h_x$, the model is in a state where the spins are polarized in the $x$ direction. 

This phase has a unique global ground state, corresponding to the ground state of sector $(+,+)$, and the energy separation to the other sectors grows linearly in the system size $L$, 
\begin{equation}
	\Delta \approx L.
	\label{eq:Delta2}
\end{equation}
The spin polarized phase can be interpreted as a Bose condensate of magnetic vortices /monopoles, and is characterized by a non-vanishing [actually, positive] value of the (non-local) order parameter $\langle X_3 \rangle$,
\begin{equation}
	\langle X_3 \rangle \neq 0,
\end{equation}
which can be interpreted as the square of the expectation value of a creation operator for a single magnetic vortex (see Sect. \ref{sec:Z2:duality}), with $\langle X_3 \rangle=1$ for the completely polarized state at $h_x\rightarrow \infty$.
Alternatively, it is characterized by an \textit{area law} of the ground state expectation value of Wilson loops,
\begin{equation}
	\langle Z[l] \rangle \approx e^{-\beta a(l)}, 
	\label{eq:WilsonA}
\end{equation}
where $a(l)$ is the area enclosed by the loop $l$, and where $\beta = \infty$ for $h_x \rightarrow \infty$, in which case $\langle Z[l] \rangle = 0$. 


\subsubsection{Continuous quantum phase transition}

For an intermediate value of $h_x$ of about $h_x = 0.3285$ \cite{Hamer2000}, the system undergoes a continuous quantum phase transition. This is in the same universality class of that of the 2D quantum Ising model (see Sect. \ref{sec:Z2:duality}) or the 3D classical Ising model, but it is special in that it separates two phases which cannot be distinguished by a local order parameter. Instead, the transition can be characterized by the non-local order parameter $\langle X_3 \rangle$, which detects the formation of a Bose condensate of magnetic vortices /monopoles, or by the scaling of Wilson loops as a function of their size, and by the presence / absence of an approximate ground state degeneracy. 

\subsection{Local symmetry}
\label{sec:DTC:symmetry}

The symmetries of the deformed toric code Hamiltonian $H_{\text{TC}}^{\text{x}}$ in Eq. \ref{eq:Hhx} play a central role in this paper. 
Apart from the $Z_2\times Z_2$ symmetry generated by operators $X_1$ and $X_2$, $H_{\text{TC}}^{\text{x}}$ is also invariant under unitary conjugation by any star operator $A_s$,
\begin{equation}
	A_s H_{\text{TC}}^{\text{x}} A_s^{\dagger} = H_{\text{TC}}^{\text{x}},~~~~~~\forall s\in \mathcal{L},
	\label{eq:localSym}
\end{equation}
which follows from recalling that $[H_{\text{TC}}^{\text{x}}, A_s] =0$ and from noticing that $A_s=A_s^\dagger = A_s^{-1}$. Therefore, the deformed toric code Hamiltonian $H_{\text{TC}}^{\text{x}}$ has a \textit{local} $Z_2$ symmetry, generated by unitary transformations of the form
\begin{equation}
	A_s: \mathbb{V}_{\text{TC}} \rightarrow \mathbb{V}_{\text{TC}},~~~~~(A_s)^2=\text{I},
\end{equation}
which simultaneously flip the four spins included in a star $s$, for any choice of star $s \in \mathcal{L}$. 

This local symmetry implies the presence of $L^2-1$ constants of motion, namely the eigenvalues $\pm 1$ of $L^2-1$ independent star operators $A_s$. In particular, a state that fulfils all the star constraints of Eq. \ref{eq:star}, $A_s\ket{\xi} = \ket{\xi}$, that is, a state that is invariant under the local symmetry, remains so under a time evolution generated by $H_{\text{TC}}^{\text{x}}$. It is important to note, however, that the vector space $\mathbb{V}_{\text{TC}}$ also contains states that are not invariant under $A_s$. For instance, we have already seen that a state with a pair of electric charges, Eq. \ref{eq:pairEC}, violates the star constraint in two sites, where it transforms as $A_s\ket{\xi} = -\ket{\xi}$.
 
\subsection{Entanglement}
\label{sec:DTC:Entanglement}

Entanglement is another aspect of the deformed toric code model that will play a key role in for subsequent discussions on how to coarse-grain the system. 

When the lattice is in a pure state $\ket{\Psi}\in \mathbb{V}_{\mathcal{L}}$, the entanglement between a block of spins and the rest of the system can be measured by means of the Von Neumann entropy $S$ of the reduced density matrix $\rho$ of the block of spins, known as \textit{entanglement entropy},
\begin{equation}
	S = -\tr \left( \rho \log_2 \rho \right), ~~~\rho \equiv \tr_{\text{rest}}\proj{\Psi}.
\end{equation}
Fig. \ref{fig:Entanglement} shows the entanglement entropy of a block of four spins as a function of the magnetic field $h_x$, when the system is in the ground state $\ket{\Phi_{+,+}}$ of $H_{\text{TC}}^{\text{x}}$. Notice that for $h_x=0$, corresponding to the (undeformed) toric code, the entanglement entropy of the block is significantly large, indicating that the ground state is very entangled. This is in sharp contrast with the entanglement entropy for large $h_x$, which tends to zero. In this second case, each spin is completely polarized in the $\hat{x}$ direction and the ground state is a product, unentangled state.

\section{$Z_2$ lattice gauge theory and the quantum Ising model}
\label{sec:Z2}
 
In this section we briefly review $Z_2$ lattice gauge theory\cite{Wegner1971,Wilson1974LGT,Kogut1979RevModPhys,Fradkin1979, Bhanot1980,Agostini1997} in two spatial dimensions and its connection to the two-dimensional quantum Ising model\cite{Kogut1979RevModPhys, Savit1980,nussinov_2010}. We also introduce a convenient graphical notation for the different Hamiltonian terms in the Hamiltonian, and pieces of the stabilizer formalism, including the transformation rules of Pauli matrices under so-called CNOT gates. The stabilizer formalism provides us with the natural language to describe the coarse-graining transformation of systems with a local symmetry, as described in Sect. \ref{sec:coarse}.
 
\subsection{$Z_2$ lattice gauge theory}
\label{sec:Z2:LGT}

The considered restriction $J_{e} \gg J_{m},h_x$ on $H_{\text{TC}}^{\text{x}}$ ensures that low energy states fulfill the star constraints of Eq. \ref{eq:star}. If, instead, we impose those constraints by truncating the vector space of the model, we are left with the vector space $\mathbb{V}_{\text{LGT}} \subseteq \mathbb{V}_{\text{TC}}$ of $Z_2$ lattice gauge theory\cite{Wegner1971,Wilson1974LGT,Kogut1979RevModPhys},
\begin{eqnarray}
	\mathbb{V}_{\text{LGT}} &\equiv& \{~\ket{\xi} \in \mathbb{V}_{\text{TC}}~:
	~ A_s\ket{\xi} = \ket{\xi}~~\forall s \in \mathcal{L} \} \label{eq:LGT} \\ 
	&\cong& \left(\mathbb{C}_2\right)^{\otimes L^2+1}.
\end{eqnarray}
When projected onto $\mathbb{V}_{\text{LGT}}$, Hamiltonian $H_{\text{TC}}^{\text{x}}$ in Eq. \ref{eq:Hhx} becomes
\begin{equation}
	H_{\text{LGT}} = -\sum_{p} B_p - h_x \sum_{j} \sigma_j^x,
	\label{eq:HLGT}
\end{equation}
where we have neglected a constant term $-L^2J_{e}$ and we have set for simplicity, without loss of generality, $J_{m}=1$. It is also understood that each term in $H_{\text{LGT}}$ is projected onto $\mathbb{V}_{\text{LGT}}$. This model has been extensively studied before \cite{Wegner1971,Kogut1979RevModPhys,Bhanot1980,Agostini1997}. 
{It was originally introduced as a model that presents two different phases that cannot be distinguished with a local order parameter. By increasing the intensity of the magnetic field one can indeed drive a transition from the deconfined phase (at small magnetic fields) to a spin polarized phase (at large magnetic field). The phase transition around $h_x\simeq 0.3$  is in the 3D Ising model universality class. The $Z_2$ lattice gauge theory has also been intensively studied because it presents the ideal testing ground for new ideas, since its relative simplicity (as compared to gauge theories with larger gauge groups) allows to obtain very precise numerical results in Monte Carlo simulations \cite{Agostini1997}. Further interest in the model comes from a conjecture that relates its critical properties with the ones of the finite temperature deconfining transition of an $SU(2)$ lattice gauge theory in $3+1$ dimensions \cite{Svetitsky:1982gs}. Finally, the existence of a duality transformation between the $Z_2$ lattice gauge theory and the quantum 2D Ising model, as we review in Sect. \ref{sec:Z2:duality}, has inspired a lot of work in finding the relevant order parameter for confinement \cite{Fradkin1978}.}

By construction, the $Z_2$ lattice gauge can be regarded as a low energy, effective model of $H_{\text{TC}}^{\text{x}}$ (for $J_{e} \gg J_{m},h_x$) and therefore has the same ground state phase diagram.

Hamiltonian $H_{\text{LGT}}$ has a  $Z_2\times Z_2$ symmetry corresponding to operators $X_1$ and $X_2$ (properly projected onto $\mathbb{V}_{\text{LGT}}$). However, it is worth emphasizing that, in contrast with Hamiltonian $H_{\text{TC}}^{\text{x}}$, Hamiltonian $H_{\text{LGT}}$ does not have a local symmetry. Indeed, $H_{\text{LGT}}$ is defined on the subspace $\mathbb{V}_{\text{LGT}} \subseteq \mathbb{V}_{\text{TC}}$ of Eq. \ref{eq:LGT}, which is made of vectors that fulfill the star constraints. That is, each star operator $A_s$ acts as the identity operator in $\mathbb{V}_{\text{LGT}}$, and therefore the assertion that $H_{\text{LGT}}$ is invariant under transformation $A_s$ is an empty statement.


The present work aims at developing a coarse-graining scheme for lattice models with a local symmetry. For concreteness, we will most of the time restrict our attention to the subspace of states that fulfill the symmetry constraints, that is, $\mathbb{V}_{\text{LGT}}$. In this case, the coarse-graining transformation can be used also to obtain a low energy, effective description of lattice gauge models, as illustrated here in the context of $Z_2$ lattice gauge theory.

\subsection{Graphical representation of Hamiltonians}
\label{sec:Z2:graphical}

\begin{figure}[t]
  \includegraphics[width=8.5cm]{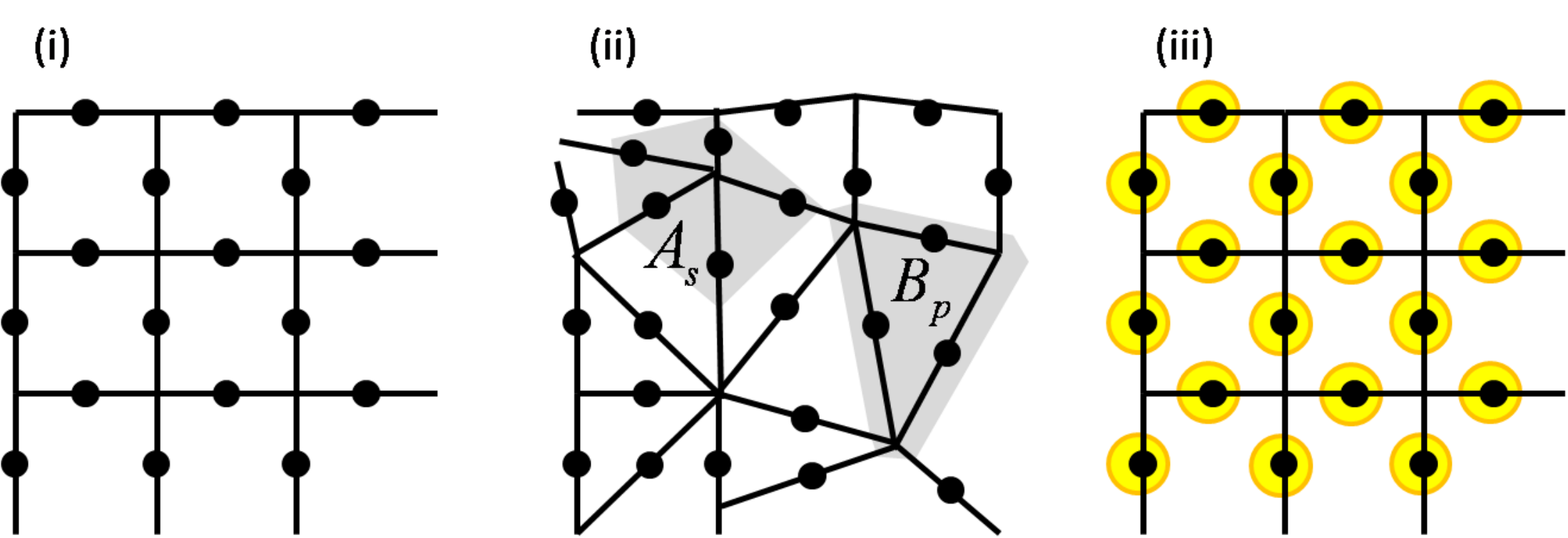}
\caption{
(i) Graphical representation of the Hamiltonian $H_{\text{TC}}$ for a $3\times 3$ lattice $\mathcal{L}$. A four-spin star operator $A_s$ acts on each site $s \in \mathcal{L}$, and a four-spin plaquette operator $B_p$ acts on each plaquette $p\in \mathcal{L}$. (ii) Hamiltonian $H_{\text{TC}}$ can also be defined on more general lattices. The example shows an irregular lattice with a five-spin star operator $A_s$ and a three spin plaquette operator $B_p$. (iii) Additionally, we represent the magnetic field of the deformed toric code Hamiltonian $H_{\text{TC}}^{\text{x}}$ by means of a yellow shade surrounding each qubit on which $-h_x\sigma^x$ acts.}
\label{fig:Hgraph}
\end{figure}

In this paper we will represent star and plaquette operators graphically by just drawing the corresponding lattice. For instance, Fig. \ref{fig:Hgraph}(i) represents the toric code Hamiltonian $H_{\text{TC}}$ of Eq. \ref{eq:HTC}. The toric code Hamiltonian $H_{\text{TC}}$ can be defined on a more general lattice by considering star operators $A_s$ and plaquette operators $B_p$ that involve a variable number of spins or qubits, see Fig. \ref{fig:Hgraph}(ii). In that case, we will also often just denote $H_{\text{TC}}$ by drawing the underlying lattice.

In addition, the presence of a magnetic field or, more generally, of related additional interactions, will be represented by a yellow shaded region enclosing all relevant qubits. For instance, Fig. \ref{fig:Hgraph}(iii) represents the deformed toric code Hamiltonian $H_{\text{TC}}^{\text{x}}$ of Eq. \ref{eq:Hhx}.   

\subsection{Qubits and CNOTs}
\label{sec:Z2:CNOT}

In order to describe our coarse-graining scheme for the deformed toric code model and the $Z_2$ lattice gauge theory, it is convenient to introduce first some basic pieces of nomenclature and formalism frequently used in the area of quantum information.

A spin-1/2 degree of freedom is an example of a quantum bit or \textit{qubit}. We will work with two preferred basis of a qubit. One basis corresponds to the eigenvectors $\ket{0}$ and $\ket{1}$ of $\sigma^z$; the second one, to the eigenvectors $\ket{+}$ and $\ket{-}$ of $\sigma^x$,
\begin{eqnarray}
	&\sigma^z\ket{0} = \ket{0},~~~~~~~~~~ &\sigma^x \ket{+} = \ket{+},\\
	&\sigma^z\ket{1} = -\ket{1},~~~~~~~~~~ &\sigma^x \ket{-} = -\ket{-},
	\label{eq:sigmaZbasis}
\end{eqnarray}
where 
\begin{equation}
	\ket{+} = \frac{\ket{0} + \ket{1}}{\sqrt{2}},~~~~~\ket{-} = \frac{\ket{0} - \ket{1}}{\sqrt{2}}.
	\label{eq:sigmaXbasis}
\end{equation}

\begin{figure}[t]
  \includegraphics[width=8.5cm]{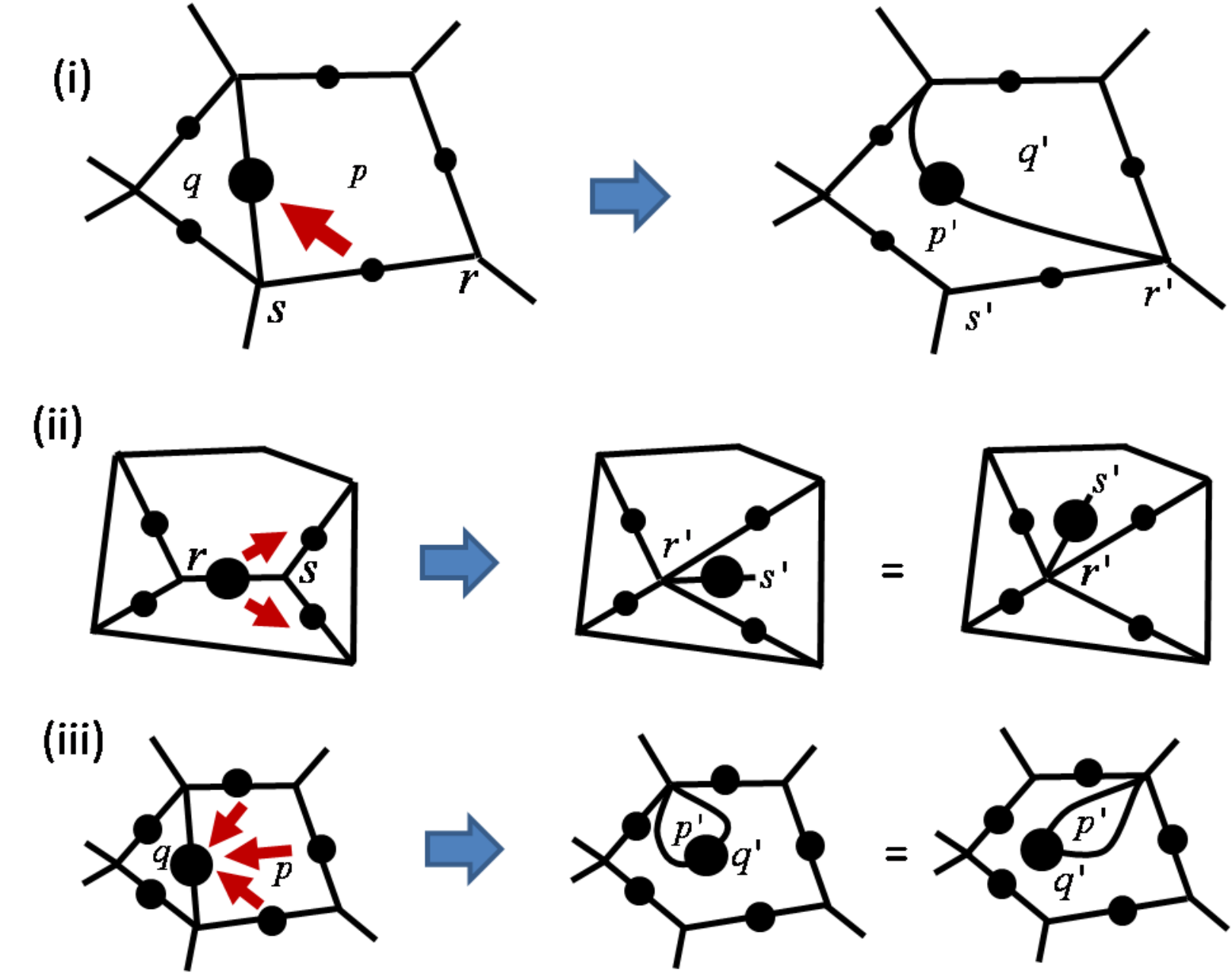}
\caption{
 (i) A CNOT is represented with an arrow from the control qubit to the target qubit. When applied on two adjacent qubits, it reconnects the target qubit. This affects the two stars $r,s$ to which the control qubits belongs, which become $r'$ and $s'$, as well as the two plaquettes $p,q$ to which the target qubit belongs, which become $p'$ and $q'$.  (ii) A sequence of CNOTs can also be used to focus a star operator $A_s$ on a single qubit, on which it acts as $A_{s'}=\sigma^x_{s'}$. As a result, the neighbouring site $r$ increases the number of qubits connected to it, becoming $r'$. Notice that the single-qubit star can be represented inside any of the plaquettes surrounding site $r'$. In this work we restrict our attention to the subspace of states that are invariant under star constraints, Eq. \ref{eq:star}. The single-qubit star $A_{s'}$ forces the qubit to be in state $\ket{+}$, and therefore unentangled with respect to the rest of the qubits. We will use this fact to remove it from the effective description of the system at low energies. (iii) A sequence of CNOTs can be used to focus a plaquette operator $B_p$ on a single qubit, on which it acts as $B_{p'}=\sigma^z_{p'}$. As a result, the neighbouring plaquette $q$ expands to $q'$. Notice that the single-qubit plaquette can be represented as connected to \textit{any} of the sites of plaquette $q'$.}
\label{fig:CNOT}
\end{figure}

A \textit{control-not} gate or CNOT gate is a unitary transformation $U_{\text{CNOT}}$ on two qubits given by 
\begin{eqnarray}
	U_{\text{CNOT}} ~&\equiv& ~~\proj{0} \otimes \text{I} ~~+ ~ \proj{1} \otimes \sigma^x\\
	&=& ~\text{I} \otimes \proj{+} ~+~  \sigma^z \otimes \proj{-},
	\label{eq:CNOT1}
\end{eqnarray}
where the first and second qubits are referred to as the \textit{control} and \textit{target} qubits, respectively. We use the stabilizer formalism \cite{Gottesman1997Thesis} to study how operators evolve under the application of a CNOT gate. Specifically, under conjugation by a CNOT, Pauli matrices on the control and target qubits are transformed according to 
\begin{eqnarray}
\text{I} \otimes \sigma^z \leftrightarrow \sigma^z\otimes \sigma^z,~~~
&&\sigma^z \otimes \text{I}  \leftrightarrow \sigma^z \otimes \text{I}, \label{eq:CNOT2}\\
\text{I} \otimes \sigma^x \leftrightarrow ~~\text{I} \otimes \sigma^x ,~~~
&&\sigma^x \otimes \text{I}  \leftrightarrow \sigma^x\otimes \sigma^x. \label{eq:CNOT3}
\end{eqnarray}

Given the toric code Hamiltonian on a given lattice, the action of a CNOT between two neighbouring qubits is to reconnect the target qubit according to Fig. \ref{fig:CNOT}(i). Notice that we represent a CNOT gate by an arrow from the control qubit to the target qubit.

Following Ref. \cite{Dennis2002}, a sequence of CNOTs involving the qubits of a star can be used to reduce the star operator to a single qubit, Fig. \ref{fig:CNOT}(ii). The star operator reduces to $\sigma^{x}$ acting on that qubit. Similarly, a sequence of CNOTs involving the qubits of a plaquette can then be used to reduce the plaquette to a single qubit, as illustrated in Fig. \ref{fig:CNOT}(iii). The plaquette operator reduces to $\sigma^z$ acting on that qubit. 

\begin{figure}[t]
  \includegraphics[width=8.5cm]{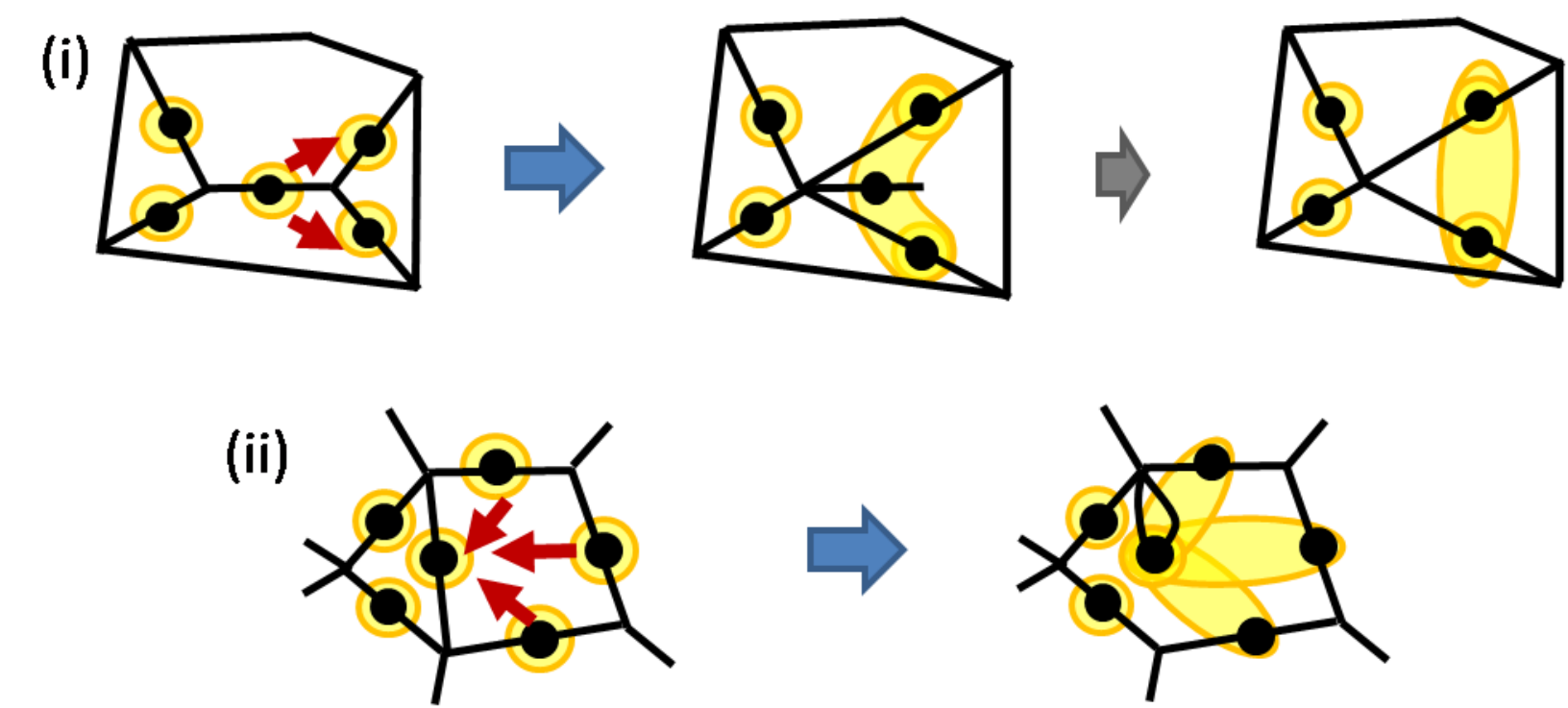}
\caption{
The magnetic field $\sigma^x$ acting on control qubits expands into a multi-qubit interaction $\sigma^x\otimes\sigma^x \otimes \cdots$ including all target qubits. (i) In concentrating a star operator on a single qubit, the magnetic field on that qubit spreads to all the qubits originally involved on that star. In the subspace of locally symmetric states, Eq. \ref{eq:star}, we can remove the control qubit (now in state $\ket{+}$ and thus unentangled from the rest of the system, see caption of Fig \ref{fig:CNOT}(ii)) from the effective description. (ii) In concentrating a plaquette operator on a single qubit, the magnetic field $\sigma^x$ of the rest of qubits on the original plaquette turn into a $\sigma^x \otimes \sigma_x$ interaction between those qubits and the target qubit.}
\label{fig:CNOT2}
\end{figure}

Of course a CNOT between two qubits will not only modify the star and plaquette operators. For instance, according to Eq. \ref{eq:CNOT3}, a magnetic field $\sigma^x$ acting on the control qubit will be transformed into a $\sigma^x\otimes\sigma^x$ interaction between control and target qubits, see Fig. \ref{fig:CNOT2}.

In the rest of the paper we refer to qubits as spins.

\begin{figure}[t]
  \includegraphics[width=8.5cm]{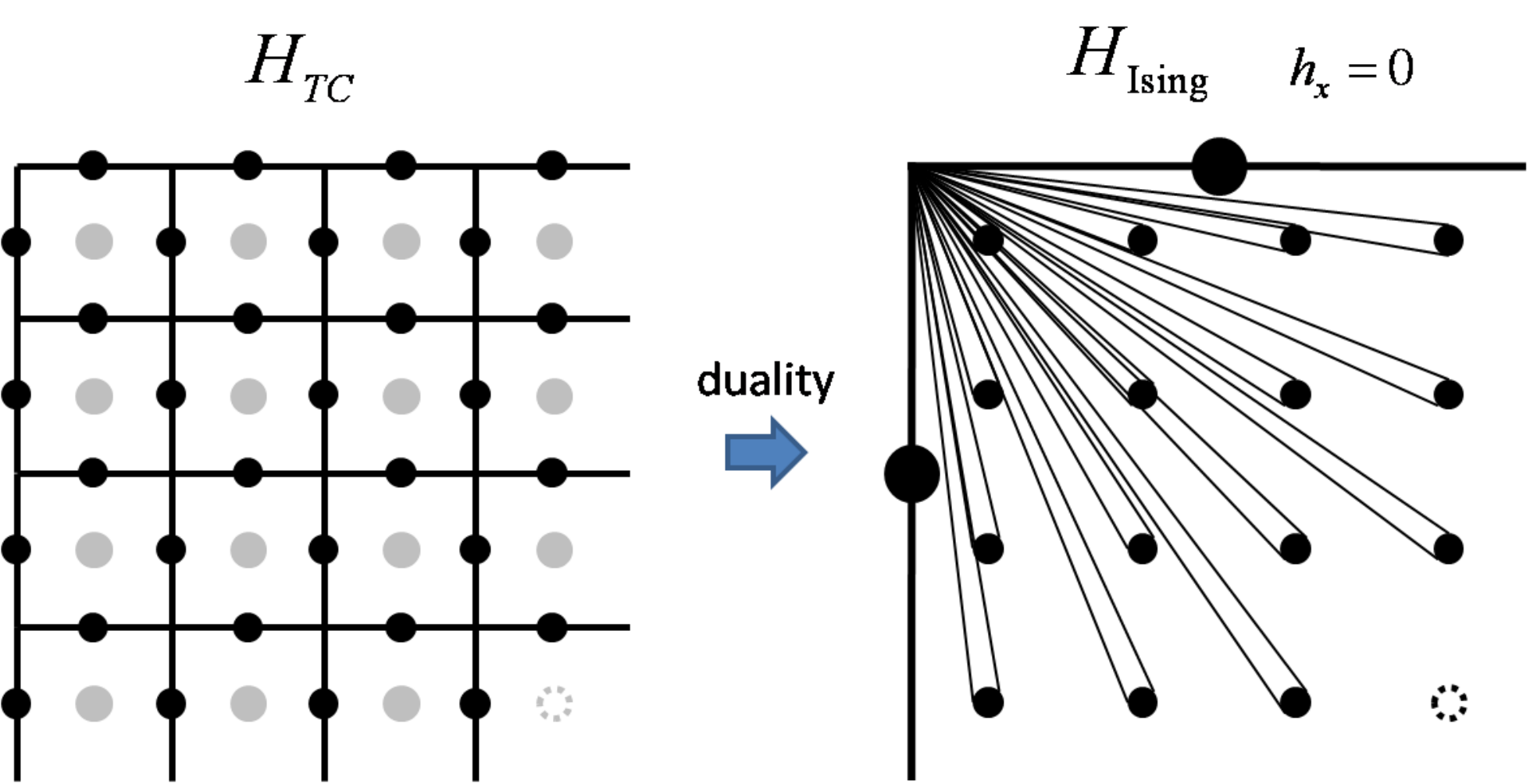}
\caption{
(i) Left: graphical representation of the toric code Hamiltonian $H_{\text{TC}}$, equivalently, $H_{\text{LGT}}$ for $h_x=0$, for a lattice $\mathcal{L}$ made of $4\times 4$ sites or $4^2 \times 2 = 16$ spins. Right: graphical representation of the dual quantum Ising Hamiltonian $H_{\text{Ising}}$. There are $4\times 4-1=15$ spins with a single-spin plaquette operator, that is, with a transverse magnetic field $-J_{m}\mu^{z}$. Notice the vacancy and two additional \textit{topological} spins which are not subject to any star or plaquette constraints. Operators $X_1$ and $X_2$ act each on one of these qubits. } 
\label{fig:Ising}
\end{figure}

\subsection{Duality transformation: quantum Ising model}
\label{sec:Z2:duality}
	
\begin{figure}[t]
  \includegraphics[width=8.5cm]{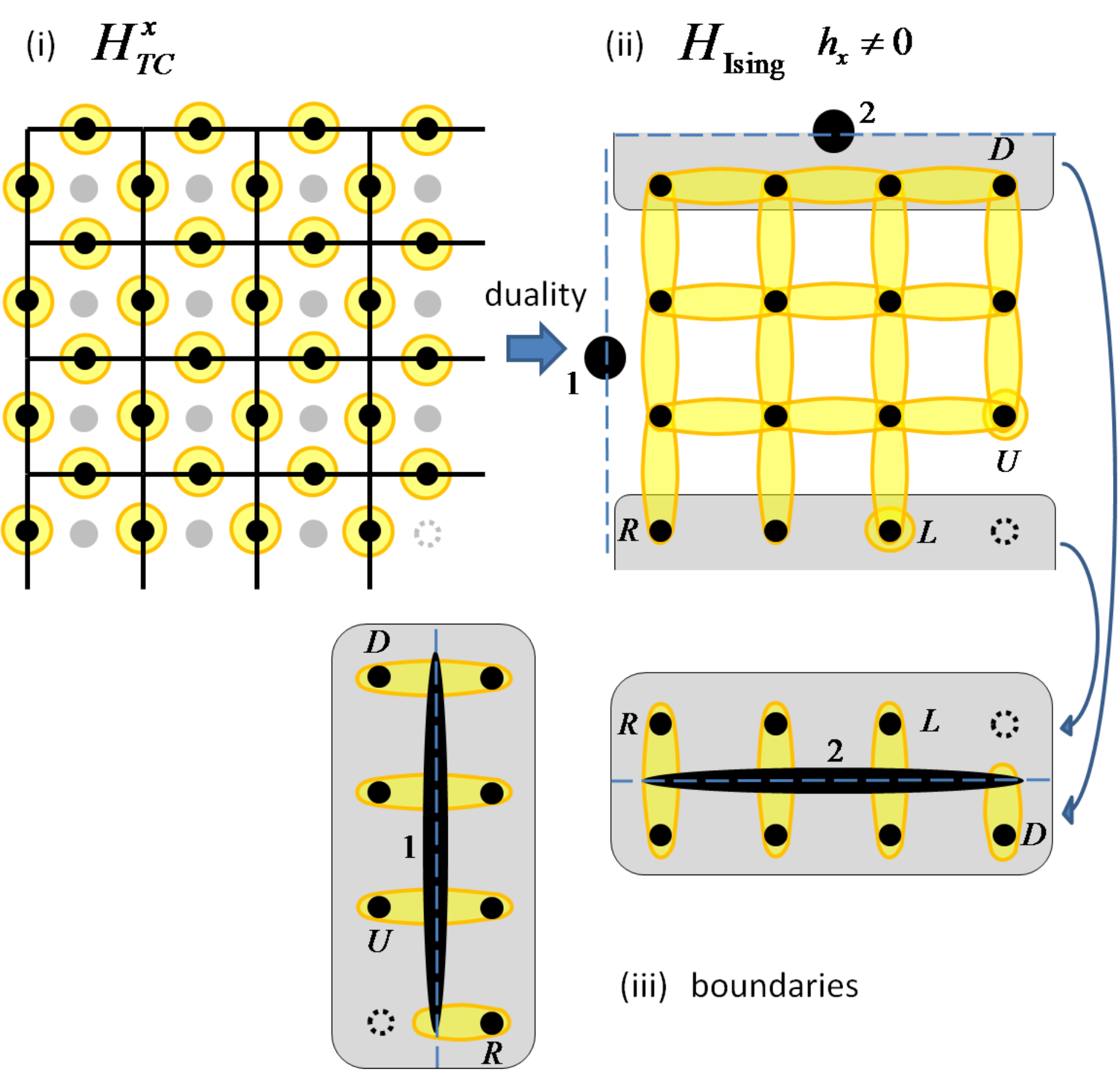}
\caption{
 (i) Graphical representation of the Hamiltonian $H_{\text{TC}}^{\text{x}}$, equivalently $H_{\text{LGT}}$, including the magnetic field $-h_x\sum_{j} \sigma^x_j$. (ii) The duality transformation maps the magnetic field to a nearest neighbour ferromagnetic Ising interaction $-h_x\mu^x\otimes \mu^x$. For simplicity, transverse magnetic field terms $-J \mu^{z}$ of $H_{\text{Ising}}$ are not depicted here. Notice that the four spins surrounding the vacancy, denoted $U$, $D$, $L$ and $R$, as well as all boundary spins, follow a different interaction pattern. On the one hand, $U$ and $L$ have one interaction term less than the rest of the spins, but have instead an additional term $-h_x\mu^x$ acting on them. (iii) On the other hand, the nearest neighbour interaction between spins $p$ and $q$ at both sides of a boundary is mediated by a coupling $-h_x\mu_p^x \otimes X_{\alpha} \otimes \mu^x_{q}$ that includes one topological spin (here, $X_{\alpha}$ corresponds to one of the non-local operators $X_1$ or $X_2$ defined in Eq. \ref{eq:X}). In the sector $(+,+)$, the topological spins are in state $\ket{+}\otimes \ket{+}$ and boundary spins are coupled with a regular Ising interaction $-h_x\mu^x_p \otimes \mu^x_q$, corresponding to periodic boundary conditions (BC) in both directions. However, in e.g. sector $(+,-)$ the topological spins are in state $\ket{+}\otimes \ket{-}$ and the coupling between the spins in the uppermost and lowermost rows becomes antiferromagnetic, $h_x\mu^x_p \otimes \mu^x_q$, corresponding to antiperiodic BC in the vertical direction, while in the horizontal direction the BC are still periodic. In general, sector $(v_1,v_2)$ the BC in the horizontal (respectively vertical) direction will be periodic/antiperiodic depending on whether $v_1$ (respectively $v_2$) is $+/-$.} 
\label{fig:Ising2}
\end{figure}

The $Z_2$ lattice gauge theory has plaquette degrees of freedom that can take two values, namely the eigenvalues $\pm 1$ of the plaquette operators $B_p$, corresponding to the absence/presence of a magnetic vortex on that plaquette. Thus plaquettes behave as spin-1/2 degree of freedom. In addition, as argued in Eq. \ref{eq:pairMV}, the operator $\sigma^x_j$ can flip the value of the two nearest neighbour plaquettes $p$ and $q$ that share the link $j$. This interpretation of the $Z_2$ lattice gauge theory as a spin model can be materialized explicitly by means of a well-known duality transformation that maps it into the quantum Ising model with a transverse magnetic field\cite{Kogut1979RevModPhys} (see appendix \ref{app:duality}), 
\begin{equation}
	H_{\text{Ising}} \equiv -h_x \sum_{\langle p,p'\rangle} \mu^{x}_p\mu^{x}_{p'} - \sum_{p} \mu^z_p, 
	\label{eq:HIsing}
\end{equation}
where the spins are placed on the sites of a $L\times L$ lattice $\mathcal{L}^{\text{dual}}$ dual to the original lattice $\mathcal{L}$ (i.e., they can be identified with the plaquettes of $\mathcal{L}$), and $\mu^x$ and $\mu^z$ are Pauli matrices, see Figs. \ref{fig:Ising}-\ref{fig:Ising2}. However, the resulting Ising model inherits two unconventional elements. On the one hand, lattice $\mathcal{L}^{\text{dual}}$ has one vacancy, i.e. there are only $L^2-1$ spins, reflecting the fact that the $Z_2$ lattice gauge theory only had $L^2-1$ independent plaquettes.  [It is worth noticing that the $\mu_p^x$ terms around the vacancy explicitly break the $Z_2$ symmetry of the Ising model.] On the other hand, there are two additional \textit{non-local} spin-1/2 degrees of freedom, associated to operators $X_1$ and $X_2$, each of which is coupled to $L$ spins along one of the two directions in $\mathcal{L}^{\text{dual}}$, see Fig. \ref{fig:Ising2}.
 

By ignoring the presence of the vacancy and non-local spins, which do not affect local observables in the thermodynamic limit, the ground state phase diagram of $H_{\text{LGT}}$ (and thus of $H_{\text{TC}}^{\text{x}}$ for $J_{e} \gg J_{m}, h_x$) can be recovered from the well-known ground state phase diagram of $H_{\text{Ising}}$ \cite{He1990,Oitmaa1991,Hamer2000,Pelissetto2002,Jordan2008,Evenbly20092DIsing,Tagliacozzo2009} by inverting the duality transformation. Recall that the $2D$ quantum Ising model at zero temperature has a disordered phase for small $h_{x}$ and an ordered phase for large $h_x$, as characterized by the vanishing (respectively non-vanishing) value of the spontaneous magnetization $m_x$, 
\begin{equation}
	m_x \equiv \frac{1}{L^2}\sum_{p} \langle \mu_p^x \rangle, 
\end{equation}
which is a local order parameter, and that the two phases are separated by a continuous quantum phase transition that occurs, indeed, at $h_x \approx 0.33$ \cite{He1990,Oitmaa1991,Hamer2000,Pelissetto2002,Jordan2008,Evenbly20092DIsing,Tagliacozzo2009}.

In particular, we can gain insight into the meaning of the nonlocal order parameter $\langle X_3 \rangle$ of the deformed toric code by noticing that operator $X_3$ in Eq. \ref{eq:X3} can be written as
\begin{equation}
	X_3 = \mu^x_{p_0} \mu^x_{p_1},
	\label{eq:X3mu}
\end{equation}
where $p_0$ and $p_1$ are two spins in $\mathcal{L}^{\text{dual}}$ (or plaquettes in $\mathcal{L}$) separated by $O(L)$ sites. In the limit $L \gg 1$ of a large lattice, where $\langle \mu^x_{p_0} \mu^x_{p_1} \rangle$ is expected to factorize into $\langle \mu^x_{p_0} \rangle \langle \mu^x_{p_1} \rangle$ since plaquettes $p_0$ and $p_1$ are far apart, this expectation value equals the square of the spontaneous magnetization $m_x$,
\begin{equation}
\langle X_3 \rangle = \langle \mu^x_{p_0} \mu^x_{p_1} \rangle \approx \langle \mu^x_{p_0} \rangle \langle \mu^x_{p_1} \rangle = m_x^2
\end{equation}
Thus, we can interpret $\langle X_3 \rangle$ as the square of the expectation value of a (fictitious) creation operator of \textit{single} magnetic vortices, and interpret a vanishing/non-vanishing value for $\langle X_3 \rangle$ as indicating the absence/presence of a Bose condensate of magnetic vortices.

\section{Entanglement renormalization}
\label{sec:ER}


In this section we briefly review one possible route to coarse-graining a lattice model in real space based on entanglement renormalization \cite{Vidal2007ER,Vidal2008MERA,Evenbly2009Alg,Vidal2009Chapter}. Entanglement renormalization is a specific real-space implementation of the renormalization group \cite{Wilson1983RG}. The coarse-graining transformation can be used to define the multi-scale entanglement renormalization ansatz \cite{Vidal2008MERA} (MERA), a variational ansatz for ground states and low energy states of a local Hamiltonian. We also review how to use entanglement renormalization and the MERA in the presence of a global internal symmetry, in such a way that the symmetry is exactly preserved and exploited to reduce computational costs. Then, in Sect. \ref{sec:coarse}, we will explore how to generalize entanglement renormalization and the MERA in the presence of a local symmetry. 

\begin{figure}[t]
  \includegraphics[width=8cm]{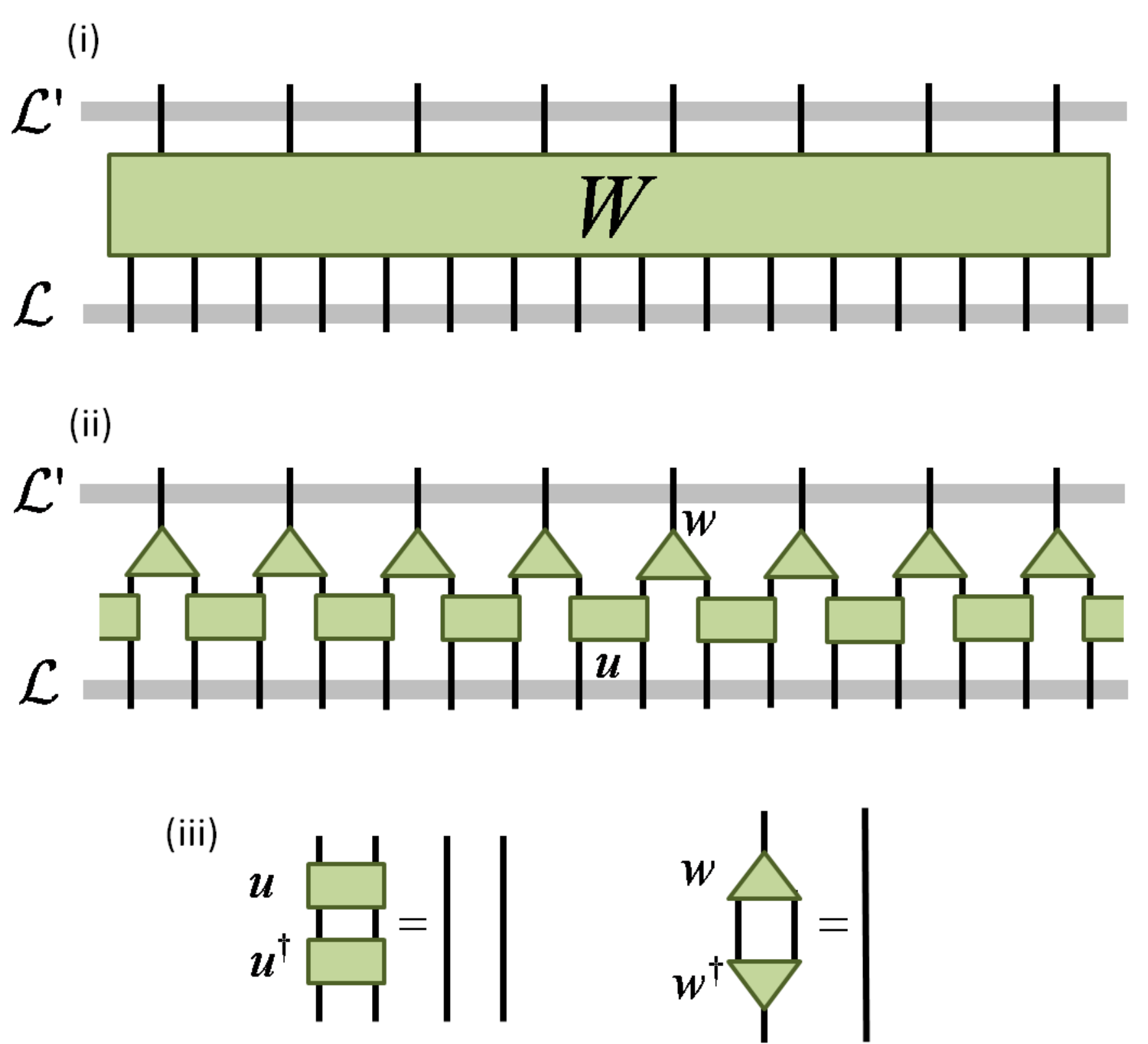}
\caption{
(i) The coarse-graining transformation $W$ defines a linear map between the space $\mathbb{V}_{\mathcal{L}'}$ of lattice $\mathcal{L}'$ and the space $\mathbb{V}_{\mathcal{L}}$ of lattice $\mathcal{L}$. (ii) Isometric transformation $W$ that decomposes as a product of disentanglers $u$ and isometries $w$. (iii) Constraints fulfilled by disentanglers and isometries, see Eq. \ref{eq:uwiso}.}
\label{fig:W}
\end{figure}

\subsection{Coarse-graining transformation}
\label{sec:ER:coarse}

\begin{figure}[t]
  \includegraphics[width=8cm]{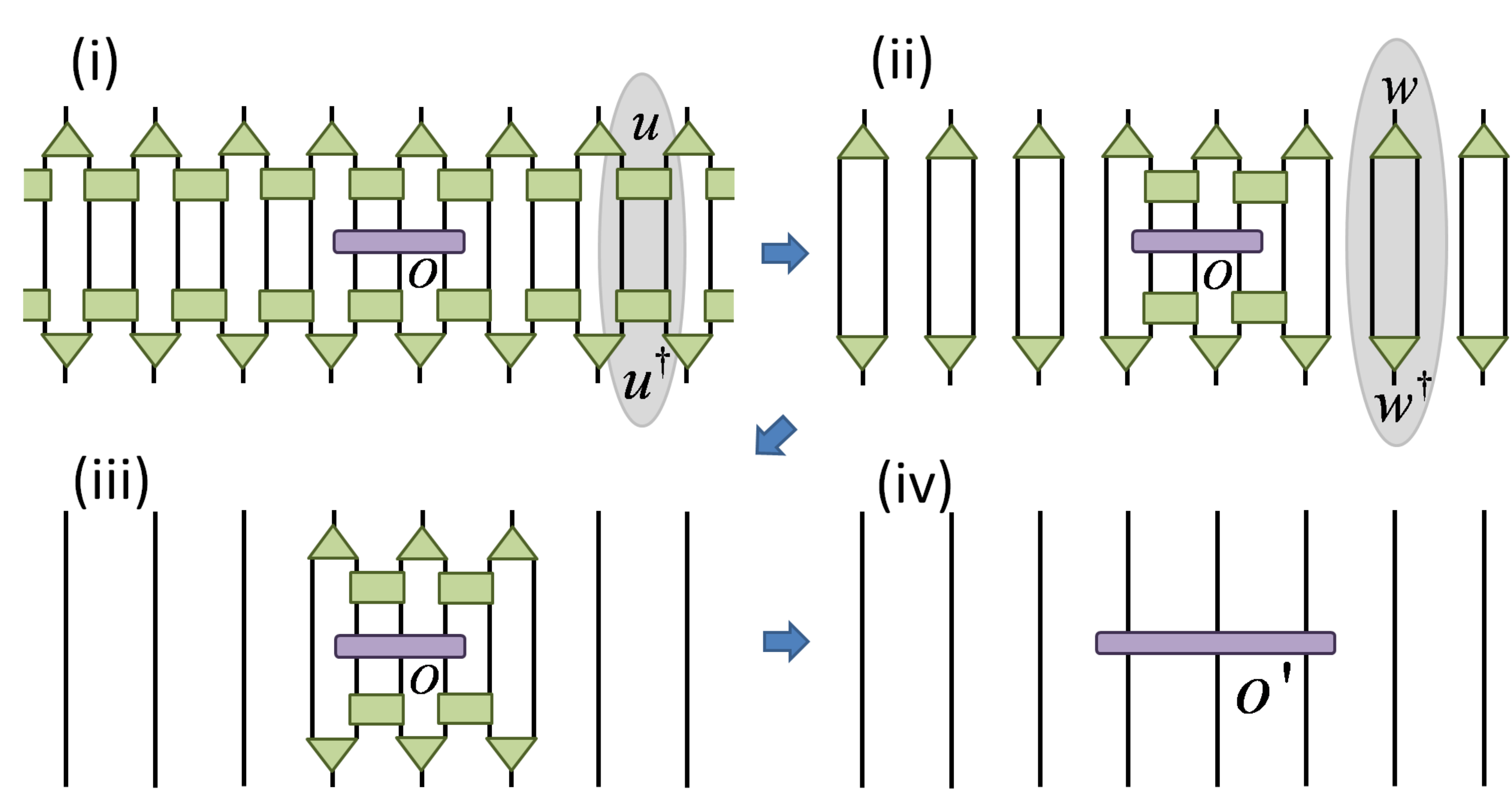}
\caption{
 Coarse-graining of a local operator $o$ acting on three contiguous sites of $\mathcal{L}$, into a local operator $o'$ acting on three contiguous sites of $\mathcal{L}'$. (i) Diagrammatic expression for $o'=W^{\dagger}oW$, where $W$ decomposes as a product of disentanglers $u$ and isometries $w$. (ii) Using that $u^{\dagger}u=I$, where $I$ is the identity on two sites of $\mathcal{L}$, most disentanglers can be removed. (iii) Using that $w^{\dagger}w=I$, where $I$ is the identity on one site of $\mathcal{L}'$, most isometries can be removed, so that (iv) $o'$ acts as the identity in all but three sites of $\mathcal{L}'$.}
\label{fig:Wlocal}
\end{figure}

\begin{figure}[t]
  \includegraphics[width=8cm]{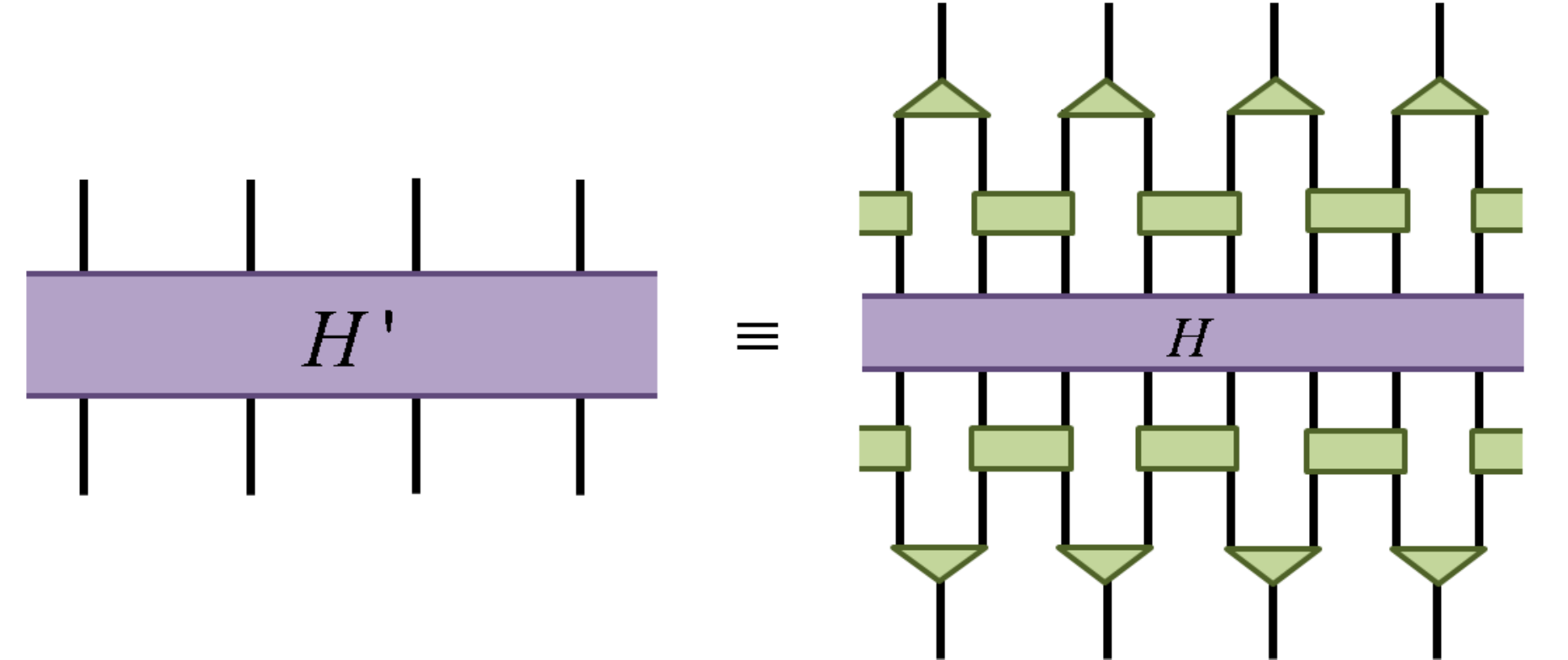}
\caption{
 Effective Hamiltonian $H' \equiv W^{\dagger} H W$ expressed in terms of the original Hamiltonian $H$ and disentanglers $u$ ans isometries $w$. If $H$ is a sum of local terms, then $H'$ can also be expressed as a sum of local terms, see Fig. \ref{fig:Wlocal}.}
\label{fig:H2}
\end{figure}

Given a lattice model, characterized by a lattice $\mathcal{L}$ and a Hamiltonian $H$, a coarse-graining transformation aims to produce a new, simplified lattice model, characterized by a coarse-grained lattice $\mathcal{L}'$ with fewer sites and an effective Hamiltonian $H'$. Here we are interested in a coarse-graining transformation $W$
\begin{equation}
	W^{\dagger} : \mathbb{V}_{\mathcal{L}} \rightarrow \mathbb{V}_{\mathcal{L}'}
\end{equation}
that defines a linear map from the space $\mathbb{V}_{\mathcal{L}}$ of the original model to the space $\mathbb{V}_{\mathcal{L}'}$ of the effective model, see Fig. \ref{fig:W}(i). [Notice that, for consistency with Ref. \onlinecite{Vidal2008MERA}, we actually regard $W$ as a map from $\mathbb{V}_{\mathcal{L}'}$ to $\mathbb{V}_{\mathcal{L}}$. ]
 
More specifically, we consider a coarse-graining transformation $W$ that maps local operators in $\mathcal{L}$ (that is operators that act non-trivially on a finite subset of neighbouring sites of $\mathcal{L}$) into local operators in $\mathcal{L}'$. This property is automatically fulfilled if $W$ is an isometric tensor,
\begin{equation}
	W^{\dagger}W=I_{\mathcal{L'}},
\end{equation}
that decomposes as the product of isometric tensors $u$ and $w$, see Figs. \ref{fig:W}(ii)-(iii),
\begin{equation}
	u^{\dagger}u=I, ~~~~ w^{\dagger}w = I,
	\label{eq:uwiso}
\end{equation}
known as \textit{disentanglers} and \textit{isometries}, which corresponds to one step of \textit{entanglement renormalization} \cite{Vidal2007ER}. Indeed, as shown in Fig. \ref{fig:Wlocal}, in this case local operators are mapped into local operators. In particular, if the Hamiltonian $H$ of the original lattice model decomposes as the sum of local terms, then the effective Hamiltonian $H'$, depicted in Fig. \ref{fig:H2}, will also decompose as a sum of local terms.

The examples of Figs. \ref{fig:W}-\ref{fig:H2invariant} used to illustrate this section correspond to one dimensional lattices for simplicity. Analogous constructions for two dimensional lattices span three dimensions, see e.g. in Ref. \onlinecite{Evenbly2009Alg,Evenbly20092DIsing}. Their graphical representation is significantly more involved and is not required in order to introduce the basic elements necessary for the present discussion. We postpone until Figs. \ref{fig:Wexact} and \ref{fig:Wnum} of Sect. \ref{sec:coarse} the explicit representation of three dimensional structures.

\subsection{Renormalization Group flow}
\label{sec:ER:RG}

\begin{figure}[t]
  \includegraphics[width=8cm]{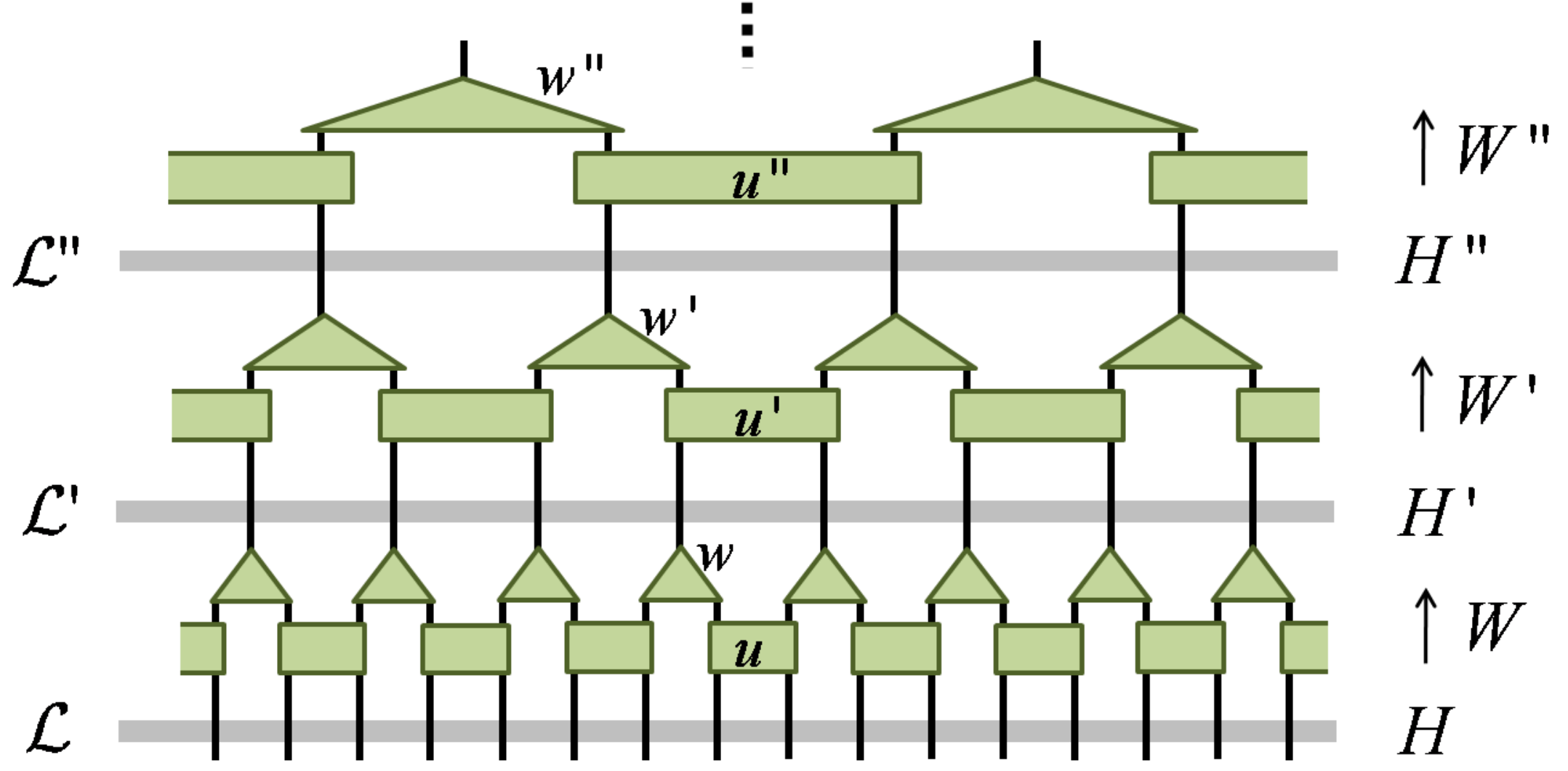}
\caption{
 By composition, coarse-graining transformations $\{W, W' ,W'', \cdots\}$ produce a sequence of increasingly coarse-grained lattices $\{\mathcal{L}, \mathcal{L}', \mathcal{L}'', \cdots\}$ with effective Hamiltonians $\{H, H', H'',\cdots\}$.}
\label{fig:RGflow}
\end{figure}

Successive applications of coarse-graining transformations $\{W, W' , W'', \cdots\}$ produce a sequence of increasingly coarse-grained lattices $\{ \mathcal{L}, \mathcal{L}', \mathcal{L}'', \cdots\}$ together with a sequence of effective, local Hamiltonians $\{ H, H', H'', \cdots\}$, see Fig. \ref{fig:RGflow}. If at each step the coarse-graining transformation projects onto the low energy subspace of the Hamiltonian, then the sequence of effective Hamiltonians defines a discrete renormalization group (RG) flow towards a fixed point model that captures the low energy/large scale properties of the original model. 

Suppose now that the original lattice $\mathcal{L}$ is made of a finite number $N$ of sites. Then the effective lattice $\mathcal{L}'$ is made of a smaller number $N'=N/b$ of sites, where $b=2$ in the example of Fig. \ref{fig:W}(ii). In particular, after $O(\log_b(N))$ applications of the coarse-graining transformation, lattice $\mathcal{L}$ is reduced to a small lattice $\mathcal{L}^{\text{top}}$ whose size is independent of $N$, and such that $H^{\text{top}}$ is amenable to exact diagonalization techniques. Notice that a state $\ket{\Psi^{\text{top}}}\in \mathbb{V}_{\mathcal{L}^{\text{top}}}$ depends on a small number of parameters that is also independent of the original system size $N$.

\subsection{MERA}
\label{sec:ER:MERA}

\begin{figure}[t]
  \includegraphics[width=8cm]{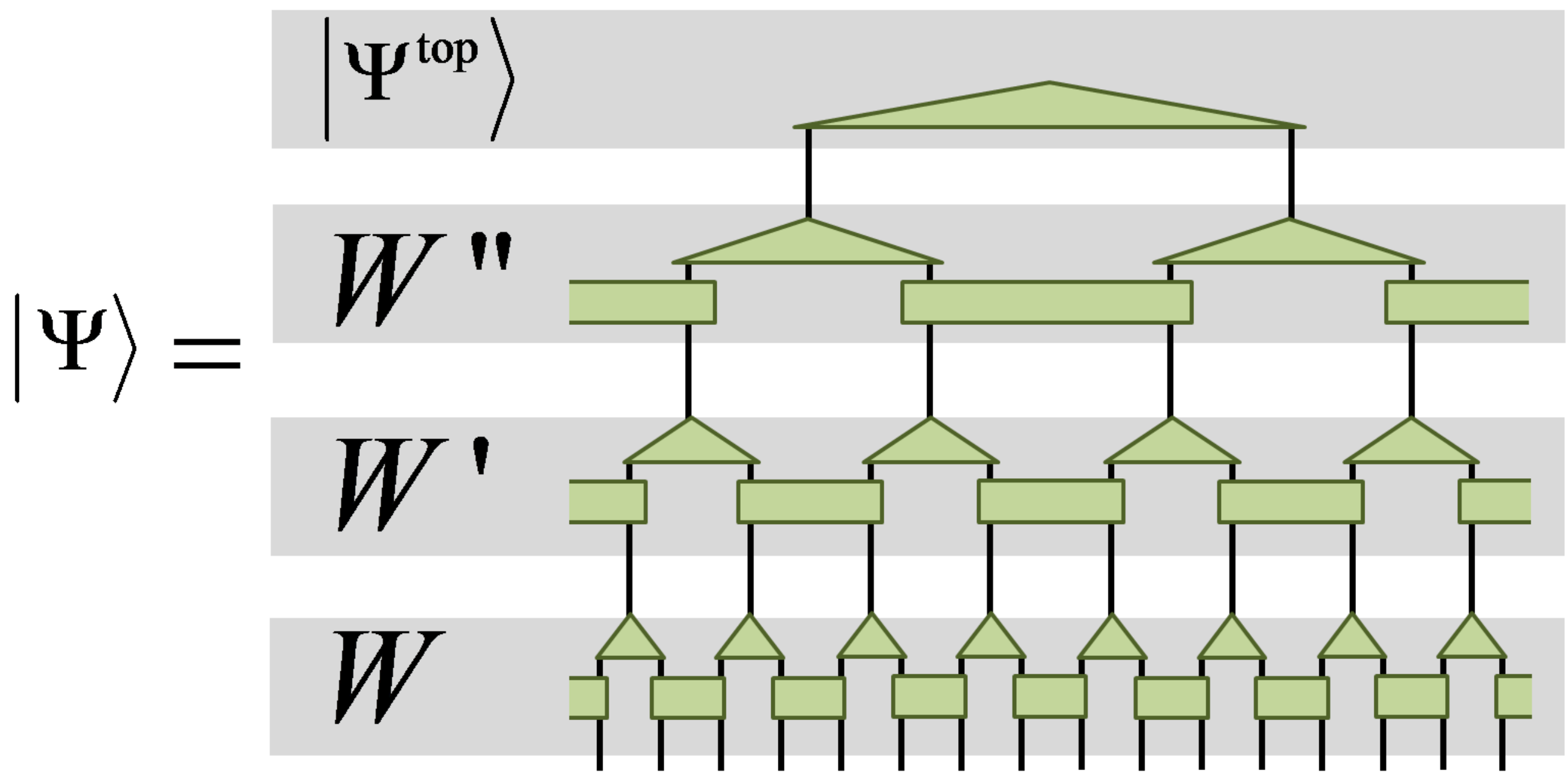}
\caption{
Example of a state of a lattice $\mathcal{L}$ made of 16 sites, $\ket{\Psi} = W W' W''\ket{\Psi^{\text{top}}}$, expressed as a MERA made of three layers of disentanglers and isometries corresponding to three coarse-graining transformations $W$, $W'$ and $W''$.}
\label{fig:MERA}
\end{figure}

A state $\ket{\Psi^{\text{top}}}\in \mathbb{V}_{\mathcal{L}^{\text{top}}}$ of this reduced lattice can be mapped into a state $\ket{\Psi}\in \mathbb{V}_\mathcal{L}$ of the original lattice by reversing the coarse-graining transformations, see Fig. \ref{fig:MERA},
\begin{equation}
	\ket{\Psi} = WW'W''\cdots \ket{\Psi^{\text{top}}}.
  \label{eq:MERA}
\end{equation}
This means that the top vector $\ket{\Psi^{\text{top}}}$ together with the sequence of transformations $\{W,W',W'',\cdots\}$, as characterized by a sequence of disentanglers and isometries $\{\{u,w\},\{u',w'\}, \{u'',w''\}, \cdots\}$, can be used as an efficient representation a state $\ket{\Psi}\in \mathbb{V}_\mathcal{L}$. This representation is the multi-scale entanglement renormalization ansatz (MERA), and can be used as a variational ansatz for e.g. the ground state of $H$. This is achieved by optimizing the coefficients in $\{\{u,w\},\{u',w'\}, \{u'',w''\}, \cdots \}$ and $\ket{\Psi^{\text{top}}}$ so as to minimize the expectation value $\bra{\Psi} H \ket{\Psi}$ using the techniques discussed in Ref. \onlinecite{Evenbly2009Alg}. More generally, by replacing $\ket{\Psi^{\text{top}}}$ with an isometry $W^{\text{top}}$ that defines a $\chi^{*}$-dimensional space $\mathbb{C}_{\chi^*}$,  a whole subspace $\mathbb{V}^* \subseteq \mathbb{V}_{\mathcal{L}}$ can be represented. If $\ket{\alpha}$ denotes an orthonormal basis in $\mathbb{C}_{\chi^*}$, then
\begin{equation}
	\ket{\Psi^{\alpha}} = WW'W''\cdots W^{\text{top}} \ket{\alpha},~~~~\alpha=1,\cdots,\chi^{*}
  \label{eq:MERAsubspace}
\end{equation}
is an orthonormal basis in $\mathbb{V}^* \subseteq \mathbb{V}_{\mathcal{L}}$. In this way, the MERA can be used to approximate e.g. a low energy subspace, including the ground state(s) and several low energy excited states of $H$, as we will do in Sect. \ref{sec:bench}.

When the state $\ket{\Psi}$ in Eq. \ref{eq:MERA} is \textit{translation invariant}, which is the case we have discussed so far, the MERA is specified with $O(log_b (N))$ parameters, namely those that characterize the $O(\log_b(N))$ pairs of disentangles and isometries $\{\{u,w\},\{u',w'\}, \{u'',w''\}, \cdots\}$.  Thus, the MERA offers an efficient description of certain states of the lattice model, whose vector space $\mathbb{V}_{\mathcal{L}}$ has dimension $O(\exp(N))$. (More generally, a generic MERA is specified by $N$ independent disentanglers and isometries, and therefore depends on $O(N)$ parameters \cite{Evenbly2009Alg}). In addition, a \textit{scale invariant} state $\ket{\Psi}\in \mathbb{V}_{\mathcal{L}}$ of the form \ref{eq:MERA} can be described using $O(1)$ parameters by choosing all disentanglers and isometries to be identical. The scale invariant MERA is useful to represent the ground state of fixed-points of the RG flow, whether corresponding to critical systems \cite{Vidal2007ER,Vidal2008MERA,Giovannetti2008,Pfeifer2009} or to systems with topological order \cite{Aguado2008TC,Konig2009}. 

In Sect. \ref{sec:bench} we will use a translation invariant coarse-graining transformation for the deformed toric code model, which will lead to a translation invariant MERA for the ground state(s) and low energy excited states of $H_{\text{TC}}^{\text{x}}$. However, for simplicity we will not attempt to use disentanglers in their full form (see Sect. \ref{sec:coarse:num} for details) as a result of which the computational cost will grow as $O(\exp(\sqrt{N}))$. The limits of small and large magnetic field of Hamiltonian $H_{\text{TC}}^{\text{x}}$, i.e. $h_{x} = 0$ and $h_{x} = \infty$, correspond to fixed points of the RG flow, namely to a topologically ordered fixed point \cite{Aguado2008TC} and to a trivial fixed point, respectively. In these two cases, we will be able to represent the ground state with a scale invariant MERA.

\subsection{Global symmetry}
\label{sec:ER:global}

\begin{figure}[t]
  \includegraphics[width=8cm]{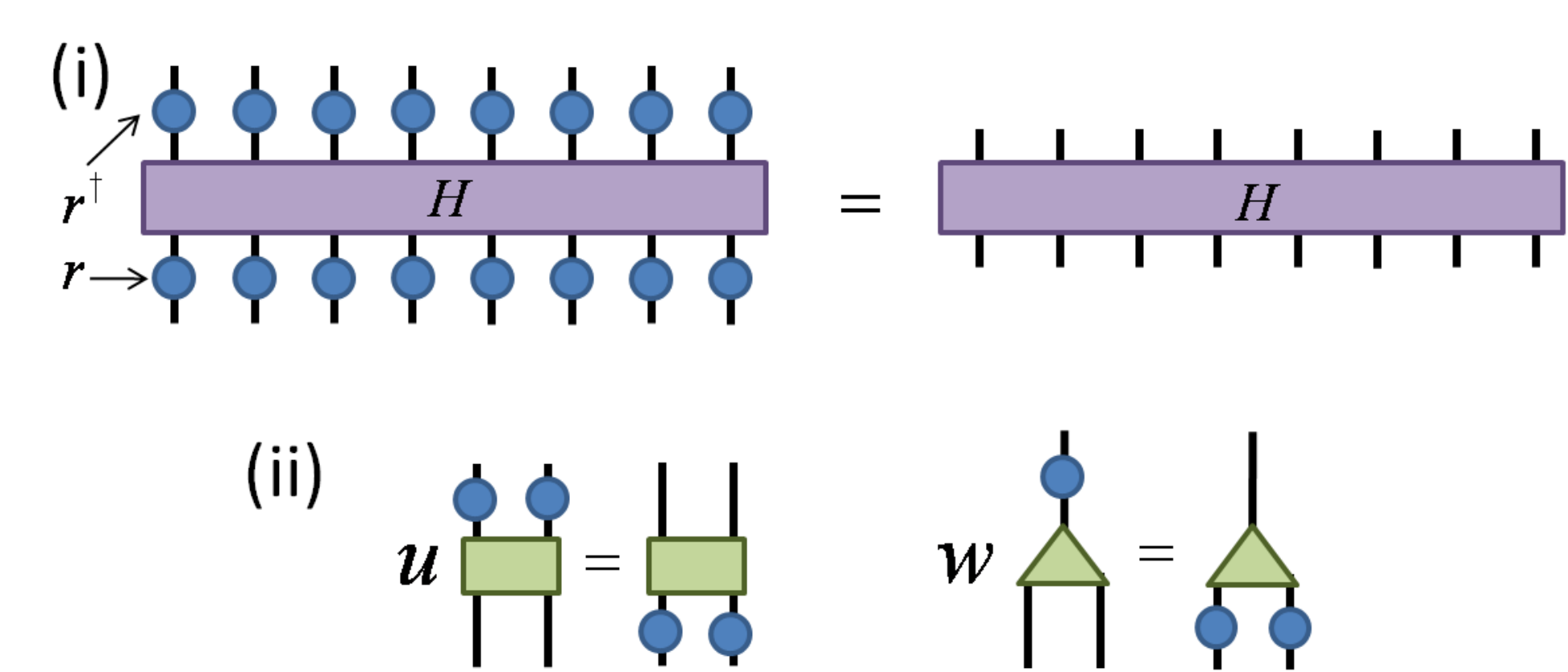}
\caption{
(i) Invariance of a Hamiltonian $H$ under a global symmetry implemented by simultaneously acting with one-site transformation $r$ on all sites of $\mathcal{L}$, see Eq. \ref{eq:RHR}. (ii) Disentangler $u$ and isometry $w$ invariant under the action of the single site transformation $r$, represented by a circle, acting on all upper/lower indices of these tensors. Notice that the upper index of the isometry $w$ is transformed according to $r'$, which might be a different representation of the symmetry group.}
\label{fig:Hinvariant}
\end{figure}

\begin{figure}[t]
  \includegraphics[width=8.5cm]{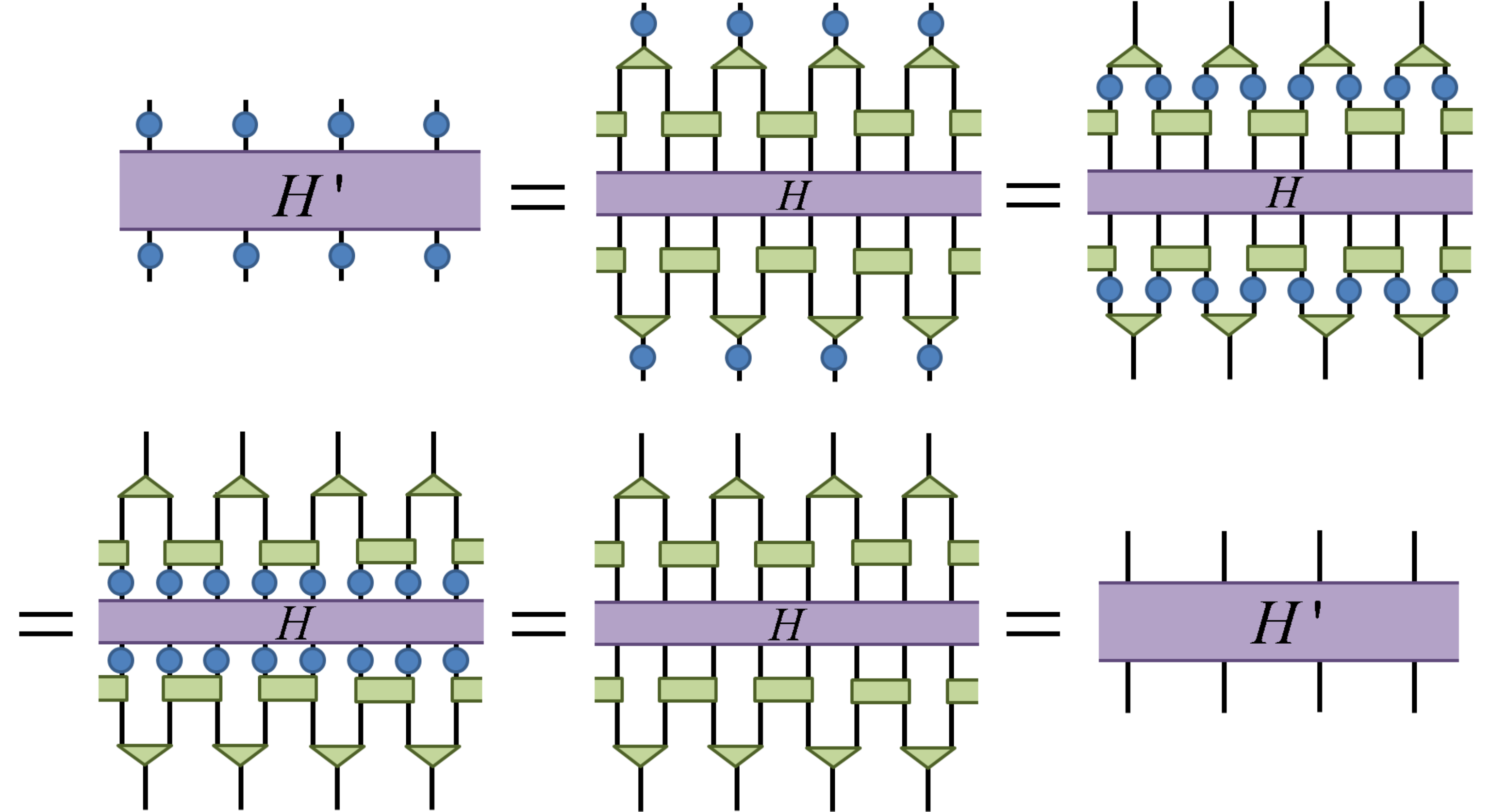}
\caption{
Sequence of equalities showing that a global symmetry $R=r^{\otimes N}$ of Hamiltonian $H$ is exactly preserved by a coarse-graining transformation $W$ that is the product of disentanglers $u$ and isometries $w$, provided the latter are also invariant under transformation $r$, Fig. \ref{fig:Hinvariant}. Indeed, the effective Hamiltonian $H' = W^{\dagger}HW$ is invariant under the global transformation $R'=(r')^{\otimes N'}$.}
\label{fig:H2invariant}
\end{figure}

Our goal is to coarse-grain a lattice model with a \textit{local} symmetry. It is instructive to first consider the simpler and better understood case of a \textit{global} symmetry\cite{Singh2009Global}. 

Let $r$ be a unitary transformation acting on one site of lattice $\mathcal{L}$, and let 
\begin{equation}
	R \equiv r^{\otimes N}
\end{equation}
be the unitary transformation that results from applying the same transformation $r$ on all sites of $\mathcal{L}$ \textit{simultaneously}. For instance, in the case where a site is described by a spin-1/2 degree of freedom, the one-site unitary transformation $r$ could correspond to a Pauli matrix, say $r=i\sigma^x$, in which case $R=(i\sigma_x)^{\otimes N}$ corresponds to simultaneously rotating all spins in the lattice by and angle $\pi$ in the $\hat{x}$ direction. 

The lattice Hamiltonian $H$ is invariant under $R$ if
\begin{equation}
	R H R^{\dagger} = H,
	\label{eq:RHR}
\end{equation}
an equation that can be represented diagrammatically as in Fig. \ref{fig:Hinvariant}(i). In this case, we say that the lattice model has a global symmetry $R$. Recall that if $R$ is a global symmetry of $H$, then $R^2$ and $R^{-1}$ are also a global symmetries of $H$ and, more generally, the global symmetries of $H$ always form a group. In the case of $r=i\sigma^x$, where $r^2 = I$, this group is $Z_2$. We also notice in passing that this particular choice $r= i \sigma^x$ is actually a global symmetry of the deformed toric code Hamiltonian $H_{\text{TC}}^{\text{x}}$, since it is included as part of its local symmetry. There, $R$ corresponds to the product of $N/2$ star operators $A_s$, chosen according to a checkerboard pattern.

The presence of a global symmetry is a fundamental property of a lattice model, since the eigenvectors of $H$ are organized according to representations of the symmetry group. It is therefore of interest to consider a coarse-graining transformation $W$ that preserves the global symmetry of a model. With exception of models with spontaneous symmetry breaking, which will not be considered here, the low energy subspace automatically inherits the symmetries of a model and, in principle, there should be no need to enforce symmetry preservation during coarse-graining. However, due to the numerical (and therefore approximate) nature of $W$, the symmetry may be lost during coarse-graining, unless some measures are put in place to explicitly protect it. 

In the case where $W$ is the product of disentanglers $u$ and isometries $w$, a global symmetry is automatically preserved if these tensors are chosen to be symmetric themselves, that is
\begin{equation}
	(r\otimes r)u(r\otimes r)^{\dagger} = u,~~~~~(r \otimes r) w (r')^{\dagger} = w,
\end{equation}
where $r'$ is a unitary matrix acting on a site of $\mathcal{L}'$, which in general will be transformed according to a different representation of the same symmetry group. The invariance of $u$ and $w$ implies that the action of the group commutes with these tensors, see Fig. \ref{fig:Hinvariant}(ii). As shown in Fig. \ref{fig:H2invariant}, in this case indeed the coarse-grained Hamiltonian $H' = W^{\dagger}HW$ will also be invariant under the global transformation $R' \equiv (r')^{\otimes N'}$.

Using symmetric tensors $u$ and $w$ does not only ensure exact preservation of the global symmetry, but it also allows for important computational savings. On the one hand, symmetric tensors depend on less parameters than generic tensors, leading to a more compact variational ansatz. On the other, manipulating symmetric tensors (for instance in order to compute the expectation value of a local observable) has lower computational cost than manipulating generic tensors.

\subsection{Local symmetry}
\label{sec:local}

\begin{figure}[t]
  \includegraphics[width=8.5cm]{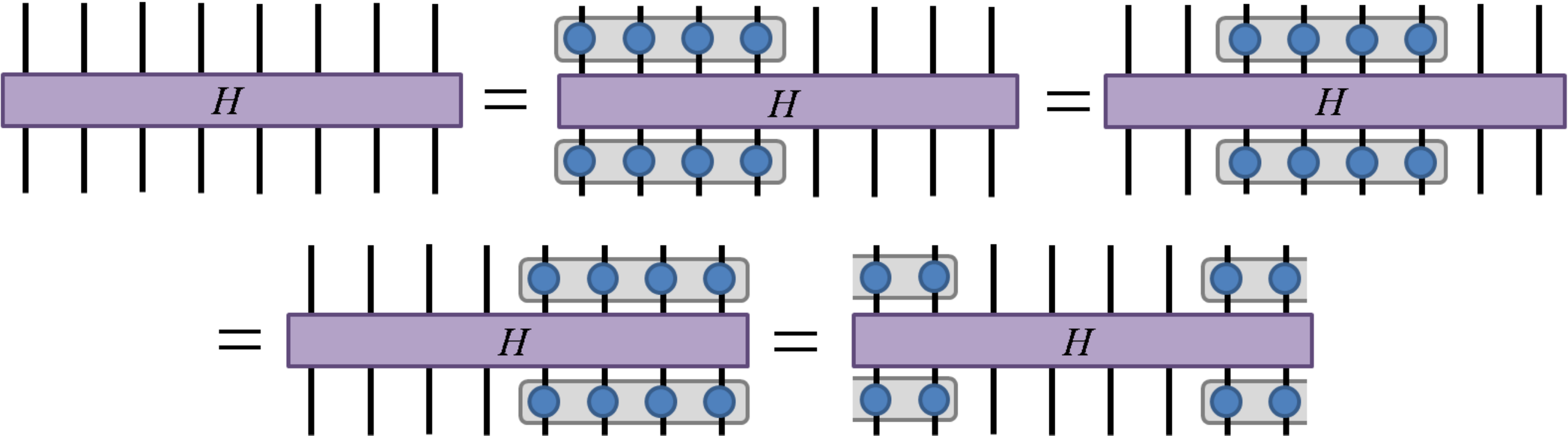}
\caption{
 Diagrammatic representation of a Hamiltonian with a local symmetry. In this example the local symmetry is implemented  by unitary transformations $r\otimes r \otimes r \otimes r$, acting on some blocks of four contigous sites.}
\label{fig:localSym}
\end{figure}

Let us now assume that the lattice model has a local symmetry. In this case the Hamiltonian $H$ is invariant under unitary transformations that act on a small subset of neighbouring sites, as exemplified by Eq. \ref{eq:localSym} for the deformed toric code Hamiltonian $H_{\text{TC}}^{\text{x}}$, that is invariant under star operators $A_s$ acting on four contiguous spins. This situation is diagrammatically represented in Fig. \ref{fig:localSym}.

How can a local symmetry be exactly preserved during coarse-graining? And how can it be exploited to obtain a more compact variational ansatz and to reduce the computational cost of simulations, as in the case of a global symmetry?
These questions  were already addressed by Fradkin et al. in Ref. \onlinecite{fradkin_real-space_1979} and before it by the seminal work of Migdal et al. in Ref. \onlinecite{Migdal:1975zg}.  In the next section we take a fresh look at them, by applying the ideas of  Entanglement Renormalization summarized in Sect. \ref{sec:ER} to this  subject.  Next section is indeed devoted to describe a coarse-graining transformation $W$ for the deformed toric code Hamiltonian $H_{\text{TC}}^{\text{x}}$ that explicitly preserves its local $Z_2$ symmetry.
 
\section{Coarse-graining of a lattice model with a local symmetry}
\label{sec:coarse}

In this section we propose a coarse-graining transformation for lattice models with a local symmetry.  For concreteness, we consider the toric code model \cite{Kitaev2003ToricCode} deformed with a magnetic field\cite{Trebst2007BreakDown} as reviewed in Sect. \ref{sec:DTC}, which has a local $Z_2$ symmetry, although it is possible to generalize the present construction to quantum double models \cite{Kitaev2003ToricCode} with a generic discrete symmetry group using the results of Refs. \onlinecite{Aguado2008TC}, as well as to the more exotic context of string-net models\cite{Levin2005SN} using the results of Ref. \onlinecite{Konig2009}.

In order to motivate our construction, we start by discussing two other possibilities, namely (i) the use of a \textit{bare} coarse-graining transformation that simply ignores the presence of the local symmetry altogether (as a result of which the local symmetry may not be preserved due to numerical errors), and (ii) the use of the (non-local) duality between the $Z_2$ lattice gauge theory and the quantum Ising model, to apply a coarse-graining transformation to the Ising model instead. Then we present an overview of the strategy considered in this work, followed by a more detailed explanation of the two parts into which the coarse-graining transformation $W$ splits: an exact transformation $W_{\text{exact}}$ and a numerical transformation $W_{\text{num}}$. Finally, we also discuss how to compose several layers of coarse-graining and how to build a variational ansatz for low energy states of the original model.

\subsection{Motivation}
\label{sec:coarse:motivation}

Let us restate our goal. We would like to build an isometric transformation $W$, 
\begin{equation}
	W^{\dagger}:\mathbb{V}_{\mathcal{L}} \rightarrow \mathbb{V}_{\mathcal{L}'}
\end{equation}
to coarse-grain the deformed toric code model,
\begin{equation}
	H_{\text{TC}}^{\text{x}} \equiv -J_{e}\sum_{s} A_s - J_{m}\sum_p B_p - h_x\sum_{j} \sigma_j^x,
	\label{eq:Hhx2}
\end{equation}
defined on a lattice $\mathcal{L}$ made of $L\times L$ sites (or $2L^{2}$ spins), as reviewed in Sect. \ref{sec:DTC}, into an effective lattice $\mathcal{L}'$. For the case $J_e \gg J_m,h_x$, the entire low energy subspace of $H_{\text{TC}}^{\text{x}}$ is made of states that are invariant under the local symmetry, Eq. \ref{eq:star}. We next discuss two simple options.

\subsubsection{Bare coarse-graining}
 
A first option is to proceed with a \textit{bare} coarse-graining scheme that ignores the presence of the local $Z_2$ symmetry or, at most, merely exploits the global $Z_2$ symmetry given by
\begin{equation}
	\left(\prod_j \sigma^{x}_j\right) H_{\text{TC}}^{\text{x}}	\left(\prod_j \sigma^{x}_j\right) =  H_{\text{TC}}^{\text{x}}.
\end{equation}
In this case, one could consider a transformation $W$ made of several types of disentanglers and isometries along the lines of the schemes used to study other two-dimensional lattice models with entanglement renormalization \cite{Cincio2008, Evenbly2009Alg, Evenbly20092DIsing, Evenbly2010Kagome}. Notice that the local symmetry, arguably a fundamental property of the model, will in general not be protected against numerical errors that may occur during a bare coarse-graining transformation. In addition, several attempts in this direction (see e.g. Fig. \ref{fig:Bare}) indicate that, while it is possible to coarse-grain the system without exploiting the local symmetry, the computational cost of applying the bare approach to the deformed toric code is prohibitively large in the deconfined phase of the model, $h_x\leq h_x^{\text{crit.}}$, due to the presence of large amounts of entanglement, as illustrated in Fig. \ref{fig:Entanglement}.

\subsubsection{Duality transformation}

It might therefore be more convenient to exploit the local $Z_2$ symmetry. The symmetry introduces local constraints on the degrees of freedom of the model, implying that the vector space $\mathbb{V}_{\mathcal{L}}$ of the theory is redundantly large. It is possible to exploit these constraints to obtain an equivalent description in a smaller vector space, and to then build the coarse-graining transformation directly on the reduced vector space. As reviewed in Sect. \ref{sec:Z2:LGT}, when $J_e \gg J_m,h_x$, projection onto the low energy subspace leads to the $Z_2$ lattice gauge theory, with Hamiltonian
\begin{equation}
	H_{\text{LGT}} = -\sum_{p} B_p - h_x \sum_{j} \sigma_j^x,
	\label{eq:HLGT2}
\end{equation}
where $J_m$ was set to $1$ without loss of generality. Recall that this model can be mapped into the quantum Ising model\cite{Kogut1979RevModPhys,Savit1980}, as reviewed in Sect. \ref{sec:Z2:duality},
\begin{equation}
	H_{\text{Ising}} \equiv -h_x \sum_{\langle p,p'\rangle} \mu^{x}_p\mu^{x}_{p'} - \sum_{p} \mu^z_p,
	\label{eq:HIsing2}
\end{equation}
where different topological sectors of $H_{\text{TC}}^{\text{x}}$ correspond to different boundary conditions for $H_{\text{Ising}}$. This is the reduced description we were looking for. 

Therefore a second option is to first transform the low energy sector of the deformed toric code $H_{\text{TC}}^{\text{x}}$ (with $J_e \gg J_m,h_x$) into the Ising model $H_{\text{Ising}}$; then to apply coarse-graining techniques to determine the ground state phase diagram of the Ising model $H_{\text{Ising}}$; and, finally, to translate the results back to the deformed toric code model by undoing the duality transformation. 

This is indeed a viable option. As a matter of fact, entanglement renormalization techniques have already been used to coarse-grain the quantum Ising model and study its ground state phase diagram \cite{Cincio2008,Evenbly20092DIsing}. In particular, with the specific layout of disentanglers and isometries proposed in Ref. \onlinecite{Evenbly20092DIsing}, it was possible to study arbitrarily large lattices. The presence of a smaller amount of entanglement in the ground state of $H_{\text{Ising}}$, as compared to the ground state of $H_{\text{TC}}^{\text{x}}$, explains why the coarse-graining strategy has a lower computational cost for the Ising model than for the deformed toric code model. 

This second approach can be extended to all those models with a local symmetry such that their low energy subspace is dual to a simpler model. This is the case, for instance, of the quantum double model\cite{Kitaev2003ToricCode} for any discrete Abelian group $Z_n$, whose low energy sector (if one allows only for certain type of excitations) corresponds to the $Z_n$ lattice gauge theory, which in turn is dual to the $n$-state Potts model with a transverse magnetic field \cite{Green1978}. It seems, however, that for non-Abelian groups, an analogous duality transformation produces a highly non-local model \cite{Rico2010priv}. 

In spite of its simplicity, this second approach also has several drawbacks. Let us discuss two of them.

\textit{Vacancy.---} The duality transformation produces an Ising model with a vacancy that breaks translation invariance. While this may not be relevant in the thermodynamic limit, the presence of a vacancy in a finite system implies that the coarse-graining transformation cannot be homogeneous. As a result, for instance, the cost of simulations with the MERA increases from $O(\log L)$ for a translation invariant system to $O(L^2)$ for an inhomogeneous system. 

\textit{Weak non-locality.---} Another, more fundamental limitation of using the duality is that this transformation is not local, as reflected by the presence of two non-local, boundary spins in the Ising model, see Fig. \ref{fig:Ising2}. We refer to the resulting model as being \textit{weakly} non-local, since its Hamiltonian is still local in the bulk. A particular topological sector of $H_{\text{TC}}^{\text{x}}$ can be studied by just fixing these boundary qubits to some product state, corresponding to fixing the boundary conditions of $H_{\text{Ising}}$, in which case the model is completely local. However, in a finite system it might be of interest to study processes that simultaneously involve several topological sectors of the model. 
These processes can still be studied with the resulting Ising model by allowing the boundary spins of Fig. \ref{fig:Ising2} to be entangled with the rest of the spins. However, each boundary spin is coupled to $O(L)$ neighbouring spins. This level of non-locality is expected to produce large amounts of entanglement, with a subsequent increase in computational costs, possibly rendering the approach too expensive.

\subsubsection{Beyond a global duality transformation}

Finally, there are two fundamental reasons why in this work we explore an alternative coarse-graining transformation that is not based on simply mapping the low energy subspace of $H_{\text{TC}}^{\text{x}}$ into $H_{\text{Ising}}$. 

One is that our ultimate goal is to develop a systematic approach that can also be applied to locally symmetric models and lattice gauge theories with a more general symmetry group, such as e.g. non-Abelian groups, and not just to cases where a duality transformation exists to a simpler model (as is the case for finite Abelian groups $Z_n$ \cite{Green1978}). In other words, we use the deformed toric code, arguably the simplest possible example, for illustrative purposes only. The coarse-graining transformation described below can be suitably generalized to quantum double models with arbitrary (possibly non-Abelian) discrete group (equivalently, to lattice gauge theories with arbitrary discrete gauge group) by using the results of Ref. \onlinecite{Aguado2008TC}, as well as to string-net models\cite{Levin2005SN} by using the results of Ref. \onlinecite{Konig2009}. 

The second reason is that we are interested in a scheme that can be generalized to models with a Hamiltonian where the local symmetry is explicitly broken (for the deformed toric code, by adding e.g. an additional magnetic field $h_y\sum_{j}\sigma_j^{y}$ to $H_{\text{TC}}^{\text{x}}$) but such that the local symmetry is still recovered at low energies. In this broader context, the map to the Ising model is strongly non-local, in that each $\sigma_y$ is transformed into a string operator, whereas the coarse-graining scheme presented here can be suitably extended in a way that locality is strictly preserved, as discussed in subsequent work.

\begin{figure}[t]
  \includegraphics[width=8cm]{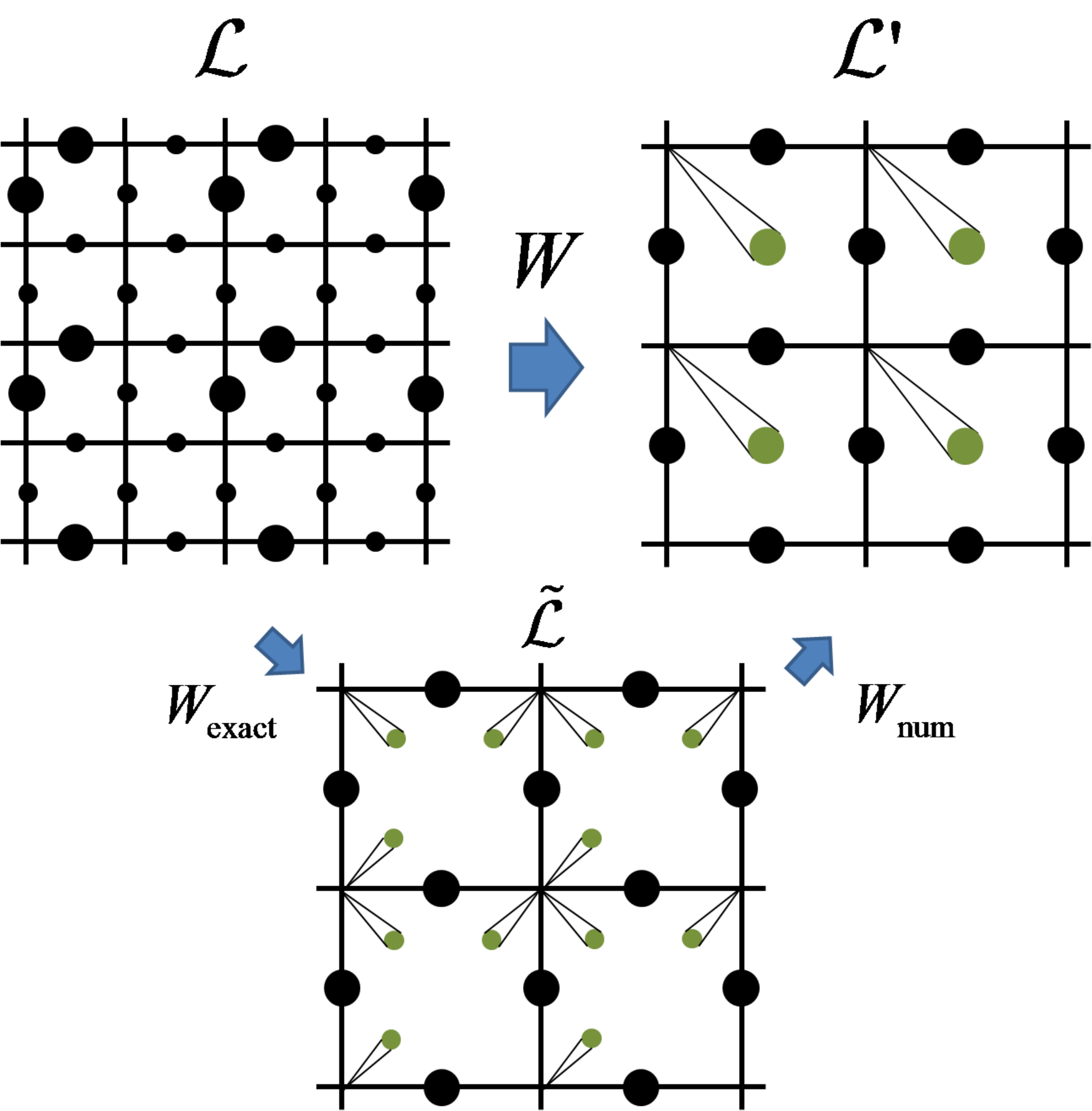}
\caption{
 Upper part: The coarse-graining transformation $W$ transforms the lattice $\mathcal{L}$ into the effective lattice $\mathcal{L}'$. Lattice $\mathcal{L}$ has spins on its edges. Lattice $\mathcal{L}'$ also has spins on its edges but, in addition, has an effective free spins sitting in the interior of each plaquette. Lower part: Transformation $W=W_{\text{exact}}W_{\text{num}}$ breaks into an exact transformation $W_{\text{exact}}$, which produces an intermediate lattice $\tilde{\mathcal{L}}$ with constrained and free spins, and a numerical transformation $W_{\text{num}}$, which coarse-grains the free spins of lattice $\tilde{\mathcal{L}}$.}
\label{fig:Wall}
\end{figure}

\subsection{The strategy}
\label{sec:coarse:strategy}

\begin{figure}[t]
  \includegraphics[width=8.5cm]{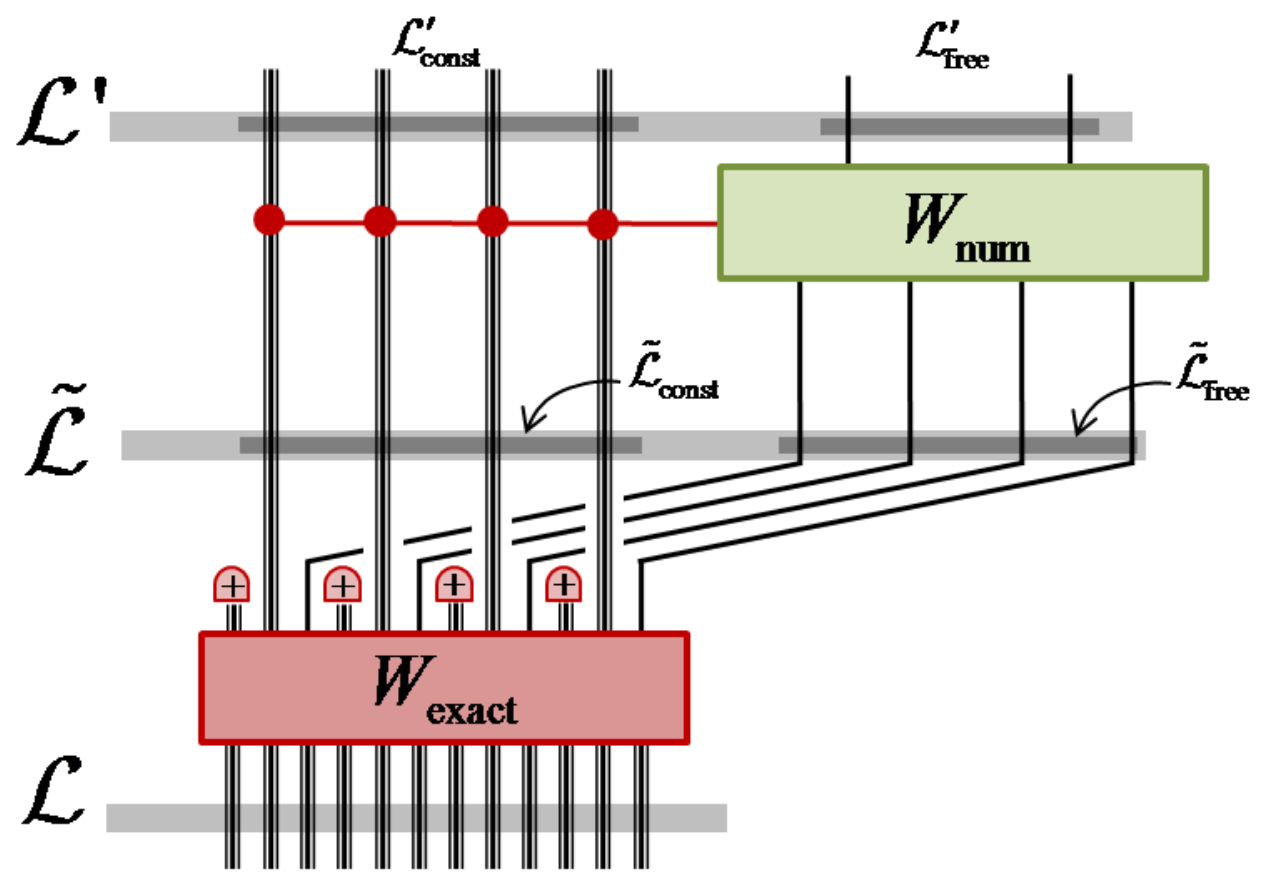}
\caption{
The proposed coarse-graining transformation $W$ breaks into two pieces $W_{\text{exact}}$ and $W_{\text{num}}$. $W$ transforms the (constrained) spins of $\mathcal{L}$ into the two types of spins present in the effective lattice $\mathcal{L}'$, namely constrained spins, represented by triple lines, and free spins, represented by single lines. $W_{\text{exact}}$ acts on constrained spins, whereas $W_{\text{num}}$ coarse-grains the free spins and acts on constrained spins diagonally on the $\sigma^x$ basis.}
\label{fig:Scheme}
\end{figure}

Next we describe the coarse-graining transformation proposed in this work. The original lattice $\mathcal{L}$ is transformed into an effective lattice $\mathcal{L}'$, where each plaquette of $\mathcal{L}'$ is obtained by coarse-graining a block of four plaquettes of $\mathcal{L}$, see Fig. \ref{fig:Wall}. Thus, if the original lattice $\mathcal{L}$ is made of $L\times L$ sites, the effective lattice $\mathcal{L}'$ is made of $L'\times L'$ sites with $L'=L/2$. The coarse-graining is implemented by an isometric transformation $W$,
\begin{equation}
	W^{\dagger}: \mathbb{V}_{\mathcal{L}} \rightarrow \mathbb{V}_{\mathcal{L}'},
\end{equation}
that maps the original Hamiltonian $H_{\text{TC}}^{\text{x}}$ into an effective Hamiltonian $H'$,
\begin{equation}
	H' \equiv W^{\dagger} H_{\text{TC}}^{\text{x}} W.
\end{equation}
What makes transformation $W$ special is that the effective $H'$ \textit{exactly} retains the local $Z_2$ symmetry of $H_{\text{TC}}^{\text{x}}$. This is accomplished by decomposing $W$ into two parts,
\begin{equation}
	W = W_{\text{exact}}W_{\text{num}},
\end{equation}
as illustrated in Figs. \ref{fig:Wall} and \ref{fig:Scheme}.
The first part, $W_{\text{exact}}$, consists of a fixed sequence of CNOT gates (see Sect. \ref{sec:Z2:CNOT}) and single-spin projections.
The second part is an isometric transformation $W_{\text{num}}$, to be determined numerically.

In order to explain the role of $W_{\text{exact}}$ and $W_{\text{num}}$, it is convenient to distinguish between two types of spins, depending on how the local $Z_2$ symmetry acts on them. We say that a spin is \textit{constrained} by the local symmetry, or just '\textit{constrained}', if it is included in the support of at least one star operator $A_s$. In other words constrained spins are transformed non-trivially by the local symmetry. We say that a spin is '\textit{free}' (that is, not constrained by the local symmetry) if no star operator $A_s$ acts on it. 

Notice that all the spins in $\mathcal{L}$ are initially constrained, since each of them belongs to the support of two star operators $A_s$. In contrast, the coarse-grained lattice $\mathcal{L}'$ will turn out to contain the above two types of degrees of freedom: \textit{constrained} spins, sitting on the edges of $\mathcal{L}'$, and \textit{free} spins, located inside the plaquettes of $\mathcal{L}'$, see Figs. \ref{fig:Wall} and \ref{fig:Scheme}.

\subsubsection{Analytic, exact coarse-graining}

$W_{\text{exact}}$ is applied only on constrained spins and its role is to transform some of these spins into free spins, while leaving other spins still constrained. It maps the original lattice $\mathcal{L}$ into an intermediate lattice $\tilde{\mathcal{L}}$, Figs. \ref{fig:Wall} and \ref{fig:Scheme},
\begin{equation}
	(W_{\text{exact}})^{\dagger}:\mathbb{V}_{\mathcal{L}}\rightarrow  \mathbb{V}_{\tilde{\mathcal{L}}},
\end{equation}
where the vector space $\mathbb{V}_{\tilde{\mathcal{L}}}$ contains both constrained spins and free spins,
\begin{equation}
	\mathbb{V}_{\tilde{\mathcal{L}}} \cong  \mathbb{V}_{\tilde{\mathcal{L}}}^{\text{const}} \otimes \mathbb{V}_{\tilde{\mathcal{L}}}^{\text{free}}.
\end{equation}
The details of how $W_{\text{exact}}$ manages to free some of the spins will be explained in Sect. \ref{sec:coarse:exact}. For now, we simply recall that the original lattice $\mathcal{L}$ contains $2L^2$ constrained spins subject to $L^2$ star operators $A_s$, Eq. \ref{eq:star} (of which $L^2-1$ are linearly independent, Eq. \ref{eq:constrained}). That is, there are only half as many star operators $A_s$ as constrained spins. It is therefore plausible that, by properly reorganizing the vector space $\mathbb{V}_{\mathcal{L}}$, transformation $W_{\text{exact}}$ can turn some of the constrained spins into free spins. $W_{\text{exact}}$ is also in charge of eliminating some of the spins of lattice $\mathcal{L}$, namely those that after the sequence of CNOT gates are forced by the local symmetry to be in a fixed, unentangled state $\ket{+}$. 

An important feature of the transformation $W_{\text{exact}}$ is that it is \textit{exact}: not only it is specified analytically (as opposed to numerically), but it can also be exactly reversed, implying that no approximation errors are introduced while squeezing the low energy, locally symmetric sector of $H_{\text{TC}}^{\text{x}}$ into the smaller vector space $\mathbb{V}_{\tilde{\mathcal{L}}}$ of lattice $\tilde{\mathcal{L}}$. 

Finally, $W_{\text{exact}}$ only depends on a few structural aspects of the spin model, such as the fact that $\mathcal{L}$ is a square lattice and how the local $Z_2$ symmetry acts on its spins. In particular, $W_{\text{exact}}$ is independent of the value of the magnetic field $h_x$ in $H_{\text{TC}}^{\text{x}}$. 

We note that $W_{\text{exact}}$ is inspired in a similar transformation proposed in Ref. \onlinecite{Aguado2008TC} for the (underformed) toric code, $h_x=0$, and that was used to show that the model is a fixed point of the RG flow. Here we will use $W_{\text{exact}}$ as part of a coarse-graining strategy valid for an arbitrary value of $h_{x}$.

\subsubsection{Numerical, approximate coarse-graining}

In contrast, transformation $W_{\text{num}}$ is mostly concerned with free spins. Its goal is to coarse-grain these \textit{free} spins into (\textit{effective}) \textit{free} spins, while acting only in a restricted way (to be explained below) on the constrained spins. It transforms the intermediate lattice $\tilde{\mathcal{L}}$ into the effective lattice $\mathcal{L}'$, see Figs. \ref{fig:Wall} and \ref{fig:Scheme},
\begin{equation}
	(W_{\text{num}})^{\dagger}: \mathbb{V}_{\tilde{\mathcal{L}}} \rightarrow \mathbb{V}_{\mathcal{L}'}, 
\end{equation}
where the vector space $\mathbb{V}_{\mathcal{L}'}$ also contains both constrained spins and free spins,
\begin{equation}
	\mathbb{V}_{\mathcal{L}'} \cong  \mathbb{V}_{\mathcal{L}'}^{\text{const}} \otimes \mathbb{V}_{\mathcal{L}'}^{\text{free}}.
\end{equation}

$W_{\text{num}}$ is determined \textit{numerically}, through some optimization procedure, and it is \textit{approximate}, in that the compression of the low energy sector of $H_{\text{TC}}^{\text{x}}$ into $\mathcal{L}'$ may include errors, whose size and implications have to be monitored. 

Finally, $W_{\text{num}}$ depends on the specific details of Hamiltonian $H_{\text{TC}}^{\text{x}}$, namely on $h_x$.
 
\subsubsection{Key properties} 
 
Before moving to a more detailed description of transformations $W_{\text{exact}}$ and $W_{\text{num}}$ in Sects. \ref{sec:coarse:exact} and \ref{sec:coarse:num}, we re-emphasize the two properties that the composite coarse-graining map $W$ is designed to fulfill:

\textit{Preservation of locality.---} $W$ transforms local operators in $\mathcal{L}$ into local operators in $\mathcal{L}'$. In particular, we will see that the effective Hamiltonian $H'$ remains local.

\textit{Preservation of the local $Z_2$ symmetry.---} The effective Hamiltonian $H'$ retains the local $Z_2$ symmetry of the original Hamiltonian $H_{\text{TC}}^{\text{x}}$. The local symmetry is preserved \textit{exactly} (in spite of the fact that $W$ contains an approximate, numerical part $W_{\text{num}}$) and exploited to obtain a significant reduction in computational costs (when compared to a bare coarse-graining strategy).

It is the combination of these two properties, namely simultaneous preservation of locality and of the local $Z_2$ symmetry, that distinguishes the present approach from the two other options mentioned earlier in Sect. \ref{sec:coarse:motivation}.

\begin{figure}[t]
  \includegraphics[width=8.5cm]{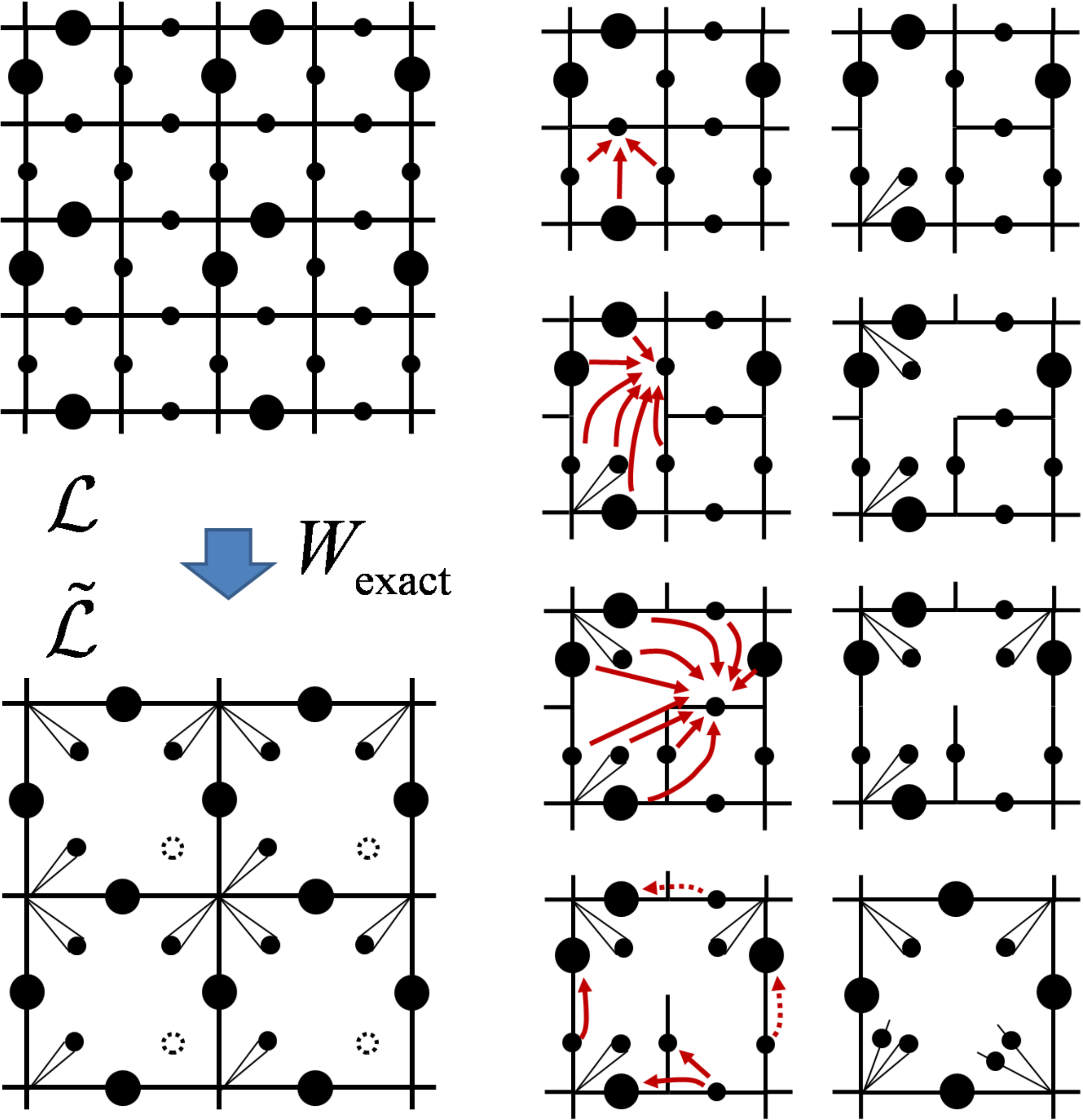}
\caption{
 Exact coarse-graining transformation $W_{\text{exact}}$. All spins of the original lattice $\mathcal{L}$ are equivalent. However, as a guide to the eye, larger circles are used to denote those spins that will remain constrained spins in $\tilde{\mathcal{L}}$. Left: a region of the original lattice $\mathcal{L}$ is mapped by $W_{\text{exact}}$ into a region of the intermediate lattice $\tilde{\mathcal{L}}$. Blocks of four plaquettes of $\mathcal{L}$ are mapped into single square plaquettes of $\tilde{\mathcal{L}}$, which contain three free spins and a vacancy in their interior. Right: detailed sequence of the CNOT gates, denoted by red arrows, applied to each block of four plaquettes in order to modify the support of the star operators $A_s$. {For clarity, dashed arrows also show two CNOTs corresponding to neighbouring unit cells.} At the end of the sequence, there are three types of spins: free spins (green), unentangled spins (in state $\ket{+}$) and constrained spins. Unentangled spins are removed by single-spin projections, and are therefore not present in $\tilde{\mathcal{L}}$.}
\label{fig:Wexact}
\end{figure}

\subsection{Exact transformation $W_{\text{exact}}$}
\label{sec:coarse:exact}

Transformation $W_{\text{exact}}$ consists of the sequence of CNOT gates and single-spin projections specified in Fig. \ref{fig:Wexact}, which maps blocks of four plaquettes of $\mathcal{L}$ into single plaquettes of the intermediate lattice $\tilde{\mathcal{L}}$. 

\subsubsection{Free spins, unentangled spins and constrained spins}

The CNOT gates are applied according to a spatially periodic pattern that has a block of four plaquettes of $\mathcal{L}$ as a unit cell, see right side of Fig. \ref{fig:Wexact}. Their goal is to modify the support of star operators $A_s$. Some of these star operators, initially acting on four spins, end up acting as the operator $\sigma^x$ on a single spin, while some other star operators end up acting on four spins of $\mathcal{L}$. This produces three types of spins: 

(i) Constrained spins that belong to the support of a star operator $A_{s'}$, where each star operator $A_{s'}$ acts on the four spins surrounding site $s'\in\tilde{\mathcal{L}}$. Constrained spins sit at the boundaries of the square plaquettes of $\tilde{\mathcal{L}}$.

(ii) Free spins, on which no star operator acts. They are depicted as green filled circles in Fig. \ref{fig:Wexact}. Free spins sit in the interior of a square plaquette of $\tilde{\mathcal{L}}$. Each free spin also belongs to a single-spin plaquette that emerges from one of the corners of a larger, square plaquette.

(iii) Unentangled spins, constrained to be an eigenstate of $\sigma^x$. They are produced by the last three CNOT gates on the right side of Fig. \ref{fig:Wexact} and can be identified as those spins sitting on an open edge (each unentangled spin is depicted next to a ket $'\ket{+}'$). Notice that any such spin must be in a product state. Indeed, if the collective state $\ket{\xi}$ of many spins is an eigenstate of $\sigma_j^x$ (with positive eigenvalue $+1$) for some specific spin $j$, then that spin $j$ cannot be entangled with the rest of the spins, but must instead be in the state $\ket{+}$ (see Eq. \ref{eq:sigmaXbasis}),
\begin{equation}
	\sigma_j^{x}\ket{\xi} = \ket{\xi}~~~~~\Rightarrow~~~~\ket{\xi} = \ket{+_j}\otimes \ket{\xi_{\text{rest}}}.
\end{equation}
{Notice that this property is common to all the ground states of the Hamiltonians  $H_{\text{TC}}^{\text{x}}$ independently of the strength of the magnetic field $h_x$ since the magnetic field commutes with the star operators.} 
Transformation $W_{\text{exact}}$ also includes projecting out any such spin, an operation that exactly preserve the state $ \ket{\xi_{\text{rest}}}$ of the rest of the spins,
\begin{equation}
\ket{\xi} \stackrel{\text{projection}}{\longrightarrow}	\braket{+_j}{\xi} = \ket{\xi_{\text{rest}}}.
\end{equation}
Therefore these unentangled spins do not appear in the intermediate lattice $\tilde{\mathcal{L}}$.

\begin{figure}[t]
  \includegraphics[width=8.5cm]{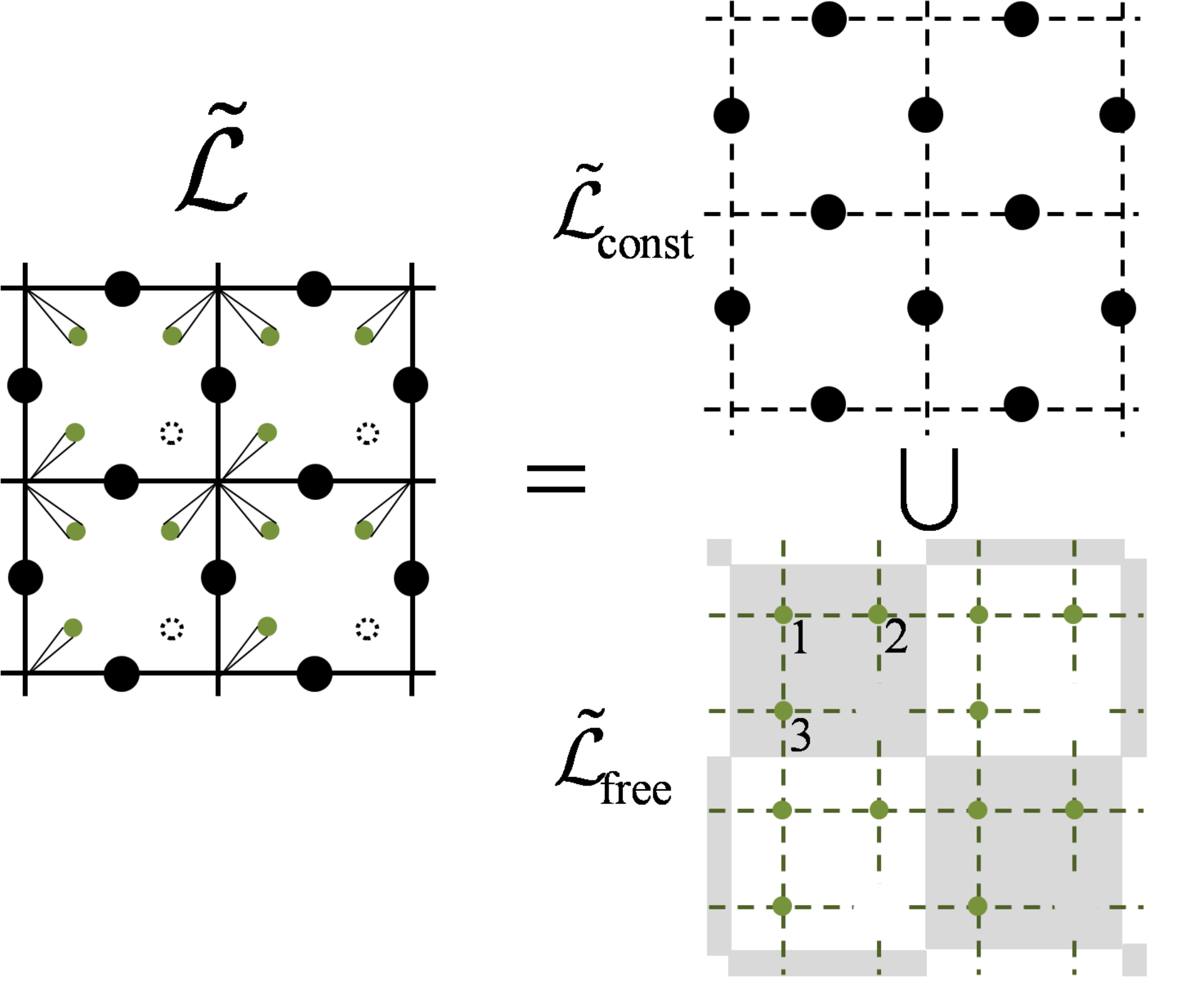}
\caption{
 The intermediate lattice $\tilde{\mathcal{L}} = \tilde{\mathcal{L}}_{\text{const}}  \cup \tilde{\mathcal{L}}_{\text{free}}$ decomposes as the union of a square sublattice $\tilde{\mathcal{L}}_{\text{const}}$ with a constrained spin on each of its edges and a square sublattice $\tilde{\mathcal{L}}_{\text{free}}$ with either a free spin or a vacancies on each of its sites. [Shading in $\tilde{\mathcal{L}}_{\text{free}}$ is used to indicate position relative to the plaquettes in $\tilde{\mathcal{L}}_{\text{const}}$]. 
}
\label{fig:Ltilde}
\end{figure}

In summary, constrained spins of the initial $\mathcal{L}$ are either mapped into constrained spins of $\tilde{\mathcal{L}}$ (sitting at the boundary of a square plaquette), or free spins of $\tilde{\mathcal{L}}$ (sitting at the interior of square plaquette), or they are removed. This occurs in the following proportions: 
\begin{equation}
	\begin{array}{c}
	8 \\
	\text{constrained}  \\
	\text{spins}\in \mathcal{L}
	 \end{array}
\longrightarrow \left\{ 
	\begin{array}{l}
	\text{2 spins} \in \tilde{\mathcal{L}}_{\text{const}}\\
	\text{3 spins} \in \tilde{\mathcal{L}}_{\text{free}} \\
  \text{3 spins removed}\\
	\end{array}
	\right.
\end{equation}
Lattice $\tilde{L}$ can be thought of as being made of two square sublattices $\tilde{\mathcal{L}} = \tilde{\mathcal{L}}_{\text{const}}  \cup \tilde{\mathcal{L}}_{\text{free}}$. The first contains constrained spins in its edges, and the second contains free spins (or vacancies) on its sites, see Fig. \ref{fig:Ltilde}.
 
\subsubsection{Transformed Hamiltonian}

\begin{figure}[t]
  \includegraphics[width=8.5cm]{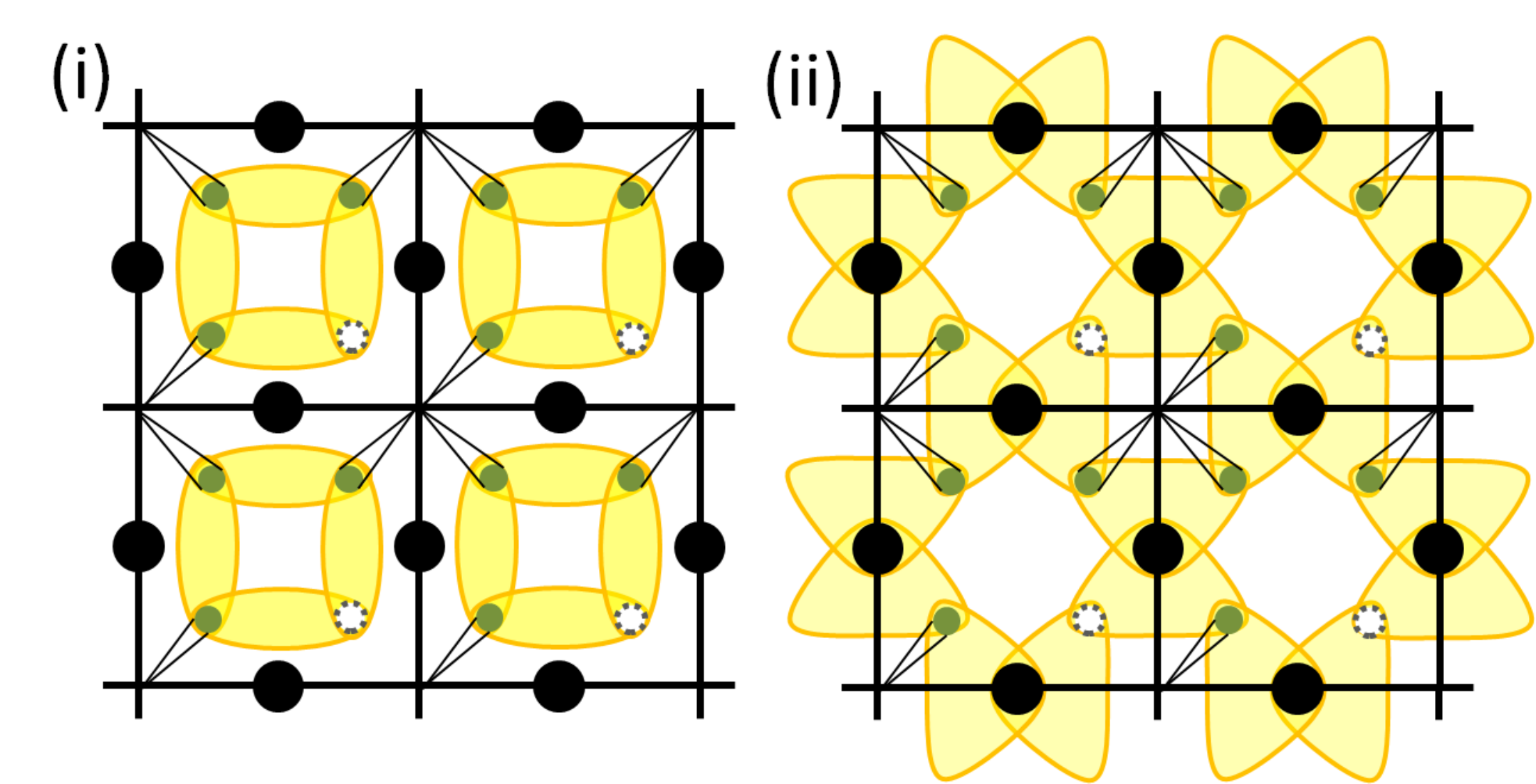}
\caption{
Depiction of the terms included in the transformed Hamiltonian $\tilde{H}$. From the structure of lattice $\tilde{\mathcal{L}}$, one can readily read the new star and plaquette operators of $\tilde{H}$ (as explained in Fig. \ref{fig:Hgraph}(ii)). 
In addition, the magnetic field $-h_x\sum_i \sigma^x_j$ in $H_{\text{TC}}^{\text{x}}$ gives rise to other interacting terms $-h_x\tilde{K}_{\text{int}}$, which are depicted as yellow shades. 
(i) Two-spin interaction $-h_x\sigma^x \otimes \sigma^x$ between nearest neighbour free spins contained in the same plaquette. Notice that in the presence of a vacancy, some of these terms become a magnetic field $-h_x \sigma^x$ acting on a free spin. 
(ii) Three-spin interaction $-h_x \sigma^x \otimes \sigma^x \otimes \sigma^x$ between two free spins on neighbouring plaquettes and a constrained spin sitting at the boundary between those plaquettes. Again, the presence of a vacancy reduces the support of some of the interaction operators, which end up coupling one free spin and one constrained spin according to $-h_x\sigma^x \otimes \sigma^x$.}
\label{fig:Wexact2}
\end{figure}

Under $W_{\text{exact}}$, the initial Hamiltonian $H_{\text{TC}}^{\text{x}}$ is transformed into a new Hamiltonian
\begin{equation}
	\tilde{H}\equiv (W_{\text{exact}})^{\dagger} H_{\text{TC}}^{\text{x}} W_{\text{exact}}
\end{equation}
that includes several types of terms,
\begin{equation}
	\tilde{H} = -J_{e}\sum_{s'\in \tilde{\mathcal{L}}} A_{s'} 
	- J_m \left( \sum_{p' \in \tilde{\mathcal{L}}}B_{p'} + \sum_{j \in \text{free}} \sigma^{z}_{j} \right) 
	-h_x \tilde{K}_{\text{int}}
	\label{eq:Htilde}
\end{equation}
All these terms are represented in Fig. \ref{fig:Wexact2}.

First, $\tilde{H}$ has terms coming from the star operators $A_s$ of $H_{\text{TC}} ^{\text{x}}$. As just mentioned, some star operators $A_s$ are simply eliminated (after being transformed into a single-spin operator $\sigma^x$), whereas others become four-spin star operators $A_{s'}$ acting on the sites of $\tilde{\mathcal{L}}$. Specifically, from every four star operators $A_s$ acting on $\mathcal{L}$, only one becomes a star operator $A_{s'}$ acting on $\tilde{\mathcal{L}}$.

Hamiltonian $\tilde{H}$ also has terms that originate in the plaquette operators $B_p$ of $H_{\text{TC}}^{\text{x}}$. For every four plaquette operators acting on $\mathcal{L}$, three are transformed into single-site operators $\sigma^z$ that act on the free spins of $\tilde{\mathcal{L}}$, whereas the fourth one becomes a seven-spin plaquette operator $B_{p'}$,
\begin{equation}
	B_{p'} = \prod_{j\in p'} \sigma_j^{z},
	\label{eq:Bpnew}
\end{equation}
where the product includes the four spins at the boundary of plaquette $p' \in \tilde{\mathcal{L}}$ and the three free spins in its interior.

Finally, $\tilde{H}$ also contains terms that result from transforming the magnetic field $-h_x\sum_{j}\sigma_j^x$ in $H_{\text{TC}}^{\text{x}}$. Under $W_{\text{exact}}$, the magnetic field becomes an interaction between free spins. There are two types of interactions: interactions within a square plaquette of $\tilde{\mathcal{L}}$, and interactions across the boundary between two plaquettes of $\tilde{\mathcal{L}}$, represented by yellow bulbs in Figs. \ref{fig:Wexact2}(i) and \ref{fig:Wexact2}(ii) respectively. Interactions within the plaquette are of the form $-h_x\sigma^x\otimes \sigma^x$ and couple pairs of first neighbour free spins. [Where the first neighbour of a free spin is a vacancy, the interaction is reduced to a magnetic field $-h_x\sigma^{x}$ acting on the free spin.] Interactions between free spins that belong to neighbouring plaquettes of $\tilde{\mathcal{L}}$ are 'mediated' by the constrained spin sitting at the boundary between the plaquettes, and it is of the form $-h_x\sigma^x\otimes \sigma^x \otimes \sigma^x$. [Again, where the interaction would involve a vacancy, it is reduced to the form $-h_x\sigma^x\otimes \sigma^x$ acting on a free spin and a constrained spin.] 

The pattern of the interaction is suggestive. The free spins indeed are subject to interactions characteristic of a magnetically charged matter field. In this case the matter field  transforms trivially under the action of the gauge group, since it belongs to $\mathbb{V}_{\text{LGT}}$ on which by definition the constraints act as the identity operator (see Sect. \ref{sec:Z2}).  It would  then be tempting  to interpret the free spins as an emergent matter field minimally coupled to the Hodge dual of the constrained spins, that play the role of the  magnetic photons. In this way the mechanism of confinement as dual superconductivity of the vacuum for the $Z_2$ lattice gauge theory becomes manifest \cite{tHooft:1981ht,Mandelstam:1974pi}.

\subsection{Local implementation of the duality transformation}
\label{sec:coarse:LocalDuality}

So far we have introduced transformation $W_{\text{exact}}$ as a sequence of CNOT gates and single-spin projections, and have described the Hamiltonian $\tilde{H}$ resulting from transforming the deformed toric code Hamiltonian $H_{\text{TC}}^{\text{x}}$. Let us now take a minute to analyse these results.

$W_{\text{exact}}$ has an interesting interpretation in relation to the duality transformation between the $Z_2$ lattice gauge theory and the quantum Ising model reviewed in Sect. \ref{sec:Z2:duality}. Indeed, looking back at Fig. \ref{fig:Wexact2} one can see that the three free spins inside a square plaquette of lattice $\tilde{\mathcal{L}}$ form a small, $2\times 2$ quantum Ising model on their own (with a vacancy), in that they interact according to the Hamiltonian
\begin{equation}
	 -h_x\sum_{\langle j,k\rangle} \sigma^{x}_j\sigma^{x}_{k} -\sum_j \sigma^z_j,
\end{equation}
[where the interactions involving a free spin and the vacancy are reduced to a term $-h_x\sigma^x$ acting on the spin.] Therefore, $W_{\text{exact}}$ can be thought of as implementing a duality transformation inside each block of four plaquettes of $\mathcal{L}$. This visualization is useful. It tells us that the mechanism used by $W_{\text{exact}}$ in order to remove degrees of freedom that are determined by the local symmetry (namely to transform some star operators $A_s$ into operators $\sigma^x$ acting on a single spin, which must consequently be in state $\ket{+}$, and to project out such spins) is the same that is used to transform the $Z_2$ lattice gauge theory into the Ising model. However, while the duality transformation is usually applied once globally on the whole system, $W_{\text{exact}}$ applies an \textit{independent} duality transformation on each block of four plaquettes of $\mathcal{L}$. This makes $W_{\text{exact}}$ a fully local transformation, namely one that maps any local operator on $\mathcal{L}$ into a local operator in $\tilde{\mathcal{L}}$.

The resulting tiny (3-spin) quantum Ising models inside the square plaquettes of $\tilde{\mathcal{L}}$, represented in Fig. \ref{fig:Wexact2}(i), are not independent, but rather patched together by further interactions represented in Fig. \ref{fig:Wexact2}(ii). These interactions,
\begin{equation}
	-h_x\sigma^x\otimes\sigma^{x}\otimes \sigma^x, 
	\label{eq:xxx}
\end{equation}
can be interpreted as Ising interactions between two free spins (one on each plaquette) that are \textit{mediated} by the constrained spin sitting at the boundary between the two plaquettes. Specifically, depending on whether the constrained spin is in state $\ket{+}$ or $\ket{-}$, the two free spins interact according to $- h_x\sigma^{x}\otimes \sigma^x$ or $h_x\sigma^{x}\otimes \sigma^x$. Again, this is reminiscent of the way the two non-local, boundary spin control the boundary conditions of the Ising model resulting from applying a global duality transformation, see Fig. \ref{fig:Ising2}.

\subsection{Disentangling power of $W_{\text{exact}}$}
\label{sec:coarse:disentangling}

Another relevant feature of $W_{\text{exact}}$ is that it removes short-range entanglement present in the low energy states of $H_{\text{TC}}^{\text{x}}$ for $J_e \gg J_m,h_x$, and which has its origin in the presence of the local symmetry. To illustrate this point, we analyse the ground state of the transformed Hamiltonian $\tilde{H}$ in the limits of small and large magnetic field $h_x$, where an analytic treatment is possible. Having established earlier that $W_{\text{exact}}$ implements some sort of duality transformation to the quantum Ising model at a local level, it is easy to anticipate that, in these two limits, the free spins of $\tilde{\mathcal{L}}$ will be in an unentangled state, since this is the case in the analogous limits of the quantum Ising model.
 
\subsubsection{Small magnetic field}

For $h_x=0$, the transformed Hamiltonian $\tilde{H}$ of Eq. \ref{eq:Htilde} becomes
\begin{equation}
	-J_{e}\sum_{s'\in \tilde{\mathcal{L}}} A_{s'} 
	- J_m \left( \sum_{p' \in \tilde{\mathcal{L}}}B_{p'} + \sum_{j \in \text{free}} \sigma^{z}_{j} \right).
	\label{eq:undeformedHtilde}
\end{equation}
Since in this case $\tilde{H}$ is obtained by transforming the undeformed toric code Hamiltonian $H_{\text{TC}}$ of Eq. \ref{eq:HTC}, where all terms commute with each other, all terms in the above expression also commute with each other. Therefore for $h_x=0$ the (four linearly independent) ground states of $\tilde{H}$ are eigenstates of each individual Hamiltonian term. In particular, a term $-J_{m}\sigma^z$ acts on each free spin in $\tilde{\mathcal{L}}$, which therefore must be in state $\ket{0}$. Thus, for $h_x=0$ any ground state of $\tilde{H}$ factorizes as
\begin{equation}
	\ket{\xi^{\text{const}}_{\tilde{\mathcal{L}}}}\otimes \ket{\xi^{\text{free}}_{\tilde{\mathcal{L}}}},~~~
\end{equation}
where $\ket{\xi^{\text{const}}_{\tilde{\mathcal{L}}}}\in \mathbb{V}^{\text{const}}_{\tilde{\mathcal{L}}}$ is an entangled state of the constrained spins whereas $\ket{\xi^{\text{free}}_{\tilde{\mathcal{L}}}}\in \mathbb{V}^{\text{free}}_{\tilde{\mathcal{L}}}$ is a product state of the free spins,
\begin{equation}
	\ket{\xi^{\text{free}}_{\tilde{\mathcal{L}}}} = \ket{0}\otimes \ket{0} \otimes \cdots \otimes \ket{0}.
\end{equation}

This factorization is not surprising --after all, the limit of small magnetic field $h_x$ corresponds to the spin polarized state of the quantum Ising model in the limit of a large transverse magnetic field (in the $\hat{z}$ direction). And yet, it is a result that has important implications for the purposes of this paper. It means that $W_{\text{exact}}$ has transformed a robustly entangled state (the ground state of the underformed toric code) into a state where a fraction of the spins are not entangled at all. This is the basis for a significant reduction in computational costs.

Following Ref. \onlinecite{Aguado2008TC}, we further notice that the state $\ket{\xi_{\text{const}}}$ for the constrained spins is again the ground state of the undeformed toric code. Indeed, after projecting out the free spins in state $\ket{\xi_{\text{free}}}$, the Hamiltonian in Eq. \ref{eq:undeformedHtilde} becomes the toric code Hamiltonian of Eq. \ref{eq:HTC} defined now over the constrained spins of lattice $\tilde{\mathcal{L}}$. In other words, the models before and after the coarse-graining transformation are locally identical, and the undeformed toric code is seen to be a fixed-point of the RG flow\cite{Aguado2008TC}. 

For a finite $h_x \neq 0$ the free spins will no longer be in a product state, but rather entangled between themselves and the constrained spins. However, for small values of $h_x$, the free spins are expected to still be only weakly entangled, an expectation confirmed by numerical simulations, see Fig. \ref{fig:Entanglement}. Thus, one of the merits of transformation $W_{\text{exact}}$ is that it reduces ground state entanglement throughout the deconfined phase of the deformed toric mode model, see Sect. \ref{sec:DTC}.

\subsubsection{Large magnetic field}

We have just seen that transformation $W_{\text{exact}}$ is capable of removing entanglement from the ground state of $H_{\text{TC}}^{\text{x}}$ in the deconfined phase, a property that the present coarse-graining scheme will exploit to obtain a simplified description. It would be unfortunate, however, if $W_{\text{exact}}$ would introduce additional entanglement in the other phase of $H_{\text{TC}}^{\text{x}}$, namely the spin polarized phase, see Sect. \ref{sec:DTC}. This does not seem to be the case, see Fig. \ref{fig:Entanglement}. In the limit of large magnetic field, this is easy to see.

Indeed, for very large $h_x$, and using that $J_e \gg J_m,h_x$, the deformed toric code Hamiltonian $H_{\text{TC}}^{\text{x}}$ can be written as
\begin{equation}
	 -J_{e}\sum_{s} A_{s} - h_x \sum_{j}\sigma_j^{x},
\end{equation}
and its (unique) ground state is the spin polarized state $\ket{+}\otimes \ket{+} \otimes \cdots \otimes \ket{+}$. It is clear from Eq. \ref{eq:CNOT1} that the sequence of CNOT gates leave this state invariant. Therefore, for large $h_x$, the ground state of $\tilde{H}$ is also an unentangled, spin polarized state. 

\subsection{Numerical transformation $W_{\text{num}}$}
\label{sec:coarse:num}

\begin{figure}[t]
  \includegraphics[width=6cm]{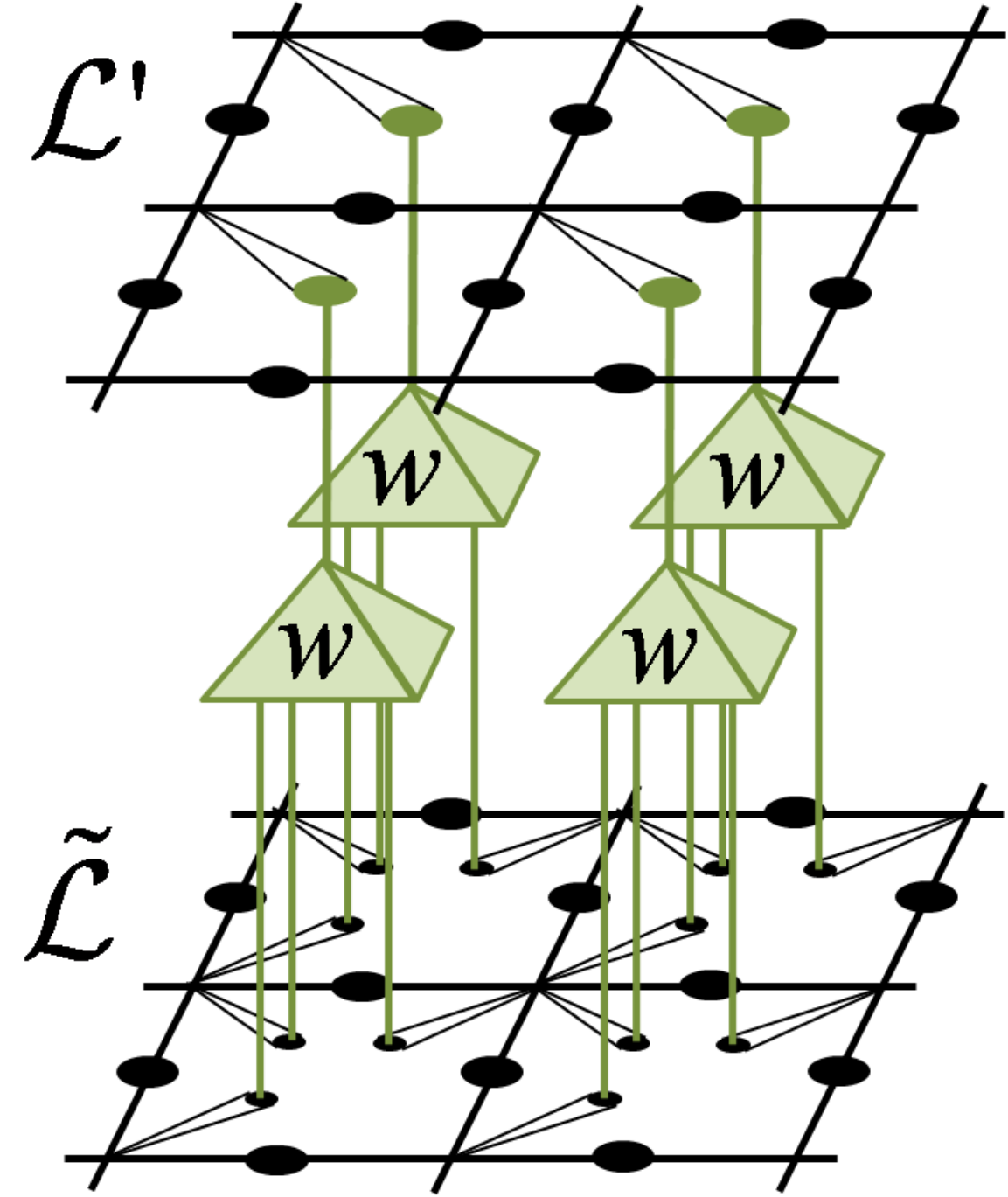}
\caption{
 This simplified choice of the numerical transformation $W_{\text{num}}$ consist of a tensor product of isometries $w$ that coarse-grain the three free spins inside a square plaquette of $\tilde{\mathcal{L}}$ into an effective free spin of $\mathcal{L}'$, see Eq. \ref{eq:w3to1}}.
\label{fig:Wnum}
\end{figure}
 
For a finite value of $h_x$, the ground state of $\tilde{H}$ will be such that the free spins are entangled with the constrained spins of $\tilde{\mathcal{L}}$. The numerical transformation $W_{\text{num}}$ coarse-grains the three free spins inside each square plaquette of the intermediate lattice $\tilde{\mathcal{L}}$ into an effective free spin of lattice $\mathcal{L}'$, see Figs. \ref{fig:Wall} and \ref{fig:Scheme}.

Transformation $W_{\text{num}}$ is required to fulfill two conditions. On the one hand it must map local operators in $\tilde{\mathcal{L}}$ into local operators in $\mathcal{L}'$. This can be accomplished, for instance, by building $W_{\text{num}}$ as a product of disentanglers and isometries, as explained in Sect. \ref{sec:ER}. On the other hand, $W_{\text{num}}$ must commute with the local $Z_2$ symmetry. This is achieved by requiring that, at most, it acts on the constrained spins by means of operators that are diagonal in the $\sigma^x$ basis. For instance, a unitary transformation
\begin{equation}
	\exp(-i\varphi \sigma^{x}\otimes\sigma^x\otimes\sigma^y),~~~~~\varphi \in [0,2\pi)
\end{equation}
where the first and third Pauli matrices act on free spins and the middle one on a constrained spin, could be used to remove part of the entanglement introduced by the coupling of Eq. \ref{eq:xxx}. Since this unitary transformation acts on the constrained spin diagonally on the $\sigma_x$ basis, it commutes with any star operator $A_{s'}$ acting on that spin, and therefore does not alter the action of the local $Z_2$ symmetry on $\tilde{\mathcal{L}}$.

In this work we use, for illustrative purposes, a simplified transformation $W_{\text{num}}$ that is made only of isometries, see Fig. \ref{fig:Wnum}. That is, we do not incorporate disentanglers in the numerical part of the coarse-graining transformation. This particular choice is clearly a limitation: local entanglement involving free spins is not removed by $W_{\text{num}}$ and, as a result, it will accumulate over successive applications of the coarse-graining transformation $W$, in a way similar to what is observed using a tree tensor network (TTN) ansatz \cite{Tagliacozzo2009}. In Sect. \ref{sec:coarse:variational} we will see that this choice of $W_{\text{num}}$ limits the system sizes that can be addressed.

On the other hand, restricting our attention to a numerical transformation $W_{\text{num}}$ that is made only of isometries also has several advantages from a pedagogical point of view. Its simplicity allows us to explicitly keep track of how the different terms in the initial Hamiltonian $H_{\text{TC}}^{\text{x}}$ are transformed. In addition, the absence of numerical disentanglers in $W_{\text{num}}$ (whose effect has already been studied in other two-dimensional systems\cite{Cincio2008,Evenbly20092DIsing,Evenbly2010Kagome}) makes the role of $W_{\text{exact}}$ in the whole coarse-graining, and in particular its disentangling power, more transparent. We refer to Ref. \onlinecite{Tagliacozzo2010prep} for an example of a more complex $W_{\text{num}}$.

\begin{figure}[t]
  \includegraphics[width=8.5cm]{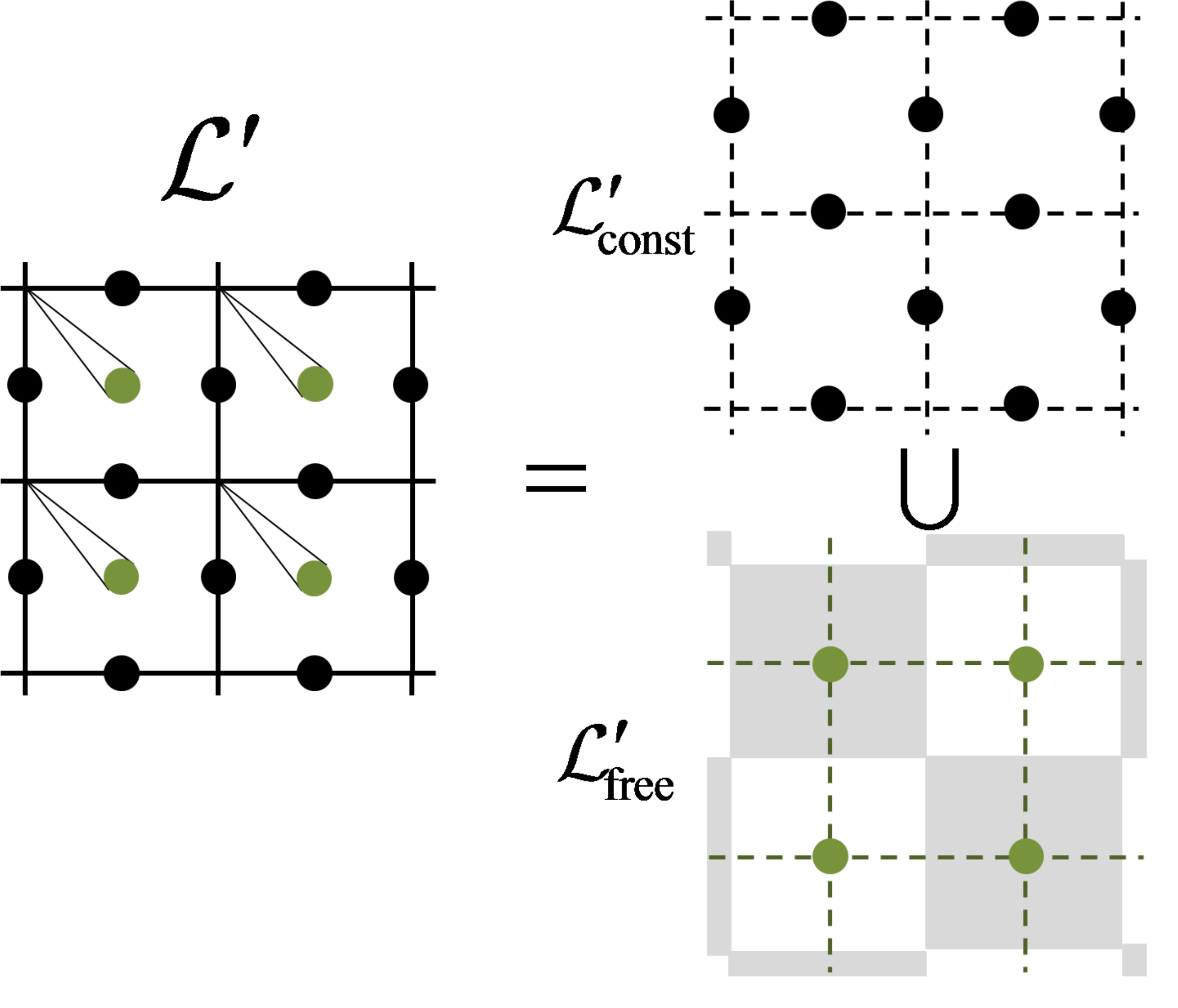}
\caption{
 The effective lattice $\mathcal{L}' = \mathcal{L}'_{\text{const}}  \cup \mathcal{L}'_{\text{free}}$ decomposes as the union of a square sublattice $\mathcal{L}'_{\text{const}}$ with a constrained spin on each of its edges and a square sublattice $\mathcal{L}'_{\text{free}}$ with an effective free spin on each of its sites. [Shading in $\mathcal{L}'_{\text{free}}$ is used to indicate position relative to the plaquettes in $\mathcal{L}'_{\text{const}}$].
}
\label{fig:Lprime}
\end{figure}

\subsubsection{Effective free spins}

Let us then see how our simplified choice of $W_{\text{num}}$, Fig. \ref{fig:Wnum}, transforms the system. Each isometry $w$ in $W_{\text{num}}$ simply maps the three free spins in the interior of a plaquette of $\tilde{\mathcal{L}}$ into an effective free spin,
\begin{equation}
	w^{\dagger}: \left( \mathbb{C}_{2}\right) ^{\otimes 3} \rightarrow \mathbb{C}_{\chi'},
	\label{eq:w3to1}
\end{equation}
where $\mathbb{C}_{2}$ is the vector space of a free spin and $\mathbb{C}_{\chi'}$ is the $\chi'$-dimensional space of an effective free spin.

The dimension $\chi'$ controls the degree of approximation introduced during coarse-graining. Ideally, it should be chosen large enough so that $\mathbb{C}_{\chi'}$ can accommodate the support of the reduced density matrix of the three free spins (see e.g. Ref. \onlinecite{Tagliacozzo2009}). Let $\chi'_{\rho}$ be the dimension of this support.  The computational cost of the approach grows as a power of $\chi'$. If, as a means to reduce the computational cost, $\chi'$ is chosen to be smaller than $\chi'_{\rho}$, then the coarse-graining transformation becomes approximate.

For $h_{x}=0$ and $h_{x}=\infty$ the free spins are in a product state and $\chi'$ can be chosen to be just $1$, and the coarse-graining transformation becomes trivial. For other values of $h_x$, the free spins are entangled and a larger value of $\chi'$ must be used.

\subsubsection{Effective Hamiltonian}

\begin{figure}[t]
  \includegraphics[width=8.5cm]{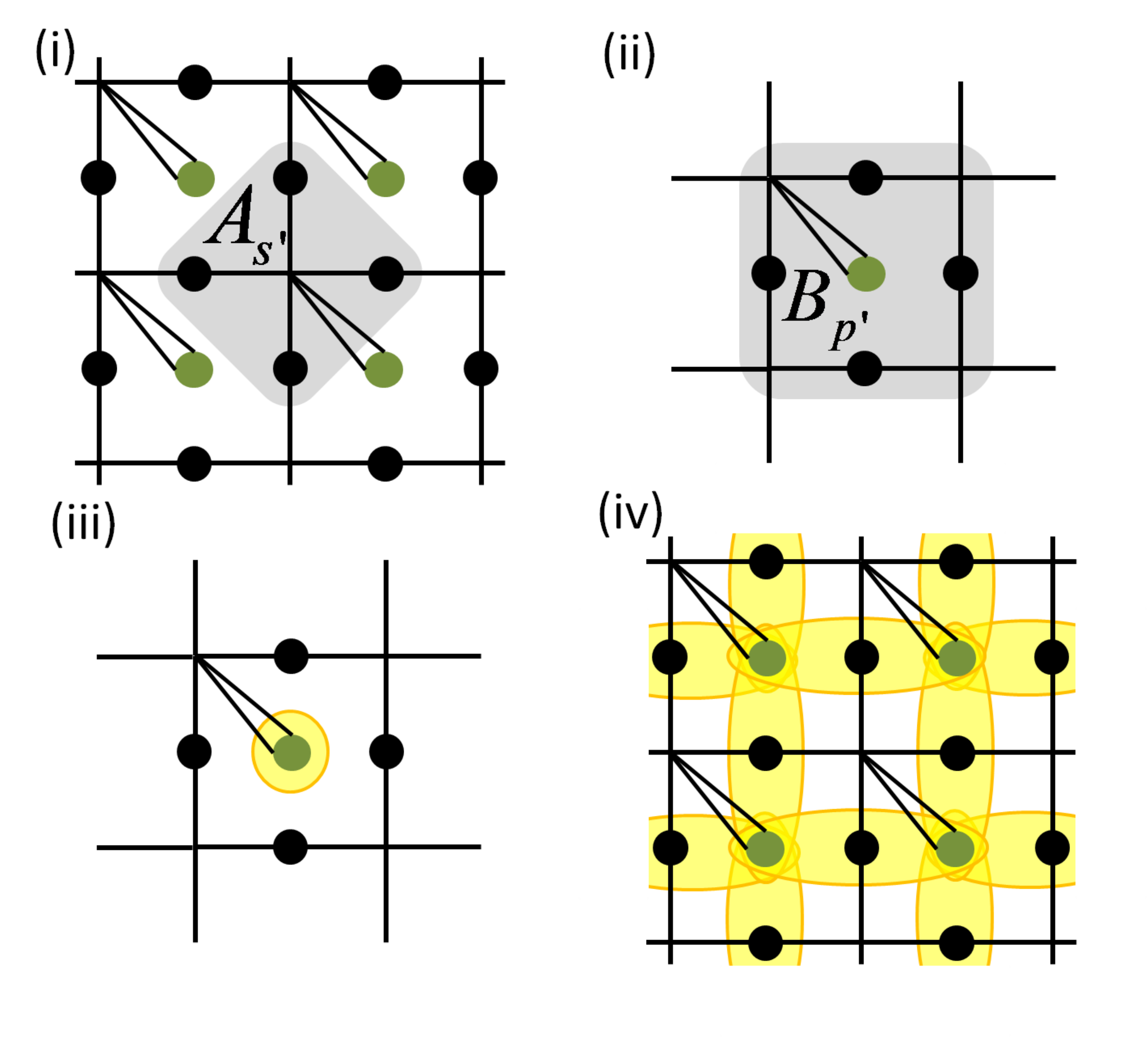}
\caption{
Terms that appear in the effective Hamiltonian $H'$, Eqs. \ref{eq:Hprime} and \ref{eq:Hprime2}. (i) Star operator $A_{s'}$ involving four constrained spins. (ii) Plaquette operator $B_{p'}$ involving four constrained spins and an effective free spin. (iii) Operators $\Sigma^z$, Eq. \ref{eq:Sigma} and $\Sigma^x$, Eq. \ref{eq:Sigmax}, acting on an effective free spin. (iv) Interactions between effective free spins on neighbouring plaquettes of $\mathcal{L}'$, controlled by the a constrained spin. }
\label{fig:Heff}
\end{figure}

Under our simplified choice of $W_{\text{num}}$, Hamiltonian $\tilde{H}$ is mapped into the effective Hamiltonian
\begin{equation}
	H' \equiv (W_{\text{num}})^{\dagger} \tilde{H} W_{\text{num}} = W^{\dagger} H_{\text{TC}}^{\text{x}} W, 
	\label{eq:Hprime}
\end{equation}
which again contains a number of terms, see Fig. \ref{fig:Heff}
\begin{eqnarray}
	H' &=& -J_{e}\sum_{s'\in \tilde{\mathcal{L}}} A_{s'} 
	- J_m \left( \sum_{p' \in \tilde{\mathcal{L}}}B_{p'} + \sum_{j \in \text{free}} \left( {\Sigma}^z_{j}+{\Sigma}^x_{j} \right) \right)	\nonumber 
\\&&-h_x K'_{\text{int}} .
	\label{eq:Hprime2}
\end{eqnarray}
In this expression operators $\Sigma^{z}$ for an effective free spin are obtained by coarse-graining the sum of the three $\sigma^z$'s acting on three free spins,
\begin{equation}
	\Sigma^{z} \equiv w^{\dagger} \left( \sigma^z_1 + \sigma^z_2 + \sigma^z_3 \right)w.
	\label{eq:Sigma}
\end{equation}
The operator $\Sigma^{x}$ for an effective free spin is obtained by coarse graining the sum of the interactions inside a block,
\begin{equation}
\Sigma^{x} \equiv w^{\dagger} \left( \sigma^x_1 \sigma^x_2 + \sigma^x_1 \sigma^x_3+ \sigma^x_2 +\sigma^x_3\right)w.
	\label{eq:Sigmax}
\end{equation}
  In both equations \ref{eq:Sigma} and \ref{eq:Sigmax} the numbering of the $\sigma^x$  and  $\sigma^z$ operators refers to the position of the free spin as reported in Fig. \ref{fig:Ltilde}.
The five-spin plaquette operator $B_{p'}$ is now the product of four $\sigma^z$'s corresponding to the constrained spins in plaquette $p'\in\mathcal{L}'$ and of $\Sigma^z$ corresponding to the effective free spin inside this plaquette. The term $-h_x K'_{\text{int}}$ collects interactions between effective free spins located in neighbouring plaquettes, which are controlled by a constrained spin. What is significant about $H'$ is that all operators are local (their support spans a small number of neighbouring spins). We will see that this property is robust under successive coarse-graining transformations.

\subsection{Exact preservation of the local symmetry}
\label{sec:coarse:preservation}

Let us discuss how the coarse-graining transformation $W$ preserves the local $Z_2$ symmetry. 

The local $Z_2$ symmetry acts on the original lattice $\mathcal{L}$ by means of the star operators $A_s$, see Sect. \ref{sec:DTC:symmetry}. As we have seen above, $W$ maps one fourth of these operators into star operators $A_{s'}$ acting on the effective lattice $\mathcal{L}'$, whereas the remaining three fourth of star operators $A_s$ have been mapped into operators $\sigma^x$ that act on spins that have been excluded from $\mathcal{L}'$. The roles played by $W_{\text{exact}}$ and $W_{\text{num}}$ in this process are markedly different. $W_{\text{exact}}$ is in charge of transforming the star operators and removing three fourth of them from the effective model. This is done analytically, so that no numerical errors are introduced, and with a transformation that is independent of the magnetic field $h_x$, so that the way in which the local $Z_2$ symmetry acts on $\mathcal{L}'$ is always the same. In contrast, $W_{\text{num}}$ coarse-grains the system numerically while avoiding to modify the way in which the local $Z_2$ symmetry acts. This is achieved by considering a transformation that commutes with all the star operators $A_{s'}$, and it implies that the local $Z_2$ symmetry will not be affected by the possible (numerical and truncation) errors that $W_{\text{num}}$ may introduce.

\subsection{Further coarse-graining}
\label{sec:coarse:further}

\begin{figure}[t]
  \includegraphics[width=8.5cm]{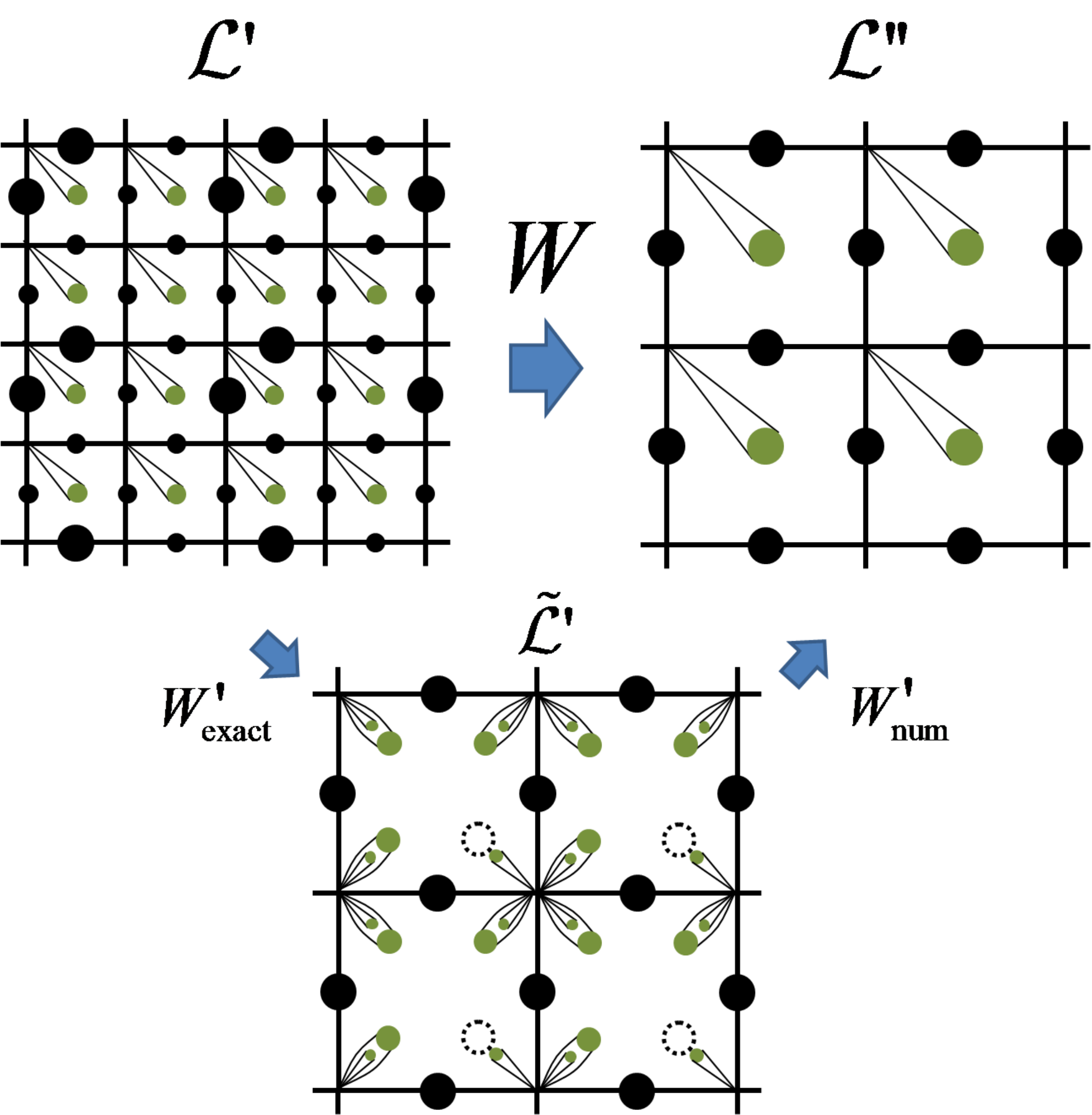}
\caption{
 The coarse-graining transformation $W'$ transforms a block of four plaquettes of the effective lattice $\mathcal{L}'$ into a single plaquette of a new effective lattice $\mathcal{L}''$. Notice that lattice $\mathcal{L}''$ has the same local composition of constrained and free spins as $\mathcal{L}'$. Transformation $W'$ again breaks into an exact transformation $W'_{\text{exact}}$, which produces an intermediate lattice $\tilde{\mathcal{L}'}$, and a numerical transformation $W_{\text{num}}$.}
\label{fig:WallSecond}
\end{figure}

\begin{figure}[t]
  \includegraphics[width=8.5cm]{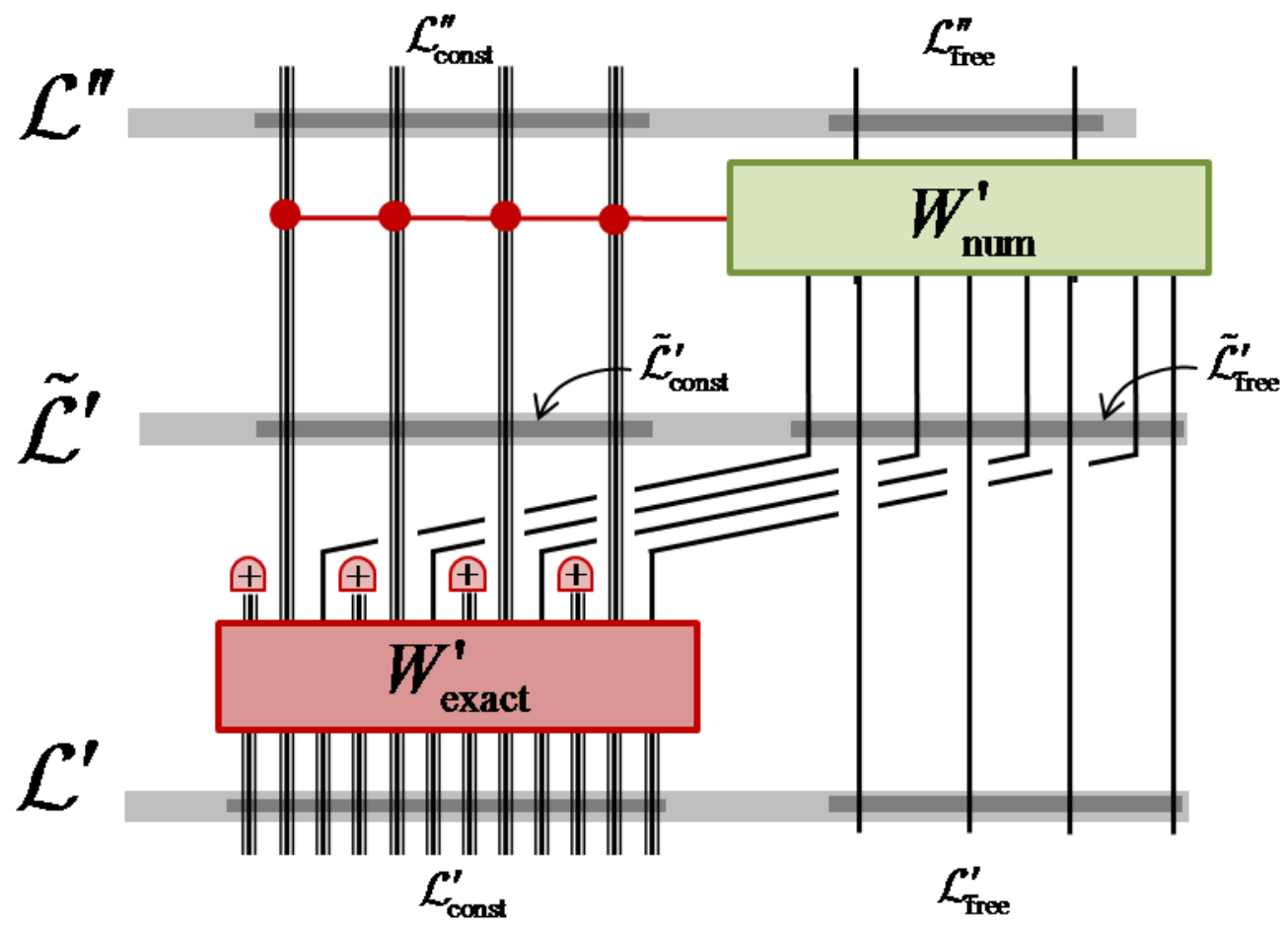}
\caption{
The coarse-graining transformation $W'$ is composed of an exact transformation $W_{\text{exact}}$ that acts only on constrained spins and a numerical transformation $W_{\text{num}}$ that coarse-graines the fresh free spins obtained from $W_{\text{exact}}$ together with the free spins in $\mathcal{L}'$ to produce the free spins of  $\mathcal{L}''$  while acting on the constrained spins diagonally in the $\sigma^x$ basis.}
\label{fig:SchemeSecond}
\end{figure}

\begin{figure}[t]
  \includegraphics[width=8cm]{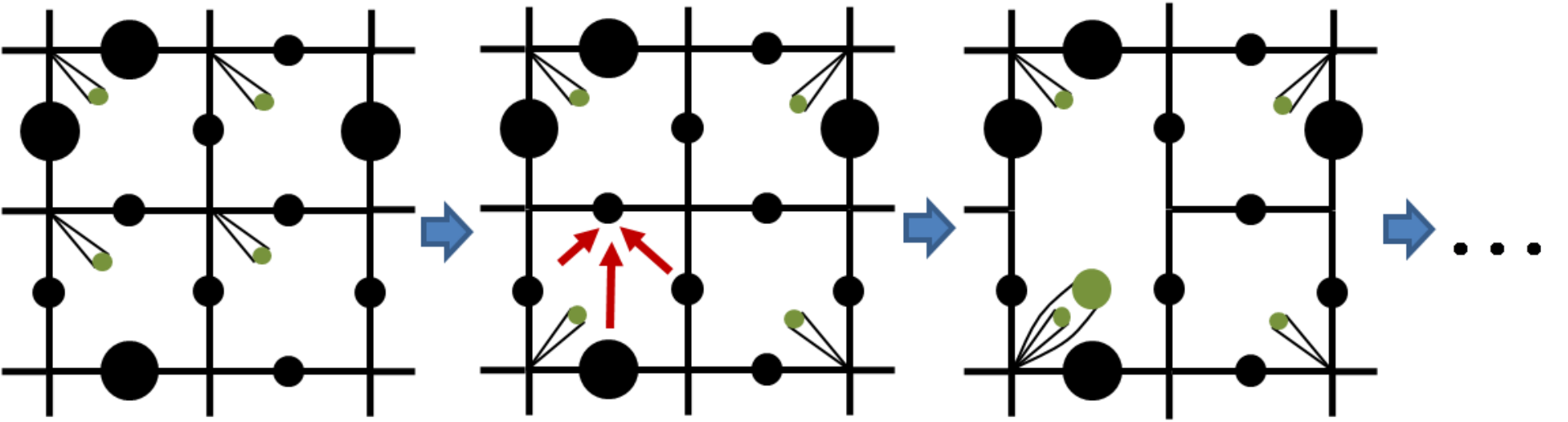}
\caption{
First round of the CNOT gates that constitute $W_{\text{exact}}$. Notice that the free spins are simply ignored by the CNOT gates. [In the first step, we have repositioned the free spins for notational convenience]. The rest of the sequence of CNOT gates proceeds as in Fig. \ref{fig:Wexact}. }
\label{fig:WexactSecond}
\end{figure}

\begin{figure}[t]
  \includegraphics[width=4cm]{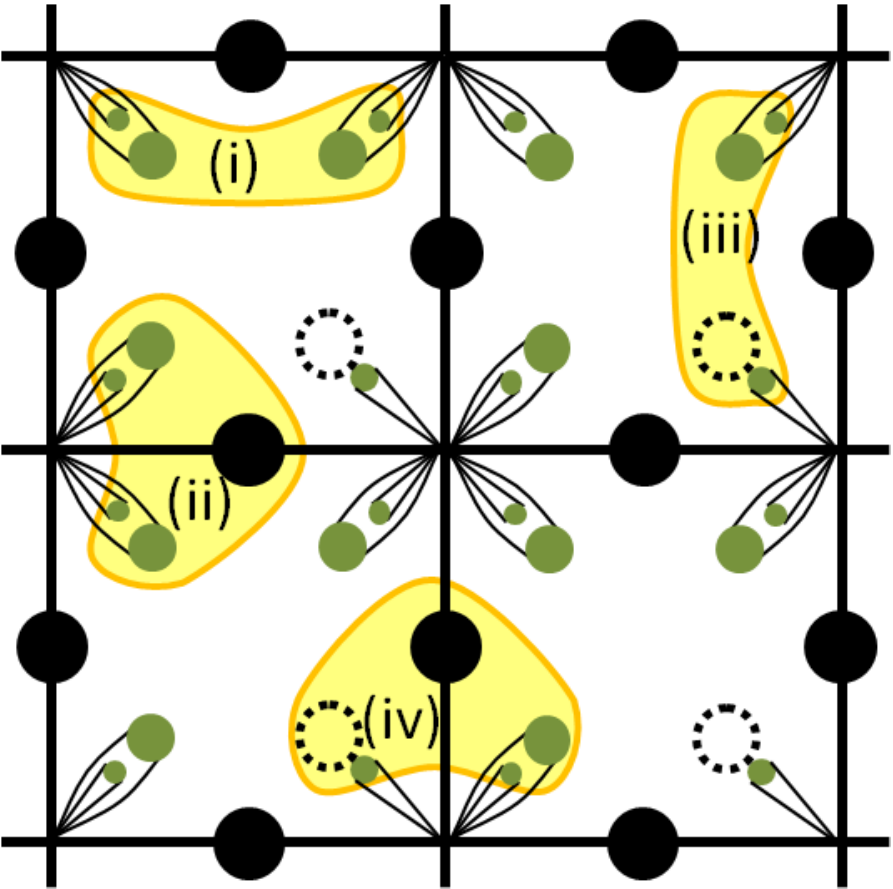}
\caption{
Examples of interactions contained in the term $(W'_{\text{exact}})^{\dagger}\left(-h_xK'_{\text{int}}\right)W'_{\text{exact}}$ of the Hamiltonian $(W'_{\text{exact}})^{\dagger} H' W'_{\text{exact}}$. (i) The interaction between two effective free spins is mediated by two fresh spins, on which it acts as $\sigma^x \otimes \sigma^x$. (ii) Analogous interaction across the boundary of a square plaquette, which involves a constrained spin. (iii)-(iv) In the presence of a vacancy, the interaction simply acts on one spin less.}
\label{fig:Wexact2Second}
\end{figure}

We can now apply a second coarse-graining transformation
\begin{equation}
	(W')^{\dagger}: \mathbb{V}_{\mathcal{L}'} \rightarrow \mathbb{V}_{\mathcal{L}''},
\end{equation}
which again breaks into an exact part $W'_{\text{exact}}$ and a numerical part $W'_{\text{num}}$, see Figs. \ref{fig:WallSecond} and \ref{fig:SchemeSecond},
\begin{equation}
	W' = W'_{\text{exact}}W'_{\text{num}}.
\end{equation}

Transformation $W'$ is very similar to transformation $W$, but it acts on a lattice made of both constrained and free spins, whereas $W$ acts on a lattice made only of constrained spins. The exact transformation $W'_{\text{exact}}$ acts on the constrained spins of $\mathcal{L'}$, and it does so by applying exactly the same sequence of CNOT gates and single-spin projections as $W_{\text{exact}}$ while simply ignoring the free spins, see Fig. \ref{fig:WexactSecond}, to produce both constrained spins and fresh free spins. Then $W'_{\text{num}}$ coarse-grains the free spins of $\mathcal{L}'$ and the freshly produced free spins while acting on constrained spins diagonally in the $\sigma^x$ basis. 

The simplified numerical transformation $W'_{\text{num}}$ considered in this work is made of isometries $w'$,
\begin{equation}
	(w')^{\dagger}: (\mathbb{C}_{\chi'})^{\otimes 4} \otimes (\mathbb{C}_{2})^{\otimes 3} \rightarrow \mathbb{C}_{\chi''}
\end{equation}
that map the free spins in the interior of a plaquette of $\mathcal{L}'$ (namely four effective free spins and three fresh free spins) into an effective free spin with vector dimension $\mathbb{C}_{\chi''}$, where again $\chi''$ is the refinement parameter of the transformation.

$W'$ maps local operators into local operators in very similar way as $W$ does. In particular, the effective Hamiltonian $H''$,
\begin{equation}
	H'' \equiv (W')^{\dagger} H' W',
\end{equation}
can be seen to contain terms analogous to those of $H'$. [To illustrate this point, Fig. \ref{fig:Wexact2Second} shows the Hamiltonian terms that are obtained in transforming $-h_xK'_{\text{int}}$ in $H'$ according to $W'_{\text{exact}}$, and which will produce $-h_xK''_{\text{int}}$.] With these observations, we can conclude that $W'$ also preserves the locality of operators and the local $Z_2$ symmetry, and that both properties will also be preserved under analogous subsequent coarse-graining transformations.

Notice in Fig. \ref{fig:WallSecond} that lattices $\mathcal{L}'$ and $\mathcal{L}''$ display an identical local pattern of constrained and free spins. Indeed, they both can be decomposed as the union of a sublattice for constrained spins and a square sublattice for free spins, see Fig. \ref{fig:Lprime}. Therefore all following coarse graining transformations $W'', W''', \cdots$ have the same structure of  $W'$. 

\begin{figure}[t]
  \includegraphics[width=8.5cm]{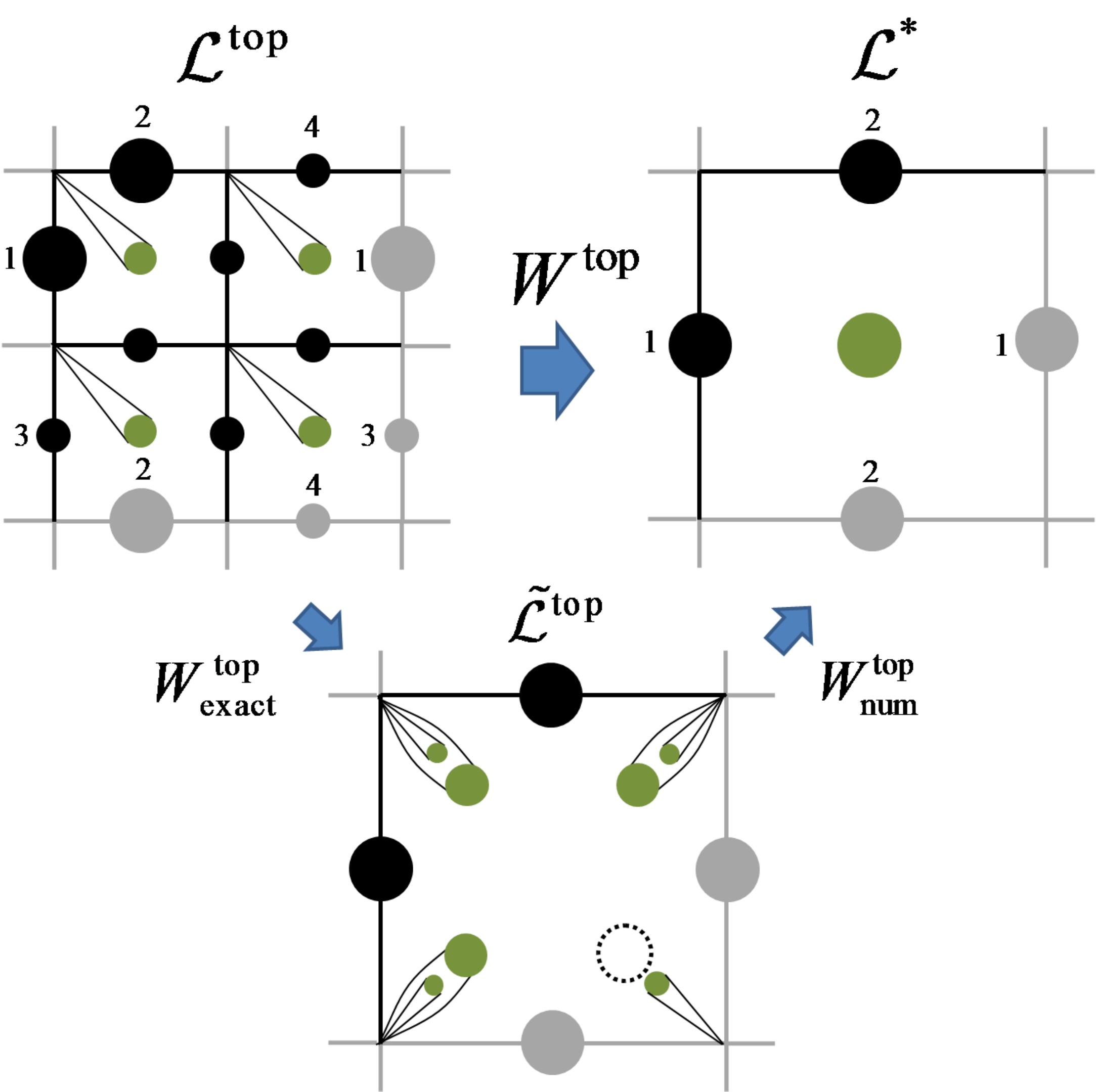}
\caption{
 Graphical representation of $\mathcal{L}^{\text{top}}$, the intermediate lattice $\tilde{\mathcal{L}}^{\text{top}}$, and the final lattice $\mathcal{L}^*$ made of just two topological spins and a free spin. Notice the toric boundary conditions.}
\label{fig:WallTop}
\end{figure}

This allow us to consider  a larger sequence of coarse-graining transformations $\{W, W', W'', \cdots \}$. Each coarse-graining transformation halves the linear size of the lattice, reducing the number of constrained spins to one fourth and, starting with $W'$, reducing the number of free spins also by one forth. Suppose that the initial lattice has $L\times L$ sites (or $2L^2$ constrained spins) with $L=2^{K}$. After $(\log_2L)-1 = K-1$ layers of coarse-graining, producing a sequence $\{\mathcal{L}, \mathcal{L}' \mathcal{L}'', \cdots \mathcal{L}^{\text{top}}\}$, the original lattice has been reduced to a small lattice $\mathcal{L}^{\text{top}}$, referred to as \textit{top} lattice, with $2\times 2$ sites or, equivalently, eight constrained spins and four free spins, see Fig. \ref{fig:WallTop}.

\begin{figure}[t]
  \includegraphics[width=8.5cm]{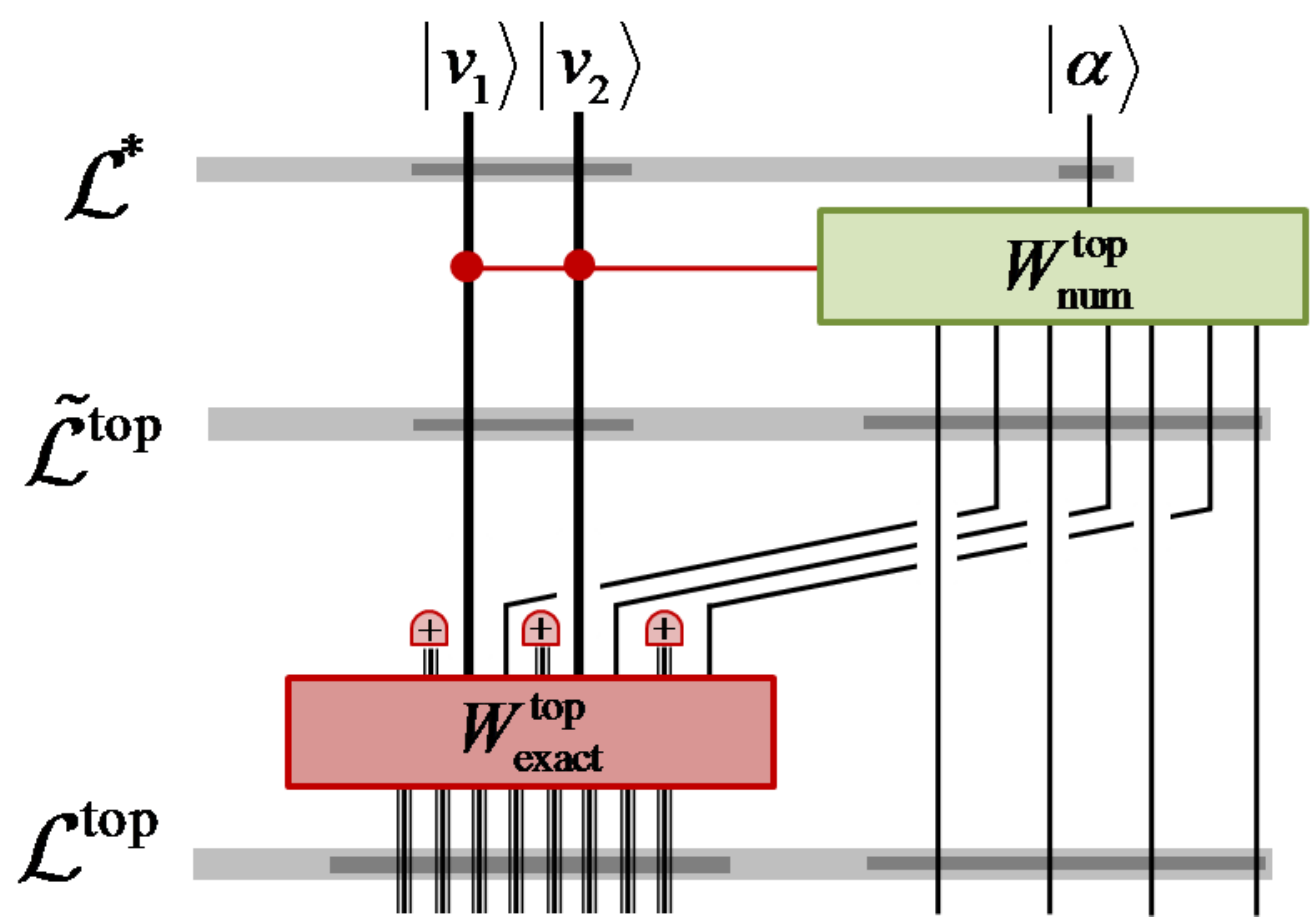}
\caption{
Transformation $W^{\text{top}}_{\text{exact}}$ produces two topological spins and three fresh new free spins. Transformation $W^{\text{top}}_{\text{num}}$ coarse-grains seven free spins into a single spin, while acting diagonally in a $\sigma^x$ basis on the two topological spins. Topological sector $(v_1,v_2)$ can be specified by setting the topological spins in the state $\ket{v_1}\otimes \ket{v_2}$. Index $\alpha$ labels an orthonormal basis $\ket{\alpha} \in \mathbb{C}_{\chi}$ of states of the final free spin. Each choice of topological sector $(v_1,v_2)$ and of state $\ket{\alpha}$ labels one of $4 \chi$ orthonormal states $\ket{\Psi^{\alpha}_{(v_1,v_2)}} \in \mathbb{V}_{\mathcal{L}}$ of the original lattice $\mathcal{L}$, which can be reconstructed by undoing the sequence of coarse-graining transformations.}
\label{fig:SchemeTop}
\end{figure}
 
\begin{figure}[t]
  \includegraphics[width=8.5cm]{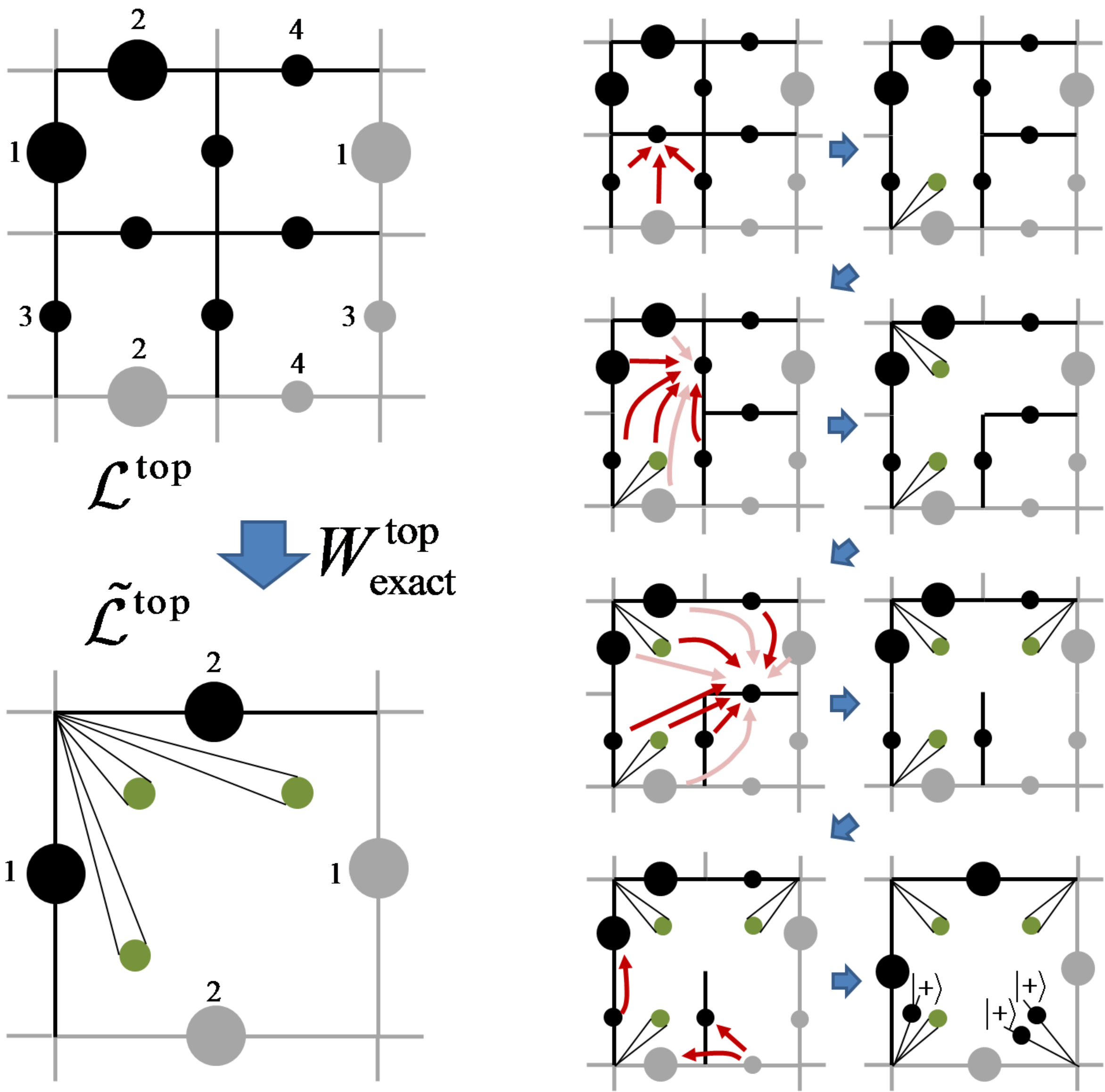}
\caption{
Lattices $\mathcal{L}^{\text{top}}$ and $\tilde{\mathcal{L}}^{\text{top}}$, with toric boundary conditions, and sequence of CNOT gates and single-spin projections included in $W^{\text{top}}_{\text{exact}}$. For simplicity, the four free spins of $\mathcal{L}^{\text{top}}$ have not been depicted. Notice that due to the boundary conditions, in occasions two CNOT gates (denoted in lighter red) are applied twice on the same control and target spins, in which case the do not need to be applied, since the square of a CNOT gate is the identity operator.}
\label{fig:WexactTop}
\end{figure}

\subsection{Top coarse-graining transformation $W^{\text{top}}$ and topological degrees of freedom}
\label{sec:coarse:top}
 
The top lattice $\mathcal{L}^{\text{top}}$ is mapped into a set of just three spins, collectively denoted $\mathcal{L}^{*}$, by a coarse-graining transformation $W^{\text{top}}$, see Figs. \ref{fig:WallTop} and \ref{fig:SchemeTop},
\begin{equation}
	(W^{\text{top}})^{\dagger}: \mathbb{V}_{\mathcal{L}^{\text{top}}} \rightarrow \mathbb{V}_{\mathcal{L}^{*}},
\end{equation}
that decomposes as usual into exact and numerical parts, $W^{\text{top}} = W^{\text{top}} _{\text{exact}} W^{\text{top}} _{\text{num}}$.
 
Transformation $W^{\text{top}}$ needs to be described explicitly, as it differs from previous transformations in a few aspects. Let us start with $W^{\text{top}}_{\text{exact}}$. The unit cell of four plaquettes on which the sequence of CNOT gates in Fig. \ref{fig:Wexact} was defined amounts now to the whole lattice $\mathcal{L}^{\text{top}}$. As shown in Fig. \ref{fig:WexactTop}, this implies that some CNOT gates are applied twice on the same control and target spins. Since a CNOT gate squares to the identity, such gates do not need to be applied. Another important difference is that in this case the sequence of CNOT gates does not produce any constrained spins. Indeed, the two spins that would otherwise be constrained (denoted as 1 and 2 in Fig. \ref{fig:WexactTop}) are now free. This is due to the fact that toric boundary conditions imply that a star operator $A_s$ acting on these spins acts twice on each of them, and since the square of operator $(\sigma^x)$ is the identity operator, the symmetry acts trivially. These two spins are very special. 
One can check that the string operators $X_1$ and $X_2$ of Eq. \ref{eq:X} acting on $\mathcal{L}$ are mapped, under coarse-graining, into operators $\sigma^{x}_1$ and $\sigma^x_2$ acting on these two spins. In other words, the state of these spins determines in which topological sector the model is. For this reason, instead of calling these spins free, we refer to them as \textit{topological} spins. Incidentally, on a state $\ket{\xi}$ invariant under the local $Z_2$ symmetry, Eq. \ref{eq:star}, operators $X_1$ and $X_2$ are equivalent to any operator obtained by multiplying them by star operators $A_s$. For instance, $A_sX_1$ is equivalent to $X_1$,
\begin{equation}
 (A_sX_1)\ket{\xi} = X_1A_s\ket{\xi} = X_1 \ket{\xi}, 
\end{equation}
where we have used that $X_1$ commutes with $A_s$ and Eq. \ref{eq:star}. Thus, by multiplication by star operators $A_s$ allows us to locally deform the original support of $X_1$ and $X_2$ (namely the non-contractible cuts $c_1$ and $c_2$ in Fig. \ref{fig:ABcc}) without changing the action of these operators. In particular, operator $A_sX_1$ above (and any such products) will still be coarse-grained by the present scheme into $\sigma^x_1$, and the same holds for deformations of the support of $X_2$. On the other hand, the operators $Z_1$ and $Z_2$ of Eq. \ref{eq:Z} acting on $\mathcal{L}$ are mapped, under the present coarse-graining, into operators $\sigma^{z}_1$ and $\sigma^{z}_2$ acting on these two spins. [Notice, however, that except in very special circumstances (e.g. when representing ground states of $H_{\text{TC}}^{\text{x}}$ for $h_x=0$, which also fulfill Eq. \ref{eq:plaquette}) multiplying $Z_1$ and $Z_2$ by plaquette operators $B_p$ will not produce an operator that will still be mapped into $\sigma^z_1$ and $\sigma^z_2$. In other words, in our scheme, the support of $X_1$ and $X_2$ can be locally deformed without changing the properties of these operators, whereas this is not true of $Z_1$ and $Z_2$, which must be supported on the specific non-contractible loops $l_1$ and $l_2$ shown in Fig. \ref{fig:ABcc} in order to be mapped to $\sigma^z$ operators on the topological spins.]

The role of transformation $W^{\text{top}}_{\text{num}}$ is to coarse-grain all free spins in $\tilde{\mathcal{L}}^{\text{top}}$ into a single free spin described by a vector space of dimension $\chi^*$, while acting on the two topological spins diagonally in the $\sigma^x$ basis. For each of four possible choices of the topological charges $(v_1,v_2)$, $W^{\text{top}}_{\text{num}}$ is characterized by an isometry $w_{v_1,v_2}$,
\begin{equation}
	\left(w_{v_1,v_2}^{\text{top}}\right)^{\dagger}: \left( \mathbb{C}_{\chi^{\text{top}}} \right)^{\otimes 4} \otimes \left( \mathbb{C}_2 \right)^{\otimes 3} \rightarrow \mathbb{C}_{\chi^*},
\end{equation}

This completes our description of the coarse-graining transformations $\{ W, W', W'', \cdots, W^{\text{top}} \}$. By composition, they reduce the original lattice $\mathcal{L}$, made of $2L^2$ constrained spins with vector space $(\mathbb{C}_2)^{\otimes 2L^2}$, to just two topological spins and one free spin with vector space
\begin{equation}
	\mathbb{V}_{\mathcal{L}^*} \cong \mathbb{C}_2 \otimes \mathbb{C}_2 \otimes \mathbb{C}_{\chi^*}.
\end{equation}

\subsection{Variational ansatz}
\label{sec:coarse:variational}

Let $\{\ket{\alpha}\}$, $\alpha=1,\cdots,\chi$, denote an orthonormal basis of $\mathbb{C}_{\chi^{*}}$. Then from states of $\mathcal{L}^*$ of the form
\begin{equation}
	\ket{v_1} \otimes \ket{v_2} \otimes \ket{\alpha} \in \mathbb{V}_{\mathcal{L}^*}
\end{equation} 
we can obtain a set of $4\chi^{*}$ orthonormal states $\ket{\Phi^{\alpha}_{v_1,v_2}} \in \mathbb{V}_{\mathcal{L}}$ of the original lattice $\mathcal{L}$ by undoing the sequence of coarse-graining transformation,
\begin{equation}
	\ket{\Phi^{\alpha}_{v_1,v_2}} \equiv WW'W'' \cdots W^{\text{top}} \ket{v_1} \otimes \ket{v_2} \otimes \ket{\alpha}. 
\end{equation}

As it is costumary within MERA algorithms (see Sect. \ref{sec:ER:MERA}), we use the set of isometric transformations $\{ W, W', W'', \cdots, W^{\text{top}} \}$ to define a variational ansatz for a $4\chi$-dimensional subspace $\mathbb{V}^{*}$ of $\mathbb{V}_{\mathcal{L}}$. The variational parameters, denoted $\vec{a}$, are encoded in the isometric tensors (namely disentanglers and isometries) that form the numerical transformations $\{ W_{\text{num}}, W'_{\text{num}}, W''_{\text{num}}, \cdots, W^{\text{top}}_{\text{num}} \}$. 
 
In order to obtain an approximation of the ground state(s) and lowest energy excited state of Hamiltonian $H^{\text{x}}_{\text{TC}}$ with this ansatz, the variational parameters $\vec{a}$ are optimized so as to minimize the expectation value $E_{\vec{a}}$ of Hamiltonian $H^{\text{x}}_{\text{TC}}$ on the subspace $\mathbb{V}^{*} \subset \mathbb{V}_{\mathcal{L}}$, namely
\begin{equation}
	\min_{\vec{a}} E_{\vec{a}},~~~~~~~~E_{\vec{a}} \equiv \tr\left( H^{\text{x}}_{\text{TC}} P_{\vec{a}}\right), 
\end{equation}
where $P_{\vec{a}}$ is a projector on subspace $\mathbb{V}^{*}$. This optimization can be performed using the optimization techniques extensively described in Refs. \onlinecite{Evenbly2009Alg, Tagliacozzo2009}.  

The ansatz used in this work is a hybrid between a MERA and a TTN. The exact transformation $W_{\text{exact}}$ contains disentanglers (sequence of CNOT gates) and isometries (single-spin projections) and therefore it conforms to the definition of MERA. However, the simplified choice of numerical transformation $W_{\text{num}}$ only has isometries, corresponding to a TTN. In the next section we refer to the ansatz as the \textit{hybrid} ansatz, to be able to distinguish it from the proper MERA used in Ref. \onlinecite{Tagliacozzo2010prep} to recover the RG fixed points.

\section{BENCHMARK RESULTS}
\label{sec:bench}

In this section we present the results of computations used to benchmark the performance of the present approach. We have used the hybrid ansatz introduced in the previous section to obtain an approximation to the ground state(s) and several excited states of the Hamiltonian $H^{\text{x}}_{\text{TC}}$ for the deformed toric code model, for lattices $\mathcal{L}$ of linear sizes $L = 4,6,8$ and $16$. 

The hybrid ansatz offers an explicit representation of the wave function of the system, from which it is possible to evaluate a number of quantities of interest. These include the expectation value of arbitrary local observables, such as the energy, but also of non-local observables, such as the disorder parameter $\langle X_3 \rangle$ and Wilson loops. In addition, it is possible to compute the overlap between different wave functions, leading to alternative tools to characterize the ground state phase diagram of the model.

For the system sizes $L=4,6,8$, it appears that the optimal choice of the numerical part of the coarse-graining is a simple TTN made of one layer with two isometries each mapping free spins into one free spin of dimension $\chi^{\text{top}}$, followed by the top numerical isometry $W^{\text{top}}_{\text{num}}$. The refinement parameter $\chi^{\text{top}}$, from now on denoted simply as $\chi$, dominates the computational cost, which scales as $O(\chi^4)$, see Ref. \onlinecite{Tagliacozzo2009}. Energy minimization proceeds until no change is observed in the first ten digits after ten optimization steps. As an example of the required computational effort, obtaining the ground state and first excited state within the topological sector $(+,+)$ for $L=4,6$ and  $8$ took, on a desktop computer with 2 processors at 2Ghz,  with 8Gb of Ram, a total of 1, 5 and 8 hours. {Unless stated otherwise all the results we present here are obtained by fixing the value of $\chi$ to $\chi=100$. In order to determine this value, we have performed a scaling analysis and found that this value is enough to produce a consistent approximation in both the topological and the spin polarized phase on the  $8 \times 8$ torus. The scaling analysis is reported in the appendix \ref{app:precision}}.

Fig. \ref{fig:Bare} illustrates one advantage the hybrid ansatz has over a TTN (resulting from a bare coarse-graining transformation, see Sect. \ref{sec:coarse:motivation}). In the example, given the same computational cost, the hybrid ansatz leads to 4 more significant figures of accuracy for the ground state energy in the deconfined phase of $H_{\text{TC}}^{\text{x}}$, which is robustly entangled, see also Fig. \ref{fig:Entanglement}.

\begin{figure}[t]
  \includegraphics[width=7.5cm]{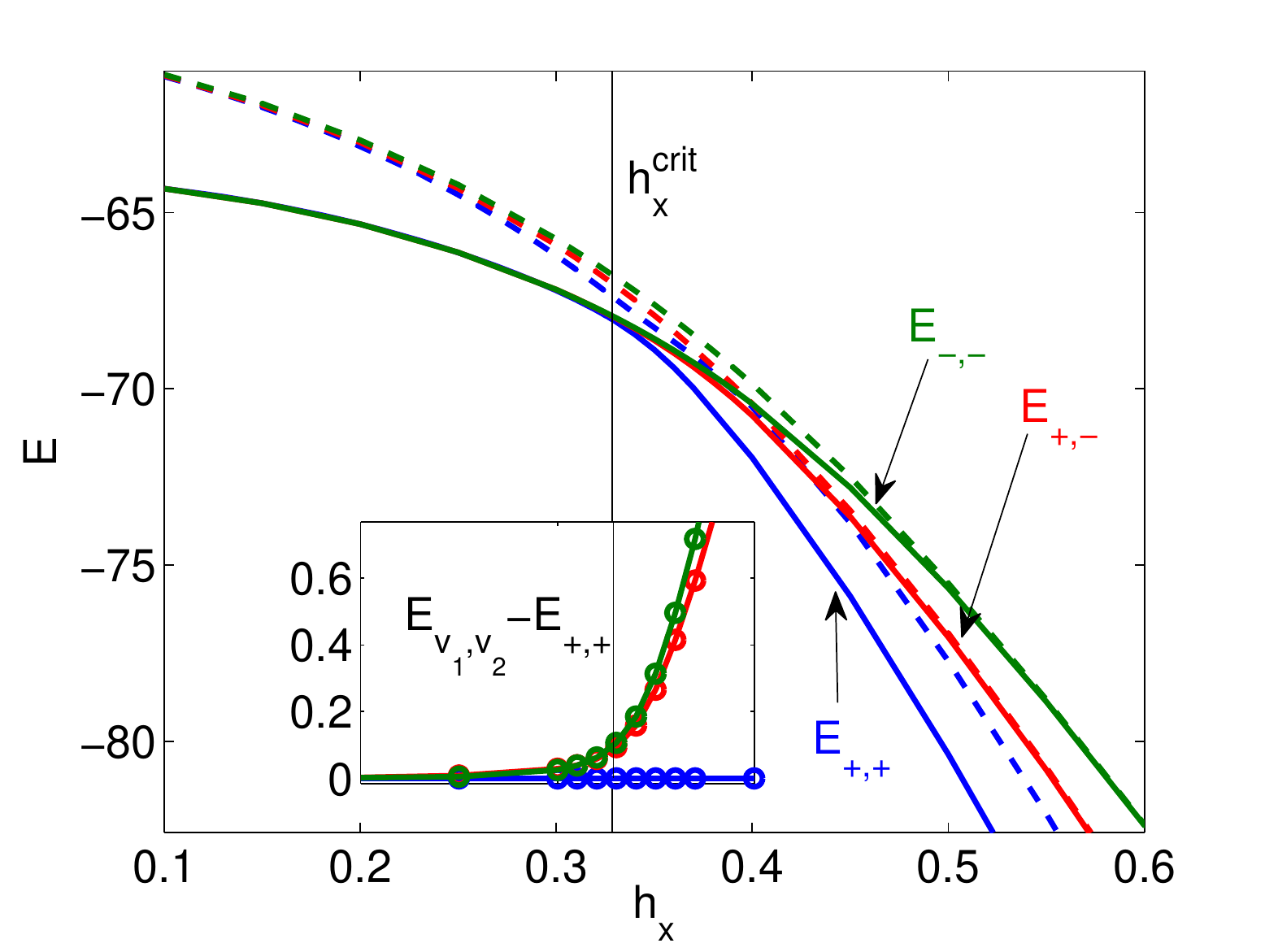}
\caption{
Lowest eigenvalues of $H^{\text{x}}_{\text{TC}}$ as a function of the magnetic field $h_x$, for a system of linear size $L=8$, corresponding to $8^2\times 2 = 128$ spins. The two lowest energies in each topological sector $(v_1,v_2)$ are plotted. {The lowest energy in each sector is represented  with a solid line while the second lowest with a dashed line. Different colours refer to different sectors. The energies in sector (-,+) are identical to the energies in sector (+,-).} The inset shows the gap between the energy of the ground states in sector $(v_1,v_2)$ and in the sector  $(+,+)$. These gaps are very small for values of the magnetic field $h_x$ smaller than some critical value $h_{x}^{\text{crit}} = 0.3285$\cite{Hamer2000} (denoted by a vertical line) and much larger for larger magnetic field $h_x$, see Fig. \ref{fig:TopoGaps}. These results were obtained with the hybrid ansatz with $\chi = 100$.}
\label{fig:Spectrum}
\end{figure}

\begin{figure}[t]
  \includegraphics[width=8.5cm]{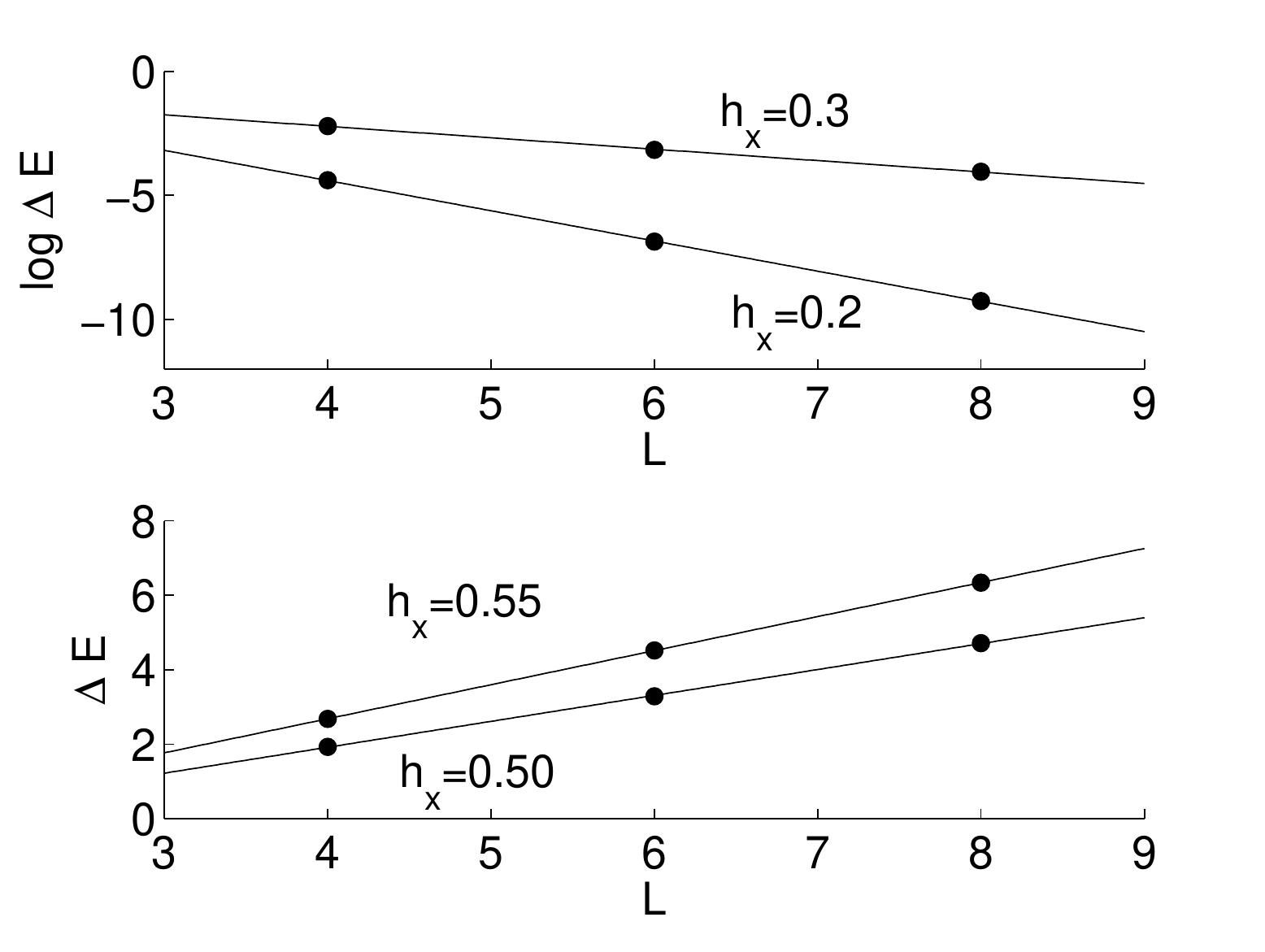}
\caption{
Upper panel: Energy gap $\Delta E$ between the ground states of the topological sectors $(+,-)$ and $(+,+)$ (see Eq. \ref{eq:DeltaE}) as a function of the linear size $L$ for $h_x = 0.2$ and $0.3$, corresponding to the deconfined phase of the deformed toric code model. In accordance with Eq. \ref{eq:Delta}, $\Delta E$ decays exponentially with $L$. These results were obtained with the hybrid ansatz with $\chi = 100$.
Lower panel: Energy gap $\Delta E$ between the ground states of the topological sectors $(+,-)$ and $(+,+)$ (see Eq. \ref{eq:DeltaE}) as a function of the linear size $L$ for $h_x = 0.5$ and $0.55$, corresponding to the spin polarized phase of the deformed toric code model. In accordance with Eq. \ref{eq:Delta2}, $\Delta E$ grows linearly with $L$. These results were obtained with the hybrid ansatz with $\chi = 100$.}
\label{fig:TopoGaps}
\end{figure}

\begin{figure}[t]
  \includegraphics[width=7.5cm]{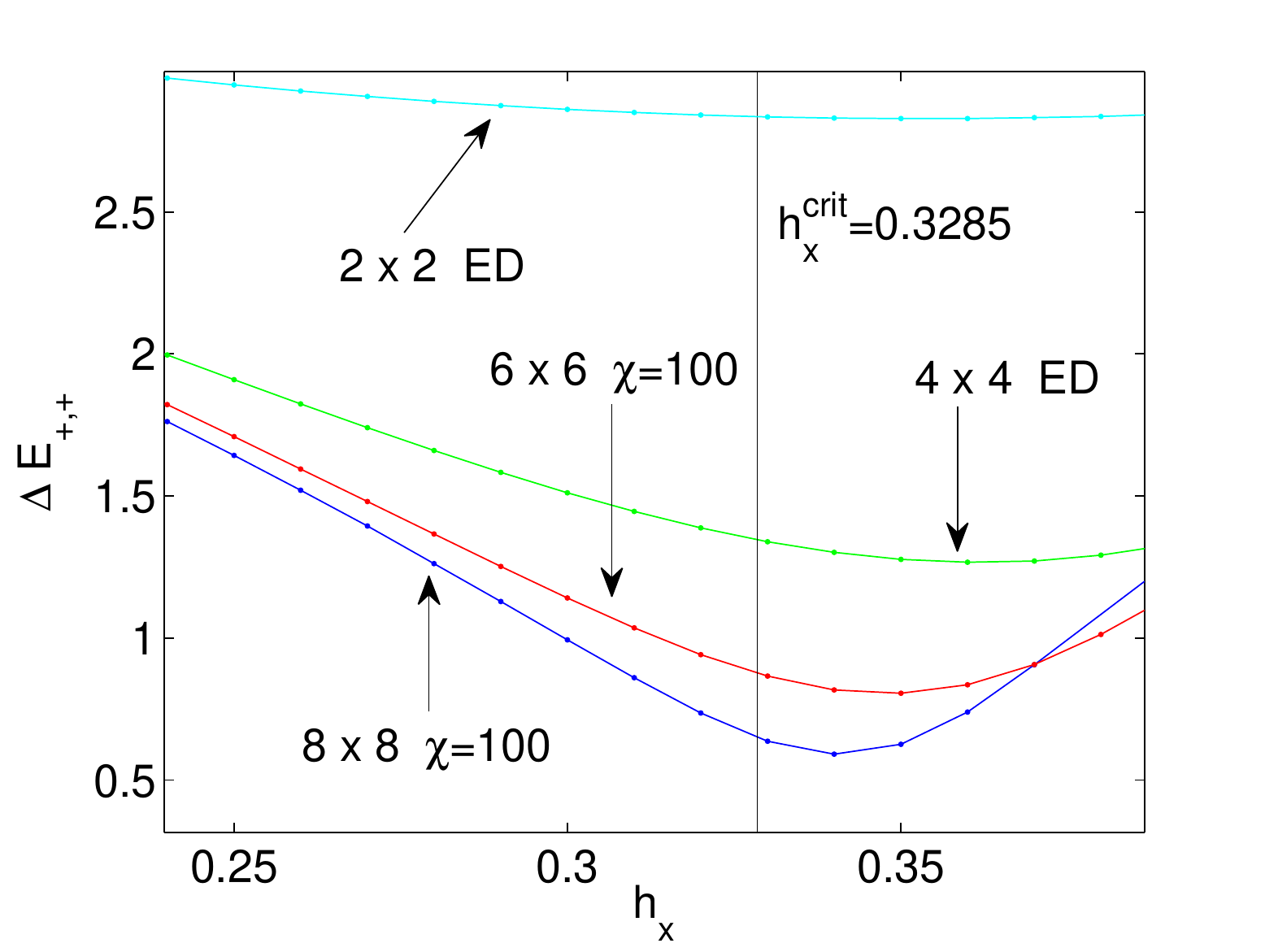}
  \includegraphics[width=7.5cm]{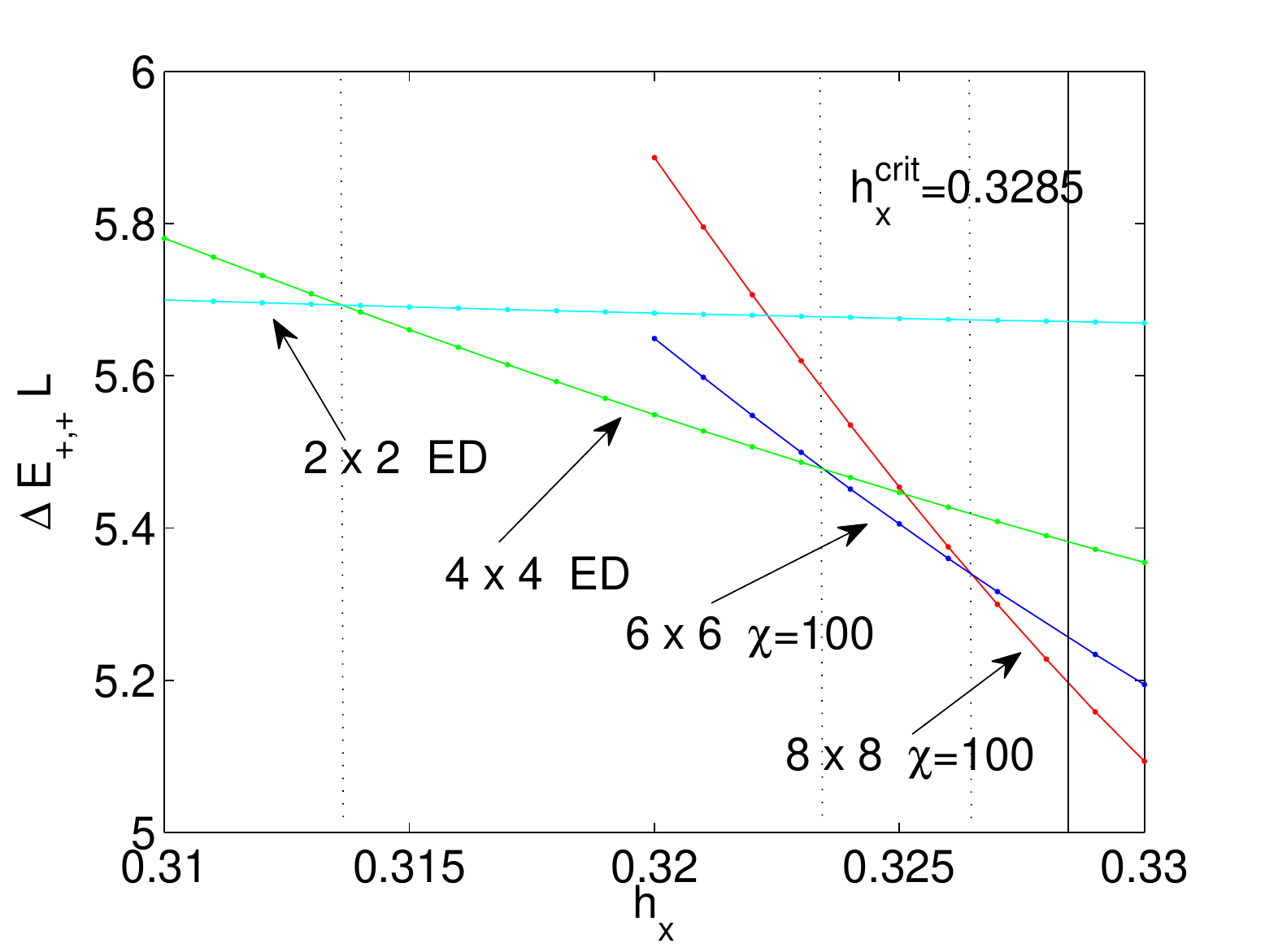}
\caption{
Upper panel: Gap $\Delta E_{+,+}$ between the ground state and first excited state of the topological sector $(+,+)$, as a function of the magnetic field $h_x$, and for system sizes $L=4,6$ and $8$.
Lower panel: Product of the systems size $L$ times the energy gap $\Delta E_{+,+}$, plotted as a function of the magnetic field $h_x$ (compare with Fig. \ref{fig:GapPlusPlus}). An estimate of $0.3267(5)$ for the critical magnetic field $h_{x}^{\text{crit}} = 0.3285(1)$\cite{Hamer2000} (denoted by a solid vertical line) is obtained by extrapolating, using the phenomenological renormalization group \cite{nightingale_scaling_1975}, the sequence of  intersections among curves obtained from systems with consecutive sizes ($2-4$, $4-6$ and $6-8$) identified by dotted vertical lines in the plot using the known $L$ dependence for the sequence \cite{Hamer2000}.
}
\label{fig:GapPlusPlus}
\end{figure}

\subsection{Local observables}
\label{sec:bench:local}

\subsubsection{Low energy spectrum} 

Fig. \ref{fig:Spectrum} shows the low energy spectrum of $H^{\text{x}}_{\text{TC}}$ as a function of the magnetic field $h_x$. It includes the energy of the ground state and first excited state of each of the four topological sectors $(v_1,v_2)$. As expected, for small $h_x$ the ground state energies in different topological sectors are very similar, whereas they depart from each other for values of the magnetic fields larger than $h_x \approx 0.3$, in which case the ground state in the sector $(+,+)$ becomes the global ground state. 

The low energy spectrum of $H^{\text{x}}_{\text{TC}}$ can be obtained by minimizing the ansatz in either the whole vector space or within each specific topological sector $(v_1,v_2)$, which is achieved by fixing the topological spins to state $\ket{v_1} \otimes \ket{v_2}$. Since the second option needs to deal with less low energy states at a time, it is generally more economical. In addition, in the spin polarized phase, where ground states in different topological sectors have markedly different energies, restricting the minimization to a single topological sector, e.g. $(+,-)$, is important in order to get an accurate representation of its ground state. Indeed, for large $h_x$ the ground state $\ket{\Phi_{+,-}}$ is no longer a low energy state of $H^{\text{x}}_{\text{TC}}$ when the whole vector space is taken into account (as shown next, $\ket{\Phi_{(+,+)}}$ has $O(L)$ lower energy), and will not be properly captured through an unrestricted energy minimization.

\subsubsection{Gaps between topological sectors}

Fig. \ref{fig:TopoGaps} studies the gap 
\begin{equation}
	\Delta E \equiv E_{+,-} - E_{+,+}
	\label{eq:DeltaE}
\end{equation}
between the ground states of the topological sectors $(+,-)$ and $(+,+)$ as a function of the system size. We obtain a clear confirmation of the exponential decay anticipated for the deconfined phase, see Eq. \ref{eq:Delta}, as well as of the linear growth of the gap with system size in the spin polarized phase, see Eq. \ref{eq:Delta2}. 

\subsubsection{Gap within the topological sector ($+,+$)}

Recall that the ground state $\ket{\Phi_{+,+}}$ is the absolute ground state of $H^{\text{x}}_{\text{TC}}$ for an arbitrary value of the magnetic field $h_x$.
Fig. \ref{fig:GapPlusPlus} shows the energy gap $\Delta E_{+,+}$ within the topological sector $(+,+)$, as a function of $h_x$, and for $L=4,6,8$. Notice that the minimum of $\Delta E_{+,+}$ as a function of $h_x$ closes roughly as $1/L$. This minimum occurs for a value of $h_x$ that diminishes with $L$ and a rough estimate of the critical value $h_{x}^{\text{crit}}$ in the thermodynamic limit can be obtained by a large $L$ extrapolation. However, a much more accurate estimate of $h_{x}^{\text{crit}}$ is obtained by drawing the curves of $L \Delta E_{+,+}$ for increasing values of $L$ in the range $L=2,4,6, 8$. The position of intersections between those curves obtained at consecutive $L$ (i.e $L=2$ with $L=4$, $L=4$ with $L=6$ and $L=6$ with $L=8$) produces a sequence of $\tilde{h}_x^{crit}(L)$ that following the seminal work of Nightingale  in Ref. \onlinecite{nightingale_scaling_1975} is expected to converge to the location of the fixed point as $\tilde{h}_x^{crit}(L)={h}_x^{crit}-A/L^4$, see Ref. \onlinecite{Hamer2000}. By fitting this expression to the sequence of intersections reported in the lower panel of Fig. \ref{fig:GapPlusPlus} we obtain an estimate of the critical point ${h}_x^{crit}=0.3267(5)$. This value is off the exact value $0.3285$ \cite{Hamer2000} by  only  $0.5 \%$. The systematic  error induced by limiting $\chi$ to $\chi=100$ indeed pushes the transition towards smaller $h_x$ since it limits the amount of entanglement in the ground state, however its effects are surprisingly small.
 
\subsection{Non-local observables}
\label{sec:bench:nonlocal}

\begin{figure}[t]
  \includegraphics[width=8.5cm]{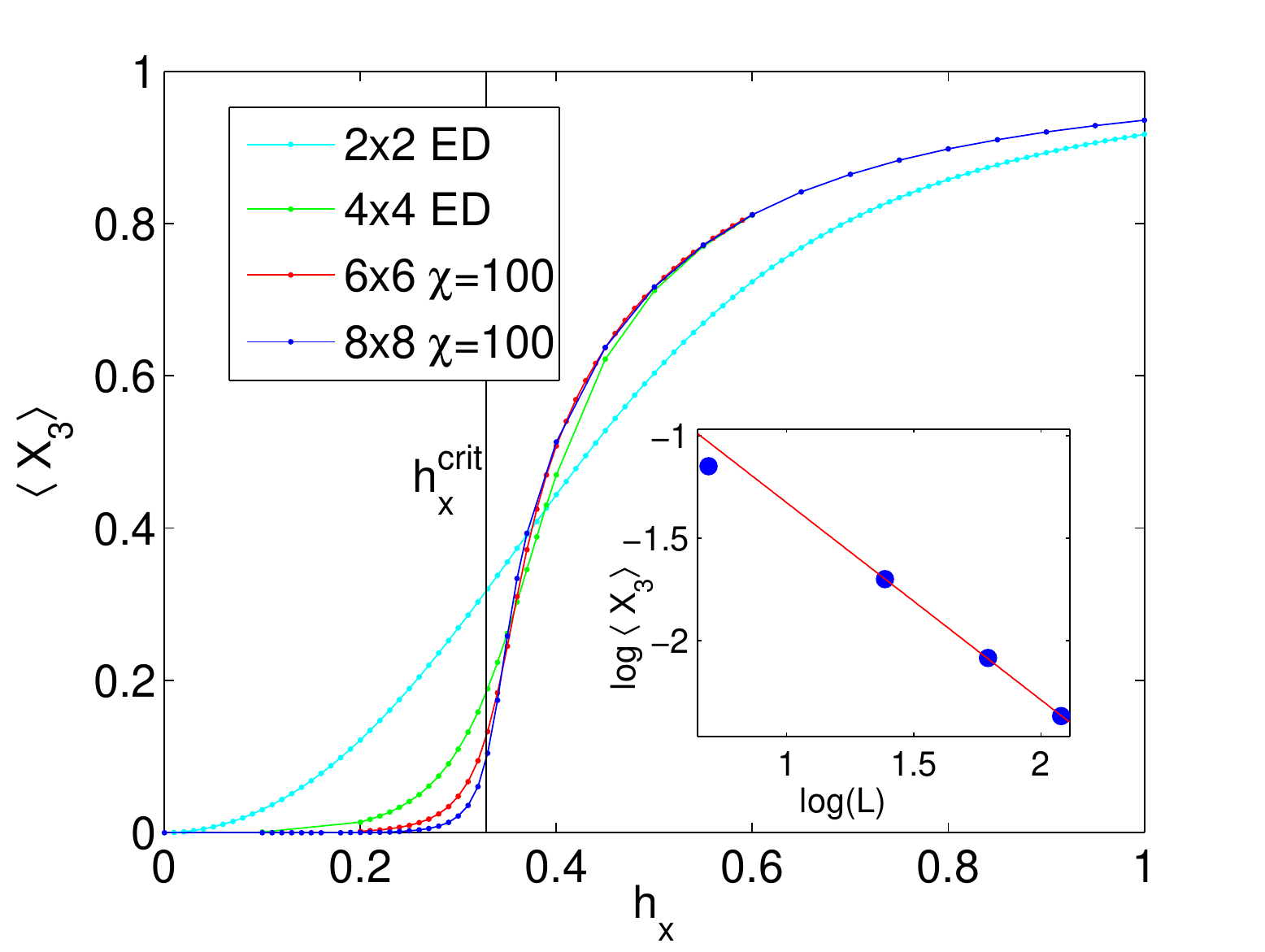}
\caption{
Disorder parameter $\langle X_3 \rangle$ as a function of the magnetic field $h_x$, for several lattice sizes. The critical magnetic field $h_{x}^{\text{crit}} = 0.3285(1)$\cite{Hamer2000} is denoted by a vertical line. The inset shows a log-log plot of $\langle X_3 \rangle$ at $h_x = h_x^{\text{crit}}$ as a function of $L$, from which we obtain an estimate of $0.96(5)$ for $2\beta/\nu$, whose current estimates are around $1.05$\cite{Pelissetto2002}.}
\label{fig:Mono}
\end{figure}

\subsubsection{Disorder parameter}

Fig. \ref{fig:Mono} shows the disorder parameter $\langle X_3 \rangle$, introduced in see Sect. \ref{sec:DTC:DTC}, which measures the formation of a condensate of magnetic vortices (or magnetic monopoles), as a function of the magnetic field $h_x$. Results for different system sizes $L=2,4,6$ and $8$ are presented. $\langle X_3 \rangle$ can clearly be used to distinguish the deconfined phase, where it vanishes, from the spin polarized phase. Notice that the drop of $\langle X_3 \rangle$ to zero for diminishing magnetic field $h_x$ near the critical point becomes sharper with increasing system size $L$. For $L=8$, we obtain an estimate $h_x^{\text{crit}} = 0.327\pm0.05$ for the critical magnetic field and $\beta=0.33\pm 0.02$ for the critical exponent $\beta$ that should be compared with the best estimates coming from Monte Carlo simulations $\beta=0.32652(15)$ \cite{Pelissetto2002}. It is important to notice that a direct measurement of this operator in Monte Carlo simulations is very challenging. Only recently, in the context of $U(1)$ LGT,   an algorithm has been proposed to directly compute these kind of operators \cite{D'Elia:2006vg}. 

\subsubsection{Wilson loops}
\begin{figure}[t]
  \includegraphics[width=8.5cm]{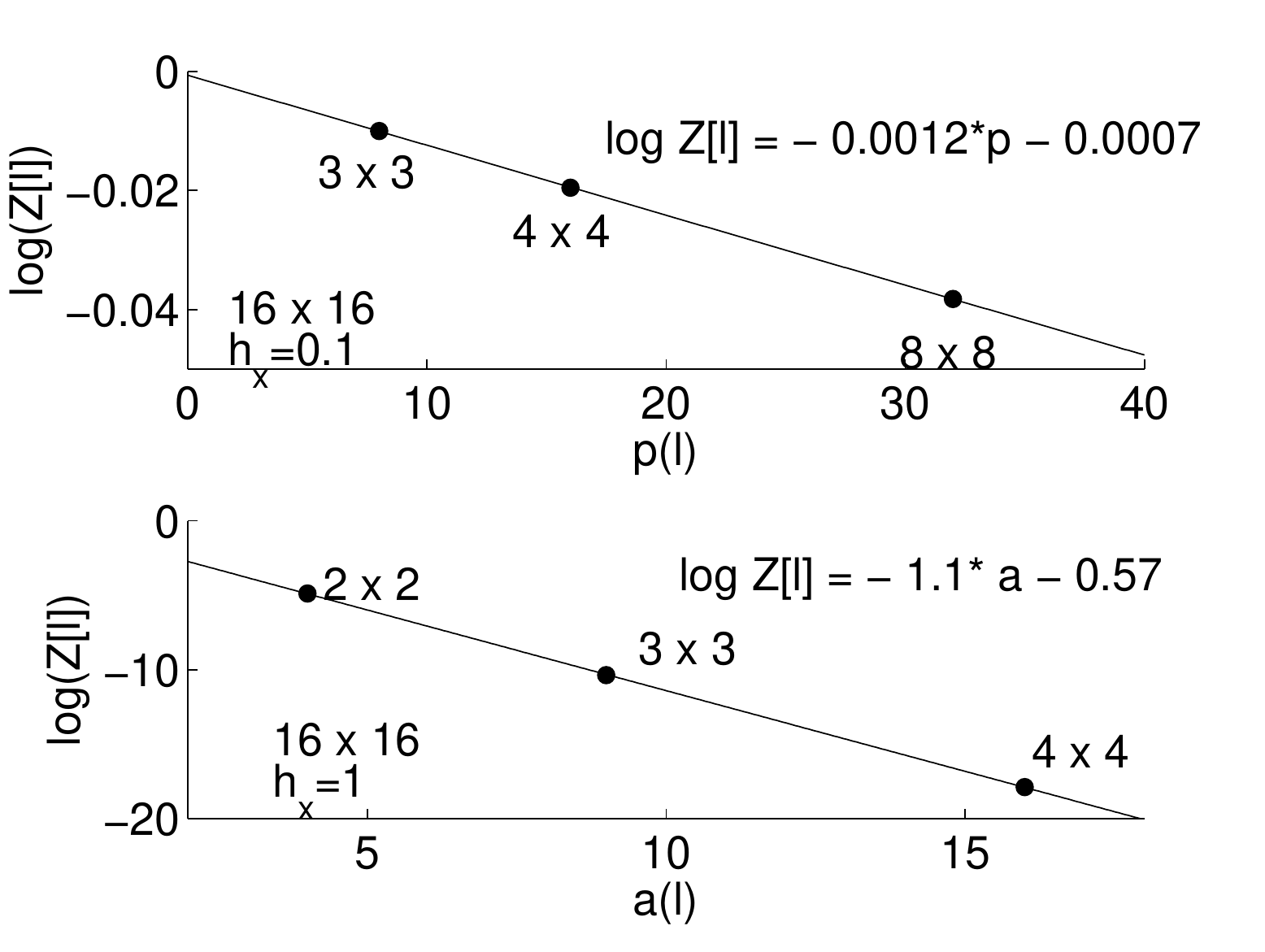}
  \caption{
Upper panel: Wilson loop $\langle Z[l]\rangle$ as a function of the perimeter $p(l)$ of loop $l$, on a $16 \times 16$ lattice with magnetic field $h_x=0.1$, corresponding to the deconfined phase. Notice the exponential decay of $\langle Z[l]\rangle$ as a function of the perimeter $p(l)$ of loop $l$, or \textit{perimeter law}, see Eq. \ref{eq:WilsonP}.
Lower panel: Wilson loop $\langle Z[l]\rangle$ as a function of the area $a(l)$ of loop $l$, on a $16 \times 16$ lattice with magnetic field $h_x=1$, corresponding to the spin polarized phase. Notice the exponential decay of $\langle Z[l]\rangle$ as a function of the area $a(l)$ of loop $l$, or \textit{area law}, see Eq. \ref{eq:WilsonA}. 
  }
  \label{fig:Wilson}
\end{figure}
Fig. \ref{fig:Wilson} shows the scaling of Wilson loops, see Sect. \ref{sec:DTC:DTC}, for the ground state of the topological sector $(+,+)$. In this case, a lattice of $16\times 16$ sites, or $16^2\times 2 = 512$, spins was considered for a magnetic field $h_x = 0.2$ (deconfined phase) and $h_x = 1$ (spin polarized phase). For these values of the magnetic field the system is away from the critical point, as reflected in a drop in the amount of ground state entanglement, see Fig. \ref{fig:Entanglement}. This allows the hybrid ansatz to reliably represent much larger lattices with just $\chi=100$.

For $h_{x} = 0.2$ our results confirm that a Wilson loop decays exponentially fast with the size of the loop according to a perimeter law, as it is characteristic of the deconfined phase. Instead, for $h_{x}=1$ the decay is exponential in the area encircled by the loop, as it is characteristic of the spin polarized phase. 

\subsection{Wave function fidelities}
\label{sec:bench:fidelity}

\begin{figure}[t]
  \includegraphics[width=8.5cm]{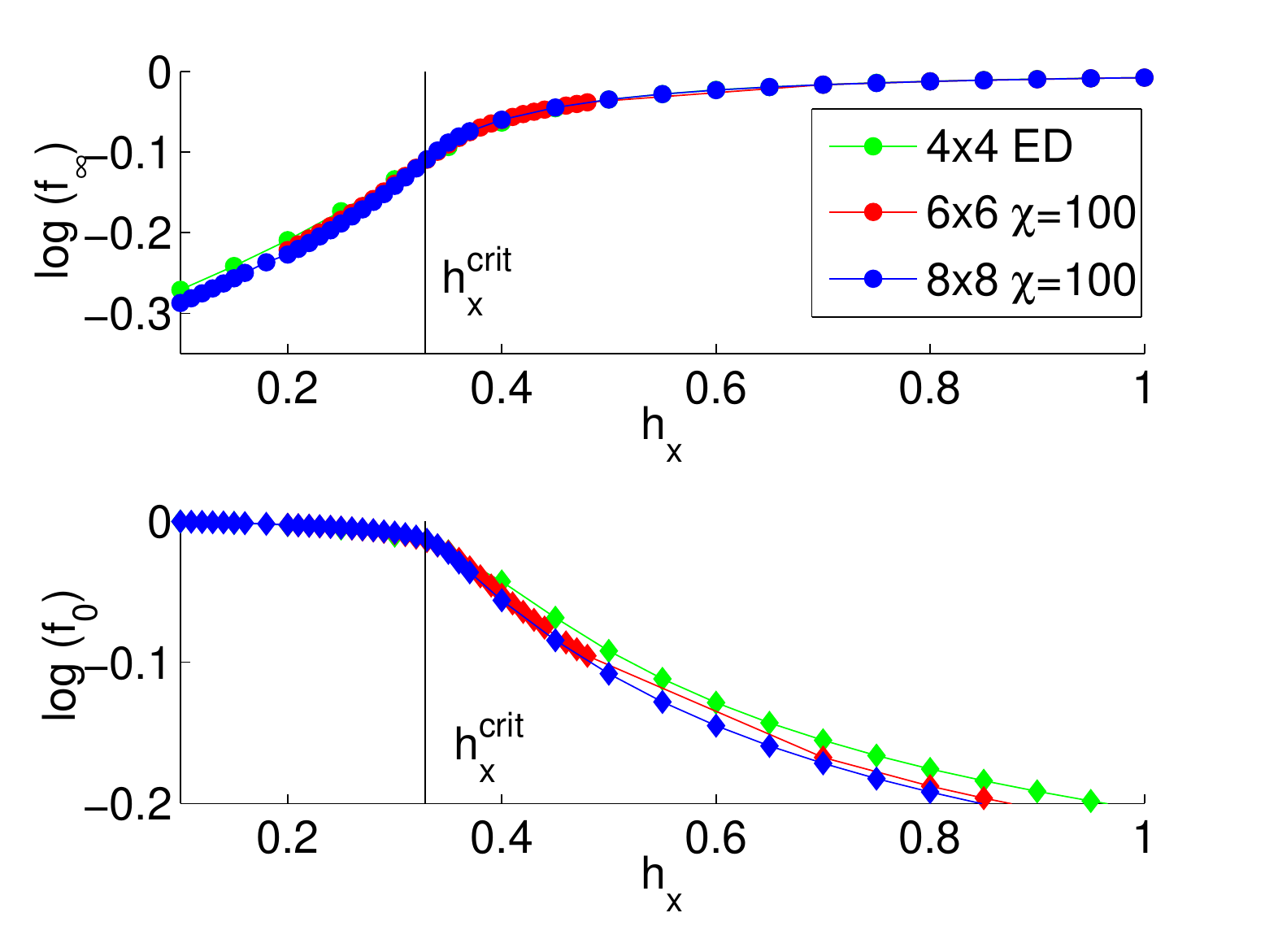}
  \caption{
Logarithm of the intensive fidelities $f_0(h_x)$ and $f_{\infty(h_x)}$ as a function of the magnetic field $h_x$, for different values $L=4,6$ and $8$ of the system size. The critical magnetic field $h_x^{\text{crit}} = 0.3285(1)$\cite{Hamer2000} is denoted by a vertical line.}
  \label{fig:Fidelity}
\end{figure}

\subsubsection{Ground state fidelities}
 
The overlap or fidelity $\braket{\Phi(h_{x})}{\Phi(h_{x}')}$ between the ground state of $H^{\text{x}}_{\text{TC}}$ for two values $h_x,h_x'$ of the magnetic field can be used as an alternative way to detect the presence of phases in the model and the location of their boundaries. Indeed, the ground state wave function is somewhat similar within a phase and experiences a radical change when the system undergoes a phase transition, with these two facts being captured by the ground state fidelity. Interestingly the characterization of phase boundaries using the fidelity does not require the knowledge of the order parameters. Here we will focus on the intensive fidelity $f(h_x,h_x')$ \cite{zanardi_2006,liu-2009,cozzini_2007,cozzini_2007b,Venuti99,buonsante_2007,you_2007,jian_2008,orus_2008} , defined through
\begin{equation}
	\log f(h_x,h_x') \equiv \frac{1}{L^2} \log \left( |\braket{\Phi(h_{x})}{\Phi(h_{x}')}| \right).
\end{equation}

The present approach yields an explicit representation of the ground state wave function, from which computing overlaps is straightforward (this is true of the hybrid ansatz used in this work but not in the more general case where $W_{\text{num}}$ also contains disentanglers\cite{Tagliacozzo2010prep}). Fig. \ref{fig:Fidelity} shows the logarithm of the intensive fidelity
\begin{equation}
	f_0(h_x) \equiv f(h_x,0)
\end{equation}
between the ground states of $H^{\text{x}}_{\text{TC}}$ in the topological sector $(+,+)$ for $h_x=0$ and another value $h_x \geq 0$; 
as well as the logarithm of the intensive fidelity
\begin{equation}
	f_\infty(h_x) \equiv f(h_x,\infty)
\end{equation}
between the ground states of $H^{\text{x}}_{\text{TC}}$ in the topological sector $(+,+)$ for $h_x=\infty$ and another value $h_x \geq 0$.
Fidelity $f_0(h_x)$ shows a markedly different behaviour for small magnetic fields roughly smaller than $h_x^{\text{crit}}$ (corresponding to the deconfined phase, to which $\ket{\Phi(0)}$ also belongs to) and for large magnetic fields, where the fidelity vanishes. An analogous behaviour is also observed for $f_{\infty}(h_x)$.

\begin{figure}[t]
  \includegraphics[width=7.5cm]{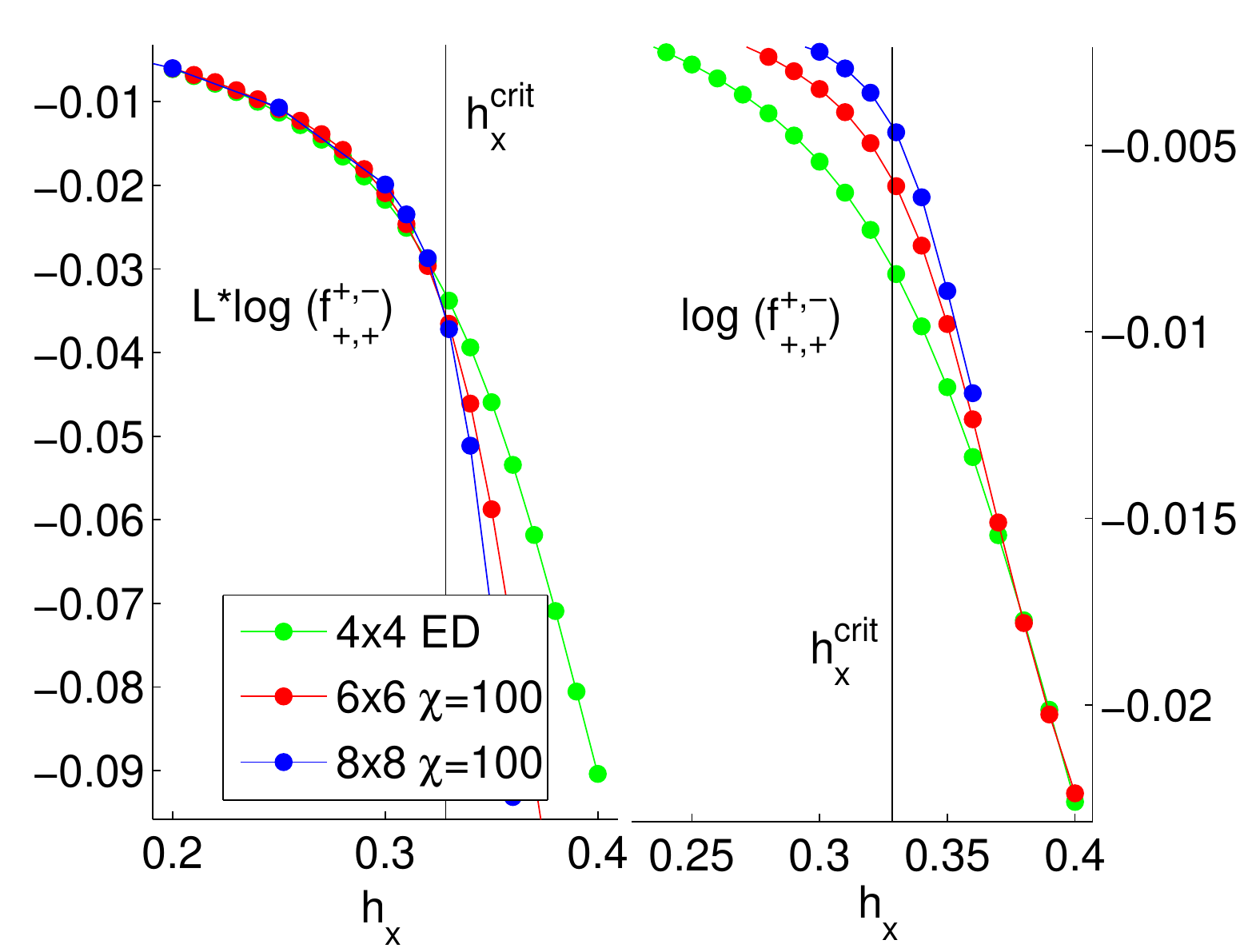}
  \caption{
Logarithm of the intensive \textit{topological} fidelity $f_{+,+}^{+,-}$ as a function of the magnetic field $h_x$. In the deconfined phase, $f_{+,+}^{+,-}$ remains large (its logarithm remains close to zero), indicating that the ground states $\ket{\Phi_{+,+}}$ and $\ket{\Phi_{+,-}}$ differ mostly on their topological spins but are otherwise very similar. This trend changes for larger magnetic fields $h_x$. As we enter the spin polarized phase, the ground states $\ket{\Phi_{+,+}}$ and $\ket{\Phi_{+,-}}$ are structurally different, and modifying the topological spins of e.g. $\ket{\Phi_{+,-}}$ does no longer produce a good approximation to $\ket{\Phi_{+,+}}$. Left: data collapse occurs in the deconfined phase when we plot $L\log(f_{+,+}^{+,-})$. The critical magnetic field $h_x^{\text{crit}} = 0.3285(1)$\cite{Hamer2000} is denoted by a vertical line. Right: instead, plotting $\log(f_{+,+}^{+,-})$ leads to data collapse in the spin polarized phase. {This behaviour is expected as explained in the main text below equation \ref{eq:TopFid2}.} }
  \label{fig:TopoFid}
\end{figure}
\subsubsection{Topological fidelities}

For $h_x=0$, the ground states in two topological sectors only differ in the expectation value of operators $X_1$ and $X_2$, see Sect. \ref{sec:DTC:TC}. One can map e.g. the ground state $\ket{\Phi_{+,+}}$ of topological sector $(+,+)$ into the ground state $\ket{\Phi_{+,-}}$ of topological sector $(+,-)$ by just applying operator $Z_2$ of Eq. \ref{eq:Z}, namely $\ket{\Phi_{+,-}} = Z_2 \ket{\Phi_{+,+}}$. Therefore $\bra{\Phi_{+,+}} Z_2 \ket{\Phi_{+,-}} = 1$ and the \textit{topological} fidelity $f_{+,+}^{+,-}$ between these two sectors, defined as
\begin{equation}
		\log(f_{+,+}^{+,-})  \equiv \frac{1}{L^2} \log\left( \bra{\Phi_{+,+}} Z_2 \ket{\Phi_{+,-}} \right), 
\end{equation}
is maximal, $f_{+,+}^{+,-}=1$, for $h_x=0$. More generally, the \textit{topological} fidelity between sectors $(v_1,v_2)$ and $(v_1',v_2')$, defined as
\begin{equation}
	\log(f_{v_1,v_2}^{v_1',v_2'})  \equiv \frac{1}{L^2} \log \bra{\Phi_{v_1,v_2}} \left(Z_1\right)^{w_1} \left(Z_2\right)^{w_2} \ket{\Phi_{v_1',v_2'}},
\end{equation}
where $w_i$ is $0$ if $v_i=v_i'$ and $1$ otherwise, also fulfils
\begin{equation}
	f_{v_1,v_2}^{v_1',v_2'} = 1~~~~~(h_x=0).
  \label{eq:TopFid}
\end{equation}

For $h_x\neq 0$ Eq. \ref{eq:TopFid} is no longer expected to hold. In particular, in the limit of large magnetic field $h_x$ a simple analytical characterization of the ground states  $\ket{\Phi_{v_1,v_2}}$ exists and can be used to show e.g. that $|\bra{\Phi_{+,+}} Z_2\ket{\Phi_{+,-}}| = L^{-1/2}$, so that $f_{+,+}^{+,-}$ vanishes for large $L$. More generally,
\begin{equation}
 f_{v_1,v_2}^{v_1',v_2'} = 0 ~~~~~(h_x = \infty, L = \infty).
\label{eq:TopFid2}
\end{equation}
For intermediate values of $h_x$, the topological fidelity is expected to remain close to one in the deconfined phase and sharply decay in the spin polarized phase. Fig. \ref{fig:TopoFid} for $\log (f_{+,+}^{+,-})$ as a function of $h_x$ shows that this is indeed the case. In addition, the data for different values of $L$ collapses into a single curve in the deconfined phase \textit{if} $\log (f_{+,+}^{+,-})$ is multiplied by $L$, whereas in the spin polarized phase the data collapses into a single curve when $\log (f_{+,+}^{+,-})$ is directly plotted. This is reminiscent of the two possible behaviour of Wilson loops, namely perimeter versus area law, see Fig. \ref{fig:Wilson}. After all, operators $Z_1$ and $Z_2$ can be understood as Wilson loops with non-contractible support.

\section{Discussion}
\label{sec:discussion}

In this paper we have proposed a coarse-graining transformation for lattice models with a local symmetry that simultaneously preserves locality and the symmetry, while exploiting the latter to significantly reduce computational costs. This coarse-graining transformation, made of an analytical part $W_{\text{exact}}$ and a numerical part $W_{\text{num}}$, gives rise to a variational ansatz for the ground state(s) and low energy states of the lattice model. Here we have focused on a simplified form of the ansatz, a hybrid between a TTN and the MERA, in which $W_{\text{num}}$ does not contain disentanglers (although $W_{\text{exact}}$ does). The computational cost of the resulting approach grows exponentially with the linear size $L$ of the lattice, severely restricting the size of systems that can be analysed, although it permits to consider systems well beyond the scope of exact diagonalization techniques. 

By adding disentanglers to $W_{\text{num}}$, it is possible to remove more entanglement from the ground state. The resulting ansatz is a proper MERA, with which much larger systems can be addressed. In addition, as explained in Ref. \onlinecite{Tagliacozzo2010prep}, under coarse-graining transformations the ground state of $H_{\text{TC}}^{\text{x}}$ for a small magnetic field $h_x$ can now be seen to flow back to the $h_x=0$ fixed point of RG flow, which has a richer, local $Z_2 \times Z_2$ symmetry, whereas the ground state for $h_x > h_x^{\text{crit}}$ flows to the spin polarized RG fixed point characterized by an infinite $h_{x}$.

\subsection{Generalizations}
\label{sec:discussion:general}

The coarse-graining transformation proposed in this paper can be generalized to more complex settings in a number of ways.

\subsubsection{Static electric charges}

Throughout the paper we have assumed that the state of the lattice model satisfies the star constraints of Eq. \ref{eq:star}, corresponding to the absence of electric charges. However, the coarse-graining scheme and related ansatz can be readily adapted to the case where one or several star constraints are violated on specific sites, provided that the state $\xi$ of the lattice is constrained to satisfy 
\begin{equation}
	A_s \ket{\xi} = -\ket{\xi} 	
\end{equation}
on those specific sites. Indeed, the only change that is required in the coarse-graining scheme is that when the star operator is eventually transformed by a sequence of CNOT gates into a single-spin operator $-\sigma^x$, the spin on which it acts, forced now to be in state $\ket{-}$ instead of $\ket{+}$, is projected out accordingly. 

In this way one could study the ground state of the system in the presence of two electric charges separated a distance $l$, and obtain the ground state energy as a function of $l$, from which the string tension can be evaluated.

\subsubsection{Magnetic field in the $xz$-plane}

If the toric code Hamiltonian is perturbed by adding a magnetic field $-\sum_{j}(h_x \sigma^{x}_j + h_z \sigma^z_j)$ on the $xz$-plane, a lattice model with local $Z_2$ symmetry can still be recovered by adding dummy spin variables, corresponding to matter field, on the sites of lattice $\mathcal{L}$, see Ref. \onlinecite{Tupitsyn2008}. In that case the local $Z_2$ symmetry can still be exploited, but in a trivial sense (the dummy spin variables are factored back out as part of the coarse-graining transformation) that appears to have no advantage to the case of an arbitrary perturbation, which is considered next.

\subsubsection{Arbitrary perturbation of the toric code} 

In this work we have focused on a perturbation of the toric code Hamiltonian $H_{\text{TC}}$ in Eq. \ref{eq:HTC} that breaks its $Z_2\times Z_2$ local symmetry for $h_x=0$, generated by both star operators $A_s$ and plaquette operators $B_p$, into a local $Z_2$ symmetry for $h_x \neq 0$ generated by star operators only. A generic perturbation of Hamiltonian $H_{\text{TC}}$ will break the local symmetry completely. However, as recently proven by Bravyi, Hastings and Michalakis in Ref. \onlinecite{Bravyi2010}, topological phases are robust under arbitrary (sufficiently weak) perturbations of the Hamiltonian.

For a perturbation that completely breaks the local $Z_2 \times Z_2$ symmetry of the toric code model, say a weak magnetic field in a direction $\hat{r}=(r_x,r_y,r_z)$ different from directions $\pm\hat{x}$ and $\pm\hat{z}$ \cite{vidal_phase_2008,vidal_self-duality_2009}, we can modify $W_{\text{exact}}$ so that it no longer projects spins into $\ket{+}$ state. The sequence of CNOT gates in Fig. \ref{fig:Wexact}, which produces six spins per plaquette on which either operator $-J_e \sigma^x$ or operator $-J_e\sigma^z$ is acting, is now followed by an extended numerical transformation $W_{\text{num}}$ that acts on all these six spins to produce an effective spin inside each plaquette of $\mathcal{L}'$. For small values of the perturbation, the sequence of CNOT gates will still map a robustly entangled ground state into a weakly entangled ground state, lowering significantly the computational cost of the approach, while the use of disentanglers as part of $W_{\text{num}}$ is expected to again produce a flow back to the RG fixed point given by the undeformed toric code Hamiltonian $H_{\text{TC}}^{\text{x}}$.

\subsubsection{Quantum double models and string-net models}
 
The analytic transformation $W_{\text{exact}}$ used in this work is, in essence, equivalent to the RG transformation proposed in Ref. \onlinecite{Aguado2008TC} to show that the toric code model is a fixed point of the RG flow. In that work, analogous analytical transformations were proposed also for generalizations of the toric code, which corresponds to the quantum double of the $Z_2$ group, to models based on the quantum double of any discrete group $G$\cite{Kitaev2003ToricCode}. In the appropriate regime, the low energy sector of the quantum double model with group $G$ corresponds to a lattice gauge theory with the same gauge group. A similar analysis was subsequently carried forward in Ref. \onlinecite{Konig2009} for string-net models \cite{Levin2005SN}. 

The analytical transformations described in Refs. \onlinecite{Aguado2008TC,Konig2009} for the fixed-point Hamiltonians of quantum double and string-net models map robustly entangled spins \textit{locally} into spins that are in a product state and can therefore be factored out (or projected out) from the coarse-grained system. They can again be used as the basis for a generalized analytical transformation $W_{\text{exact}}$, which needs to be supplemented with a numerical transformation $W_{\text{num}}$ when the fixed-point Hamiltonian is deformed with an arbitrary perturbation. In this way, we obtain a coarse-graining scheme for deformed quantum double models (equivalent in the appropriate regime to lattice gauge theories) and string-net models, as well as corresponding tensor network ans\"atze for their ground state wave functions.

\subsubsection{Renormalization of PEPS} 
 
The use of an analytical transformation $W_{\text{exact}}$ as part of a coarse-graining procedure can be also exported to the domain of approaches based on PEPS (also referred to as TPS). This can occur in two different contexts. 

On the one hand, given a PEPS representation for a state $\ket{\xi}$ (e.g. the ground state of Hamiltonian $H_{\text{TC}}^{x}$) with a local $Z_2$ symmetry, an important task is to evaluate the tensor network corresponding to the scalar product $\braket{\xi}{\xi}$ (and related quantities), which plays a central role in the computation of expectation values of observables and various optimization algorithms. The tensor network for $\braket{\xi}{\xi}$ can be evaluated by a number of different methods, such as corner transfer matrix (CTM) methods\cite{nishino_1996,orus_2009}, MPS techniques\cite{Verstraete2004, Jordan2008} or the tensor entanglement renormalization group\cite{Gu2008, Xie2009, Chen2009} (TERG). All these methods proceed by coarse-graining the tensor network for the scalar product $\braket{\xi}{\xi}$ (as opposed to coarse-graining a tensor network for the state $\ket{\xi}$), with a cost that depends on the amount of entanglement in the system. This cost can again be significantly reduced if the coarse-graining incorporates analytic local pre-processing along the lines of the transformation $W_{\text{exact}}$ discussed in this work.

On the other hand, a PEPS representation for the same state $\ket{\xi}$ can also be coarse-grained as a wave function. This process, termed \textit{wavefunction renormalization} in Ref. \onlinecite{Chen2010}, is thoroughly equivalent to \textit{entanglement renormalization}\cite{Vidal2007ER,Evenbly2009Alg} and produces a MERA\cite{Vidal2007ER} as a result. Therefore, the exact transformation $W_{\text{exact}}$ discussed in this paper can also be used in the context of wavefunction renormalization.

\subsection{Conclusions}

Tensor network algorithms offer a variational approach to strongly interacting systems on a lattice that is free of the sign problem and can therefore be applied to systems that are beyond the reach of Monte Carlo sampling techniques, such as frustrated antiferromagnets and interacting fermions. In this paper we have explored the use of tensor network techniques in the context of lattice gauge theory and, more broadly, systems with topological order. We have explain how to incorporate a local symmetry into a coarse-graining scheme and the resulting variational ansatz, for the simplest non-trivial case of the $Z_2$ lattice gauge theory.

We envisage that proper generalizations of the results presented here, possibly along the lines of those discussed in Sect. \ref{sec:discussion:general}, will constitute the basis for future numerical simulations of lattice gauge theories and more general models with topological order.

We thank M. Aguado, P. Calabrese, P. Corboz, A. Di Giacomo, M. D'Elia, G. Evenbly, F. Gliozzi, A. Hamma, J.I. Latorre and G. Sierra for useful discussions and comments. The authors acknowledge financial support from the Australian Research Council (FF0668731, DP0878830, DP1092513).


\begin{appendix}

\section{Duality transformation}
\label{app:duality}

In this appendix we review the well-known duality transformation\cite{Wegner1971,Kogut1979RevModPhys} between $Z_2$ lattice gauge theory and the quantum Ising model in two spatial dimensions as described in Sect. \ref{sec:Z2:duality}, see Fig.\ref{fig:Ising}. The transformation is expressed in terms of $CNOT$ gates in order to highlight its similarities and differences with the analytical transformation $W_{\text{exact}}$ used in this work.

\begin{figure}
\begin{center}
\includegraphics[width=8cm]{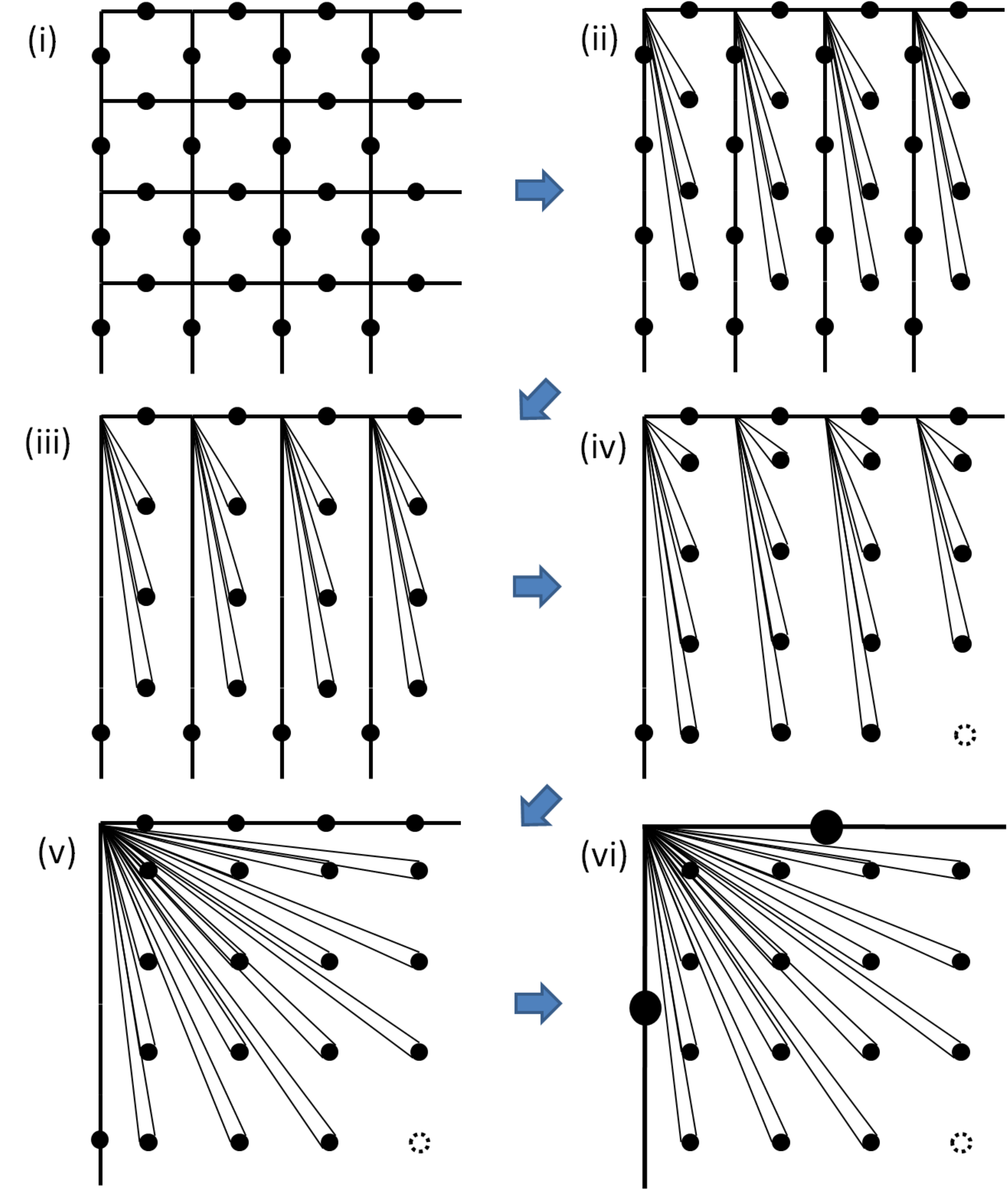}
\caption{
Duality transformation between the $Z_2$ lattice gauge theory and the quantum Ising model, broken into six stages ($i$)-($vi$).} 
\label{fig:Duality}
\end{center}
\end{figure}

\begin{figure}
\begin{center}
\includegraphics[width=7cm]{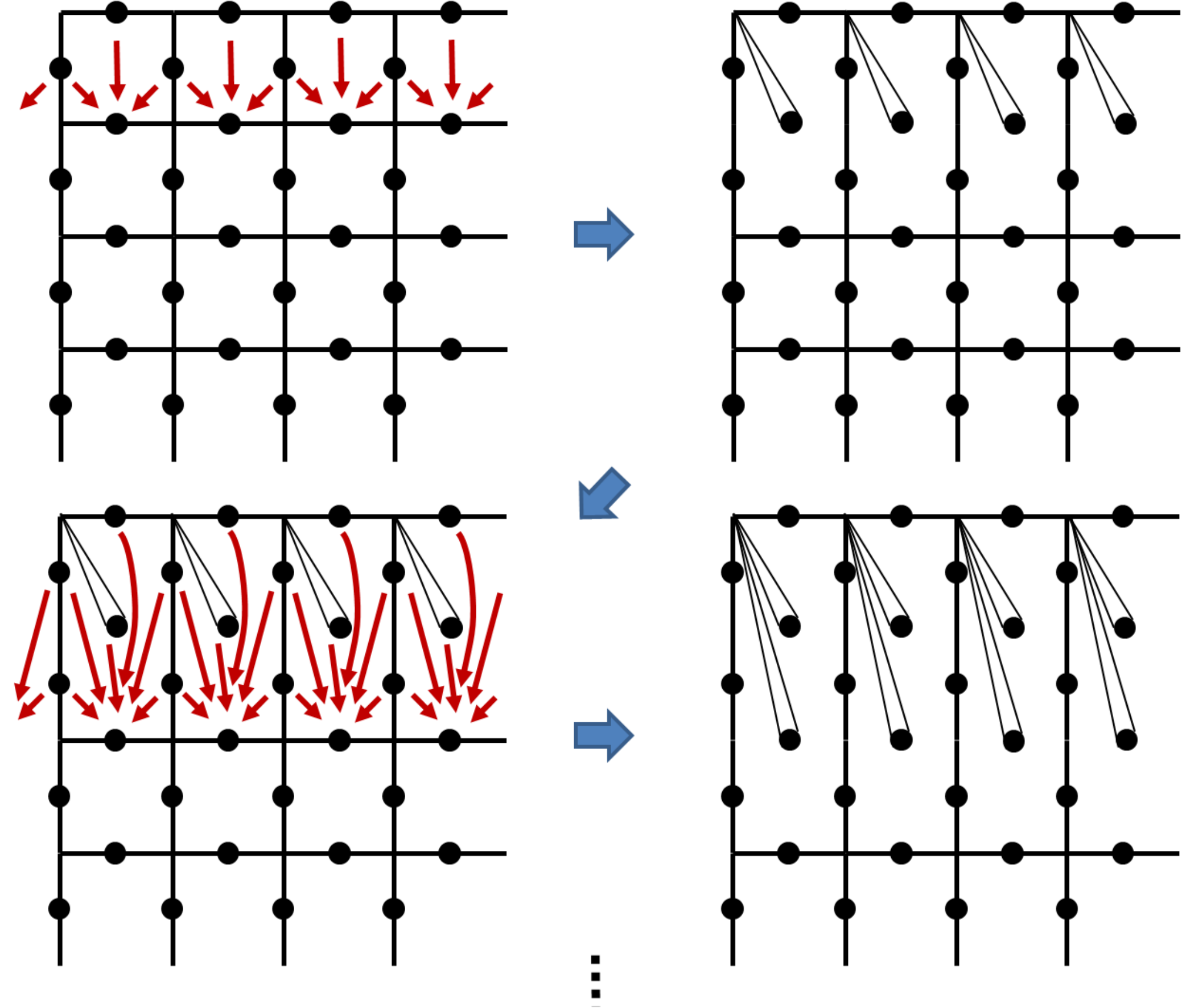}
\caption{
 CNOT gates that transform stage ($i$) into stage ($ii$). The sequence proceeds from top to bottom. Only two of the three required steps are depicted. In a $L\times L$ lattice, $L-1$ such steps are required.} 
\label{fig:Duality2}
\end{center}
\end{figure}

\begin{figure}
\begin{center}
\includegraphics[width=7cm]{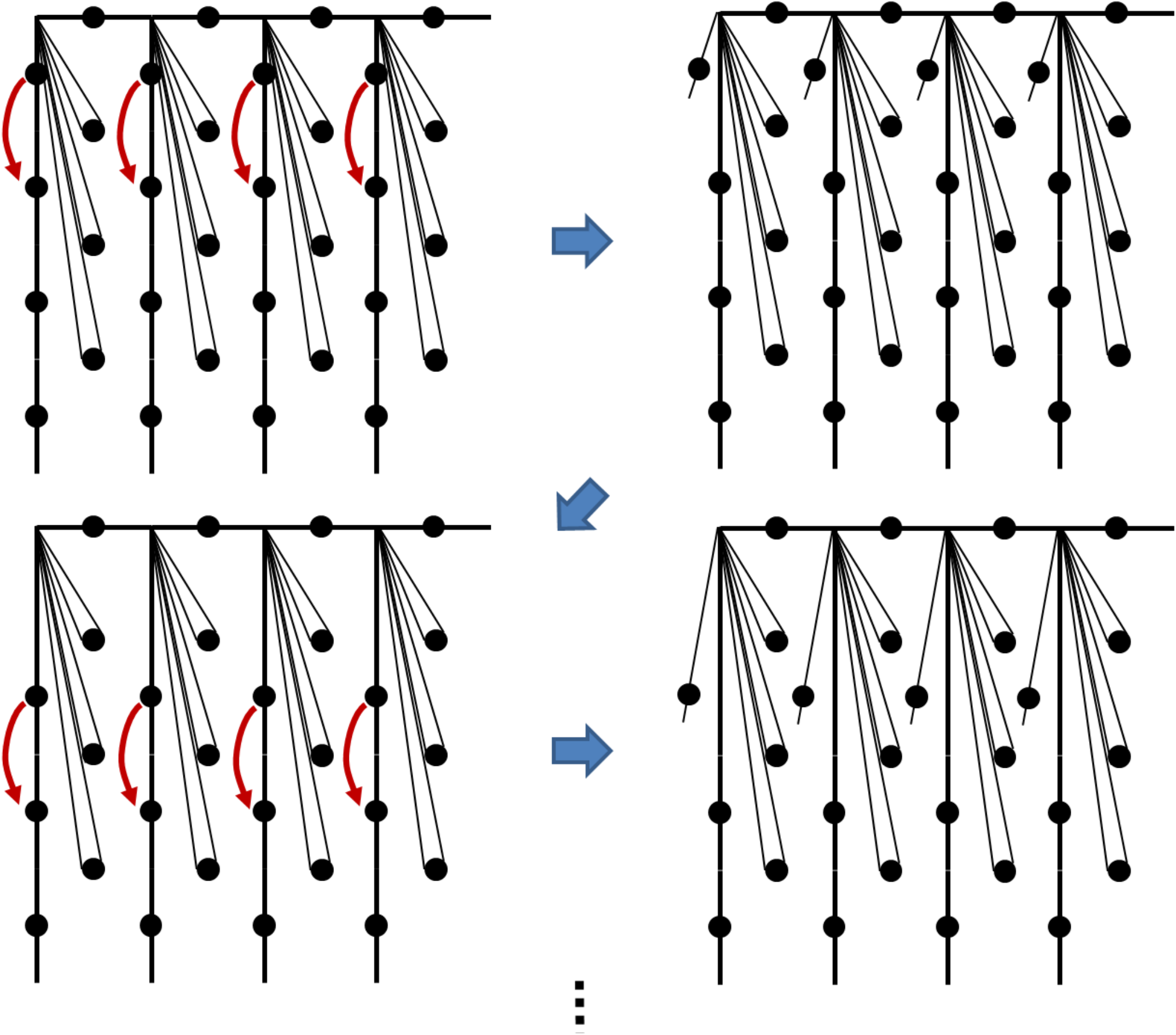}
\caption{
 CNOT gates that transform stage ($ii$) into stage ($iii$). The sequence proceeds from top to bottom. Only two of the three required steps are depicted. In a $L\times L$ lattice, $L-1$ such steps are required.} 
\label{fig:Duality3}
\end{center}
\end{figure}

\begin{figure}
\begin{center}
\includegraphics[width=7cm]{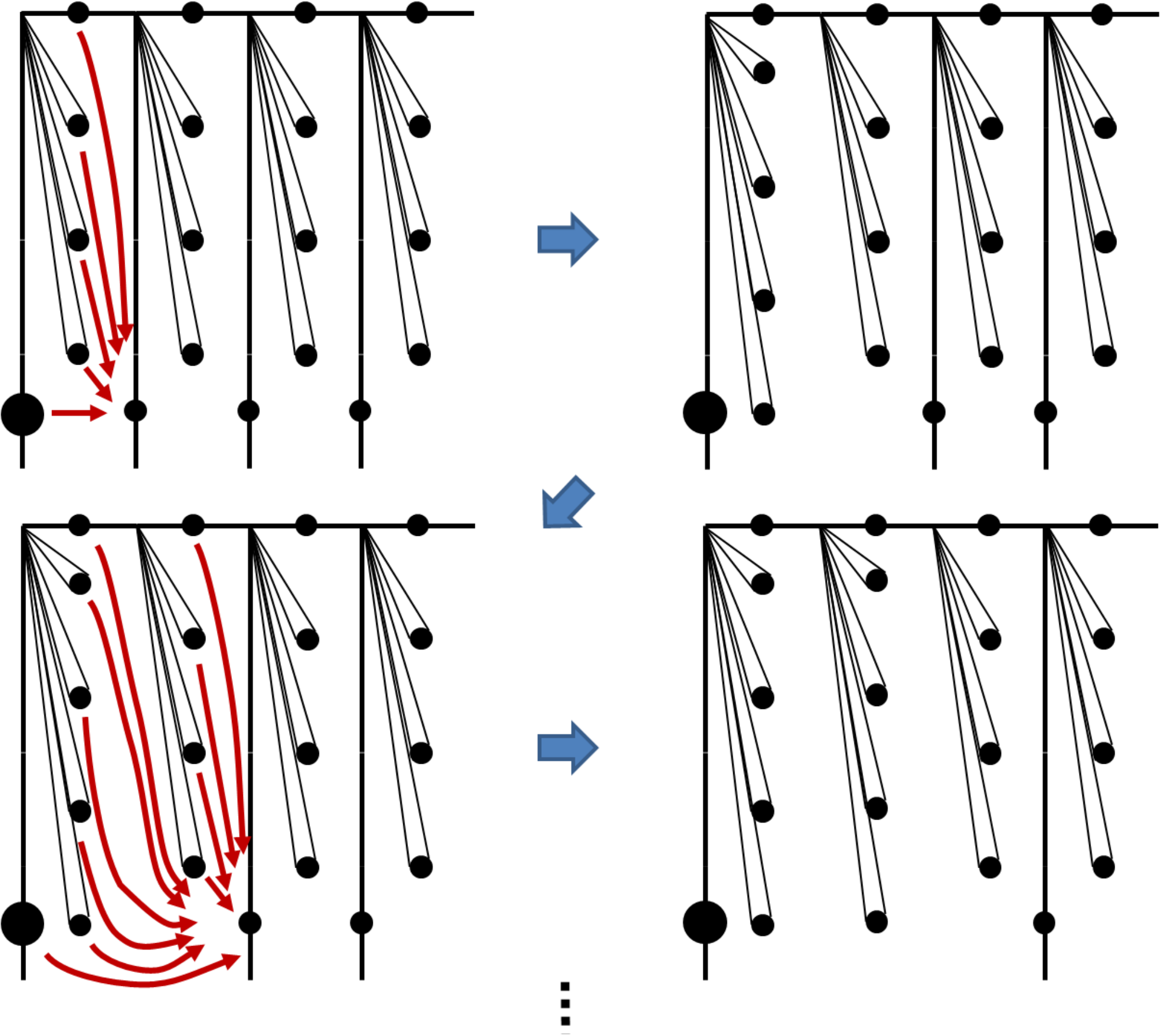}
\caption{
CNOT gates that transform stage ($iii$) into stage ($iv$). Notice that stage ($iii$) contains four columns of single-spin plaquettes, with three spins on each column. Instead, stage ($iv$) contains four columns of single-spin plaquettes with four spins on each column, except for the rightmost column, which only contains three spins and a vacancy. The sequence proceeds from left to right. Only two of the three required steps are depicted. In a $L\times L$ lattice, $L-1$ such steps are required.} 
\label{fig:Duality4}
\end{center}
\end{figure}

\begin{figure}
\begin{center}
\includegraphics[width=7cm]{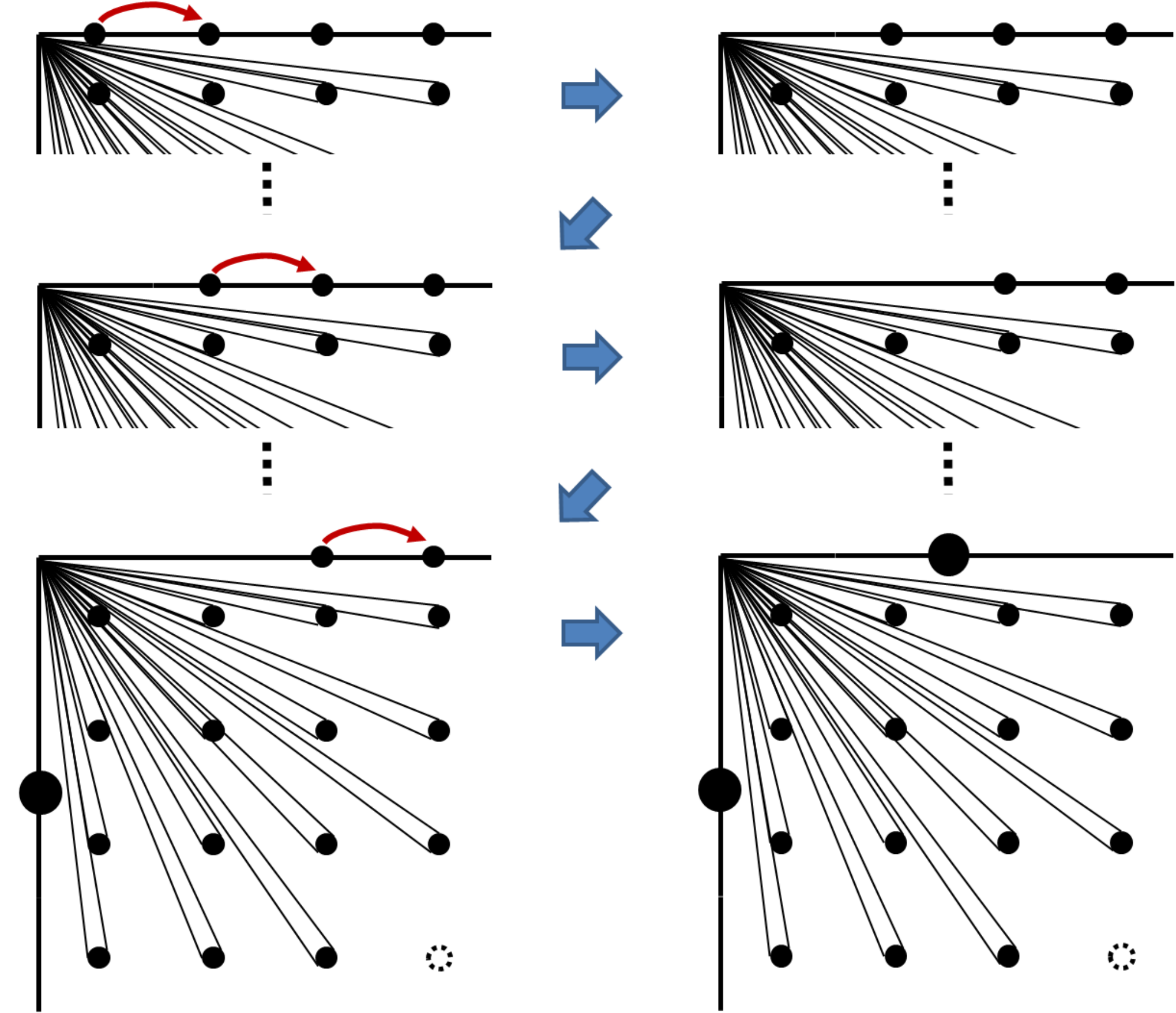}
\caption{
CNOT gates that transform stage ($v$) into stage ($vi$). Three required steps are depicted. In a $L\times L$ lattice, $L-1$ such steps are required.} 
\label{fig:Duality5}
\end{center}
\end{figure}

As indicated in Figs. \ref{fig:Duality}-\ref{fig:Duality5} for the case of a lattice of $4\times 4$ sites with periodic boundary conditions in both directions, this transformation can be implemented as a sequence of CNOT gates as indicated. Fig. \ref{fig:Duality} represents six stages ($i$)-($vi$) of this transformation, with stages ($i$) and ($vi$) corresponding to the $Z_2$ lattice gauge theory and the quantum Ising model respectively.

Fig. \ref{fig:Duality2} indicates the sequence of CNOT gates that transform stage ($i$) into stage ($ii$). This sequence progresses from top to bottom and can be divided into four columns of CNOT gates, where gates in different columns commute. 

Fig. \ref{fig:Duality3} indicates the sequence of CNOT gates that transform stage ($ii$) into stage ($iii$). Again, the sequence progresses from top to bottom and can be divided into four columns of CNOTs gates, where gates in different columns commute.

Fig. \ref{fig:Duality4} explains how to transform stage ($iii$) into stage ($iv$). Notice that stages ($iv$) and ($v$) are equivalent, since they only differ in where the single-spin plaquettes are connected to the rest of the lattice -- that is, they correspond exactly to the same Hamiltonian.

Finally, Fig. \ref{fig:Duality5} shows how to transform stage ($v$) into stage ($vi$.), corresponding to the Ising model.  

\section{Assessment of the precision of the numerical results}
\label{app:precision}

Section \ref{sec:bench} presented a number of numerical results for the deformed toric code Hamiltonian $H_{\text{TC}}^{\text{x}}$ on a torus of linear size $L=4,6$ and $8$, obtained with a hybrid tensor network with a fixed value of the refinement parameter $\chi=100$. Those results successfully reproduced both quantitatively and qualitatively the main properties of the ground state of the system in the deconfined and spin polarized phases and of the second order phase transition between them. In this appendix we perform a scaling analysis with respect to $\chi$, to show that the particular choice $\chi=100$ used in Sect. \ref{sec:bench} did not have a significant effects on the numerical results.

We start by reminding that the hybrid tensor network ansatz is, as far as its numerical part $W_{\text{num}}$ is concerned, a tree tensor network TTN, namely one used to approximate the ground state of the lattice model after being transformed according to $W_{\text{exact}}$. As discussed in Ref. \onlinecite{Tagliacozzo2009}, a sufficiently large value of $\chi$ will reproduce the ground state to arbitrary accuracy, but this value will have to be exponentially large in the linear size $L$ of the lattice.

\begin{figure}
\begin{center}
i) \includegraphics[width=7cm]{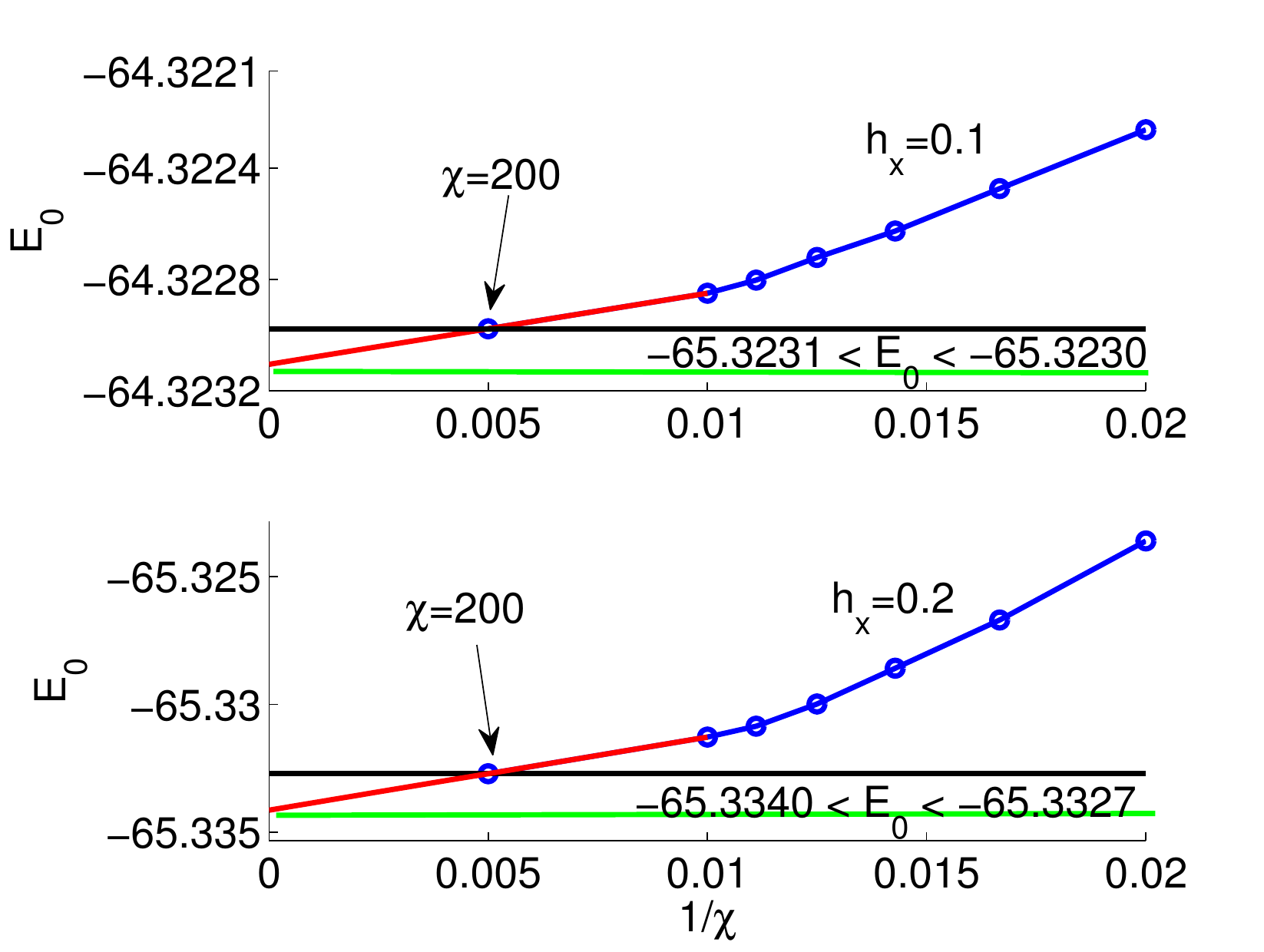} 

ii) \includegraphics[width=7cm]{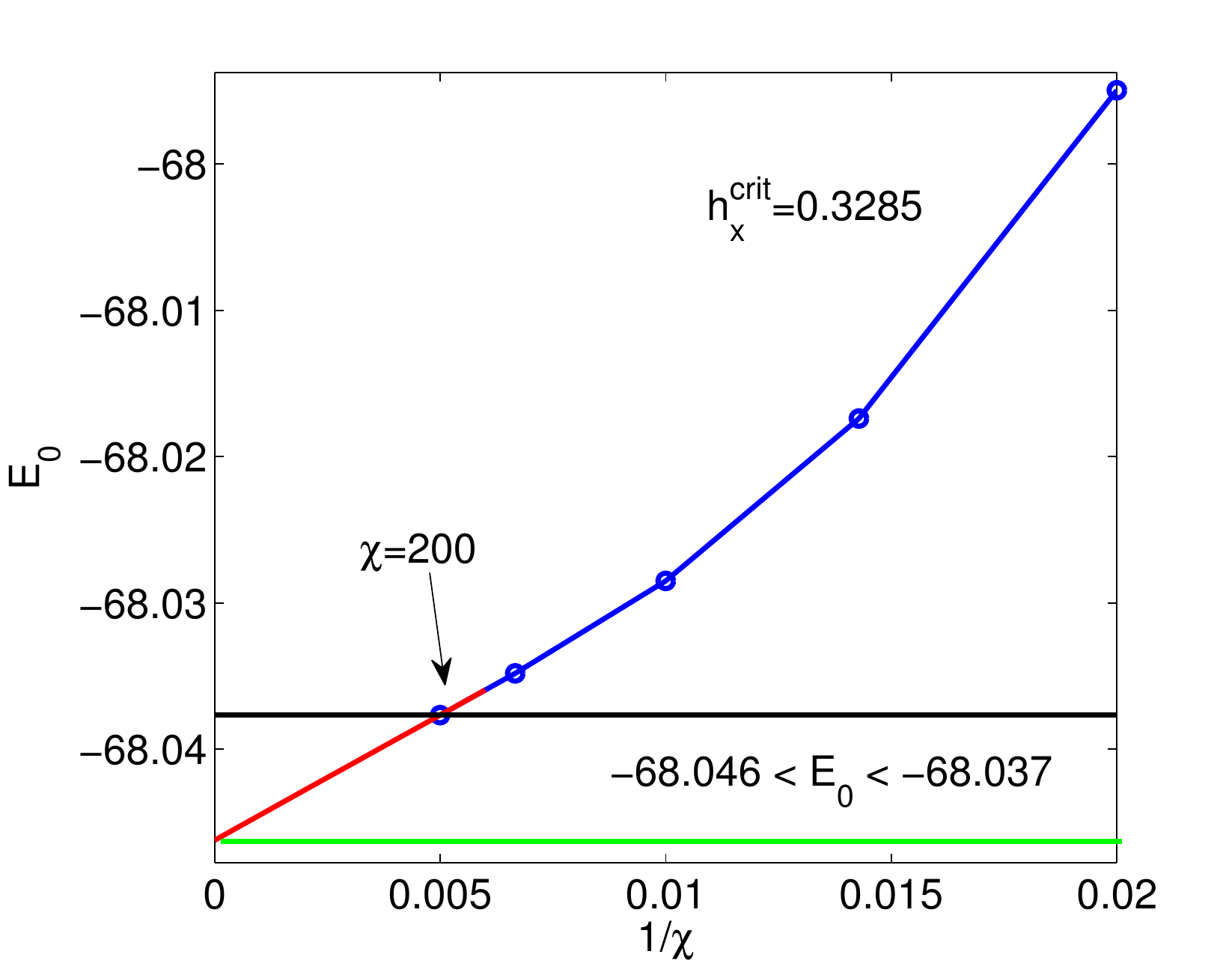}

iii) \includegraphics[width=7cm]{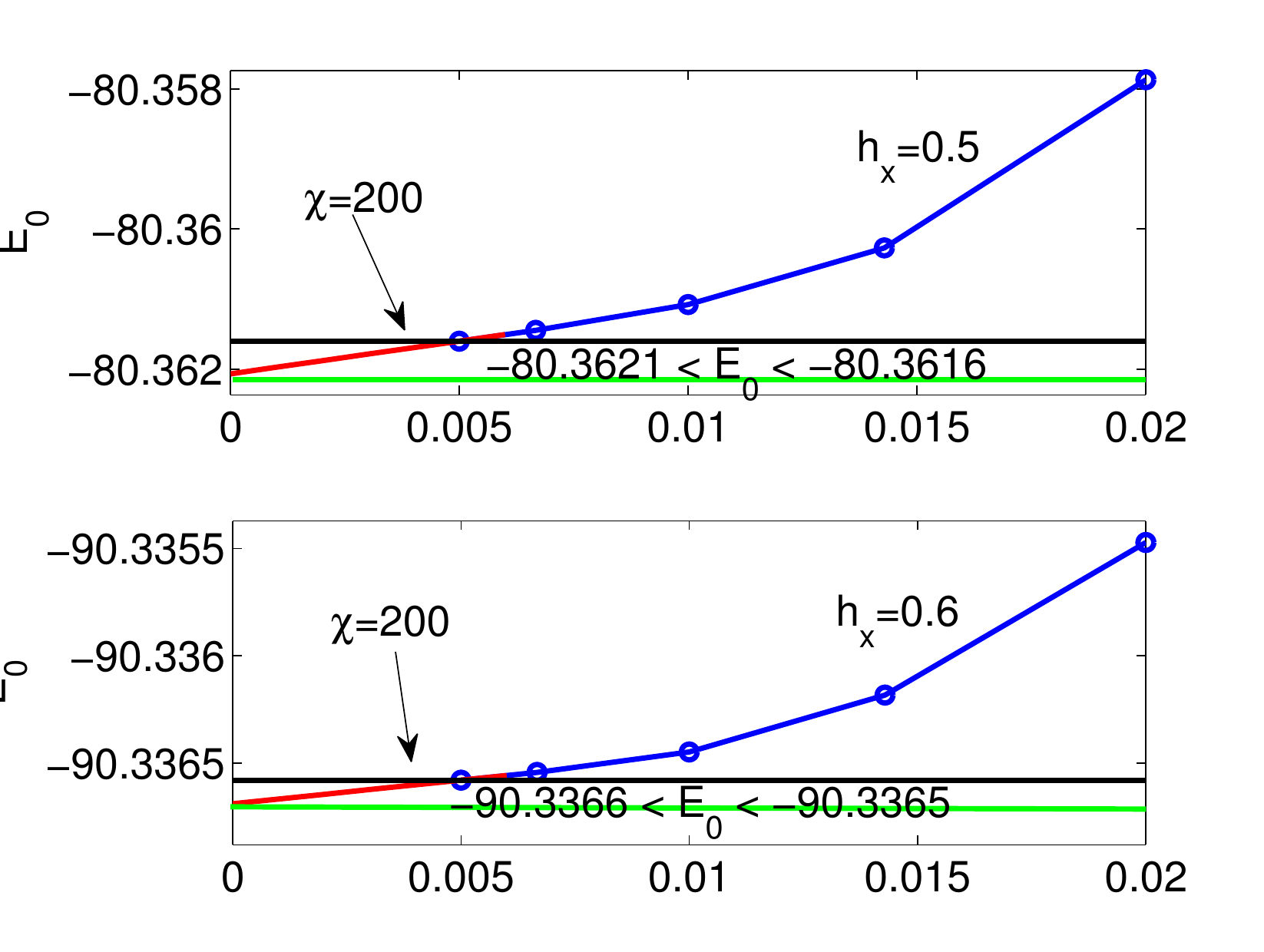}
\caption{
Ground state energy $E_0$ for the deformed toric code on a $8 \times 8$ torus as a function of the refinement parameter $1/ \chi$ for  different values of the magnetic field, i) $h_x=0.1$ and $h_x=0.2$ in the topological phase, ii) $h_{x}^{crit}=0.3285$ at the critical point, iii) $h_x=0.5$ and  $h_x=0.6$ in the spin polarized phase.
Results of the simulations in the range of refinement parameter $50 \le \chi \le 200$ are shown with blue circles. We estimate upper and lower bounds for the ground state energy using the analysis of Ref. \onlinecite{Tagliacozzo2009}. The upper bound indicated by a black horizontal line is the value of $E_0$ at $\chi=200$ while the lower bound is shown by a green line. In the plots we also write the exact numerical values for both upper and lower bounds. They provide narrow windows (in general around $0.01\%$ or less  of the exact value) in which the exact ground state energy is contained.} 
\label{fig:gde}
\end{center}
\end{figure}

Our scaling analysis consists of two parts. First we study, in a torus of linear size $L=8$, the convergence of the ground state energy as a function of $1/\chi$, see Fig. \ref{fig:gde}, for specific values of the magnetic field $h_x$, namely $h_x = \{0.1,0.2\}$ (deconfined phase), $h_x =\{0.5,0.6\}$ (spin polarized phase) and $h=0.33 \approx h_x^{\text{crit}}$ (critical point). From these results, lower and upper bounds to the ground state energy are obtained by observing that the ground state energy is a monotonic function of $1/\chi$, $E_0(1/\chi)$, with positive, monotonically increasing derivative \cite{Tagliacozzo2009}. Assuming that this trend continues all the way to $1/\chi=0$, this implies that $E_0(\chi=200)$ provides us with the best upper bound to the ground state energy, while a lower bound can be found with a linear extrapolation of $E_0(\chi=150)$ and $E_0(\chi=200)$ (the two best results) to $1/\chi =0$. For all the values of the magnetic field considered these bounds provide narrow windows (around $0.01\%$ of the actual value).

\begin{figure}
\begin{center}
\includegraphics[width=7cm]{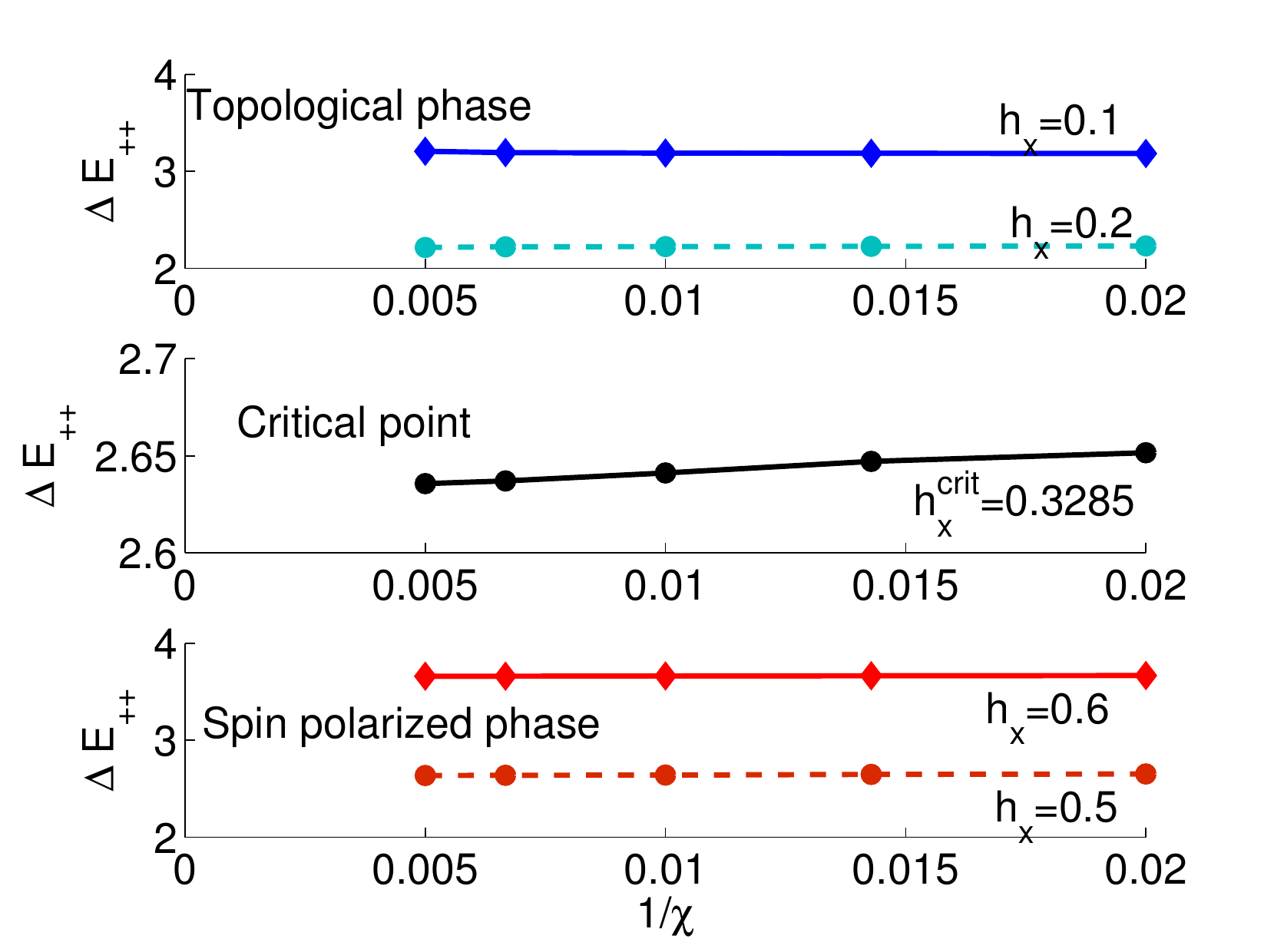} 
\caption{
Scaling of the gap $\Delta E_{++}$ for the deformed toric code on the $8 \times 8$ torus as a function of $1 / \chi$ for different values of the magnetic field in the topological phase, $h_x=0.1$ and $h_x=0.2$ (top panel), at the critical point , $h_x^{crit}=0.3285$ (central panel), and in the spin polarized phase, $h_x=0.5$ and $h_x=0.6$ (lower panel). The dependence of the gap as a function of $1 / \chi$ is not uniform as in the case of the ground state energy and thus we cannot extract rigid bounds on its value. However due to the small variations of the gap as a function of $\chi$ in the range  $50\le  \chi \le 200$ (in the worst case of the critical point of about $1\%$ of its value) compared  with its large variation as a function of the magnetic field $h_x$, we can neglect the dependence on $\chi$ of the gap provided we accept an uncertainty of about $1\%$ on the actual gap value. }
\label{fig:dgap}
\end{center}
\end{figure}

\begin{figure}
\begin{center}
\includegraphics[width=7cm]{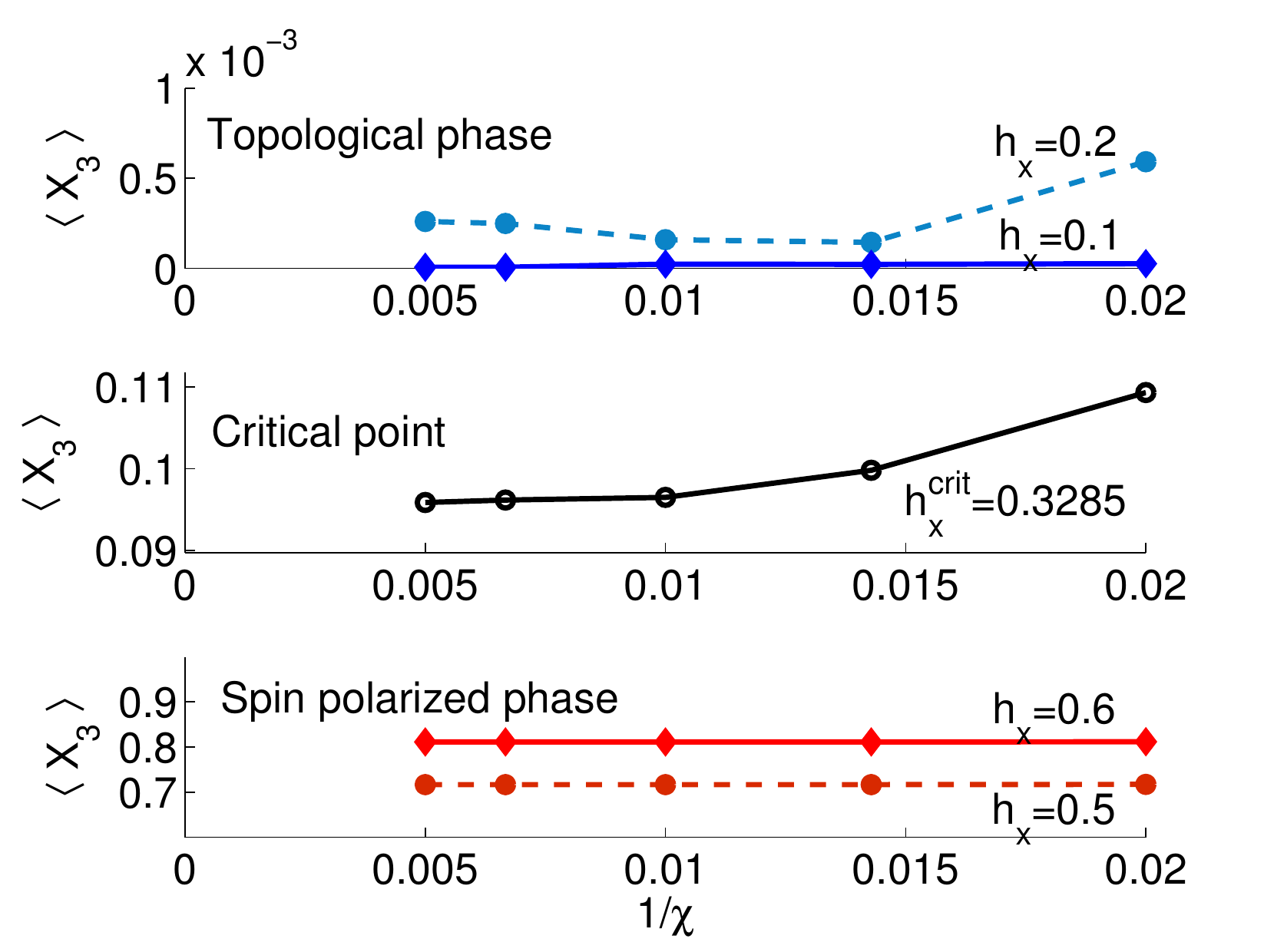}
 \caption{
Scaling of the disorder parameter $\langle X_3 \rangle$ for the deformed toric code on the $8 \times 8$ torus as a function of $1 / \chi$ for different values of the magnetic field in the topological phase, $h_x=0.1$ and $h_x=0.2$ (top panel), at the critical point, $h_x^{crit}=0.3285$ (central panel), and in the spin polarized phase, $h_x=0.5$ and $h_x=0.6$ (lower panel). 
On the one hand, the values in the deconfined phase (upper panel) are compatible with the theoretical expectation that the disorder parameter vanishes. However, they fluctuate substantially as a function of $\chi$ (we appreciate variations of around $30\%$ around the value of disorder parameter at $h_x=0.2$). We can still distinguish the results at $h_x=0.1$, where the disorder parameter fluctuates around $10^{-5}$, from those at $h_x=0.2$ where it fluctuates around $10^{-4}$. We believe that the origin of these large fluctuations is due not as much to the finite size of $\chi$ as to a lack of convergence of the iterative procedure used to compute the ground state, combined with the small values of the disordered paramter. Our iterative optimization procedure is stopped when the ground state energy has converged to about 10 digits. This criterion could be insufficient for a disorder parameter if the relevant critical exponents imply a much slower convergence than the ground state energy \cite{Sandvik2010}. On the other hand, both at the critical point (central panel) and in the spin polarized phase (lower panel), we extract again very accurate results and we can neglect the dependence on $\chi$ of the disorder parameter by introducing a systematic error lower than $1\%$ of its actual value.}
\label{fig:gdorpar}
\end{center}
\end{figure}

Once we have established the level of convergence of the ground state energy, we look at the dependence of other observables with $\chi$. Figs. \ref{fig:dgap} and \ref{fig:gdorpar} show the disorder parameter $\langle X_3 \rangle$ and the gap $\Delta E_{(+,+)}$ in the $(+,+)$ sector, respectively. In contrast with the ground state energy, which was monotonic in $1/\chi$, we see that $\langle X_3 \rangle$ and $\Delta E_{(+,+)}$ behave more erratically as a function of $1/\chi$. In this case, we can use the variations in value as a function of $1/\chi$ as an estimate of the error introduced by finite $\chi$. Importantly, this error is much smaller than the variation of $\langle X_3 \rangle$ and $\Delta E_{(+,+)}$ as a function of the magnetic field $h_x$. Further description is left to the captions of the figures.

\end{appendix}

\input{rev_MERAgauge2.bbl}

\begin{thebibliography}{100}%
\makeatletter
\providecommand \@ifxundefined [1]{%
 \ifx #1\undefined \expandafter \@firstoftwo
 \else \expandafter \@secondoftwo
\fi
}%
\providecommand \@ifnum [1]{%
 \ifnum #1\expandafter \@firstoftwo
 \else \expandafter \@secondoftwo
\fi
}%
\providecommand \enquote [1]{``#1''}%
\providecommand \bibnamefont  [1]{#1}%
\providecommand \bibfnamefont [1]{#1}%
\providecommand \citenamefont [1]{#1}%
\providecommand\href[0]{\@sanitize\@href}%
\providecommand\@href[1]{\endgroup\@@startlink{#1}\endgroup\@@href}%
\providecommand\@@href[1]{#1\@@endlink}%
\providecommand \@sanitize [0]{\begingroup\catcode`\&12\catcode`\#12\relax}%
\@ifxundefined \pdfoutput {\@firstoftwo}{%
 \@ifnum{\z@=\pdfoutput}{\@firstoftwo}{\@secondoftwo}%
}{%
 \providecommand\@@startlink[1]{\leavevmode\special{html:<a href="#1">}}%
 \providecommand\@@endlink[0]{\special{html:</a>}}%
}{%
 \providecommand\@@startlink[1]{%
  \leavevmode
  \pdfstartlink
   attr{/Border[0 0 1 ]/H/I/C[0 1 1]}%
   user{/Subtype/Link/A<</Type/Action/S/URI/URI(#1)>>}%
  \relax
 }%
 \providecommand\@@endlink[0]{\pdfendlink}%
}%
\providecommand \url  [0]{\begingroup\@sanitize \@url }%
\providecommand \@url [1]{\endgroup\@href {#1}{\urlprefix}}%
\providecommand \urlprefix [0]{URL }%
\providecommand \Eprint[0]{\href }%
\@ifxundefined \urlstyle {%
  \providecommand \doi [1]{doi:\discretionary{}{}{}#1}%
}{%
  \providecommand \doi [0]{doi:\discretionary{}{}{}\begingroup
  \urlstyle{rm}\Url }%
}%
\providecommand \doibase [0]{http://dx.doi.org/}%
\providecommand \Doi[1]{\href{\doibase#1}}%
\providecommand \bibAnnote [3]{%
  \BibitemShut{#1}%
  \begin{quotation}\noindent
    \textsc{Key:}\ #2\\\textsc{Annotation:}\ #3%
  \end{quotation}%
}%
\providecommand \bibAnnoteFile [2]{%
  \IfFileExists{#2}{\bibAnnote {#1} {#2} {\input{#2}}}{}%
}%
\providecommand \typeout [0]{\immediate \write \m@ne }%
\providecommand \selectlanguage [0]{\@gobble}%
\providecommand \bibinfo [0]{\@secondoftwo}%
\providecommand \bibfield [0]{\@secondoftwo}%
\providecommand \translation [1]{[#1]}%
\providecommand \BibitemOpen[0]{}%
\providecommand \bibitemStop [0]{}%
\providecommand \bibitemNoStop [0]{.\EOS\space}%
\providecommand \EOS [0]{\spacefactor3000\relax}%
\providecommand \BibitemShut [1]{\csname bibitem#1\endcsname}%
\bibitem{Misner1973}%
  \BibitemOpen
  \bibfield{author}{%
  \bibinfo {author} {\bibfnamefont{C.~W.}\ \bibnamefont{Misner}}, \bibinfo
  {author} {\bibfnamefont{K.~S.}\ \bibnamefont{Thorne}},\ and\ \bibinfo
  {author} {\bibfnamefont{J.~A.}\ \bibnamefont{Wheeler}},\ }%
  \emph{\bibinfo {title} {{Gravitation}}}\ (\bibinfo {address} {San
  Francisco},\ \bibinfo {year} {1973})%
  \bibAnnoteFile{NoStop}{Misner1973}%
\bibitem{Peskin1995}%
  \BibitemOpen
  \bibfield{author}{%
  \bibinfo {author} {\bibfnamefont{M.~E.}\ \bibnamefont{Peskin}}\ and\ \bibinfo
  {author} {\bibfnamefont{D.~V.}\ \bibnamefont{Schroeder}},\ }%
  \emph{\bibinfo {title} {{An Introduction to Quantum Field Theory}}}\
  (\bibinfo {publisher} {Addison-Wesley Advanced Book Program},\ \bibinfo
  {year} {1995})%
  \bibAnnoteFile{NoStop}{Peskin1995}%
\bibitem{Wilson1974LGT}%
  \BibitemOpen
  \bibfield{author}{%
  \bibinfo {author} {\bibfnamefont{K.~G.}\ \bibnamefont{Wilson}},\ }%
  \bibfield{journal}{%
  \bibinfo {journal} {Phys. Rev. D}\ }%
  \textbf{\bibinfo {volume} {10}},\ \bibinfo {pages} {2445} (\bibinfo {year}
  {1974})%
  \bibAnnoteFile{NoStop}{Wilson1974LGT}%
\bibitem{bali_qcd_2000}%
  \BibitemOpen
  \bibfield{author}{%
  \bibinfo {author} {\bibfnamefont{G.~S.}\ \bibnamefont{Bali}},\ }%
  \bibfield{journal}{%
  \bibinfo {journal} {{Phys. Rept.} 343 (2001) 1-136}}%
   (\bibinfo {year} {2001})%
  \bibAnnoteFile{NoStop}{bali_qcd_2000}%
\bibitem{jlqcd_collaboration_determination_2009}%
  \BibitemOpen
  \bibfield{author}{%
  \bibinfo {author} {\bibfnamefont{J.}~\bibnamefont{collaboration}}, \bibinfo
  {author} {\bibfnamefont{H.}~\bibnamefont{Fukaya}}, \bibinfo {author}
  {\bibfnamefont{S.}~\bibnamefont{Aoki}}, \bibinfo {author}
  {\bibfnamefont{S.}~\bibnamefont{Hashimoto}}, \bibinfo {author}
  {\bibfnamefont{T.}~\bibnamefont{Kaneko}}, \bibinfo {author}
  {\bibfnamefont{J.}~\bibnamefont{Noaki}}, \bibinfo {author}
  {\bibfnamefont{T.}~\bibnamefont{Onogi}},\ and\ \bibinfo {author}
  {\bibfnamefont{N.}~\bibnamefont{Yamada}},\ }%
  \bibfield{journal}{%
  \Doi{doi:10.1103/PhysRevLett.104.122002}{\bibinfo {journal}
  {Phys.Rev.Lett.}}\ }%
  \textbf{\bibinfo {volume} {104}},\ \bibinfo {pages} {122002} (\bibinfo
  {month} {nov}\ \bibinfo {year} {2009}),\ \bibinfo {note} {2010}%
  \bibAnnoteFile{NoStop}{jlqcd_collaboration_determination_2009}%
\bibitem{bazavov_full_2009}%
  \BibitemOpen
  \bibfield{author}{%
  \bibinfo {author} {\bibfnamefont{A.}~\bibnamefont{Bazavov}}, \bibinfo
  {author} {\bibfnamefont{C.}~\bibnamefont{Bernard}}, \bibinfo {author}
  {\bibfnamefont{C.}~\bibnamefont{{DeTar}}}, \bibinfo {author}
  {\bibfnamefont{S.}~\bibnamefont{Gottlieb}}, \bibinfo {author}
  {\bibfnamefont{U.~M.}\ \bibnamefont{Heller}}, \bibinfo {author}
  {\bibfnamefont{J.~E.}\ \bibnamefont{Hetrick}}, \bibinfo {author}
  {\bibfnamefont{J.}~\bibnamefont{Laiho}}, \bibinfo {author}
  {\bibfnamefont{L.}~\bibnamefont{Levkova}}, \bibinfo {author}
  {\bibfnamefont{P.~B.}\ \bibnamefont{Mackenzie}}, \bibinfo {author}
  {\bibfnamefont{M.~B.}\ \bibnamefont{Oktay}}, \bibinfo {author}
  {\bibfnamefont{R.}~\bibnamefont{Sugar}}, \bibinfo {author}
  {\bibfnamefont{D.}~\bibnamefont{Toussaint}},\ and\ \bibinfo {author}
  {\bibfnamefont{R.~S.~V.}\ \bibnamefont{de~Water}},\ }%
  \bibfield{journal}{%
  \bibinfo {journal} {0903.3598}}%
   (\bibinfo {month} {Mar.}\ \bibinfo {year} {2009}),\ \url{arXiv:0903.3598}%
  \bibAnnoteFile{NoStop}{bazavov_full_2009}%
\bibitem{ruester_phase_2005}%
  \BibitemOpen
  \bibfield{author}{%
  \bibinfo {author} {\bibfnamefont{S.~B.}\ \bibnamefont{Ruester}}, \bibinfo
  {author} {\bibfnamefont{V.}~\bibnamefont{Werth}}, \bibinfo {author}
  {\bibfnamefont{M.}~\bibnamefont{Buballa}}, \bibinfo {author}
  {\bibfnamefont{I.~A.}\ \bibnamefont{Shovkovy}},\ and\ \bibinfo {author}
  {\bibfnamefont{D.~H.}\ \bibnamefont{Rischke}},\ }%
  \bibfield{journal}{%
  \Doi{10.1103/PhysRevD.72.034004}{\bibinfo {journal} {Phys. Rev.}}\ }%
  \textbf{\bibinfo {volume} {D72}},\ \bibinfo {pages} {034004} (\bibinfo {year}
  {2005})%
  \bibAnnoteFile{NoStop}{ruester_phase_2005}%
\bibitem{Affleck1988}%
  \BibitemOpen
  \bibfield{author}{%
  \bibinfo {author} {\bibfnamefont{I.}~\bibnamefont{Affleck}}, \bibinfo
  {author} {\bibfnamefont{T.}~\bibnamefont{Kennedy}}, \bibinfo {author}
  {\bibfnamefont{E.~H.}\ \bibnamefont{Lieb}},\ and\ \bibinfo {author}
  {\bibfnamefont{H.}~\bibnamefont{Tasaki}},\ }%
  \bibfield{journal}{%
  \bibinfo {journal} {Commun. Math. Phys.}\ }%
  \textbf{\bibinfo {volume} {115}},\ \bibinfo {pages} {477 } (\bibinfo {year}
  {1988})%
  \bibAnnoteFile{NoStop}{Affleck1988}%
\bibitem{Fannes1992}%
  \BibitemOpen
  \bibfield{author}{%
  \bibinfo {author} {\bibfnamefont{M.}~\bibnamefont{Fannes}}, \bibinfo {author}
  {\bibfnamefont{B.}~\bibnamefont{Nachtergaele}},\ and\ \bibinfo {author}
  {\bibfnamefont{R.~F.}\ \bibnamefont{Werner}},\ }%
  \bibfield{journal}{%
  \bibinfo {journal} {Comm. Math. Phys.}\ }%
  \textbf{\bibinfo {volume} {144}},\ \bibinfo {pages} {443} (\bibinfo {year}
  {1992})%
  \bibAnnoteFile{NoStop}{Fannes1992}%
\bibitem{Ostlund1995}%
  \BibitemOpen
  \bibfield{author}{%
  \bibinfo {author} {\bibfnamefont{S.}~\bibnamefont{\"{O}stlund}}\ and\
  \bibinfo {author} {\bibfnamefont{S.}~\bibnamefont{Rommer}},\ }%
  \bibfield{journal}{%
  \bibinfo {journal} {Phys. Rev. Lett.}\ }%
  \textbf{\bibinfo {volume} {75}},\ \bibinfo {pages} {3537} (\bibinfo {year}
  {1995})%
  \bibAnnoteFile{NoStop}{Ostlund1995}%
\bibitem{White1992}%
  \BibitemOpen
  \bibfield{author}{%
  \bibinfo {author} {\bibfnamefont{S.~R.}\ \bibnamefont{White}},\ }%
  \bibfield{journal}{%
  \bibinfo {journal} {Phys. Rev. Lett.}\ }%
  \textbf{\bibinfo {volume} {69}},\ \bibinfo {pages} {2863} (\bibinfo {year}
  {1992})%
  \bibAnnoteFile{NoStop}{White1992}%
\bibitem{White1993}%
  \BibitemOpen
  \bibfield{author}{%
  \bibinfo {author} {\bibfnamefont{S.~R.}\ \bibnamefont{White}},\ }%
  \bibfield{journal}{%
  \bibinfo {journal} {Phys. Rev. B}\ }%
  \textbf{\bibinfo {volume} {48}},\ \bibinfo {pages} {10345} (\bibinfo {year}
  {1993})%
  \bibAnnoteFile{NoStop}{White1993}%
\bibitem{Schollwock2005}%
  \BibitemOpen
  \bibfield{author}{%
  \bibinfo {author} {\bibfnamefont{U.}~\bibnamefont{Schollw\"{o}ck}},\ }%
  \bibfield{journal}{%
  \bibinfo {journal} {Rev. Mod. Phys.}\ }%
  \textbf{\bibinfo {volume} {77}},\ \bibinfo {pages} {259} (\bibinfo {year}
  {2005})%
  \bibAnnoteFile{NoStop}{Schollwock2005}%
\bibitem{Verstraete2004}%
  \BibitemOpen
  \bibfield{author}{%
  \bibinfo {author} {\bibfnamefont{F.}~\bibnamefont{Verstraete}}\ and\ \bibinfo
  {author} {\bibfnamefont{J.~I.}\ \bibnamefont{Cirac}},\ }%
  \bibfield{journal}{%
  \bibinfo {journal} {arXiv:cond-mat/0407066v1}}%
   (\bibinfo {year} {2004})%
  \bibAnnoteFile{NoStop}{Verstraete2004}%
\bibitem{Jordan2008}%
  \BibitemOpen
  \bibfield{author}{%
  \bibinfo {author} {\bibfnamefont{J.}~\bibnamefont{Jordan}}, \bibinfo {author}
  {\bibfnamefont{R.}~\bibnamefont{Or\'{u}s}}, \bibinfo {author}
  {\bibfnamefont{G.}~\bibnamefont{Vidal}}, \bibinfo {author}
  {\bibfnamefont{F.}~\bibnamefont{Verstraete}},\ and\ \bibinfo {author}
  {\bibfnamefont{J.~I.}\ \bibnamefont{Cirac}},\ }%
  \bibfield{journal}{%
  \bibinfo {journal} {Phys. Rev. Lett.}\ }%
  \textbf{\bibinfo {volume} {101}},\ \bibinfo {pages} {250602} (\bibinfo {year}
  {2008})%
  \bibAnnoteFile{NoStop}{Jordan2008}%
\bibitem{Verstraete2008}%
  \BibitemOpen
  \bibfield{author}{%
  \bibinfo {author} {\bibfnamefont{F.}~\bibnamefont{Verstraete}}, \bibinfo
  {author} {\bibfnamefont{V.}~\bibnamefont{Murg}},\ and\ \bibinfo {author}
  {\bibfnamefont{J.~I.}\ \bibnamefont{Cirac}},\ }%
  \bibfield{journal}{%
  \bibinfo {journal} {Advance in Physics}\ }%
  \textbf{\bibinfo {volume} {57}},\ \bibinfo {pages} {143 } (\bibinfo {year}
  {2008})%
  \bibAnnoteFile{NoStop}{Verstraete2008}%
\bibitem{Gu2008}%
  \BibitemOpen
  \bibfield{author}{%
  \bibinfo {author} {\bibfnamefont{Z.-C.}\ \bibnamefont{Gu}}, \bibinfo {author}
  {\bibfnamefont{M.}~\bibnamefont{Levin}},\ and\ \bibinfo {author}
  {\bibfnamefont{X.-G.}\ \bibnamefont{Wen}},\ }%
  \bibfield{journal}{%
  \bibinfo {journal} {Phys. Rev. B}\ }%
  \textbf{\bibinfo {volume} {78}},\ \bibinfo {pages} {205116} (\bibinfo {year}
  {2008})%
  \bibAnnoteFile{NoStop}{Gu2008}%
\bibitem{Gu2009}%
  \BibitemOpen
  \bibfield{author}{%
  \bibinfo {author} {\bibfnamefont{Z.-C.}\ \bibnamefont{Gu}}\ and\ \bibinfo
  {author} {\bibfnamefont{X.-G.}\ \bibnamefont{Wen}},\ }%
  \bibfield{journal}{%
  \bibinfo {journal} {Phys. Rev. B}\ }%
  \textbf{\bibinfo {volume} {80}},\ \bibinfo {pages} {155131} (\bibinfo {year}
  {2009})%
  \bibAnnoteFile{NoStop}{Gu2009}%
\bibitem{Murg2009}%
  \BibitemOpen
  \bibfield{author}{%
  \bibinfo {author} {\bibfnamefont{V.}~\bibnamefont{Murg}}, \bibinfo {author}
  {\bibfnamefont{F.}~\bibnamefont{Verstraete}},\ and\ \bibinfo {author}
  {\bibfnamefont{J.~I.}\ \bibnamefont{Cirac}},\ }%
  \bibfield{journal}{%
  \bibinfo {journal} {Phys. Rev. B}\ }%
  \textbf{\bibinfo {volume} {79}},\ \bibinfo {pages} {195119} (\bibinfo {year}
  {2009})%
  \bibAnnoteFile{NoStop}{Murg2009}%
\bibitem{Xie2009}%
  \BibitemOpen
  \bibfield{author}{%
  \bibinfo {author} {\bibfnamefont{Z.~Y.}\ \bibnamefont{Xie}}, \bibinfo
  {author} {\bibfnamefont{H.~C.}\ \bibnamefont{Jiang}}, \bibinfo {author}
  {\bibfnamefont{Q.~N.}\ \bibnamefont{Chen}}, \bibinfo {author}
  {\bibfnamefont{Z.~Y.}\ \bibnamefont{Weng}},\ and\ \bibinfo {author}
  {\bibfnamefont{T.}~\bibnamefont{Xiang}},\ }%
  \bibfield{journal}{%
  \bibinfo {journal} {Phys. Rev. Lett.}\ }%
  \textbf{\bibinfo {volume} {103}},\ \bibinfo {pages} {160601} (\bibinfo {year}
  {2009})%
  \bibAnnoteFile{NoStop}{Xie2009}%
\bibitem{Chen2009}%
  \BibitemOpen
  \bibfield{author}{%
  \bibinfo {author} {\bibfnamefont{P.}~\bibnamefont{Chen}}, \bibinfo {author}
  {\bibfnamefont{C.-Y.}\ \bibnamefont{Lai}},\ and\ \bibinfo {author}
  {\bibfnamefont{M.-F.}\ \bibnamefont{Yang}},\ }%
  \bibfield{journal}{%
  \bibinfo {journal} {J. Stat. Mech.: Theor. Exp.}\ }%
  \textbf{\bibinfo {volume} {2009}},\ \bibinfo {pages} {P10001} (\bibinfo
  {year} {2009})%
  \bibAnnoteFile{NoStop}{Chen2009}%
\bibitem{Zhao2010}%
  \BibitemOpen
  \bibfield{author}{%
  \bibinfo {author} {\bibfnamefont{H.~H.}\ \bibnamefont{Zhao}}, \bibinfo
  {author} {\bibfnamefont{Z.~Y.}\ \bibnamefont{Xie}}, \bibinfo {author}
  {\bibfnamefont{Q.~N.}\ \bibnamefont{Chen}}, \bibinfo {author}
  {\bibfnamefont{Z.~C.}\ \bibnamefont{Wei}}, \bibinfo {author}
  {\bibfnamefont{J.~W.}\ \bibnamefont{Cai}},\ and\ \bibinfo {author}
  {\bibfnamefont{T.}~\bibnamefont{Xiang}},\ }%
  \bibfield{journal}{%
  \bibinfo {journal} {Phys. Rev. B}\ }%
  \textbf{\bibinfo {volume} {81}},\ \bibinfo {pages} {174411} (\bibinfo {year}
  {2010})%
  \bibAnnoteFile{NoStop}{Zhao2010}%
\bibitem{Kraus2010}%
  \BibitemOpen
  \bibfield{author}{%
  \bibinfo {author} {\bibfnamefont{C.~V.}\ \bibnamefont{Kraus}}, \bibinfo
  {author} {\bibfnamefont{N.}~\bibnamefont{Schuch}}, \bibinfo {author}
  {\bibfnamefont{F.}~\bibnamefont{Verstraete}},\ and\ \bibinfo {author}
  {\bibfnamefont{J.~I.}\ \bibnamefont{Cirac}},\ }%
  \bibfield{journal}{%
  \bibinfo {journal} {Phys. Rev. A}\ }%
  \textbf{\bibinfo {volume} {81}},\ \bibinfo {pages} {52338} (\bibinfo {year}
  {2010})%
  \bibAnnoteFile{NoStop}{Kraus2010}%
\bibitem{Corboz2010fPEPS}%
  \BibitemOpen
  \bibfield{author}{%
  \bibinfo {author} {\bibfnamefont{P.}~\bibnamefont{Corboz}}, \bibinfo {author}
  {\bibfnamefont{R.}~\bibnamefont{Or\'{u}s}}, \bibinfo {author}
  {\bibfnamefont{B.}~\bibnamefont{Bauer}},\ and\ \bibinfo {author}
  {\bibfnamefont{G.}~\bibnamefont{Vidal}},\ }%
  \bibfield{journal}{%
  \bibinfo {journal} {Phys. Rev. B}\ }%
  \textbf{\bibinfo {volume} {81}},\ \bibinfo {pages} {165104} (\bibinfo {year}
  {2010})%
  \bibAnnoteFile{NoStop}{Corboz2010fPEPS}%
\bibitem{Barthel2009}%
  \BibitemOpen
  \bibfield{author}{%
  \bibinfo {author} {\bibfnamefont{T.}~\bibnamefont{Barthel}}, \bibinfo
  {author} {\bibfnamefont{C.}~\bibnamefont{Pineda}},\ and\ \bibinfo {author}
  {\bibfnamefont{J.}~\bibnamefont{Eisert}},\ }%
  \bibfield{journal}{%
  \bibinfo {journal} {Phys. Rev. A}\ }%
  \textbf{\bibinfo {volume} {80}},\ \bibinfo {pages} {42333} (\bibinfo {year}
  {2009})%
  \bibAnnoteFile{NoStop}{Barthel2009}%
\bibitem{sierra_density_1998}%
  \BibitemOpen
  \bibfield{author}{%
  \bibinfo {author} {\bibfnamefont{G.}~\bibnamefont{Sierra}}\ and\ \bibinfo
  {author} {\bibfnamefont{M.~A.}\ \bibnamefont{{Martin-Delgado}}},\ }%
  \bibfield{journal}{%
  \bibinfo {journal} {arXiv:cond-mat/9811170}}%
   (\bibinfo {month} {Nov.}\ \bibinfo {year} {1998})%
  \bibAnnoteFile{NoStop}{sierra_density_1998}%
\bibitem{maeshima_vertical_2001}%
  \BibitemOpen
  \bibfield{author}{%
  \bibinfo {author} {\bibfnamefont{N.}~\bibnamefont{Maeshima}}, \bibinfo
  {author} {\bibfnamefont{Y.}~\bibnamefont{Hieida}}, \bibinfo {author}
  {\bibfnamefont{Y.}~\bibnamefont{Akutsu}}, \bibinfo {author}
  {\bibfnamefont{T.}~\bibnamefont{Nishino}},\ and\ \bibinfo {author}
  {\bibfnamefont{K.}~\bibnamefont{Okunishi}},\ }%
  \bibfield{journal}{%
  \bibinfo {journal} {cond-mat/0101360}}%
   (\bibinfo {year} {2001}),\ \bibinfo {note} {{Phys.Rev.E64:016705,2001}}%
  \bibAnnoteFile{NoStop}{maeshima_vertical_2001}%
\bibitem{nishio_tensor_2004}%
  \BibitemOpen
  \bibfield{author}{%
  \bibinfo {author} {\bibfnamefont{Y.}~\bibnamefont{Nishio}}, \bibinfo {author}
  {\bibfnamefont{N.}~\bibnamefont{Maeshima}}, \bibinfo {author}
  {\bibfnamefont{A.}~\bibnamefont{Gendiar}},\ and\ \bibinfo {author}
  {\bibfnamefont{T.}~\bibnamefont{Nishino}},\ }%
  \bibfield{journal}{%
  \bibinfo {journal} {arXiv:cond-mat/0401115}}%
   (\bibinfo {year} {2004})%
  \bibAnnoteFile{NoStop}{nishio_tensor_2004}%
\bibitem{Vidal2007ER}%
  \BibitemOpen
  \bibfield{author}{%
  \bibinfo {author} {\bibfnamefont{G.}~\bibnamefont{Vidal}},\ }%
  \bibfield{journal}{%
  \bibinfo {journal} {Phys. Rev. Lett.}\ }%
  \textbf{\bibinfo {volume} {99}},\ \bibinfo {pages} {220405} (\bibinfo {year}
  {2007})%
  \bibAnnoteFile{NoStop}{Vidal2007ER}%
\bibitem{Vidal2008MERA}%
  \BibitemOpen
  \bibfield{author}{%
  \bibinfo {author} {\bibfnamefont{G.}~\bibnamefont{Vidal}},\ }%
  \bibfield{journal}{%
  \bibinfo {journal} {Phys. Rev. Lett.}\ }%
  \textbf{\bibinfo {volume} {101}},\ \bibinfo {pages} {110501} (\bibinfo {year}
  {2008})%
  \bibAnnoteFile{NoStop}{Vidal2008MERA}%
\bibitem{Cincio2008}%
  \BibitemOpen
  \bibfield{author}{%
  \bibinfo {author} {\bibfnamefont{L.}~\bibnamefont{Cincio}}, \bibinfo {author}
  {\bibfnamefont{J.}~\bibnamefont{Dziarmaga}},\ and\ \bibinfo {author}
  {\bibfnamefont{M.~M.}\ \bibnamefont{Rams}},\ }%
  \bibfield{journal}{%
  \bibinfo {journal} {Phys. Rev. Lett.}\ }%
  \textbf{\bibinfo {volume} {100}},\ \bibinfo {pages} {240603} (\bibinfo {year}
  {2008})%
  \bibAnnoteFile{NoStop}{Cincio2008}%
\bibitem{Evenbly20092DIsing}%
  \BibitemOpen
  \bibfield{author}{%
  \bibinfo {author} {\bibfnamefont{G.}~\bibnamefont{Evenbly}}\ and\ \bibinfo
  {author} {\bibfnamefont{G.}~\bibnamefont{Vidal}},\ }%
  \bibfield{journal}{%
  \bibinfo {journal} {Phys. Rev. Lett.}\ }%
  \textbf{\bibinfo {volume} {102}},\ \bibinfo {pages} {180406} (\bibinfo {year}
  {2009})%
  \bibAnnoteFile{NoStop}{Evenbly20092DIsing}%
\bibitem{Evenbly2009Alg}%
  \BibitemOpen
  \bibfield{author}{%
  \bibinfo {author} {\bibfnamefont{G.}~\bibnamefont{Evenbly}}\ and\ \bibinfo
  {author} {\bibfnamefont{G.}~\bibnamefont{Vidal}},\ }%
  \bibfield{journal}{%
  \bibinfo {journal} {Phys. Rev. B}\ }%
  \textbf{\bibinfo {volume} {79}},\ \bibinfo {pages} {144108} (\bibinfo {year}
  {2009})%
  \bibAnnoteFile{NoStop}{Evenbly2009Alg}%
\bibitem{Giovannetti2009}%
  \BibitemOpen
  \bibfield{author}{%
  \bibinfo {author} {\bibfnamefont{V.}~\bibnamefont{Giovannetti}}, \bibinfo
  {author} {\bibfnamefont{S.}~\bibnamefont{Montangero}}, \bibinfo {author}
  {\bibfnamefont{M.}~\bibnamefont{Rizzi}},\ and\ \bibinfo {author}
  {\bibfnamefont{R.}~\bibnamefont{Fazio}},\ }%
  \bibfield{journal}{%
  \bibinfo {journal} {Phys. Rev. A}\ }%
  \textbf{\bibinfo {volume} {79}},\ \bibinfo {pages} {52314} (\bibinfo {year}
  {2009})%
  \bibAnnoteFile{NoStop}{Giovannetti2009}%
\bibitem{Evenbly2010Kagome}%
  \BibitemOpen
  \bibfield{author}{%
  \bibinfo {author} {\bibfnamefont{G.}~\bibnamefont{Evenbly}}\ and\ \bibinfo
  {author} {\bibfnamefont{G.}~\bibnamefont{Vidal}},\ }%
  \bibfield{journal}{%
  \bibinfo {journal} {Phys. Rev. Lett.}\ }%
  \textbf{\bibinfo {volume} {104}},\ \bibinfo {pages} {187203} (\bibinfo {year}
  {2010})%
  \bibAnnoteFile{NoStop}{Evenbly2010Kagome}%
\bibitem{Corboz2009fMERA}%
  \BibitemOpen
  \bibfield{author}{%
  \bibinfo {author} {\bibfnamefont{P.}~\bibnamefont{Corboz}}\ and\ \bibinfo
  {author} {\bibfnamefont{G.}~\bibnamefont{Vidal}},\ }%
  \bibfield{journal}{%
  \bibinfo {journal} {Phys. Rev. B}\ }%
  \textbf{\bibinfo {volume} {80}},\ \bibinfo {pages} {165129} (\bibinfo {year}
  {2009})%
  \bibAnnoteFile{NoStop}{Corboz2009fMERA}%
\bibitem{Pineda2010}%
  \BibitemOpen
  \bibfield{author}{%
  \bibinfo {author} {\bibfnamefont{C.}~\bibnamefont{Pineda}}, \bibinfo {author}
  {\bibfnamefont{T.}~\bibnamefont{Barthel}},\ and\ \bibinfo {author}
  {\bibfnamefont{J.}~\bibnamefont{Eisert}},\ }%
  \bibfield{journal}{%
  \bibinfo {journal} {Phys. Rev. A}\ }%
  \textbf{\bibinfo {volume} {81}},\ \bibinfo {pages} {50303} (\bibinfo {year}
  {2010})%
  \bibAnnoteFile{NoStop}{Pineda2010}%
\bibitem{Mezzacapo2010}%
  \BibitemOpen
  \bibfield{author}{%
  \bibinfo {author} {\bibfnamefont{F.}~\bibnamefont{Mezzacapo}}\ and\ \bibinfo
  {author} {\bibfnamefont{J.~I.}\ \bibnamefont{Cirac}},\ }%
  \bibfield{journal}{%
  \bibinfo {journal} {arXiv:1006.4480v2}}%
   (\bibinfo {year} {2010})%
  \bibAnnoteFile{NoStop}{Mezzacapo2010}%
\bibitem{Li2010}%
  \BibitemOpen
  \bibfield{author}{%
  \bibinfo {author} {\bibfnamefont{S.-H.}\ \bibnamefont{Li}}, \bibinfo {author}
  {\bibfnamefont{Q.-Q.}\ \bibnamefont{Shi}},\ and\ \bibinfo {author}
  {\bibfnamefont{H.-Q.}\ \bibnamefont{Zhou}},\ }%
  \bibfield{journal}{%
  \bibinfo {journal} {arXiv:1001.3343v1}}%
   (\bibinfo {year} {2010})%
  \bibAnnoteFile{NoStop}{Li2010}%
\bibitem{Wegner1971}%
  \BibitemOpen
  \bibfield{author}{%
  \bibinfo {author} {\bibfnamefont{F.}~\bibnamefont{Wegner}},\ }%
  \bibfield{journal}{%
  \bibinfo {journal} {J. Math. Phys.}\ }%
  \textbf{\bibinfo {volume} {12}},\ \bibinfo {pages} {2259} (\bibinfo {year}
  {1971})%
  \bibAnnoteFile{NoStop}{Wegner1971}%
\bibitem{Horn1979}%
  \BibitemOpen
  \bibfield{author}{%
  \bibinfo {author} {\bibfnamefont{D.}~\bibnamefont{Horn}}, \bibinfo {author}
  {\bibfnamefont{M.}~\bibnamefont{Weinstein}},\ and\ \bibinfo {author}
  {\bibfnamefont{S.}~\bibnamefont{Yankielowicz}},\ }%
  \bibfield{journal}{%
  \bibinfo {journal} {Phys. Rev. D}\ }%
  \textbf{\bibinfo {volume} {19}},\ \bibinfo {pages} {3715} (\bibinfo {year}
  {1979})%
  \bibAnnoteFile{NoStop}{Horn1979}%
\bibitem{Kogut1979RevModPhys}%
  \BibitemOpen
  \bibfield{author}{%
  \bibinfo {author} {\bibfnamefont{J.~B.}\ \bibnamefont{Kogut}},\ }%
  \bibfield{journal}{%
  \bibinfo {journal} {Reviews of Modern Physics}\ }%
  \textbf{\bibinfo {volume} {51}},\ \bibinfo {pages} {659} (\bibinfo {year}
  {1979})%
  \bibAnnoteFile{NoStop}{Kogut1979RevModPhys}%
\bibitem{wen_topological_1995}%
  \BibitemOpen
  \bibfield{author}{%
  \bibinfo {author} {\bibfnamefont{X.}~\bibnamefont{Wen}},\ }%
  \bibfield{journal}{%
  \bibinfo {journal} {Advances in Physics}\ }%
  \textbf{\bibinfo {volume} {44}},\ \bibinfo {pages} {405} (\bibinfo {year}
  {1995}),\ ISSN \bibinfo {issn} {0001-8732}%
  \bibAnnoteFile{NoStop}{wen_topological_1995}%
\bibitem{Kitaev2003ToricCode}%
  \BibitemOpen
  \bibfield{author}{%
  \bibinfo {author} {\bibfnamefont{A.~Y.}\ \bibnamefont{Kitaev}},\ }%
  \bibfield{journal}{%
  \bibinfo {journal} {Ann. Phys.}\ }%
  \textbf{\bibinfo {volume} {303}},\ \bibinfo {pages} {2} (\bibinfo {year}
  {2003})%
  \bibAnnoteFile{NoStop}{Kitaev2003ToricCode}%
\bibitem{Sachdev_1991}%
  \BibitemOpen
  \bibfield{author}{%
  \bibinfo {author} {\bibfnamefont{N.}~\bibnamefont{Read}}\ and\ \bibinfo
  {author} {\bibfnamefont{S.}~\bibnamefont{Sachdev}},\ }%
  \bibfield{journal}{%
  \Doi{10.1103/PhysRevLett.66.1773}{\bibinfo {journal} {Phys. Rev. Lett.}}\ }%
  \textbf{\bibinfo {volume} {66}},\ \bibinfo {pages} {1773} (\bibinfo {month}
  {Apr}\ \bibinfo {year} {1991})%
  \bibAnnoteFile{NoStop}{Sachdev_1991}%
\bibitem{Wen_1991}%
  \BibitemOpen
  \bibfield{author}{%
  \bibinfo {author} {\bibfnamefont{X.~G.}\ \bibnamefont{Wen}},\ }%
  \bibfield{journal}{%
  \Doi{10.1103/PhysRevB.44.2664}{\bibinfo {journal} {Phys. Rev. B}}\ }%
  \textbf{\bibinfo {volume} {44}},\ \bibinfo {pages} {2664} (\bibinfo {month}
  {Aug}\ \bibinfo {year} {1991})%
  \bibAnnoteFile{NoStop}{Wen_1991}%
\bibitem{Sachdev_1991b}%
  \BibitemOpen
  \bibfield{author}{%
  \bibinfo {author} {\bibfnamefont{N.}~\bibnamefont{Read}}\ and\ \bibinfo
  {author} {\bibfnamefont{S.}~\bibnamefont{Sachdev}},\ }%
  \bibfield{journal}{%
  \bibinfo {journal} {International Journal of Modern Physics B 5, 219}}%
   (\bibinfo {year} {1991}),\ \url{arXiv:cond-mat/0402109v1}%
  \bibAnnoteFile{NoStop}{Sachdev_1991b}%
\bibitem{Sachdev_2008}%
  \BibitemOpen
  \bibfield{author}{%
  \bibinfo {author} {\bibfnamefont{S.}~\bibnamefont{Sachdev}},\ }%
  \bibfield{journal}{%
  \bibinfo {journal} {Lectures at the Les Houches School on "Modern theories of
  correlated electron systems", France, May 2009}}%
   (\bibinfo {year} {2009}),\ \url{arXiv:1002.3823v2}%
  \bibAnnoteFile{NoStop}{Sachdev_2008}%
\bibitem{Trebst2007BreakDown}%
  \BibitemOpen
  \bibfield{author}{%
  \bibinfo {author} {\bibfnamefont{S.}~\bibnamefont{Trebst}}, \bibinfo {author}
  {\bibfnamefont{P.}~\bibnamefont{Werner}}, \bibinfo {author}
  {\bibfnamefont{M.}~\bibnamefont{Troyer}}, \bibinfo {author}
  {\bibfnamefont{K.}~\bibnamefont{Shtengel}},\ and\ \bibinfo {author}
  {\bibfnamefont{C.}~\bibnamefont{Nayak}},\ }%
  \bibfield{journal}{%
  \bibinfo {journal} {Phys. Rev. Lett.}\ }%
  \textbf{\bibinfo {volume} {98}},\ \bibinfo {pages} {70602} (\bibinfo {year}
  {2007})%
  \bibAnnoteFile{NoStop}{Trebst2007BreakDown}%
\bibitem{hamma_2008}%
  \BibitemOpen
  \bibfield{author}{%
  \bibinfo {author} {\bibfnamefont{A.}~\bibnamefont{Hamma}}\ and\ \bibinfo
  {author} {\bibfnamefont{D.~A.}\ \bibnamefont{Lidar}},\ }%
  \bibfield{journal}{%
  \Doi{10.1103/PhysRevLett.100.030502}{\bibinfo {journal} {Phys. Rev. Lett.}}\
  }%
  \textbf{\bibinfo {volume} {100}},\ \bibinfo {pages} {030502} (\bibinfo
  {month} {Jan}\ \bibinfo {year} {2008})%
  \bibAnnoteFile{NoStop}{hamma_2008}%
\bibitem{Sugihara2005}%
  \BibitemOpen
  \bibfield{author}{%
  \bibinfo {author} {\bibfnamefont{T.}~\bibnamefont{Sugihara}},\ }%
  \bibfield{journal}{%
  \bibinfo {journal} {JHEP}\ }%
  \textbf{\bibinfo {volume} {2005}},\ \bibinfo {pages} {22} (\bibinfo {year}
  {2005})%
  \bibAnnoteFile{NoStop}{Sugihara2005}%
\bibitem{Schuch2010}%
  \BibitemOpen
  \bibfield{author}{%
  \bibinfo {author} {\bibfnamefont{N.}~\bibnamefont{Schuch}}, \bibinfo {author}
  {\bibfnamefont{I.}~\bibnamefont{Cirac}},\ and\ \bibinfo {author}
  {\bibfnamefont{D.}~\bibnamefont{Perez-Garcia}},\ }%
  \bibfield{journal}{%
  \bibinfo {journal} {arXiv:1001.3807v2}}%
   (\bibinfo {year} {2010})%
  \bibAnnoteFile{NoStop}{Schuch2010}%
\bibitem{Swingle2010}%
  \BibitemOpen
  \bibfield{author}{%
  \bibinfo {author} {\bibfnamefont{B.}~\bibnamefont{Swingle}}\ and\ \bibinfo
  {author} {\bibfnamefont{X.-G.}\ \bibnamefont{Wen}},\ }%
  \bibfield{journal}{%
  \bibinfo {journal} {arXiv:1001.4517v1}}%
   (\bibinfo {year} {2010})%
  \bibAnnoteFile{NoStop}{Swingle2010}%
\bibitem{Chen2010Symmetry}%
  \BibitemOpen
  \bibfield{author}{%
  \bibinfo {author} {\bibfnamefont{X.}~\bibnamefont{Chen}}, \bibinfo {author}
  {\bibfnamefont{B.}~\bibnamefont{Zeng}}, \bibinfo {author}
  {\bibfnamefont{Z.-C.}\ \bibnamefont{Gu}}, \bibinfo {author}
  {\bibfnamefont{I.~L.}\ \bibnamefont{Chuang}},\ and\ \bibinfo {author}
  {\bibfnamefont{X.-G.}\ \bibnamefont{Wen}},\ }%
  \bibfield{journal}{%
  \bibinfo {journal} {arXiv:1003.1774v1}}%
   (\bibinfo {year} {2010})%
  \bibAnnoteFile{NoStop}{Chen2010Symmetry}%
\bibitem{Hamer2000}%
  \BibitemOpen
  \bibfield{author}{%
  \bibinfo {author} {\bibfnamefont{C.~J.}\ \bibnamefont{Hamer}},\ }%
  \bibfield{journal}{%
  \bibinfo {journal} {J. Phys. A: Math. Gen.}\ }%
  \textbf{\bibinfo {volume} {33}},\ \bibinfo {pages} {6683} (\bibinfo {year}
  {2000})%
  \bibAnnoteFile{NoStop}{Hamer2000}%
\bibitem{mcculloc_2002}%
  \BibitemOpen
  \bibfield{author}{%
  \bibinfo {author} {\bibfnamefont{I.~P.}\ \bibnamefont{McCulloch}}\ and\
  \bibinfo {author} {\bibfnamefont{M.}~\bibnamefont{Gulacsi}},\ }%
  \bibfield{journal}{%
  \bibinfo {journal} {EPL (Europhysics Letters)}\ }%
  \textbf{\bibinfo {volume} {57}},\ \bibinfo {pages} {852} (\bibinfo {year}
  {2002})%
  \bibAnnoteFile{NoStop}{mcculloc_2002}%
\bibitem{Sierra1997505}%
  \BibitemOpen
  \bibfield{author}{%
  \bibinfo {author} {\bibfnamefont{G.}~\bibnamefont{Sierra}}\ and\ \bibinfo
  {author} {\bibfnamefont{T.}~\bibnamefont{Nishino}},\ }%
  \bibfield{journal}{%
  \Doi{DOI: 10.1016/S0550-3213(97)00217-4}{\bibinfo {journal} {Nuclear Physics
  B}}\ }%
  \textbf{\bibinfo {volume} {495}},\ \bibinfo {pages} {505 } (\bibinfo {year}
  {1997}),\ ISSN \bibinfo {issn} {0550-3213}%
  \bibAnnoteFile{NoStop}{Sierra1997505}%
\bibitem{Buividovich2008141}%
  \BibitemOpen
  \bibfield{author}{%
  \bibinfo {author} {\bibfnamefont{P.}~\bibnamefont{Buividovich}}\ and\
  \bibinfo {author} {\bibfnamefont{M.}~\bibnamefont{Polikarpov}},\ }%
  \bibfield{journal}{%
  \bibinfo {journal} {Physics Letters B}\ }%
  \textbf{\bibinfo {volume} {670}},\ \bibinfo {pages} {141 } (\bibinfo {year}
  {2008}),\ ISSN \bibinfo {issn} {0370-2693}%
  \bibAnnoteFile{NoStop}{Buividovich2008141}%
\bibitem{kadanoff_1971}%
  \BibitemOpen
  \bibfield{author}{%
  \bibinfo {author} {\bibfnamefont{L.~P.}\ \bibnamefont{Kadanoff}}\ and\
  \bibinfo {author} {\bibfnamefont{H.}~\bibnamefont{Ceva}},\ }%
  \bibfield{journal}{%
  \Doi{10.1103/PhysRevB.3.3918}{\bibinfo {journal} {Phys. Rev. B}}\ }%
  \textbf{\bibinfo {volume} {3}},\ \bibinfo {pages} {3918} (\bibinfo {month}
  {Jun}\ \bibinfo {year} {1971})%
  \bibAnnoteFile{NoStop}{kadanoff_1971}%
\bibitem{Frohlich:1987er}%
  \BibitemOpen
  \bibfield{author}{%
  \bibinfo {author} {\bibfnamefont{J.}~\bibnamefont{Frohlich}}\ and\ \bibinfo
  {author} {\bibfnamefont{P.~A.}\ \bibnamefont{Marchetti}},\ }%
  \bibfield{journal}{%
  \Doi{10.1007/BF01217817}{\bibinfo {journal} {Commun. Math. Phys.}}\ }%
  \textbf{\bibinfo {volume} {112}},\ \bibinfo {pages} {343} (\bibinfo {year}
  {1987})%
  \bibAnnoteFile{NoStop}{Frohlich:1987er}%
\bibitem{digiacomo-1995-349}%
  \BibitemOpen
  \bibfield{author}{%
  \bibinfo {author} {\bibfnamefont{L.~D. D.~A.}\ \bibnamefont{{Di Giacomo}}}\
  and\ \bibinfo {author} {\bibfnamefont{G.}~\bibnamefont{Paffuti}},\ }%
  \bibfield{journal}{%
  \bibinfo {journal} {Physics Letters B}\ }%
  \textbf{\bibinfo {volume} {349}},\ \bibinfo {pages} {513} (\bibinfo {year}
  {1995}),\ \url{arXiv:hep-lat/9403013}%
  \bibAnnoteFile{NoStop}{digiacomo-1995-349}%
\bibitem{Montvay:1994cy}%
  \BibitemOpen
  \bibfield{author}{%
  \bibinfo {author} {\bibfnamefont{I.}~\bibnamefont{Montvay}}\ and\ \bibinfo
  {author} {\bibfnamefont{G.}~\bibnamefont{Munster}}\ }%
  \bibinfo {note} {cambridge, UK: Univ. Pr. (1994) 491 p. (Cambridge monographs
  on mathematical physics)}%
  \bibAnnoteFile{NoStop}{Montvay:1994cy}%
\bibitem{Kogut:1974ag}%
  \BibitemOpen
  \bibfield{author}{%
  \bibinfo {author} {\bibfnamefont{J.~B.}\ \bibnamefont{Kogut}}\ and\ \bibinfo
  {author} {\bibfnamefont{L.}~\bibnamefont{Susskind}},\ }%
  \bibfield{journal}{%
  \Doi{10.1103/PhysRevD.11.395}{\bibinfo {journal} {Phys. Rev.}}\ }%
  \textbf{\bibinfo {volume} {D11}},\ \bibinfo {pages} {395} (\bibinfo {year}
  {1975})%
  \bibAnnoteFile{NoStop}{Kogut:1974ag}%
\bibitem{Kogut:1980sg}%
  \BibitemOpen
  \bibfield{author}{%
  \bibinfo {author} {\bibfnamefont{J.~B.}\ \bibnamefont{Kogut}},\ }%
  \bibfield{journal}{%
  \Doi{10.1016/0370-1573(80)90081-2}{\bibinfo {journal} {Phys. Rept.}}\ }%
  \textbf{\bibinfo {volume} {67}},\ \bibinfo {pages} {67} (\bibinfo {year}
  {1980})%
  \bibAnnoteFile{NoStop}{Kogut:1980sg}%
\bibitem{Marchesini:1981kt}%
  \BibitemOpen
  \bibfield{author}{%
  \bibinfo {author} {\bibfnamefont{G.}~\bibnamefont{Marchesini}}\ and\ \bibinfo
  {author} {\bibfnamefont{E.}~\bibnamefont{Onofri}},\ }%
  \bibfield{journal}{%
  \Doi{10.1007/BF02827437}{\bibinfo {journal} {Nuovo Cim.}}\ }%
  \textbf{\bibinfo {volume} {A65}},\ \bibinfo {pages} {298} (\bibinfo {year}
  {1981})%
  \bibAnnoteFile{NoStop}{Marchesini:1981kt}%
\bibitem{Fradkin1979}%
  \BibitemOpen
  \bibfield{author}{%
  \bibinfo {author} {\bibfnamefont{E.}~\bibnamefont{Fradkin}}\ and\ \bibinfo
  {author} {\bibfnamefont{S.~H.}\ \bibnamefont{Shenker}},\ }%
  \bibfield{journal}{%
  \bibinfo {journal} {Phys. Rev. D}\ }%
  \textbf{\bibinfo {volume} {19}},\ \bibinfo {pages} {3682} (\bibinfo {year}
  {1979})%
  \bibAnnoteFile{NoStop}{Fradkin1979}%
\bibitem{Bhanot1980}%
  \BibitemOpen
  \bibfield{author}{%
  \bibinfo {author} {\bibfnamefont{G.}~\bibnamefont{Bhanot}}\ and\ \bibinfo
  {author} {\bibfnamefont{M.}~\bibnamefont{Creutz}},\ }%
  \bibfield{journal}{%
  \bibinfo {journal} {Phys. Rev. D}\ }%
  \textbf{\bibinfo {volume} {21}},\ \bibinfo {pages} {2892} (\bibinfo {year}
  {1980})%
  \bibAnnoteFile{NoStop}{Bhanot1980}%
\bibitem{Agostini1997}%
  \BibitemOpen
  \bibfield{author}{%
  \bibinfo {author} {\bibfnamefont{V.}~\bibnamefont{Agostini}}, \bibinfo
  {author} {\bibfnamefont{G.}~\bibnamefont{Carlino}}, \bibinfo {author}
  {\bibfnamefont{M.}~\bibnamefont{Caselle}},\ and\ \bibinfo {author}
  {\bibfnamefont{M.}~\bibnamefont{Hasenbusch}},\ }%
  \bibfield{journal}{%
  \bibinfo {journal} {Nuclear Physics B}\ }%
  \textbf{\bibinfo {volume} {484}},\ \bibinfo {pages} {331} (\bibinfo {year}
  {1997})%
  \bibAnnoteFile{NoStop}{Agostini1997}%
\bibitem{Savit1980}%
  \BibitemOpen
  \bibfield{author}{%
  \bibinfo {author} {\bibfnamefont{R.}~\bibnamefont{Savit}},\ }%
  \bibfield{journal}{%
  \bibinfo {journal} {Rev. Mod. Phys.}\ }%
  \textbf{\bibinfo {volume} {52}},\ \bibinfo {pages} {453} (\bibinfo {year}
  {1980})%
  \bibAnnoteFile{NoStop}{Savit1980}%
\bibitem{nussinov_2010}%
  \BibitemOpen
  \bibfield{author}{%
  \bibinfo {author} {\bibfnamefont{E.}~\bibnamefont{Cobanera}}, \bibinfo
  {author} {\bibfnamefont{G.}~\bibnamefont{Ortiz}},\ and\ \bibinfo {author}
  {\bibfnamefont{Z.}~\bibnamefont{Nussinov}},\ }%
  \bibfield{journal}{%
  \Doi{10.1103/PhysRevLett.104.020402}{\bibinfo {journal} {Phys. Rev. Lett.}}\
  }%
  \textbf{\bibinfo {volume} {104}},\ \bibinfo {pages} {020402} (\bibinfo
  {month} {Jan}\ \bibinfo {year} {2010})%
  \bibAnnoteFile{NoStop}{nussinov_2010}%
\bibitem{Svetitsky:1982gs}%
  \BibitemOpen
  \bibfield{author}{%
  \bibinfo {author} {\bibfnamefont{B.}~\bibnamefont{Svetitsky}}\ and\ \bibinfo
  {author} {\bibfnamefont{L.~G.}\ \bibnamefont{Yaffe}},\ }%
  \bibfield{journal}{%
  \Doi{10.1016/0550-3213(82)90172-9}{\bibinfo {journal} {Nucl. Phys.}}\ }%
  \textbf{\bibinfo {volume} {B210}},\ \bibinfo {pages} {423} (\bibinfo {year}
  {1982})%
  \bibAnnoteFile{NoStop}{Svetitsky:1982gs}%
\bibitem{Fradkin1978}%
  \BibitemOpen
  \bibfield{author}{%
  \bibinfo {author} {\bibfnamefont{E.}~\bibnamefont{Fradkin}}\ and\ \bibinfo
  {author} {\bibfnamefont{L.}~\bibnamefont{Susskind}},\ }%
  \bibfield{journal}{%
  \Doi{10.1103/PhysRevD.17.2637}{\bibinfo {journal} {Phys. Rev. D}}\ }%
  \textbf{\bibinfo {volume} {17}},\ \bibinfo {pages} {2637} (\bibinfo {month}
  {May}\ \bibinfo {year} {1978})%
  \bibAnnoteFile{NoStop}{Fradkin1978}%
\bibitem{Gottesman1997Thesis}%
  \BibitemOpen
  \bibfield{author}{%
  \bibinfo {author} {\bibfnamefont{D.}~\bibnamefont{Gottesman}},\ }%
  \bibfield{journal}{%
  \bibinfo {journal} {Ph.D. thesis, Caltech}}%
   (\bibinfo {year} {1997})%
  \bibAnnoteFile{NoStop}{Gottesman1997Thesis}%
\bibitem{Dennis2002}%
  \BibitemOpen
  \bibfield{author}{%
  \bibinfo {author} {\bibfnamefont{E.}~\bibnamefont{Dennis}}, \bibinfo {author}
  {\bibfnamefont{A.}~\bibnamefont{Kitaev}}, \bibinfo {author}
  {\bibfnamefont{A.}~\bibnamefont{Landahl}},\ and\ \bibinfo {author}
  {\bibfnamefont{J.}~\bibnamefont{Preskill}},\ }%
  \bibfield{journal}{%
  \bibinfo {journal} {J. Math. Phys.}\ }%
  \textbf{\bibinfo {volume} {43}},\ \bibinfo {pages} {4452} (\bibinfo {year}
  {2002})%
  \bibAnnoteFile{NoStop}{Dennis2002}%
\bibitem{He1990}%
  \BibitemOpen
  \bibfield{author}{%
  \bibinfo {author} {\bibfnamefont{H.~X.}\ \bibnamefont{He}}, \bibinfo {author}
  {\bibfnamefont{C.~J.}\ \bibnamefont{Hamer}},\ and\ \bibinfo {author}
  {\bibfnamefont{J.}~\bibnamefont{Oitmaa}},\ }%
  \bibfield{journal}{%
  \bibinfo {journal} {J. Phys. A: Math. Gen.}\ }%
  \textbf{\bibinfo {volume} {23}},\ \bibinfo {pages} {1775} (\bibinfo {year}
  {1990})%
  \bibAnnoteFile{NoStop}{He1990}%
\bibitem{Oitmaa1991}%
  \BibitemOpen
  \bibfield{author}{%
  \bibinfo {author} {\bibfnamefont{J.}~\bibnamefont{Oitmaa}}, \bibinfo {author}
  {\bibfnamefont{C.~J.}\ \bibnamefont{Hamer}},\ and\ \bibinfo {author}
  {\bibfnamefont{Z.}~\bibnamefont{Weihong}},\ }%
  \bibfield{journal}{%
  \bibinfo {journal} {J. Phys. A: Math. Gen.}\ }%
  \textbf{\bibinfo {volume} {24}},\ \bibinfo {pages} {2863} (\bibinfo {year}
  {1991})%
  \bibAnnoteFile{NoStop}{Oitmaa1991}%
\bibitem{Pelissetto2002}%
  \BibitemOpen
  \bibfield{author}{%
  \bibinfo {author} {\bibfnamefont{A.}~\bibnamefont{Pelissetto}}\ and\ \bibinfo
  {author} {\bibfnamefont{E.}~\bibnamefont{Vicari}},\ }%
  \bibfield{journal}{%
  \bibinfo {journal} {Phys. Rep.}\ }%
  \textbf{\bibinfo {volume} {368}},\ \bibinfo {pages} {549} (\bibinfo {year}
  {2002})%
  \bibAnnoteFile{NoStop}{Pelissetto2002}%
\bibitem{Tagliacozzo2009}%
  \BibitemOpen
  \bibfield{author}{%
  \bibinfo {author} {\bibfnamefont{L.}~\bibnamefont{Tagliacozzo}}, \bibinfo
  {author} {\bibfnamefont{G.}~\bibnamefont{Evenbly}},\ and\ \bibinfo {author}
  {\bibfnamefont{G.}~\bibnamefont{Vidal}},\ }%
  \bibfield{journal}{%
  \bibinfo {journal} {Phys. Rev. B}\ }%
  \textbf{\bibinfo {volume} {80}},\ \bibinfo {pages} {235127} (\bibinfo {year}
  {2009})%
  \bibAnnoteFile{NoStop}{Tagliacozzo2009}%
\bibitem{Vidal2009Chapter}%
  \BibitemOpen
  \bibfield{author}{%
  \bibinfo {author} {\bibfnamefont{G.}~\bibnamefont{Vidal}},\ }%
  \bibfield{journal}{%
  \bibinfo {journal} {arXiv:0912.1651v2}}%
   (\bibinfo {year} {2009})%
  \bibAnnoteFile{NoStop}{Vidal2009Chapter}%
\bibitem{Wilson1983RG}%
  \BibitemOpen
  \bibfield{author}{%
  \bibinfo {author} {\bibfnamefont{K.~G.}\ \bibnamefont{Wilson}},\ }%
  \bibfield{journal}{%
  \bibinfo {journal} {Rev. Mod. Phys.}\ }%
  \textbf{\bibinfo {volume} {55}},\ \bibinfo {pages} {583} (\bibinfo {year}
  {1983})%
  \bibAnnoteFile{NoStop}{Wilson1983RG}%
\bibitem{Giovannetti2008}%
  \BibitemOpen
  \bibfield{author}{%
  \bibinfo {author} {\bibfnamefont{V.}~\bibnamefont{Giovannetti}}, \bibinfo
  {author} {\bibfnamefont{S.}~\bibnamefont{Montangero}},\ and\ \bibinfo
  {author} {\bibfnamefont{R.}~\bibnamefont{Fazio}},\ }%
  \bibfield{journal}{%
  \bibinfo {journal} {Phys. Rev. Lett.}\ }%
  \textbf{\bibinfo {volume} {101}},\ \bibinfo {pages} {180503} (\bibinfo {year}
  {2008})%
  \bibAnnoteFile{NoStop}{Giovannetti2008}%
\bibitem{Pfeifer2009}%
  \BibitemOpen
  \bibfield{author}{%
  \bibinfo {author} {\bibfnamefont{R.~N.~C.}\ \bibnamefont{Pfeifer}}, \bibinfo
  {author} {\bibfnamefont{G.}~\bibnamefont{Evenbly}},\ and\ \bibinfo {author}
  {\bibfnamefont{G.}~\bibnamefont{Vidal}},\ }%
  \bibfield{journal}{%
  \bibinfo {journal} {Phys. Rev. A}\ }%
  \textbf{\bibinfo {volume} {79}},\ \bibinfo {pages} {40301} (\bibinfo {year}
  {2009})%
  \bibAnnoteFile{NoStop}{Pfeifer2009}%
\bibitem{Aguado2008TC}%
  \BibitemOpen
  \bibfield{author}{%
  \bibinfo {author} {\bibfnamefont{M.}~\bibnamefont{Aguado}}\ and\ \bibinfo
  {author} {\bibfnamefont{G.}~\bibnamefont{Vidal}},\ }%
  \bibfield{journal}{%
  \bibinfo {journal} {Phys. Rev. Lett.}\ }%
  \textbf{\bibinfo {volume} {100}},\ \bibinfo {pages} {70404} (\bibinfo {year}
  {2008})%
  \bibAnnoteFile{NoStop}{Aguado2008TC}%
\bibitem{Konig2009}%
  \BibitemOpen
  \bibfield{author}{%
  \bibinfo {author} {\bibfnamefont{R.}~\bibnamefont{K\"{o}nig}}, \bibinfo
  {author} {\bibfnamefont{B.~W.}\ \bibnamefont{Reichardt}},\ and\ \bibinfo
  {author} {\bibfnamefont{G.}~\bibnamefont{Vidal}},\ }%
  \bibfield{journal}{%
  \bibinfo {journal} {Phys. Rev. B}\ }%
  \textbf{\bibinfo {volume} {79}},\ \bibinfo {pages} {195123} (\bibinfo {year}
  {2009})%
  \bibAnnoteFile{NoStop}{Konig2009}%
\bibitem{Singh2009Global}%
  \BibitemOpen
  \bibfield{author}{%
  \bibinfo {author} {\bibfnamefont{S.}~\bibnamefont{Singh}}, \bibinfo {author}
  {\bibfnamefont{R.~N.~C.}\ \bibnamefont{Pfeifer}},\ and\ \bibinfo {author}
  {\bibfnamefont{G.}~\bibnamefont{Vidal}},\ }%
  \bibfield{journal}{%
  \bibinfo {journal} {arXiv:0907.2994v1}}%
   (\bibinfo {year} {2009})%
  \bibAnnoteFile{NoStop}{Singh2009Global}%
\bibitem{fradkin_real-space_1979}%
  \BibitemOpen
  \bibfield{author}{%
  \bibinfo {author} {\bibfnamefont{E.}~\bibnamefont{Fradkin}}\ and\ \bibinfo
  {author} {\bibfnamefont{S.}~\bibnamefont{Raby}},\ }%
  \bibfield{journal}{%
  \Doi{10.1103/PhysRevD.20.2566}{\bibinfo {journal} {Physical Review D}}\ }%
  \textbf{\bibinfo {volume} {20}},\ \bibinfo {pages} {2566} (\bibinfo {month}
  {Nov.}\ \bibinfo {year} {1979})%
  \bibAnnoteFile{NoStop}{fradkin_real-space_1979}%
\bibitem{Migdal:1975zg}%
  \BibitemOpen
  \bibfield{author}{%
  \bibinfo {author} {\bibfnamefont{A.~A.}\ \bibnamefont{Migdal}},\ }%
  \bibfield{journal}{%
  \bibinfo {journal} {Sov. Phys. JETP}\ }%
  \textbf{\bibinfo {volume} {42}},\ \bibinfo {pages} {413} (\bibinfo {year}
  {1975})%
  \bibAnnoteFile{NoStop}{Migdal:1975zg}%
\bibitem{Levin2005SN}%
  \BibitemOpen
  \bibfield{author}{%
  \bibinfo {author} {\bibfnamefont{M.~A.}\ \bibnamefont{Levin}}\ and\ \bibinfo
  {author} {\bibfnamefont{X.-G.}\ \bibnamefont{Wen}},\ }%
  \bibfield{journal}{%
  \bibinfo {journal} {Phys. Rev. B}\ }%
  \textbf{\bibinfo {volume} {71}},\ \bibinfo {pages} {45110} (\bibinfo {year}
  {2005})%
  \bibAnnoteFile{NoStop}{Levin2005SN}%
\bibitem{Green1978}%
  \BibitemOpen
  \bibfield{author}{%
  \bibinfo {author} {\bibfnamefont{M.~B.}\ \bibnamefont{Green}},\ }%
  \bibfield{journal}{%
  \bibinfo {journal} {Nucl. Phys. B}\ }%
  \textbf{\bibinfo {volume} {144}},\ \bibinfo {pages} {473} (\bibinfo {year}
  {1978})%
  \bibAnnoteFile{NoStop}{Green1978}%
\bibitem{Rico2010priv}%
  \BibitemOpen
  \bibfield{author}{%
  \bibinfo {author} {\bibfnamefont{E.}~\bibnamefont{Rico}},\ }%
  \bibfield{journal}{%
  \bibinfo {journal} {private communication}}%
   (\bibinfo {year} {2010})%
  \bibAnnoteFile{NoStop}{Rico2010priv}%
\bibitem{tHooft:1981ht}%
  \BibitemOpen
  \bibfield{author}{%
  \bibinfo {author} {\bibfnamefont{G.}~\bibnamefont{'t~Hooft}},\ }%
  \bibfield{journal}{%
  \Doi{10.1016/0550-3213(81)90442-9}{\bibinfo {journal} {Nucl. Phys.}}\ }%
  \textbf{\bibinfo {volume} {B190}},\ \bibinfo {pages} {455} (\bibinfo {year}
  {1981})%
  \bibAnnoteFile{NoStop}{tHooft:1981ht}%
\bibitem{Mandelstam:1974pi}%
  \BibitemOpen
  \bibfield{author}{%
  \bibinfo {author} {\bibfnamefont{S.}~\bibnamefont{Mandelstam}},\ }%
  \bibfield{journal}{%
  \Doi{10.1016/0370-1573(76)90043-0}{\bibinfo {journal} {Phys. Rept.}}\ }%
  \textbf{\bibinfo {volume} {23}},\ \bibinfo {pages} {245} (\bibinfo {year}
  {1976})%
  \bibAnnoteFile{NoStop}{Mandelstam:1974pi}%
\bibitem{Tagliacozzo2010prep}%
  \BibitemOpen
  \bibfield{author}{%
  \bibinfo {author} {\bibfnamefont{L.}~\bibnamefont{Tagliacozzo}}\ and\
  \bibinfo {author} {\bibfnamefont{G.}~\bibnamefont{Vidal}},\ }%
  \bibfield{journal}{%
  \bibinfo {journal} {in preparation}}%
   (\bibinfo {year} {2010})%
  \bibAnnoteFile{NoStop}{Tagliacozzo2010prep}%
\bibitem{nightingale_scaling_1975}%
  \BibitemOpen
  \bibfield{author}{%
  \bibinfo {author} {\bibfnamefont{M.~P.}\ \bibnamefont{Nightingale}},\ }%
  \bibfield{journal}{%
  \Doi{10.1016/0378-4371(75)90021-7}{\bibinfo {journal} {Physica A: Statistical
  and Theoretical Physics}}\ }%
  \textbf{\bibinfo {volume} {83}},\ \bibinfo {pages} {561} (\bibinfo {year}
  {1975}),\ ISSN \bibinfo {issn} {0378-4371}%
  \bibAnnoteFile{NoStop}{nightingale_scaling_1975}%
\bibitem{D'Elia:2006vg}%
  \BibitemOpen
  \bibfield{author}{%
  \bibinfo {author} {\bibfnamefont{M.}~\bibnamefont{D'Elia}}\ and\ \bibinfo
  {author} {\bibfnamefont{L.}~\bibnamefont{Tagliacozzo}},\ }%
  \bibfield{journal}{%
  \Doi{10.1103/PhysRevD.74.114510}{\bibinfo {journal} {Phys. Rev.}}\ }%
  \textbf{\bibinfo {volume} {D74}},\ \bibinfo {pages} {114510} (\bibinfo {year}
  {2006}),\
  \Eprint{http://arxiv.org/abs/hep-lat/0609018}{arXiv:hep-lat/0609018}%
  \bibAnnoteFile{NoStop}{D'Elia:2006vg}%
\bibitem{zanardi_2006}%
  \BibitemOpen
  \bibfield{author}{%
  \bibinfo {author} {\bibfnamefont{P.}~\bibnamefont{Zanardi}}\ and\ \bibinfo
  {author} {\bibfnamefont{N.}~\bibnamefont{Paunkovi\ifmmode~\acute{c}\else
  \'{c}\fi{}}},\ }%
  \bibfield{journal}{%
  \Doi{10.1103/PhysRevE.74.031123}{\bibinfo {journal} {Phys. Rev. E}}\ }%
  \textbf{\bibinfo {volume} {74}},\ \bibinfo {pages} {031123} (\bibinfo {month}
  {Sep}\ \bibinfo {year} {2006})%
  \bibAnnoteFile{NoStop}{zanardi_2006}%
\bibitem{liu-2009}%
  \BibitemOpen
  \bibfield{author}{%
  \bibinfo {author} {\bibfnamefont{J.-H.}\ \bibnamefont{Liu}}, \bibinfo
  {author} {\bibfnamefont{Q.-Q.}\ \bibnamefont{Shi}}, \bibinfo {author}
  {\bibfnamefont{H.-L.}\ \bibnamefont{Wang}}, \bibinfo {author}
  {\bibfnamefont{J.}~\bibnamefont{Links}},\ and\ \bibinfo {author}
  {\bibfnamefont{H.-Q.}\ \bibnamefont{Zhou}},\ }%
  \bibfield{journal}{%
  \bibinfo {journal} {arXiv.org:0909.3031}}%
   (\bibinfo {year} {2009})%
  \bibAnnoteFile{NoStop}{liu-2009}%
\bibitem{cozzini_2007}%
  \BibitemOpen
  \bibfield{author}{%
  \bibinfo {author} {\bibfnamefont{M.}~\bibnamefont{Cozzini}}, \bibinfo
  {author} {\bibfnamefont{P.}~\bibnamefont{Giorda}},\ and\ \bibinfo {author}
  {\bibfnamefont{P.}~\bibnamefont{Zanardi}},\ }%
  \bibfield{journal}{%
  \Doi{10.1103/PhysRevB.75.014439}{\bibinfo {journal} {Phys. Rev. B}}\ }%
  \textbf{\bibinfo {volume} {75}},\ \bibinfo {pages} {014439} (\bibinfo {month}
  {Jan}\ \bibinfo {year} {2007})%
  \bibAnnoteFile{NoStop}{cozzini_2007}%
\bibitem{cozzini_2007b}%
  \BibitemOpen
  \bibfield{author}{%
  \bibinfo {author} {\bibfnamefont{M.}~\bibnamefont{Cozzini}}, \bibinfo
  {author} {\bibfnamefont{R.}~\bibnamefont{Ionicioiu}},\ and\ \bibinfo {author}
  {\bibfnamefont{P.}~\bibnamefont{Zanardi}},\ }%
  \bibfield{journal}{%
  \Doi{10.1103/PhysRevB.76.104420}{\bibinfo {journal} {Phys. Rev. B}}\ }%
  \textbf{\bibinfo {volume} {76}},\ \bibinfo {pages} {104420} (\bibinfo {month}
  {Sep}\ \bibinfo {year} {2007})%
  \bibAnnoteFile{NoStop}{cozzini_2007b}%
\bibitem{Venuti99}%
  \BibitemOpen
  \bibfield{author}{%
  \bibinfo {author} {\bibfnamefont{L.}~\bibnamefont{Campos~Venuti}}\ and\
  \bibinfo {author} {\bibfnamefont{P.}~\bibnamefont{Zanardi}},\ }%
  \bibfield{journal}{%
  \Doi{10.1103/PhysRevLett.99.095701}{\bibinfo {journal} {Phys. Rev. Lett.}}\
  }%
  \textbf{\bibinfo {volume} {99}},\ \bibinfo {pages} {095701} (\bibinfo {month}
  {Aug}\ \bibinfo {year} {2007})%
  \bibAnnoteFile{NoStop}{Venuti99}%
\bibitem{buonsante_2007}%
  \BibitemOpen
  \bibfield{author}{%
  \bibinfo {author} {\bibfnamefont{P.}~\bibnamefont{Buonsante}}\ and\ \bibinfo
  {author} {\bibfnamefont{A.}~\bibnamefont{Vezzani}},\ }%
  \bibfield{journal}{%
  \Doi{10.1103/PhysRevLett.98.110601}{\bibinfo {journal} {Phys. Rev. Lett.}}\
  }%
  \textbf{\bibinfo {volume} {98}},\ \bibinfo {pages} {110601} (\bibinfo {month}
  {Mar}\ \bibinfo {year} {2007})%
  \bibAnnoteFile{NoStop}{buonsante_2007}%
\bibitem{you_2007}%
  \BibitemOpen
  \bibfield{author}{%
  \bibinfo {author} {\bibfnamefont{W.-L.}\ \bibnamefont{You}}, \bibinfo
  {author} {\bibfnamefont{Y.-W.}\ \bibnamefont{Li}},\ and\ \bibinfo {author}
  {\bibfnamefont{S.-J.}\ \bibnamefont{Gu}},\ }%
  \bibfield{journal}{%
  \Doi{10.1103/PhysRevE.76.022101}{\bibinfo {journal} {Phys. Rev. E}}\ }%
  \textbf{\bibinfo {volume} {76}},\ \bibinfo {pages} {022101} (\bibinfo {month}
  {Aug}\ \bibinfo {year} {2007})%
  \bibAnnoteFile{NoStop}{you_2007}%
\bibitem{jian_2008}%
  \BibitemOpen
  \bibfield{author}{%
  \bibinfo {author} {\bibfnamefont{S.-J.}\ \bibnamefont{Gu}}, \bibinfo {author}
  {\bibfnamefont{H.-M.}\ \bibnamefont{Kwok}}, \bibinfo {author}
  {\bibfnamefont{W.-Q.}\ \bibnamefont{Ning}},\ and\ \bibinfo {author}
  {\bibfnamefont{H.-Q.}\ \bibnamefont{Lin}},\ }%
  \bibfield{journal}{%
  \Doi{10.1103/PhysRevB.77.245109}{\bibinfo {journal} {Phys. Rev. B}}\ }%
  \textbf{\bibinfo {volume} {77}},\ \bibinfo {pages} {245109} (\bibinfo {month}
  {Jun}\ \bibinfo {year} {2008})%
  \bibAnnoteFile{NoStop}{jian_2008}%
\bibitem{orus_2008}%
  \BibitemOpen
  \bibfield{author}{%
  \bibinfo {author} {\bibfnamefont{H.-Q.}\ \bibnamefont{Zhou}}, \bibinfo
  {author} {\bibfnamefont{R.}~\bibnamefont{Or\'us}},\ and\ \bibinfo {author}
  {\bibfnamefont{G.}~\bibnamefont{Vidal}},\ }%
  \bibfield{journal}{%
  \Doi{10.1103/PhysRevLett.100.080601}{\bibinfo {journal} {Phys. Rev. Lett.}}\
  }%
  \textbf{\bibinfo {volume} {100}},\ \bibinfo {pages} {080601} (\bibinfo
  {month} {Feb}\ \bibinfo {year} {2008})%
  \bibAnnoteFile{NoStop}{orus_2008}%
\bibitem{Tupitsyn2008}%
  \BibitemOpen
  \bibfield{author}{%
  \bibinfo {author} {\bibfnamefont{I.~S.}\ \bibnamefont{Tupitsyn}}, \bibinfo
  {author} {\bibfnamefont{A.}~\bibnamefont{Kitaev}}, \bibinfo {author}
  {\bibfnamefont{N.~V.}\ \bibnamefont{Prokof'ev}},\ and\ \bibinfo {author}
  {\bibfnamefont{P.~C.~E.}\ \bibnamefont{Stamp}},\ }%
  \bibfield{journal}{%
  \bibinfo {journal} {arXiv:0804.3175}}%
   (\bibinfo {year} {2008})%
  \bibAnnoteFile{NoStop}{Tupitsyn2008}%
\bibitem{Bravyi2010}%
  \BibitemOpen
  \bibfield{author}{%
  \bibinfo {author} {\bibfnamefont{S.}~\bibnamefont{Bravyi}}, \bibinfo {author}
  {\bibfnamefont{M.}~\bibnamefont{Hastings}},\ and\ \bibinfo {author}
  {\bibfnamefont{S.}~\bibnamefont{Michalakis}},\ }%
  \bibfield{journal}{%
  \bibinfo {journal} {arXiv:1001.0344v1}}%
   (\bibinfo {year} {2010})%
  \bibAnnoteFile{NoStop}{Bravyi2010}%
\bibitem{vidal_phase_2008}%
  \BibitemOpen
  \bibfield{author}{%
  \bibinfo {author} {\bibfnamefont{J.}~\bibnamefont{Vidal}}, \bibinfo {author}
  {\bibfnamefont{S.}~\bibnamefont{Dusuel}},\ and\ \bibinfo {author}
  {\bibfnamefont{K.~P.}\ \bibnamefont{Schmidt}},\ }%
  \bibfield{journal}{%
  \bibinfo {journal} {arXiv:0807.0487}}%
   (\bibinfo {month} {Jul.}\ \bibinfo {year} {2008})%
  \bibAnnoteFile{NoStop}{vidal_phase_2008}%
\bibitem{vidal_self-duality_2009}%
  \BibitemOpen
  \bibfield{author}{%
  \bibinfo {author} {\bibfnamefont{J.}~\bibnamefont{Vidal}}, \bibinfo {author}
  {\bibfnamefont{R.}~\bibnamefont{Thomale}}, \bibinfo {author}
  {\bibfnamefont{K.~P.}\ \bibnamefont{Schmidt}},\ and\ \bibinfo {author}
  {\bibfnamefont{S.}~\bibnamefont{Dusuel}},\ }%
  \bibfield{journal}{%
  \bibinfo {journal} {arXiv:0902.3547}}%
   (\bibinfo {month} {Feb.}\ \bibinfo {year} {2009}),\ \bibinfo {note} {phys.
  Rev. B 80, 081104 (2009)}%
  \bibAnnoteFile{NoStop}{vidal_self-duality_2009}%
\bibitem{nishino_1996}%
  \BibitemOpen
  \bibfield{author}{%
  \bibinfo {author} {\bibfnamefont{T.}~\bibnamefont{Nishino}}\ and\ \bibinfo
  {author} {\bibfnamefont{K.}~\bibnamefont{Okunishi}},\ }%
  \bibfield{journal}{%
  \Doi{10.1143/JPSJ.65.891}{\bibinfo {journal} {Journal of the Physical Society
  of Japan}}\ }%
  \textbf{\bibinfo {volume} {65}},\ \bibinfo {pages} {891} (\bibinfo {year}
  {1996})%
  \bibAnnoteFile{NoStop}{nishino_1996}%
\bibitem{orus_2009}%
  \BibitemOpen
  \bibfield{author}{%
  \bibinfo {author} {\bibfnamefont{R.}~\bibnamefont{Or\'us}}\ and\ \bibinfo
  {author} {\bibfnamefont{G.}~\bibnamefont{Vidal}},\ }%
  \bibfield{journal}{%
  \Doi{10.1103/PhysRevB.80.094403}{\bibinfo {journal} {Phys. Rev. B}}\ }%
  \textbf{\bibinfo {volume} {80}},\ \bibinfo {pages} {094403} (\bibinfo {month}
  {Sep}\ \bibinfo {year} {2009})%
  \bibAnnoteFile{NoStop}{orus_2009}%
\bibitem{Chen2010}%
  \BibitemOpen
  \bibfield{author}{%
  \bibinfo {author} {\bibfnamefont{X.}~\bibnamefont{Chen}}, \bibinfo {author}
  {\bibfnamefont{Z.-C.}\ \bibnamefont{Gu}},\ and\ \bibinfo {author}
  {\bibfnamefont{X.-G.}\ \bibnamefont{Wen}},\ }%
  \bibfield{journal}{%
  \bibinfo {journal} {arXiv:1004.3835v1}}%
   (\bibinfo {year} {2010})%
  \bibAnnoteFile{NoStop}{Chen2010}%
\bibitem{Sandvik2010}%
  \BibitemOpen
  \bibfield{author}{%
  \bibinfo {author} {\bibfnamefont{C.}~\bibnamefont{Liu}}, \bibinfo {author}
  {\bibfnamefont{L.}~\bibnamefont{Wang}}, \bibinfo {author}
  {\bibfnamefont{A.~W.}\ \bibnamefont{Sandvik}}, \bibinfo {author}
  {\bibfnamefont{Y.-C.}\ \bibnamefont{Su}},\ and\ \bibinfo {author}
  {\bibfnamefont{Y.-J.}\ \bibnamefont{Kao}},\ }%
  \bibfield{journal}{%
  \Doi{10.1103/PhysRevB.82.060410}{\bibinfo {journal} {Phys. Rev. B}}\ }%
  \textbf{\bibinfo {volume} {82}},\ \bibinfo {pages} {060410} (\bibinfo {month}
  {Aug}\ \bibinfo {year} {2010})%
  \bibAnnoteFile{NoStop}{Sandvik2010}%
\end{thebibliography}%
\end{document}